\documentclass{elsart}

\usepackage[square,comma]{natbib}
\usepackage{graphicx}

\usepackage{amssymb}

\begin{document}

\thispagestyle{empty}
\begin{Large}
\textbf{DEUTSCHES ELEKTRONEN-SYNCHROTRON}

\textbf{\large{in der HELMHOLTZ-GEMEINSCHAFT}\\}
\end{Large}

DESY 06-037

April 2006

\begin{eqnarray}
\nonumber &&\cr \nonumber && \cr \nonumber &&\cr
\end{eqnarray}
\begin{eqnarray}
\nonumber
\end{eqnarray}
\begin{center}
\begin{Large}
\textbf{Statistical Optics approach to the design of beamlines for
Synchrotron Radiation}
\end{Large}
\begin{eqnarray}
\nonumber &&\cr \nonumber && \cr
\end{eqnarray}

\begin{large}
Gianluca Geloni, Evgeni Saldin, Evgeni Schneidmiller and Mikhail
Yurkov
\end{large}
\textsl{\\Deutsches Elektronen-Synchrotron DESY, Hamburg}
\begin{eqnarray}
\nonumber
\end{eqnarray}
\begin{eqnarray}
\nonumber
\end{eqnarray}
\begin{eqnarray}
\nonumber
\end{eqnarray}
ISSN 0418-9833
\begin{eqnarray}
\nonumber
\end{eqnarray}
\begin{large}
\textbf{NOTKESTRASSE 85 - 22607 HAMBURG}
\end{large}
\end{center}
\clearpage
\newpage

\begin{frontmatter}



\title{Statistical Optics approach to the design of beamlines for Synchrotron Radiation}


\author[DESY]{Gianluca Geloni}
\author[DESY]{Evgeni Saldin}
\author[DESY]{Evgeni Schneidmiller}
\author[DESY]{Mikhail Yurkov}

\address[DESY]{Deutsches Elektronen-Synchrotron (DESY), Hamburg,
Germany}

\begin{abstract}
In this paper we analyze the image formation problem for undulator
radiation through an optical system, accounting for the influence
of the electron beam emittance. On the one hand, image formation
with Synchrotron Radiation is governed by the laws of Statistical
Optics. On the other hand,  the widely used Gaussian-Shell model
cannot be applied to describe the coherence properties of X-ray
beams from third generation Synchrotron Radiation sources. As a
result, a more rigorous analysis of coherence properties is
required. We propose a technique to explicitly calculate the
cross-spectral density of an undulator source, that we
subsequently propagate through an optical imaging system. At first
we focus on the case
of an ideal lens with a non-limiting pupil aperture. 
Our theory, which makes consistent use of dimensionless analysis,
also allows treatment and physical understanding of many
asymptotes of the parameter space, together with their
applicability region. Particular emphasis is given to the
asymptotic situation when the horizontal emittance is much larger
than the radiation wavelength, which is relevant for third
generation Synchrotron Radiation sources. First principle
calculations of undulator radiation characteristics (i.e.
ten-dimensional integrals) are then reduced to one-dimensional
convolutions of analytical functions with universal functions
specific for undulator radiation sources. We also consider the
imaging problem for a non-ideal lens in presence of aberrations
and a limiting pupil aperture, which increases the dimension of
the convolution from one to three. In particular we give emphasis
to cases when the intensity at the observation plane can be
presented as a convolution of an impulse response function and the
intensity from an ideal lens. Our results may be used in practical
cases as well as in benchmarks for numerical methods.
\end{abstract}


\end{frontmatter}


\clearpage

\tableofcontents

\newpage

\section{\label{sec:intro} Introduction}

The majority of experiments based on the use of X-rays are carried
out at Synchrotron Radiation facilities, as very high brilliance
is achievable by means of undulator devices installed in storage
rings. Guiding the photons from the exit of an undulator to the
specimen position requires the development of optical beamlines
whose main task is to re-image the undulator source to any plane
of interest. To deal with the image formation problem, one should
account for the fact that Synchrotron Radiation constitutes a
random stochastic process. In fact, the shot noise in the electron
beam causes fluctuations of the electron beam current density.
These fluctuations are random both in space and time. As a result,
the radiation field produced by the electron beam can be described
in terms of a phasor with random amplitudes and phases and, in all
generality, the laws of Statistical Optics must be applied to
solve the image formation problem. In this paper we study the
image formation problem with undulator radiation beams based on
Statistical Optics. In this framework, the basic quantity
characterizing Synchrotron Radiation sources is the second order
correlation function of the fields at two observation points on a
given transverse plane identified by the coordinate $z_o$ along
the optical beamline. Once such a plane is fixed, the two points,
$P_1$ and $P_2$, are fully characterized by their transverse
coordinates $\vec{r}_{\bot 1}$ and $\vec{r}_{\bot 2}$
respectively. Our presentation will be given in the frequency
domain. Due to the limited temporal resolution of detectors in
Synchrotron Radiation experiments, the analysis in frequency
domain is much more natural than that in the time domain. As a
consequence of this choice, and without restrictive assumptions on
the system, we study the spatial correlation between Fourier
transforms of the electric field \footnote{\label{primaf}Since our
analysis deals with the electric field in frequency domain, we
will sometimes refer to the "Fourier transform of the electric
field" simply as "the field", when this does not generate
confusion.} at a fixed frequency $\omega$, that is the
cross-spectral density

\begin{equation}
G = \left\langle \bar{E}\left(z_o,\vec{r}_{\bot 1},\omega\right)
\bar{E}^*\left(z_o,\vec{r}_{\bot 2},\omega\right) \right\rangle~.
\label{G12}
\end{equation}
In Eq. (\ref{G12}) $\bar{E}$ is the complex amplitude of the
Fourier transform of a given Cartesian component of the electric
field at the space-frequency point $(z_o,\vec{r}_{\bot},\omega)$,
the asterisk denotes complex conjugation, and brackets $<...>$
indicate an ensemble average over electron bunches. Since we study
an ultra-relativistic system, the paraxial approximation can
always be enforced so that, here, the electric field is understood
to obey the paraxial wave equation \cite{OUR1}.  The
cross-spectral density carries all information about the
transverse characteristics of undulator radiation. A fully general
study of the cross-spectral density is not a trivial one.
Difficulties arise when one tries to include simultaneously the
effect of intrinsic divergence of the radiation, due to the
presence of the undulator, of the electron beam size and of the
electron beam divergence. In \cite{OURS} a technique was
described, based on Statistical Optics, to calculate the
cross-spectral density from undulator sources in the most general
case \footnote{With the point of view of the source parameters.},
at any position after the undulator but still without optical
elements (i.e. in free space). Although self-contained, the
present study relies on that work. Expressions from \cite{OURS}
will be taken as a starting point to proceed along the optical
beamline towards the specimen position.

In general, as we will see, undulator radiation can be thought as
originating from an equivalent source localized on a transverse
plane at a given longitudinal position. By definition, such source
has the following property: it produces a field which coincides
with that from the undulator at any distance from the exit of the
undulator. In particular, we localize the equivalent source in the
center of the undulator, that will be conventionally taken as the
beamline origin $z=0$. Since in this case the equivalent source
does not reproduce the real electromagnetic field distribution in
the center of the undulator, we refer to it as \textit{virtual
source}. Further on, throughout this paper we assume that the beta
functions of the electron beam have their minimal value in the
center of the undulator. By this, as we will see, the virtual
source exhibits particular properties which simplify our
treatment. Once the concept of virtual source is introduced, the
problem of describing radiation characteristics at a certain
observation plane after a given optical element is twofold. First,
one has to characterize the cross-spectral density at the virtual
source and, second, one has to propagate the cross-spectral
density along the optical beamline to the observation plane.

Let us first consider the problem of characterizing the source. An
important simplified model which admits an analytical description
without loss of essential information about the source features is
obtained by letting both horizontal and vertical electron beam
emittances be much larger than the radiation wavelength
($\epsilon_{x,y} \gg \lambda/(2\pi)$). This is a good assumption
for second generation Synchrotron Radiation sources. The kind of
virtual source obtained for $\epsilon_{x,y} \gg \lambda/(2\pi)$
belongs to the wider class of quasi-homogeneous ones. These are
characterized by the fact that the cross-spectral density at the
virtual source plane (i.e. at $z=0$) can be written as:

\begin{eqnarray}
{G}({\vec{r}}_{\bot 1},{\vec{r}}_{\bot 2}, \omega) =
I\left({\vec{r}}_{\bot 1}, \omega\right) g({\vec{r}}_{\bot
2}-{\vec{r}}_{\bot 1}, \omega ) ~,\label{introh}
\end{eqnarray}
where

\begin{equation}
I\left({\vec{r}}_{\bot 1}, \omega\right) = \left \langle \left|
E\left({\vec{r}}_{\bot 1}, \omega\right) \right|^2
\right\rangle\label{intint}
\end{equation}
is the field intensity distribution and $g({\vec{r}}_{\bot
2}-{\vec{r}}_{\bot 1}, \omega )$ is the spectral degree of
coherence (normalized, by definition, so that $g(0,\omega) = 1$).
The definition of quasi-homogeneity amounts to a factorization of
the cross-spectral density as the product of the field intensity
distribution and the spectral degree of coherence, which contains
information about the spatial correlation. A set of necessary and
sufficient conditions for such factorization to be possible
follows: \textit{(a)} the radiation intensity at the virtual
source varies very slowly with the position across the source on
the scale of the field correlation length and \textit{(b)} the
spectral degree of coherence depends on the positions across the
source only through the difference ${\vec{r}}_{\bot
2}-{\vec{r}}_{\bot 1}$.

There are situations when the Statistical Optics description is
not the only one possible. The asymptotic limit for large electron
beam emittances ($\epsilon_{x,y} \gg \lambda/(2\pi)$) is one of
these. In this limit, the Statistical Optics description of the
source coincides with the Geometrical Optics (or Hamiltonian)
description of the source, where a photon-beam phase space is
defined and can be described in terms of rays specified by
position-angle coordinates. In the Geometrical Optics approach,
based on the uncertainty principle, only the radiation originating
by a photon-beam phase space area of order $[\lambda/(2\pi)]^2$ is
spatially coherent. When the emittance is much larger than the
wavelength, $2 \pi \epsilon_{x,y}/\lambda \gg 1$, the divergence
of the electron beam is much larger than the diffraction angle of
undulator radiation, and the transverse size of the electron beam
is much larger than the diffraction size of undulator radiation.
As a result one can completely neglect diffraction effects, and
the Geometrical Optics approach can always be applied. Since
Geometrical Optics describes a limiting situation of Statistical
Optics there must be a relation between the fundamental
Geometrical Optics quantity, the phase space distribution, and the
fundamental Statistical Optics quantity, the cross-spectral
density. It can be shown \cite{MAND} that the radiant intensity of
the field generated in free space by a quasi-homogeneous source in
the direction of a unit vector $\vec{s}$ can be expressed as

\begin{equation}
\mathcal{I}(\vec{s},\omega) \propto \Gamma(\vec{s},\omega)
~,\label{inte}
\end{equation}
$\Gamma(\vec{s},\omega)$ being the two-dimensional spatial Fourier
transform of the degree of transverse coherence $g({\vec{r}}_{\bot
2}-{\vec{r}}_{\bot 1}, \omega )$:

\begin{equation}
\Gamma(\vec{s},\omega) = \int g({\vec{\rho'}},\omega)
\exp\left[\frac{i\omega}{c} \vec{s}\cdot\vec{\rho'}\right]
d{\vec{\rho'}}~.\label{ftran}
\end{equation}
The expression for the phase space distribution is given by the
product of the intensity distribution of the source and the
radiant intensity

\begin{equation}
\Phi \left(\vec{s},\vec{r}_\bot \right) = I\left({\vec{r}}_{\bot }
,\omega\right)\mathcal{I}(\vec{s},\omega) \propto
I\left({\vec{r}}_{\bot }, \omega
\right)\Gamma\left(\vec{s},\omega\right)~, \label{phsp}
\end{equation}
where the variables $(\vec{r}_\bot,\vec{s})$ characterize a ray in
phase space. A comparison between Eq. (\ref{phsp}) and Eq.
(\ref{introh}) shows that cross-spectral density and phase space
distribution contain the same information in the limiting case
$\epsilon_{x,y} \gg \lambda/(2\pi)$.

In other words, if a source has a large angular divergence
(compared with the diffraction angle) and a large transverse size
(compared with the diffraction size), one can completely neglect
diffraction effects and treat the problem of the characterization
of the source by means of Geometrical
Optics\footnote{\label{bendm} Condition $2 \pi
\epsilon_{x,y}/\lambda \gg 1$ is sufficient, but not necessary. In
general, we cannot say that Geometrical Optics is never applicable
for electron beam emittances smaller than the radiation wavelength
$\lambda$. We will treat this subject in a more extensive fashion
in Section \ref{sub:imfoge}. There we will see that there are
situations when Geometrical Optics can be applied to describe the
(virtual) source even when the electron beam emittance is smaller
than $\lambda$. We will find that a sufficient (less restrictive,
but still not necessary) condition for the applicability of
Geometrical Optics to the description of a given undulator source
is that such source can be characterized in terms of a
quasi-homogeneous virtual source. Moreover, as it will also be
discussed in Section \ref{sub:imfoge},  our comparison of the
emittance with the radiation wavelength is done under the
assumption that the electron beam beta function is comparable with
the radiation formation length at wavelength $\lambda$. This is
often, but not always,  the case for undulator sources, since the
radiation formation length is the undulator length, which is at
least a few meters. However, it is not the case for bending magnet
radiation.}.

Despite the previous discussion, the possibilities of using
Geometrical Optics to describe undulator sources are quite limited
in many realistic situations. Applications of Synchrotron
Radiation make use of a very wide range of wavelengths which span
over four order of magnitude, from $0.1 \mathrm{\AA}$ to $10^3
\mathrm{\AA}$. For third generation light sources, either planned
or in operation, the horizontal electron beam emittance
$\epsilon_x = \sigma_x\sigma_{x'}$ is of order of $1 \div 3$ nm.
The vertical emittance is given by $\epsilon_y =
\sigma_y\sigma_{y'} = \chi \epsilon_x$, $\chi$ being the so called
coupling factor. Typical values of $\chi$ for third generation
light sources are of order $\chi \sim 0.01$, corresponding to
vertical emittances of order $0.1 \div 0.3 \mathrm{\AA}$. These
values are always near or within the diffraction limit for
wavelength ranges up to the hard X-rays in the vertical direction,
and Geometrical Optics descriptions fail. In particular, in the
VUV wavelength range, both vertical and horizontal emittances are
much smaller than the radiation wavelength ($\epsilon_{x,y} \ll
\lambda/(2\pi)$). One recovers, then, the perfectly coherent
situation when the source is diffraction limited in both
horizontal and vertical directions. This is another situation when
the Statistical Optics description is not the only one possible.
In this case, deterministic Wave Optics may be used as well. As
the wavelength becomes shorter, in the soft X-ray range, one
obtains $\epsilon_{y} \ll \lambda/(2\pi)$, but $\epsilon_{x} \gg
\lambda/(2\pi)$. At wavelengths of about $1\mathrm{\AA}$ the
vertical emittance reaches the same order of magnitude of the
wavelength $\epsilon_{y} \sim \lambda/(2\pi)$, while $\epsilon_{x}
\gg \lambda/(2\pi)$. Finally, in the hard X-ray region, at a
wavelength of about $0.1 \mathrm{\AA}$, both emittances are much
larger than the wavelength ($\epsilon_{x,y} \gg \lambda/(2\pi)$),
and Geometrical Optics can be used alongside Statistical Optics.
It follows that, for third generation light sources, only the
limiting cases for wavelengths around $100$ nm and $0.1
\mathrm{\AA}$ can be treated, respectively, my means of Wave
Optics or Geometrical Optics. The intermediate situation can be
treated in a rigorous way only with the help of Statistical
Optics, which includes both Wave Optics and Geometrical Optics as
asymptotic cases.

Strictly related to the problem of source characterization, but
separate from it, is the issue of propagating the photon beam
through the optical beamline to the observation plane. In the case
of quasi-homogeneous virtual sources, if diffraction effects from
the optical elements can be neglected, Geometrical Optics can be
taken advantage of. The virtual source can be described in terms
of phase space distribution, and interactions with optical media
can be conveniently modelled in terms of symplectic
transformations, very much like electron beams in storage rings
optics.  Several computer codes (e.g. SHADOW \cite{SHAD}), usually
referred to as ray-tracing codes, have been developed and are
standard tools used to carry out Geometrical Optics-based
calculations. However, this approach is not always possible as the
virtual source may not be quasi-homogeneous or diffraction effects
may not be neglected in the optical beamline. A rigorous analysis
of the object-image coherence relationship is of fundamental
importance in the context of several coherence-based techniques
like fluctuation correlation dynamics, phase imaging, coherent
X-ray diffraction and X-ray holography, whose development has been
fostered by the high flux of coherent X-rays provided by
state-of-the-art third generation facilities. It should be noted
that, in the case of partially coherent wavefronts, even the
calculation of the intensity distribution at the specimen position
should involve Statistical Optics techniques. In fact, to obtain
the intensity at the specimen position as some optical element is
present one first needs to track the cross-spectral density
through the beamline i.e. one needs to study the evolution of the
partially coherent wavefront.

Computer codes have been written \cite{CHUB} in order to deal with
beamline design in the case of partially coherent radiation. These
are devoted to the solution of the image formation problem
starting from first principles. Results may in fact be obtained
using numerical techniques alone, starting from the
Lienard-Wiechert expressions for the electromagnetic field and
applying the definition of the field correlation function without
any analytical manipulation. Yet, a first-principle calculation of
the field correlation function between two generic points or, in
particular, calculation of the intensity at a single point
involves very complicated and time-expensive numerical
evaluations. To be specific, one needs to perform two integrations
along the undulator device and four integrations over the
electron-beam phase space distribution to solve the problem in
free space. Then, modelling the optical beamline as a single
convergent lens, other four integrations are needed to
characterize coherence properties on the image plane, for a total
of ten integrations. The development of a universal code for any
experimental setup is then likely to be problematic. A more
conservative approach may suggest the use of computer codes based
on some analytical manipulation of first principle equations
suited for specific experimental setups. From this viewpoint our
most general expressions may be used as reliable basis for the
development of numerical methods. Yet, computer codes can
calculate properties for a given set of parameters, but can hardly
improve physical understanding, which is particularly important in
the stage of planning experiments. Our theory will allow treatment
and physical understanding of many asymptotes of the parameter
space and their applicability region with the help of a consistent
use of dimensional analysis. In the most general asymptotic cases
treated here, this will allow to reduce first principle
calculations (i.e. ten-dimensional integrals) to one-dimensional
convolutions of analytical functions with universal functions
specific for the undulator source case, and still to retain a
certain degree of generality. It is also worth to underline that
our asymptotic results may also be used as a benchmark for
numerical methods.

One of the main difficulties in applying a Statistical Optics
approach to Synchrotron Radiation sources stems from the fact that
Statistical Optics has principally developed in connection with
problems involving thermal light. Solutions to all these problems
share approximations that allow major simplifications, but are
specific of thermal sources only. For instance, thermal sources
can be modelled as perfectly incoherent, and the cross-spectral
density assumes the form

\begin{equation}
G({\vec{r}}_{\bot 1},{\vec{r}}_{\bot 2}, \omega) \propto
I\left({\vec{r}}_{\bot 1}, \omega\right)
\delta\left({\vec{r}}_{\bot 2}-{\vec{r}}_{\bot
1}\right)~,\label{therm}
\end{equation}
where $I$ is the source intensity distribution and $\delta$ is the
two-dimensional Dirac $\delta$-function. However, there is a close
connection between the state of coherence of the source and the
angular distribution of the radiant intensity (see Eq.
(\ref{inte})). The physical interpretation of Eq. (\ref{therm}) is
that the source is correlated over the minimal possible distance
(which is of order of the wavelength). This has the consequence
that the radiant intensity is distributed over a solid angle of
order $2 \pi$. This is correct for thermal sources, but is in
contradiction with the fact that any Synchrotron Radiation source
is confined within a narrow cone in the forward direction. The
high directionality of Synchrotron Radiation rules out the use of
Eq. (\ref{therm}) as a model for Synchrotron Radiation sources.
However, such high directionality is not in contrast with the poor
coherence which characterizes the quasi-homogeneous limit.
Quasi-homogeneous sources are only locally coherent over a
distance of many wavelengths but, by definition of
quasi-homogeneity, the linear dimension of the source is much
larger than the correlation distance. Even though a
quasi-homogeneous source can be described with Geometrical Optics
techniques, a coherence distance of many wavelengths rules out the
use of Eq. (\ref{therm}) as a model for Synchrotron Radiation
sources. A more precise knowledge of the cross-spectral density
(that is equivalent to the knowledge of the correct phase space
density) is necessary to solve the image formation problem. For
instance, suppose that a light source is placed at arbitrary
distance in front of a lens. If the source is perfectly incoherent
(thermal light case) the area of the light incident on the lens is
always the area of the lens. In the case of Synchrotron Radiation
source though, such area may be smaller than the lens. Even in the
limit for a large beam emittance (compared with the radiation
wavelength), information about the small angular distribution must
be printed in the wavefront at the exit of the undulator leading
once more to the same conclusion: Eq. (\ref{therm}) cannot be used
in order to model Synchrotron Radiation sources. In \cite{OURS} we
treated, among other cases, the asymptote for a large electron
beam size and divergence. The expression for the cross-spectral
density of the source in free space simplifies and a particular
quasi-homogeneous model can be given. In the same work, we
specified also the region of applicability of such model, and we
showed that it cannot be applied outside the limit for a large
electron beam size and divergence.

In relation with these remarks it should be mentioned that an
attempt to follow the path proposed in this paper is described in
\cite{ATT2}. To our knowledge, \cite{ATT2} constitutes the first
remarkable attempt to use Statistical Optics techniques in order
to characterize the evolution of partially coherent X-ray beams
through optical systems. In that work, as well as in \cite{ATT1},
the beamline optics from the undulator to the specimen can be
modelled as a critical illumination system \cite{GOOD}, the
beamline behaving as the condenser. After this, Statistical Optics
techniques are consistently used to calculate coherent properties
on the image plane. However, the authors of \cite{ATT2} reduced
the general ten-dimensional integrals to four-dimensional
integrals by postulating that the cross-spectral density
distribution at the exit of the undulator can be written as Eq.
(\ref{therm}), i.e. a perfectly incoherent source is assumed at
the exit of the undulator. As we have just seen though, this
assumption is always inconsistent in the case of Synchrotron
Radiation, even in the Geometrical Optics limit and, \textit{a
fortiori}, in the case treated by the authors (the undulator
beamline 12 at ALS), where $\epsilon_y \simeq 0.1 \lambda/(2\pi)$
and $\epsilon_x \simeq 3 \lambda/(2\pi)$, which is highly
spatially coherent.

We organize our work as follows. Besides this Introduction, in
Section \ref{sec:elem} we describe the  optical system under study
and some concepts from Statistical Optics that will be widely used
in the following Sections. In Section \ref{sec:fila} we review
some general expressions pertaining undulator radiation from a
single particle. In particular, following \cite{OURS} we present
an analytical expression for the Fourier transform of the electric
field generated by a single electron with offset and deflection
which is valid at any distance from the exit of the undulator. We
also present the analytical solution of the imaging problem for a
deterministic model of undulator radiation (absence of electron
beam emittance). In Section \ref{sec:emit} we give a derivation of
the cross-spectral density for undulator radiation based sources.
Subsequently we analyze the evolution of the cross-spectral
density function through the optical system with particular
attention to the focal and to the image plane. The following two
Sections \ref{sec:qhso} and \ref{sec:nong} describe
quasi-homogeneous sources, respectively Gaussian and non-Gaussian,
in the ideal case when the lens is aberration-free and the pupil
aperture is non-limiting. A digression is then taken in Section
\ref{sub:imfoge}, where we analyze in detail the relation between
Geometrical Optics and quasi-homogeneous sources. Such Section may
therefore be skipped in a first reading,  without interrupting the
main logical stream of our work. The next Section \ref{par:ngdgb}
describes the effects of a finite aperture size on the radiation
characteristics from quasi-homogeneous sources at the image plane,
and is followed by Section \ref{sec:abe} that assumes a
quasi-homogeneous source as well and deals with the consequences
of lens aberration on the intensity at the image plane. In Section
\ref{sec:cam} we introduce a particular setup, a pinhole camera,
capable of producing images of the sources in the absence of
lenses. The study of this particular setup is of particular
interest because it allows the reader to recognize mathematical
analogies between cases otherwise physically different and serves
as a junction between the previous Section \ref{sec:abe}  and
Section \ref{sec:focim} treating physical cases when one obtains,
surprisingly, the image of the source on the focal plane. In the
following Section \ref{sec:imany} we extend the treatment for the
focal plane to any plane of interest. In Section \ref{sec:dof} we
discuss the depth of focus, including the case of a large
non-limiting aperture and the effects of aperture size. In the
next Section \ref{sec:nonh} solutions for the image formation
problem in non-homogeneous cases relevant for third generation
Synchrotron Radiation sources are given. Before conclusions, in
Section \ref{sec:supp}, we discuss the accuracy of
quasi-homogenous source asymptotes. Finally, in Section
\ref{sec:conc}, we come to conclusions.

\section{\label{sec:elem} Elements and definitions of image formation theory}

\subsection{\label{sub:wavfr} Wave propagation in free space}

Let us indicate with ${E}_\bot(z,\vec{r}_{\bot },t)$ any fixed
polarization component (along the direction $x$ or $y$) of the
electric field at time $t$ calculated on a plane at position $z$
down the beamline at a certain transverse location
$\vec{r}_{\bot}$. ${E}_\bot(z,\vec{r}_{\bot },t)$ obeys, in free
space, the homogeneous wave equation:

\begin{equation}
c^2 \nabla^2 {E}_\bot - \frac{\partial^2 {E}_\bot}{\partial t^2} =
0~. \label{omog}
\end{equation}
Let us now introduce   the Fourier transform
$\bar{E}_\bot(z,\vec{r}_{\bot },\omega)$ of the electric field
${E}_\bot(z,\vec{r}_{\bot },t)$:

\begin{equation}
\bar{{E}}_\bot(\omega) = \int_{-\infty}^{\infty} dt {{E}}_\bot(t)
e^{i \omega t}~, \label{ftran}
\end{equation}
so that
\begin{equation}
{{E}}_\bot(t) = \frac{1}{{2\pi}} \int_{-\infty}^{\infty} d\omega
\bar{{E}}_\bot(\omega) e^{-i \omega t}~.
 \label{fanti}
\end{equation}
As already remarked in footnote \ref{primaf}, we will sometimes
refer to $\bar{E}_\bot$ as "the field", understanding that we are
working in the frequency domain.

Let us consider the field propagation problem. To this purpose, we
first introduce the complex envelope of the field:

\begin{equation}
{\widetilde{E}} ={\bar{E}}_{\bot} \exp[-i \omega z/c]~.
\label{tildaE}
\end{equation}
It is always possible to give such a definition. However, its
utility is restricted to the case when $\tilde{E}$ is a slowly
varying function of $z$ with respect to the radiation wavelength
$\lambda$. When the paraxial approximation is applicable (i.e.
always, for Synchrotron Radiation sources), this condition is
fulfilled \cite{OUR1}.

In paraxial approximation and in free space, the following
parabolic equation holds for the complex envelope
${\widetilde{E}}$ of the Fourier transform of the electric field
along a fixed polarization component:

\begin{equation}
\left({ {\nabla}_\bot}^2 + \frac{2 i \omega}{c}
\frac{\partial}{{\partial
 {z}}}\right) {\widetilde{E}} = 0 ~.\label{parabo}
\end{equation}
The derivatives in the Laplacian operator ${ {\nabla}_\bot}^2$ are
taken with respect to the transverse coordinates. One has to solve
Eq. (\ref{parabo}) with a given
initial condition at ${z}$, which is a Cauchy problem. 
Indicating with $\vec{ {r}}_o$ the transverse coordinate of an
observation point on a plane at longitudinal position ${z}_o$ we
have

\begin{equation}
{ \widetilde{E}}( {z}_o,\vec{ {r}_o}) = \frac{i \omega}{2 \pi c(
{z}_o- {z})} \int d \vec{ {r}'}~ \tilde{E}(z,\vec{r'})
\exp{\left[\frac{i \omega \left|{\vec{ {r}}}_o-\vec{
{r}'}\right|^2}{2 c ( {z}_o- {z})}\right]}~, \label{fieldpropback}
\end{equation}
where the integral is performed over the transverse plane.

Next to the propagation equation for the field in free space, Eq.
(\ref{fieldpropback}), we can discuss a propagation equation for
the spatial Fourier transform of the field, which can also be
derived from Eq. (\ref{parabo}) and will be useful in the
following parts of this work. We will indicate the spatial Fourier
transform of $\tilde{E}(z,\vec{r'})$  with $\mathrm{F}(z,
\vec{u})$\footnote{For the sake of completeness we explicitly
write the definitions of the two-dimensional Fourier transform and
inverse transform of a function $g(\vec{r})$ in agreement with the
notations used in this paper. The Fourier transform and inverse
transform pair reads:

\begin{eqnarray}
\nonumber \tilde{g}(\vec{k}) = \int \d \vec{r} ~g(\vec{r})
\exp\left[i \vec{r} \cdot \vec{k} \right]~;~~{g}(\vec{r}) =
\frac{1}{4\pi^2} \int \d \vec{k} ~\tilde{g}(\vec{k}) \exp\left[-i
\vec{r} \cdot \vec{k} \right] , \label{spatgft}
\end{eqnarray}
the integration being understood over the entire plane. If $g$ is
circular symmetric we can introduce the Fourier-Bessel transform
and inverse transform pair:

\begin{eqnarray}
\nonumber \tilde{g}({k}) = 2\pi \int_0^{\infty} d{r}~ r g(r) J_o(k
r)~;~~ {g}({r}) = \frac{1}{2\pi} \int_0^{\infty} d{k}~ k
\tilde{g}(k) J_o(k r)~,\label{bessspatgft}
\end{eqnarray}
$r$ and $k$ indicating the modulus of the vectors $\vec{r}$ and
$\vec{k}$ respectively, and $J_o$ being the zero-th order Bessel
function of the first kind.}:

\begin{equation}
\mathrm{F}\left(z,\vec{u}\right) = \int d \vec{r'}
\tilde{E}(z,\vec{r'}) \exp{\left[i \vec{r'}\cdot\vec{u}\right]}~.
\label{ftfield}
\end{equation}
Eq. (\ref{parabo}) can be rewritten in terms of $\mathrm{F}$ as

\begin{equation}
\left({ {\nabla}_\bot}^2 + \frac{2 i \omega}{c}
\frac{\partial}{{\partial {z}}}\right) \left\{\int d\vec{u}~
\mathrm{F}\left(z,\vec{u}\right) \exp{\left[-i
\vec{r}\cdot\vec{u}\right]}\right\} = 0 ~.\label{paraboft}
\end{equation}
Eq. (\ref{paraboft}) requires that

\begin{equation}
\left(-\left|\vec{u}\right|^2+ \frac{2 i \omega}{c}
\frac{\partial}{{\partial {z}}}\right)
\mathrm{F}\left(z,\vec{u}\right) = 0 ~.\label{paraboft2}
\end{equation}
Solution of Eq. (\ref{paraboft2}) can be presented as

\begin{eqnarray}
\mathrm{F}\left(z,\vec{u}\right) =
\mathrm{F}\left(0,\vec{u}\right) \exp{\left[-\frac{i c |\vec{
u}|^2 z}{2 \omega}\right]}~. \label{fieldpropback35}
\end{eqnarray}
It should be noted that the definition of $\mathrm{F}(0, \vec{u})$
is a matter of initial conditions. In many practical cases,
including the totality of the situation treated in this paper,
$\mathrm{F}$ (or $\tilde{E}$) may have no direct physical meaning
at $z=0$. For instance, in all cases considered in this paper,
$z=0$ is in the center of the undulator, well within the radiation
formation length. However, $\mathrm{F}(0, \vec{u})$ can be
considered as the spatial Fourier transform of the field produced
by a \textit{virtual} source. Such a source is defined by the fact
that, supposedly placed at $z=0$, it would produce, at any
distance from the undulator, the same field as the real source
does. The result in Eq. (\ref{fieldpropback35}) is very general.
On the one hand, the spatial Fourier transform of the electric
field exhibits an almost trivial behavior in $z$, since
$|\mathrm{F}(z)|^2 = \mathrm{const}$. On the other hand, the
behavior of the electric field itself is not trivial at all (see
Section \ref{sub:impc} and Fig.s \ref{S1}-\ref{S10}). These
properties follow directly from the propagation equation for the
field and its Fourier transform.

Let us now discuss the physical meaning of Eq.
(\ref{fieldpropback35}). The spatial Fourier transform of the
field, $\mathrm{F}(z,\vec{u})$, may be interpreted as a
superposition of plane waves (the so-called \textit{angular
spectrum}).  Once the frequency $\omega$ is fixed, the wave number
$k=\omega/c$ is fixed as well, and a given value of the transverse
component of the wave vector $\vec{k}_\bot=\vec{u}$ corresponds to
a given angle of propagation of a plane wave. Different
propagation directions correspond to different distances travelled
to get to a certain observation point. Therefore, they also
correspond to different phase shifts, which depend on the position
along the $z$ axis (see, for example, reference \cite{GOOD2}
Section 3.7). Free space basically acts as a Fourier
transformation. This means that the field in the far zone is,
phase factor and proportionality factor aside, the spatial Fourier
transform of the field at any position $z$. To show this fact, we
first recall that if we know the field at a given position
$(z,\vec{r'})$ we may use Eq. (\ref{fieldpropback}) to calculate
the field at another position $(z_o,\vec{r_o})$ . Let us now
consider the limit  $z_o \longrightarrow \infty$, with finite
ratio $\vec{r}_o/z_o$. In this case, the exponential function in
Eq. (\ref{fieldpropback}) can be expanded giving

\begin{eqnarray}
{ \widetilde{E}}( {z}_o,\vec{ {r}_o}) &=& \frac{i \omega}{2 \pi c
{z}_o} \int d \vec{ {r}'}~ \tilde{E}(z,\vec{r'})
\exp{\left[\frac{i \omega |\vec{ {r}}_o|^2}{2 c {z}_o}-\frac{i
\omega( \vec{ {r}}_o\cdot\vec{ {r}'}) }{ c {z}_o}+\frac{i \omega z
|\vec{ {r}}_o|^2}{2 c {z}_o^2}\right]}~.\cr &&
\label{fieldpropback2}
\end{eqnarray}
Letting $\vec{\theta} = \vec{r}_o/z_o$ we have

\begin{eqnarray}
{ \widetilde{E}}( {z}_o,\vec{r}_o)&=& \frac{i \omega}{2 \pi c
{z}_o} \exp{\left[\frac{i \omega |\vec{ \theta}|^2}{2
c}(z_o+z)\right]} \mathrm{F}\left(z,-\frac{\omega
\vec{\theta}}{c}\right)~.  \label{fieldpropback3}
\end{eqnarray}
With the help of Eq. (\ref{fieldpropback35}), Eq.
(\ref{fieldpropback3}) may be presented as

\begin{eqnarray}
{ \widetilde{E}}( {z}_o,\vec{r}_o)&=& \frac{i \omega}{2 \pi c
{z}_o} \exp{\left[\frac{i \omega |\vec{ \theta}|^2}{2
c}z_o\right]} \mathrm{F}\left(0,-\frac{\omega
\vec{\theta}}{c}\right)~ . \label{fieldpropback3tris}
\end{eqnarray}
Eq. (\ref{fieldpropback3tris}) shows what we wanted to
demonstrate: free space basically acts as a Fourier
transformation.

\subsection{\label{sub:imcoh} Image formation with coherent light}

\begin{figure}
\begin{center}
\includegraphics*[width=110mm]{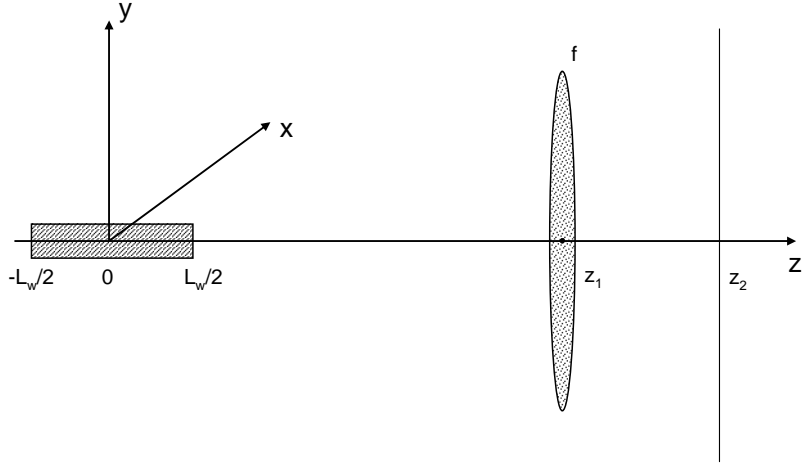}
\caption{\label{setup} Single lens imaging system with an
undulator source as object. }
\end{center}
\end{figure}
As has been remarked in \cite{ATT2}, any beamline optics used to
re-image undulator radiation to an observation plane of interest
can be modelled as a critical illumination system \cite{GOOD}.
Therefore, the setup considered in this paper can be sketched as
in Fig. \ref{setup}. It consists of an undulator of length $L_w$
centered at ${z}=0$, a convergent lens positioned at ${z} =
{z}_1$, characterized by a focusing strength $f$, and a plane of
observation at position ${z} = {z}_2$.  In principle, ${z}_1$,
${z}_2$ and ${f}$ are unrelated parameters. However our main case
of interest is a critical illumination system. Therefore, a given
source plane at coordinate $z=z_s$ - the object - is imaged at a
particular observation position $z=z_2$, that defines the image
plane. Using Ray Optics we can calculate the distance $z_2$ along
the axis behind the lens where the image is formed. This gives the
well-known lens-maker equation

\begin{equation}
\frac{1}{{f}} = \frac{1}{z_1 - z_s}+\frac{1}{ {z}_2- {z}_1}~.
\label{lens}
\end{equation}
The size of the image is magnified by a factor $|\mathrm{M}|$ and
a real image is inverted with respect to the object because

\begin{equation}
\mathrm{M} = - \frac{ {z}_2- {z}_1}{z_1-z_s}<0 ~,\label{Mdefined}
\end{equation}
as expected from Ray Optics. We also define a scale factor
$\mathrm{m}$, that is the inverse of the magnification power
$|\mathrm{M}|$ of the lens:

\begin{equation}
\mathrm{m} = \frac{1}{|\mathrm{M}|} = \frac{z_1-z_s}{ {z}_2-
{z}_1}~. \label{lensmag}
\end{equation}
Once the lens position is fixed at $z=z_1$, the position $z_s$ of
the source is a matter of choice. Such a choice fixes the position
$z_i = z_2$ of the image plane in agreement with Eq. (\ref{lens}).
For instance, with \cite{ATT2} and \cite{ATT1}, one may set the
critical illumination system to image the radiation at the
undulator exit, in which case the choice $z_s = L_w/2$ is made.
However, one is not obliged to do so. In particular, in this
paper, we will make the choice $z_s = 0$. Then, the critical
illumination system images the center of the undulator, in the
sense that the image plane $z_i = z_2$ obeys Eq. (\ref{lens}) with
$z_s = 0$.

From a mathematical viewpoint, specifying the source is equivalent
to fixing the initial conditions for Maxwell equations in terms of
a field distribution on a certain transverse plane. When a certain
field distribution (in the frequency domain) is fixed on a plane
at position $z$, Maxwell equations automatically set the way
radiation propagates in free space, and the source is univocally
defined. The choice made in \cite{ATT2} and \cite{ATT1}
corresponds to the choice of a real source. The denomination
"real" is justified by the fact that the initial condition for
Maxwell equations  amounts to the specification of a field
distribution which is actually present, and in principle
measurable, at the exit of the undulator. On the contrary, our
choice $z_s = 0$ corresponds to the position down the z-axis in
the middle of the undulator, well within the radiation formation
length. Although it makes sense to talk about the distribution of
the electromagnetic field in the middle of the undulator, it does
not make any sense to identify such distribution with the initial
condition for Maxwell equations, i.e. with the source. However, it
makes sense to define a \textit{virtual} source as in Section
\ref{sub:wavfr}. There is a particular reason for the choice $z=0$
as the position for the virtual source. In the case of a coherent
undulator source (that is being treated in the present Section),
the wavefront of the radiation at a virtual source located at
$z=0$ is plane. In other words, the far zone field from an
undulator has spherical wavefronts centered in the middle of the
undulator. This fact alone makes the center of the undulator a
privileged point with respect to others. Moreover, in the more
general case of partially coherent radiation we will assume (as it
is often verified in practice) that the beta functions of the
electron beam have their minima (in both horizontal and vertical
directions) in the center of the undulator. Subject to this
assumption, as we will see, the center of the undulator
constitutes a privileged point of interest in this case as well.
As a result of the previous discussion we set

\begin{equation}
z_s =  0~. \label{image}
\end{equation}
Let us assume that the position $z_1$ and the focal length $f$  of
the lens are set. Throughout this paper we will be particularly
interested in the radiation characteristics at two privileged
positions down the beamline:

\begin{itemize}
\item{ the image plane, at position $z_2$ identified by Eq.
(\ref{lens}).}
\item{the focal plane, at position $z_2$ identified by the
equation $ {z}_2-z_1 = {f}$.} \end{itemize}
As it will be seen in the next Sections, both image and focal
planes have special properties that can be expressed in terms of
Fourier Optics. These properties are valid for any wavefront. We
will first take advantage of them in Section \ref{sec:fila}, where
we will deal with wavefronts generated by an electron beam with
zero emittance and further on in Section \ref{sec:emit}, where
emittance effects will be discussed in the realm of Statistical
Optics. In the present Section, after having set the configuration
under study, we will limit ourselves to describe these properties
and to define basic quantities to be used in the Statistical
Optics formulation of the image formation problem.

Consider the problem of mapping the plane  \textit{immediately in
front} of the lens onto the plane at longitudinal position
$z_2>z_1$ \textit{behind} the lens (see Fig. \ref{setup}). If one
knows $\tilde{E}$ at position $z= z_1$ \textit{immediately in
front} of the lens, one can also obtain the expression for
$\tilde{E}$ \textit{immediately behind} the lens multiplying by
the transmission function:

\begin{equation}
{T}\left(\vec{ {r}'}\right) = P\left(\vec{ {r}'}\right)
\exp\left[- i \omega \frac{\left|\vec{ {r}'}\right|^2}{2 c
{f}}\right]~. \label{transm}
\end{equation}
For simplicity of notation we consider here identical focal
distances in the horizontal and in the vertical direction, i.e.
$f=f_x=f_y$. Results for more complicated optical systems (e.g. a
combination of cylindrical mirrors) are found substituting
consistently Eq. (\ref{transm}) with its straightforward
generalization. We assume with \cite{GOOD} that the complex pupil
function $P$ is zero outside the lens aperture. Its phase accounts
for aberrations and its modulus may vary along the lens to
describe apodizations: the simplest possible study case is for
$|P| = 1$ and $\arg{(P)}=0$ within the lens aperture. Accounting
for Eq. (\ref{transm}), the propagation equation for any field ${
\widetilde{E}}  ( {z}_1,\vec{ {r}'}) $ \textit{immediately in
front} of the lens to the point $(z_2,\vec{r}_2)$ on the
observation plane \textit{behind} the lens can be written as:

\begin{eqnarray}
{ \tilde{E}}( {z}_2,\vec{ {r}}_2) &=& \frac{i \omega}{2 \pi c (
{z}_2- {z}_1)}  \exp{\left[ \frac{i \omega | \vec{r}_2 |^2}{2 c (
{z}_2-z_1)} \right]}\int d \vec{ {r}'}~ \Bigg\{{ \widetilde{E}} (
{z}_1,\vec{ {r}'}) P(\vec{r'})\cr &&\times \exp{\left[\frac{i
\omega}{c} \left(\frac{1}{2( {z}_2-z_1)}-\frac{1}{2f}\right)|
\vec{r'} |^2 \right]}\Bigg\}\exp{\left[- \frac{i \omega
(\vec{r}_2\cdot\vec{r'})}{c \left(
{z}_2-z_1\right)}\right]}~.\label{fieldpropexp}
\end{eqnarray}
Since Synchrotron Radiation is highly collimated, it is
practically relevant to discuss the case when the area of the spot
of the  incident radiation  is small compared with the area of the
lens. It is also simpler and more natural to start with this
situation. Then, effects from a finite pupil dimension can be
neglected. Note that this is not the case for thermal sources:
since these are emitting into a solid angle of $2 \pi$ (they are
perfectly incoherent) the finite pupil dimensions cannot be
ignored and result in the so-called \textit{vignetting} effect
\cite{GOOD}. Considering a perfect lens with no aberrations too,
Eq. (\ref{fieldpropexp}) assumes a particularly simple form at the
focal and on the image plane.
Initially we will consider situations when the pupil presence can
be neglected. Later on we discuss how to include the effects due
to the presence of the pupil.

\subsubsection{\label{subsub:negl1} Large non-limiting aperture}

We will now specialize the result in Eq. (\ref{fieldpropexp}) in
the asymptote for a large non-limiting aperture and in the case of
the focal and of the image plane. Let us denote with $(
{z}_f,\vec{ {r}_f})$ a point on the focal plane, and with $(
{z}_i,\vec{ {r}_i})$ a point on the image plane. From Eq.
(\ref{fieldpropexp}), on the focal plane we have

\begin{eqnarray}
{ \tilde{E}}( {z}_f,\vec{ {r}_f}) &=& \frac{i \omega }{2 \pi c f}
\exp{\left[ \frac{i \omega | \vec{r}_f |^2}{2 c f} \right]}\int d
\vec{ {r}'}~ { \widetilde{E}}  ( {z}_1,\vec{ {r}'}) \exp{\left[-
\frac{i \omega(\vec{r}_f\cdot\vec{r'})}{c f }\right]}~.
\label{fieldpropexpfoc}
\end{eqnarray}
With the help of Eq. (\ref{ftfield}) we can write Eq.
(\ref{fieldpropexpfoc}) as

\begin{eqnarray}
{ \tilde{E}}( {z}_f,\vec{ {r}}_f) &=& \frac{i \omega}{2 \pi c f}
\exp{\left[ \frac{i \omega | \vec{r}_f |^2}{2 c f} \right]}
\mathrm{F}\left(z_1,-\frac{\omega \vec{r}_f}{c f}\right)~.
\label{cpact1}
\end{eqnarray}
Substitution of Eq. (\ref{fieldpropback35}) in Eq. (\ref{cpact1})
gives

\begin{eqnarray}
{ \tilde{E}}( {z}_f,\vec{ {r}}_f) &=& \frac{i \omega}{2 \pi c f }
\exp{\left[ \frac{i \omega | \vec{r}_f |^2}{2 c f} \right]}
\exp{\left[-\frac{i \omega z_1 |\vec{r}_f|^2}{2 c f^2}\right]}
\mathrm{F}\left(0,-\frac{\omega \vec{r}_f}{c
f}\right)~.\label{cpact3}
\end{eqnarray}
For the image plane, remembering that

\begin{equation}
\frac{1}{{f}} = \frac{1}{z_1}+\frac{1}{ {z}_i- {z}_1}~
\label{lensagain}
\end{equation}
and that

\begin{equation}
d_i = z_i-z_1 = \frac{z_1}{\mathrm{m}} ~, \label{magn2}
\end{equation}
we obtain, from Eq. (\ref{fieldpropexp})

\begin{eqnarray}
{ \tilde{E}}( {z}_i,\vec{ {r}_i}) &=& \frac{i \omega \mathrm{m}}{2
\pi c z_1}  \exp{\left[ \frac{i \omega \mathrm{m} | \vec{r}_i |^2
}{2 c z_1} \right]}\int d \vec{ {r}'}~\Bigg\{ { \widetilde{E}} (
{z}_1,\vec{ {r}'}) \cr &&\exp{\left[-\frac{i \omega | \vec{r'}
|^2}{2 c z_1} \right]}\Bigg\}\exp{\left[- \frac{i \omega
\mathrm{m} (\vec{r}_i\cdot\vec{r'})}{ c z_1}\right]}~.
\label{fieldpropexpima}
\end{eqnarray}
On the image plane, according to Eq. (\ref{fieldpropexpima}), we
have to calculate the Fourier transform of the product of two
factors: $\exp{[-{i \omega | \vec{r'} |^2}/({2 c z_1})]}$,
representing the phase of a spherical wave in paraxial
approximation, and ${ \widetilde{E}} ( {z}_1,\vec{ {r}'})$. A
direct calculation shows that the Fourier transform of the phase
factor is

\begin{eqnarray}
\int d \vec{ {r}'}~\exp{\left[-\frac{i \omega | \vec{r'} |^2}{2 c
z_1} \right]}\exp{\left[- \frac{i \omega \mathrm{m}
(\vec{r}_i\cdot\vec{r'})}{ c z_1}\right]}=-4 i z_1
\exp{\left[\frac{i \mathrm{m}^2 \omega |\vec{r}_i|^2}{2 c z_1}
\right]}~. \label{fieldpropexpimabbbb}
\end{eqnarray}
Since the Fourier transform of a product is equal to the
convolution of the Fourier transforms of each separate factor,
from Eq. (\ref{fieldpropexpima}) one obtains

\begin{eqnarray}
{ \tilde{E}}( {z}_i,\vec{ {r}}_i) &=& \frac{\mathrm{m}}{4\pi^2}
\exp{\left[ \frac{i \omega \mathrm{m}| \vec{r}_i |^2}{2 c z_1}
\right]}\cr &&\times\int d \vec{u}~
\mathrm{F}\left(z_1,\vec{u}\right)\exp{\left[\frac{i c z_1 }{2
\omega} \left(-\frac{\mathrm{m} \omega \vec{r}_i}{c
z_1}-\vec{u}\right)^2 \right]}~. \label{cpact2}
\end{eqnarray}
Substitution of Eq. (\ref{fieldpropback35}) in Eq. (\ref{cpact2})
gives

\begin{eqnarray}
{ \tilde{E}}( {z}_i,\vec{ {r}}_i) &=& \frac{\mathrm{m}}{4\pi^2}
\exp{\left[ \frac{i \omega \mathrm{m} | \vec{r}_i |^2}{2 c z_1}
\right]} \exp{\left[\frac{i \omega \mathrm{m}^2 |\vec{r}_i|^2}{2 c
z_1}\right]}\cr && \times\int d \vec{u}~ \mathrm{F}(0,\vec{u})
\exp{\left[i \mathrm{m} \vec{r}_i\cdot\vec{u} \right]}~, \cr &&
\label{cpact4}
\end{eqnarray}
that is

\begin{eqnarray}
{ \tilde{E}}( {z}_i,\vec{ {r}}_i) &=& \mathrm{m}  \exp{\left[
\frac{i \omega \mathrm{m} | \vec{r}_i |^2}{2 c z_1}
\right]}\exp{\left[\frac{i \omega \mathrm{m}^2 |\vec{r}_i|^2}{2 c
z_1}\right]}{ \tilde{E}}\left( 0,-{ \mathrm{m} \vec{r}_i}\right)~.
\label{cpact5}
\end{eqnarray}
The phase factor in the Fourier transform of the electric field,
given in Eq. (\ref{fieldpropback35}), cancels the quadratic phase
factor in $|\vec{u}|^2$ in Eq. (\ref{cpact2}). Therefore, the
convolution integral in Eq. (\ref{cpact2}) transforms to a Fourier
integral. As a result, in the image plane we always obtain (aside
for a scaling and a net phase factor) the inverted field
distribution in the virtual source plane. More in general, at any
observation plane located at $z=z_2$ behind the lens and the focal
plane, one observes (aside, again, for a scaling and a net phase
factor) the inverted field distribution on an object plane located
at $z = z_s$, where $z_s$ satisfies the lens condition Eq.
(\ref{lens}).

Eq. (\ref{cpact3}) and Eq. (\ref{cpact5}) are reflections of
well-known theorems of Fourier Optics. Neglecting the effects from
a finite pupil dimension and assuming a perfect lens with no
aberrations, the focal plane has the following property
\cite{GOOD2}:

\begin{itemize}
\item{For any position of the object in front of the lens, the field distribution (amplitude and phase) on the focal plane differs from the spatial Fourier transform of the field distribution on the object plane by a scale factor $-\omega/(c f)$ and a net phase factor.}
\end{itemize}
At the image plane, instead, the following property applies
\cite{GOOD2}:

\begin{itemize}
\item{For any position of the object in front of the lens, the field distribution (amplitude and phase) on the image plane differs from the field distribution on the object plane by a scale factor $-\mathrm{m}$ and a net phase factor \footnote{The prefactor $\mathrm{m}(z_i)$ is a consequence of the conservation of the total energy associated with the propagating field.}. }
\end{itemize}

These properties can be interpreted in terms  of intensity
distributions. The first tells that the intensity profile on the
focal plane has the same shape of that on a distant plane and is
obtained taking, essentially, the square modulus of the Fourier
transform of the field on the object plane. The second tells that
the intensity profile of the object is  inverted and magnified by
the lens on the image plane. For the image plane we just obtained,
for perfectly coherent light, the same result which is obtained in
Geometrical Optics in the perfectly incoherent limit.

\subsubsection{\label{subsub:incl1} Effect of aperture size}

Pupil effects are taken into account, from a general standpoint,
in Eq. (\ref{fieldpropexp}). Eq. (\ref{fieldpropexp}) takes a
specific form in the focal and in the image plane. One has

\begin{eqnarray}
{ \tilde{E}}( {z}_f,\vec{ {r}}_f) &=& \frac{i \omega}{2 \pi c f}
\exp{\left[ \frac{i \omega | \vec{r}_f |^2}{2 c f} \right]}\int d
\vec{ {r}'}~ { \widetilde{E}} ( {z}_1,\vec{ {r}'}) P(\vec{r'})
\exp{\left[- \frac{i \omega (\vec{r}_f\cdot\vec{r'})}{c
f}\right]}~\cr&&\label{fieldpropexpf}
\end{eqnarray}
and

\begin{eqnarray}
{ \tilde{E}}( {z}_i,\vec{ {r}}_i) &=& \frac{i \omega \mathrm{m}}{2
\pi c {z}_1}  \exp{\left[ \frac{i \omega \mathrm{m} | \vec{r}_i
|^2}{2 c z_1} \right]}\int d \vec{ {r}'}~ \Bigg\{{ \widetilde{E}}
( {z}_1,\vec{ {r}'}) P(\vec{r'})\exp{\left[-\frac{i \omega
|\vec{r'} |^2 }{2c z_1} \right]}\Bigg\}\cr &&\times \exp{\left[-
\frac{i \omega \mathrm{m}(\vec{r}_i\cdot\vec{r'})}{c
z_1}\right]}~.\label{fieldpropexpi}
\end{eqnarray}
Eq. (\ref{fieldpropexpf}) and Eq. (\ref{fieldpropexpi}) are formal
extensions of Eq. (\ref{fieldpropexpfoc}) and Eq.
(\ref{fieldpropexpima}), respectively. Use of the convolution
theorem on Eq. (\ref{fieldpropexpf}) and Eq. (\ref{fieldpropexpi})
allows to write analogous extensions of Eq. (\ref{cpact3}) and Eq.
(\ref{cpact5}). To this purpose we define

\begin{eqnarray}
\mathcal{P}(\vec{u}) &=& \int d\vec{r'} P(\vec{r'}) \exp{\left[-i
\vec{r'}\cdot \vec{u}\right]}~. \label{PFT}
\end{eqnarray}
Then, indicating with $\tilde{E}_P$ the field in the presence of
the pupil, on the focal plane we have

\begin{eqnarray}
\tilde{E}_P(z_f,\vec{r}_f) &=& \exp\left[\frac{i \omega
|\vec{r}_f|^2}{2 c f} \right] \cr &&\times\int d \vec{u}~
\mathcal{P}\left(\frac{\omega \vec{r}_f}{c f}-\vec{u}\right)\cdot
\exp{\left[- \frac{i c f | \vec{u} |^2}{2 \omega}
\right]}\tilde{E}\left(z_f,\frac{c f \vec{u}}{\omega}\right)
\label{EFP}
\end{eqnarray}
and on the image plane

\begin{eqnarray}
\tilde{E}_P(z_i,\vec{r}_i) &=& \exp\left[\frac{i \omega \mathrm{m}
|\vec{r}_i|^2}{2 c z_1} \right] \cr &&\times\int d \vec{u}~
\mathcal{P}\left(\frac{\omega \mathrm{m} \vec{r}_i}{c
z_1}-\vec{u}\right)\cdot\exp{\left[- \frac{i c z_1 | \vec{u}
|^2}{2 \omega \mathrm{m}} \right]} \tilde{E}\left(z_i,\frac{c z_1
\vec{u}}{\omega \mathrm{m}}\right) ~. \label{EIP}
\end{eqnarray}
One may apply the following  mnemonic rule to include the effects
of the pupil in Eq. (\ref{cpact3}) or Eq. (\ref{cpact5}). First,
divide Eq. (\ref{cpact3}) or Eq. (\ref{cpact5}) by the first phase
factor, corresponding to the phase factor outside the integral
sign in Eq. (\ref{fieldpropexp}). Second, convolve with
$\mathcal{P}$. Third, put the phase factor back.

It should be noted that, in the limit for large apertures,
$\mathcal{P}$ can be substituted by a $\delta$-Dirac function in
both Eq. (\ref{EFP}) and Eq. (\ref{EIP}). In this case, from Eq.
(\ref{EFP}) we recover $\tilde{E}_P(z_f,\vec{r}_f) =
\tilde{E}(z_f,\vec{r}_f)$, given in Eq. (\ref{cpact3}). Form Eq.
(\ref{EIP}) instead, we have $\tilde{E}_P(z_i,\vec{r}_i) =
\tilde{E}(z_i,\vec{r}_i)$, given in Eq. (\ref{cpact5}).

Unless particular conditions are met, the phase factors under the
integral signs in Eq. (\ref{EFP}) and Eq. (\ref{EIP}) compensate
only partially the phase of $\tilde{E}$, which can be found in Eq.
(\ref{cpact3}) and Eq. (\ref{cpact5}) respectively. This fact
complicates the evaluation of the convolution integrals. Let us
restrict our attention to the image plane. We can treat
analytically the case when the phase factors under integral in Eq.
(\ref{EIP}) completely compensate the phase factor in Eq.
(\ref{cpact5}), i.e. when we can neglect the second phase factor
in Eq. (\ref{cpact5}). This happens when we are in the far field
limit. The far field limit of Eq. (\ref{cpact5})  can be obtained
by substitution of Eq. (\ref{fieldpropback3tris}) in Eq.
(\ref{fieldpropexpima}). After the inverse Fourier transform in
Eq. (\ref{fieldpropexpima}) is calculated one obtains

\begin{eqnarray}
{ \tilde{E}}( {z}_i,\vec{ {r}}_i) &=& \mathrm{m}  \exp{\left[
\frac{i \omega \mathrm{m} | \vec{r}_i |^2}{2 c z_1} \right]}{
\tilde{E}}\left( 0,-{ \mathrm{m} \vec{r}_i}\right)~,
\label{fieldpropexpimafar}
\end{eqnarray}
that can also be obtained directly from Eq. (\ref{cpact5})
neglecting the second phase factor on the right hand side. This is
possible when

\begin{equation} \frac{\omega \mathrm{m}^2 |\vec{r}_i|^2}{2 c z_1}
\ll 1 \label{phasesmall}
\end{equation}
for any point $\vec{r}_i$ on the image pattern.

Let us indicate with $\sigma_i$ the characteristic size of the
image. Condition (\ref{phasesmall}) can be interpreted as the
following requirement for $z_1$:

\begin{equation}
z_1 \gg \frac{\omega \mathrm{m}^2 \sigma_i^2}{c}~.
\label{phasesmallc1}
\end{equation}
Since we are interested in the parametric dependence only, a
factor $2$ has been neglected in Eq. (\ref{phasesmallc1}). From
Eq. (\ref{cpact5}) follows that $\mathrm{m} \sigma_i$ is the
characteristic size of the virtual source. As such it is
independent of the position of the lens and of the magnification
factor $|M|=\mathrm{m}^{-1}$ as well. Condition
(\ref{phasesmallc1}) is often met in practice and means that the
radiation spot size on the lens, $c z_1/(\omega
\mathrm{m}\sigma_i)$, is much larger than the characteristic size
of the virtual source, $\mathrm{m} \sigma_i$. Then, the lens is
placed in the far zone with respect to the virtual undulator
source by definition of far zone $\lambda z_1/(2 \pi \sigma_o)\gg
\sigma_o$, $\sigma_o$ being the source size. We conclude that the
condition for the lens to be placed in the far zone is equivalent
to the condition that Eq. (\ref{cpact5}) can be reduced to Eq.
(\ref{fieldpropexpimafar}). This result will be of importance in
what follows. However, this kind of reasoning is only valid on the
image plane. On the focal plane the second phase factor in Eq.
(\ref{cpact3}) follows directly from the phase factor in Eq.
(\ref{fieldpropback35}), that is related with the propagation of
the angular spectrum: in this case one concludes that plane waves
with different directions of propagation lead to an increasing
phase difference as the distance $z_1$ increases. Therefore, in
the focal plane, some simplification may be obtained in the near
field only, as $z_1$ is small enough that the phase difference
between different plane wave components is negligible. We will not
investigate this situation further. Going back to the image plane
one can see that a term of the expansion of the phase factor under
the integral sign in Eq. (\ref{cpact2}) cancels the phase factor
in Eq. (\ref{fieldpropback35}). It follows that the second phase
factor in Eq. (\ref{cpact5}) is not related with the propagation
of the angular spectrum. Therefore, in the far field region, when
condition (\ref{phasesmallc1}) holds, we have that the pupil
effects can be accounted for by means of a simpler convolution. In
the following parts of this paper we will restrict to this
particular case when treating pupil effects. It is interesting to
remark that pupil effects due to finite pupil dimension are
especially important in the far region. In this limit the
radiation spot size on the lens is much larger than the size of
the radiation spot at the virtual source, and is often larger that
the size of the pupil. In the near field instead, the radiation
spot size on the lens is of order of the radiation spot size at
the virtual source. Therefore, in this limit, one can neglect
effects from any finite pupil aperture larger than the radiation
spot size at the virtual source.

In conclusion, explicit substitution of Eq.
(\ref{fieldpropexpimafar}) in Eq. (\ref{EIP}) yields the following
far field limit expression on the image plane:

\begin{eqnarray}
\tilde{E}_P(z_i,\vec{r}_i) &=& \exp\left[\frac{i \omega \mathrm{m}
|\vec{r}_i|^2}{2 c z_1} \right] \int d \vec{u}~
\mathcal{P}\left(\frac{\omega \mathrm{m} \vec{r}_i}{c
z_1}-\vec{u}\right)\cdot{ \tilde{E}}\left( 0,-{ \frac{c z_1
\vec{u}}{\omega }}\right) ~. \label{EIPfar}
\end{eqnarray}

\subsection{\label{sub:propar} Propagation of partially coherent light in free space}

Let us now consider the stochastic nature of the Synchrotron
Radiation field in general terms. Synchrotron Radiation is a
Gaussian stochastic process. As has been discussed in detail in
reference \cite{OURS}, the electromagnetic signal at any position
down the beamline is completely characterized, from a statistical
viewpoint, by the knowledge of the second order field correlation
function in space-frequency domain

\begin{equation}
\Gamma_{\omega}(z_o,\vec{r}_{o1},\vec{r}_{  o2},\omega,\omega') =
\left< {\bar{E}}_\bot (z_o,\vec{r}_{
o1},\omega){\bar{E}}^*_\bot(z_o,\vec{r}_{ o2},\omega') \right>~.
\label{gamma}
\end{equation}
In this paper, the averaging brackets $\langle ... \rangle$ will
always indicate an ensemble average over bunches. As it will be
better explained in the following Section \ref{sec:fila}, we will
restrict ourselves to the treatment of radiation from planar
undulators in resonance with the fundamental harmonic. Therefore
we can neglect vertically polarized radiation components and
consider $\bar{E}$ as a scalar quantity. The shot noise in the
electron beam is responsible for random fluctuations of the beam
density, both in space and time. As a result, the temporal Fourier
transform of the Synchrotron Radiation pulse at a fixed frequency
and a fixed point in space is a sum of a great many independent
contributions:

\begin{equation}
{\bar{E}}_\bot(z_o,\vec{r}_{ o},\omega)=\sum_{k=1}^{N}
\bar{E}_{s\bot}(\vec{\eta}_k,\vec{l}_k,z_o,\vec{r}_{ o},\omega)
\exp{(i\omega t_k)} ~, \label{total}
\end{equation}
where $N$ is the number of electrons in the bunch. Here
$\vec{\eta}_k,\vec{l}_k$ and $t_k$ are random variables describing
random angular direction, position and arrival time of an electron
at the reference position $z_o = 0$. As has been demonstrated in
\cite{OURS}, under the assumption - generally verified for X-ray
beams and third generation light sources - that the radiation
wavelengths of interest is much shorter than the bunch length we
can write Eq. (\ref{gamma}) as

\begin{eqnarray}
\Gamma_{\omega}(z_o,\vec{r}_{ o1},\vec{r}_{ o2}, \omega, \omega')
& = &  N {F}_{\omega}( \omega- \omega')  \cr && \times \Bigg
\langle \bar{E}_{s\bot} (\vec{\eta},\vec{l},z_o,\vec{r}_{ o1},
\omega) \bar{E}^*_{s\bot}(\vec{\eta},\vec{l},z_o,\vec{r}_{ o2},
\omega') \Bigg\rangle_{\vec{\eta},\vec{l}} ~, \label{gamma5}
\end{eqnarray}
where $F(\omega)$ is the Fourier transform of the bunch
longitudinal profile function $F_{t}(t_k)$, that is

\begin{equation}
\langle \exp{(i \omega t_k)} \rangle_{t} = \int_{-\infty}^{\infty}
d t_k F_{t}(t_k) e^{i\omega t_k} = {F}_{\omega}(\omega)~.
\label{FTlong}
\end{equation}
Note that the ensemble average on the right hand side of Eq.
(\ref{gamma5}) is done over the product of the electric field
produced by the same electron. In other words each electron is
correlated only with itself.

If the dependence of $\bar{E}_{s\bot}$ on $ \omega$ and $ \omega'$
is slow enough, so that $\bar{E}_{s\bot}$ does not vary
appreciably on the characteristic scale of ${F}_{\omega}$ we can
substitute $\bar{E}^*_{s\bot}(\vec{\eta},\vec{l},z_o,\vec{r}_{
o2}, \omega')$ with
$\bar{E}^*_{s\bot}(\vec{\eta},\vec{l},z_o,\vec{r}_{ o2}, \omega)$
in Eq. (\ref{gamma5}) thus obtaining:

\begin{eqnarray}
\Gamma_{\omega}(z_o,\vec{r}_{ o1},\vec{r}_{ o2}, \omega, \omega')
 = N {F}_{\omega}( \omega- \omega') G(z_o,\vec{r}_{o1},\vec{r}_{o2}, \omega)
\label{gamma6prima}
\end{eqnarray}
where

\begin{equation}
G(z_o,\vec{r}_{ o1},\vec{r}_{ o2}, \omega)= \Bigg\langle
\bar{E}_{s\bot} (\vec{\eta},\vec{l},z_o,\vec{r}_{ o1}, \omega)
\bar{E}^*_{s\bot}(\vec{\eta},\vec{l},z_o,\vec{r}_{ o2}, \omega)
\Bigg\rangle_{\vec{\eta},\vec{l}} ~.\label{coore}
\end{equation}
As has been shown in \cite{OURS} this assumption is by no means a
restrictive one.

From now on we will be concerned with the calculation of the
correlation function $G(z_o,\vec{r}_{ o1},\vec{r}_{ o2}, \omega)$,
while the correlation in frequency (which may be complicated by
other factors describing, for instance, the presence of a
monochromator) can be dealt with separately.

On the one hand, the cross-spectral density as is defined in Eq.
(\ref{coore}) includes the product of fields which obey the free
space propagation relation Eq. (\ref{fieldpropback}). On the other
hand, the averaging over random variables commutes with all
operations involved in the calculation of the field propagation.
More explicitly, introducing the notation
$\tilde{E}(z_o)=\mathcal{O}[\tilde{E}(z)]$ as a shortcut for Eq.
(\ref{fieldpropback}) one can write

\begin{eqnarray}
G(z_o) &=& \left\langle
\bar{E}_{s\bot}(z_o)\bar{E}_{s\bot}^*(z_o)\right\rangle =
\left\langle
\mathcal{O}\left[\bar{E}_{s\bot}(z)\right]\mathcal{O}^*\left[\bar{E}_{s\bot}^*(z)\right]\right\rangle
= \cr &&
\mathcal{O}\cdot\mathcal{O}^*\left[\left\langle\bar{E}_{s\bot}(z)\bar{E}_{s\bot}^*(z)\right\rangle\right]
= \mathcal{O}\cdot\mathcal{O}^*\left[G(z)\right] ~. \label{comm}
\end{eqnarray}
Note that $\mathcal{O}$ may represent, more in general, any linear
operator.

As a result, one can obtain a law for the propagation of the
cross-spectral density in free space in analogy with Eq.
(\ref{fieldpropback}) from position ${z}$ to position ${z}_o$:

\begin{eqnarray}
{G}( {z}_o,\vec{ {r}}_{o1},\vec{{r}}_{o2}) &=&
\frac{\omega^2}{4\pi^2 c^2( {z}_o- {z})^2} \int d \vec{{r}'}_1 d
\vec{{r}'}_2~ { {G}}({z},\vec{{r}'}_1,\vec{ {r}'}_2)\cr && \times
\exp{\left[\frac{i \omega }{2 c ( {z}_o-{z})} \left(\left|\vec{
{r}}_{o1}-\vec{{r}'}_1\right|^2- \left|\vec{{r}}_{o2}-\vec{
{r}'}_2\right|^2\right)\right]}~, \label{crspecpropo}
\end{eqnarray}
where the integral is performed  in four dimensions. We may now
proceed in parallel with Section \ref{sub:wavfr}. In analogy with
Eq. (\ref{ftfield}) let us first define

\begin{eqnarray}
{\mathcal{G}}\left( {z},\vec{ u}_1, \vec{ u}_2\right) &=& \int d
\vec{  {r}'}_1 d \vec{ {r}'}_2~ {G}\left(  {z},\vec{ {r}'}_1,
\vec{ {r}'}_2\right)  \exp \left[i  \left(\vec{ u}_1\cdot\vec{
{r}'}_1 - \vec{ u}_2\cdot\vec{ {r}'}_2\right)\right]~. \label{trG}
\end{eqnarray}
Eq. (\ref{trG}) is a second-order correlation function between
spatial Fourier transforms of the field. In the following, for
simplicity, and with some abuse of language, we will denote
$\mathcal{G}$ as the "Fourier transform of $G$". The spatial
Fourier transform of $\tilde{E}$ depends on the position along the
beamline through a phase factor only. Moreover, as already said,
the operation of ensemble average commutes with the operation of
Fourier transform. It follows that also $\mathcal{G}$ depends on
the position along the beamline through a phase factor only. To be
specific, the analogous of Eq. (\ref{fieldpropback35}) is given by
%
%
\begin{eqnarray}
{\mathcal{G}}\left( {z},\vec{ u}_1, \vec{ u}_2\right) =
{\mathcal{G}}\left(0,\vec{ u}_1, \vec{ u}_2\right) \exp \left[-
\frac{i c }{2 \omega} \left(\left|\vec{ u}_1\right|^2-\left|\vec{
u}_2\right|^2\right) {z}\right]~. \label{factprop}
\end{eqnarray}
Continuing in analogy with Section \ref{sub:wavfr} we find Eq.
(\ref{fieldpropback3tris}), which relates the far field expression
for $\tilde{E}(z_o, \vec{r}_o)$ (in the limit $z_o \longrightarrow
\infty$ and for a finite ratio $\vec{r}_o/z_o$) to the spatial
Fourier transform $F$. We can take advantage of Eq.
(\ref{fieldpropback3tris}) to obtain, with the help of Eq.
(\ref{coore}) and Eq. (\ref{trG}), a useful relation between the
cross-spectral density in the far field and the Fourier transform
of $G$ at the virtual-source position. In the limit $z_o
\longrightarrow \infty$ and for finite ratios $\vec{\theta}_{1} =
\vec{r}_{o1}/z_o$, $\vec{\theta}_{2} = \vec{r}_{o2}/z_o$ we have

%
\begin{eqnarray}
G\left(  {z}_o,\vec{  r}_{o1}, \vec{ r}_{o2}\right) =
\frac{\omega^2}{4 \pi^2 c^2 z_o^2} \exp\left[{\frac{i \omega z_o
}{2
c}\left(\left|\vec{\theta}_1\right|^2-\left|\vec{\theta}_2\right|^2\right)}
\right]{\mathcal{G}}\left( 0,-\frac{\omega\vec{ {\theta}}_1}{c},
-\frac{\omega\vec{ {\theta}}_2}{c}\right)~.\cr &&
\label{maintrick}
\end{eqnarray}
This expression will be very useful later on. In \cite{OURS} we
obtained an explicit expression for the cross-spectral density of
the undulator source in free space at any distance from the
undulator. In particular we can calculate the cross-spectral
density in the far field, which assumes a simplified form.
Consequently, the use of Eq. (\ref{maintrick}) allows to calculate
the Fourier transform of the cross-spectral density at the
virtual-source position. As a result we can characterize the
virtual source and operate with it. Our starting point, here as in
\cite{OURS} is the electron beam in the undulator device. In
contrast with this, previous literature dealing with application
of Statistical Optics to undulator radiation assumes \textit{a
priori} the validity of a postulated expression for the
cross-spectral density.

\subsection{\label{sub:impc} Image formation with partially coherent
light}

In analogy with Section \ref{sub:imcoh} we will now consider the
problem of propagating the cross-spectral density
\textit{immediately in front} of the lens through the optical
system up to the screen at $z=z_2$ \textit{behind} the lens.
Suppose that we know $ {{G}}({z}_1)$ \textit{immediately in front}
of the lens. The cross spectral density \textit{immediately
behind} the lens, $ {G}_l({z}_1)$, is related to ${G}( {z}_1)$ by

\begin{equation}
{G}_l\left( {z}_1,\vec{ {r}'}_1, \vec{ {r}'}_2\right) = {G}\left(
{z}_1,\vec{ {r}'}_1, \vec{ {r}'}_2\right) {T}\left(\vec{
{r}'_1}\right) {T}^*\left(\vec{ {r}'_2}\right)~,\label{after}
\end{equation}
where the transmission function $T$ is defined by Eq.
(\ref{transm}).
%
One can obtain  a law for the propagation of the cross-spectral
density in free space using Eq. (\ref{crspecpropo}), in agreement
with \cite{GOOD}, from position ${z}_1$ \textit{immediately
behind} the lens to the image plane at position ${z}_2$, that is

\begin{eqnarray}
{G}( {z}_2,\vec{ {r}}_1,\vec{{r}}_2) &=& \frac{\omega^2}{4\pi^2
c^2( {z}_2- {z}_1)^2} \int d \vec{{r}'}_1 d \vec{{r}'}_2~ {
{G}}_{l}({z}_1,\vec{{r}'}_1,\vec{ {r}'}_2)\cr && \times
\exp{\left[\frac{i \omega }{2 c ( {z}_2-{z}_1)} \left(\left|\vec{
{r}}_1-\vec{{r}'}_1\right|^2- \left|\vec{{r}}_2-\vec{
{r}'}_2\right|^2\right)\right]}~. \label{crspecprop}
\end{eqnarray}
Substituting Eq. (\ref{after}) in Eq. (\ref{crspecprop}) and
remembering Eq. (\ref{transm}) one finds

\begin{eqnarray}
{G}( {z}_2,\vec{ {r}}_1,\vec{{r}}_2) &=& \frac{\omega^2}{4 \pi^2
c^2 ({z}_2-{z}_1)^2} \int d \vec{{r}'}_1 d \vec{{r}'}_2~ {G}\left(
{z}_1,\vec{ {r}'}_1, \vec{{r}'}_2\right)
 P\left(\vec{ {r}'_1}\right) P^*\left(\vec{{r}'_2}\right)
\cr && \times \exp\left[\frac{i \omega}{2 c {f}}
\left({\left|\vec{ {r}'_2}\right|^2-\left|\vec{
{r}'_1}\right|^2}\right)+\frac{i \omega}{2 c( {z}_2- {z}_1)}
\left(\left|\vec{ {r}}_1-\vec{ {r}'}_1\right|^2 \right. \right.\cr
&& \left.\left. - \left|\vec{ {r}}_2-\vec{
{r}'}_2\right|^2\right)\right]~. \label{crspecprop2}
\end{eqnarray}
Manipulation of the argument in the exponential function under
integral allows the more suggestive representation

\begin{eqnarray}
{G}(  {z}_2,\vec{  {r}}_1,\vec{  {r}}_2) &=&
\frac{\omega^2}{4\pi^2 c^2(  {z}_2-  {z}_1)^2} \exp\left[\frac{i
\omega} {2  c (  {z}_2-  {z}_1)} \left(\left|\vec{
{r}}_1\right|^2- \left|\vec{  {r}}_2\right|^2\right)\right]\cr &&
\times \int d \vec{  {r}'}_1 d \vec{  {r}'}_2~ \Bigg\{ {G}\left(
{z}_1,\vec{ {r}'}_1, \vec{  {r}'}_2\right)
 P\left(\vec{  {r}'_1}\right) P^*\left(\vec{  {r}'_2}\right)
\cr && \times \exp\left[\frac{i \omega}{c}  \left(\frac{1}{2 {f}}-
\frac{1}{2( {z}_2-  {z}_1)} \right) \left({\left|\vec{
{r}'_2}\right|^2-\left|\vec{
{r}'_1}\right|^2}\right)\right]\Bigg\} \cr && \times \exp
\left[\frac{i \omega}{c (   {z}_2-  {z}_1)}\left(-\vec{
{r}}_1\cdot\vec{ {r}'}_1 + \vec{  {r}}_2\cdot\vec{
{r}'}_2\right)\right]~, \label{crspecprop3}
\end{eqnarray}
that is analogous to Eq. (\ref{fieldpropexp}). The quantity in
brackets $\{...\}$ is basically Fourier-transformed. As mentioned
before, similarities between the way $ {G}$ and $\tilde{E}$ evolve
through the beamline have to be ascribed to the fact that the
average over random variables commutes with all other operations
in the calculation of the cross-spectral density. As a result, the
reason why Eq. (\ref{crspecprop3}) is basically a Fourier
transformation is due to the particular way $\tilde{E}$ evolves.

Similarly as has been explained above for the fields, the image
and the focal plane are privileged planes for which the
cross-spectral density assumes particularly simple forms, that can
be found in terms of Fourier Optics.

\subsubsection{\label{subsub:negl2} Large non-limiting aperture}

We will now proceed in analogy with Section \ref{subsub:negl1}. On
the focal plane, similarly to Eq. (\ref{cpact1}) we have

\begin{eqnarray}
{G}(  {z}_f,\vec{  {r}}_{1f},\vec{  {r}}_{2f}) &=&
\frac{\omega^2}{4\pi^2 c^2 {f}^2} \exp\left[\frac{i \omega} {2
c{f}} \left(\left|\vec{  {r}}_{1f}\right|^2- \left|\vec{
{r}}_{2f}\right|^2\right)\right] \cr && \times {\mathcal{G}}\left(
{z}_1,-\frac{\omega\vec{{r}}_{1f}}{c {f}}, -\frac{\omega\vec{
{r}}_{2f}}{c {f}}\right)~,\label{crspecpropfoc}
\end{eqnarray}
while using Eq. (\ref{factprop}) we obtain, similarly to Eq.
(\ref{cpact3}),

\begin{eqnarray}
{G}(  {z}_f,\vec{  {r}}_{1f},\vec{  {r}}_{2f}) &=&
\frac{\omega^2}{4\pi^2 c^2 {f}^2} \exp\left[\frac{i \omega} {2
c{f}} \left(\left|\vec{  {r}}_{1f}\right|^2- \left|\vec{
{r}}_{2f}\right|^2\right)\right]\cr && \times\exp\left[-\frac{i
\omega z_1} {2 c{f^2}} \left(\left|\vec{  {r}}_{1f}\right|^2-
\left|\vec{ {r}}_{2f}\right|^2\right)\right]
{\mathcal{G}}\left(0,-\frac{\omega\vec{ {r}}_{1f}}{c {f}},
-\frac{\omega\vec{ {r}}_{2f}}{c {f}}\right)~.\cr &&\label{crsfin}
\end{eqnarray}
In analogy with Eq. (\ref{fieldpropexpima}), for the image plane
we can write

\begin{eqnarray}
{G}(  {z}_i,\vec{  {r}}_{1i},\vec{  {r}}_2) &=&
\left(\frac{\mathrm{m} \omega}{2\pi c {z}_1}\right)^2
\exp\left[\frac{i \omega \mathrm{m}} {2 c{z}_1} \left(\left|\vec{
{r}}_{1i}\right|^2- \left|\vec{ {r}}_{2i}\right|^2\right)\right]
\cr && \times \int d \vec{ {r}'}_1 d \vec{ {r}'}_2~ { {G}}(
{z}_1,\vec{ {r}'}_1,\vec{ {r}'}_2)\exp{\left[\frac{i \omega}{2 c
{z}_1} \left(\left|\vec{ {r}'}_2\right|^2- \left|\vec{
{r}'}_1\right|^2\right)\right]}\cr && \times \exp \left[-\frac{i
\mathrm{m} \omega}{c  {z}_1 }\left(\vec{{r}}_{1i}\cdot\vec{
{r}'}_1 - \vec{ {r}}_{2i}\cdot\vec{ {r}'}_2\right)\right]~.
\label{crspecpropim0}
\end{eqnarray}
Using the convolution theorem in analogy with Eq. (\ref{cpact2})
we obtain

\begin{eqnarray}
{G}(  {z}_i,\vec{  {r}}_{1i},\vec{  {r}}_{2i}) &=&
\left(\frac{\mathrm{m} }{4\pi^2 }\right)^2 \exp\left[\frac{i
\omega \mathrm{m}} {2 c{z}_1} \left(\left|\vec{
{r}}_{1i}\right|^2- \left|\vec{ {r}}_{2i}\right|^2\right)\right]
\int d \vec{ {u}} ~d \vec{ {v}}~ {\mathcal{G}}\left( {z}_1,\vec{
{u}},\vec{ {v}}\right) \cr && \times \exp\left\{\frac{i c {z}_1}{2
\omega} \left[ \left(\frac{\mathrm{m} \omega \vec{ {r}}_{1i}}{c
{z}_1}+\vec{ {u}}\right)^2- \left(\frac{\mathrm{m} \omega \vec{
{r}}_{2i}}{ c {z}_1} + \vec{ {v}}\right)^2\right] \right\}~.\cr &&
\label{crspecpropim}
\end{eqnarray}
Finally, taking advantage of Eq. (\ref{factprop}) we have

\begin{eqnarray}
{G}(  {z}_i,\vec{  {r}}_{1i},\vec{  {r}}_{2i}) &=&
\left(\frac{\mathrm{m} }{4\pi^2 }\right)^2 \exp\left[\frac{i
 \omega \mathrm{m}} {2 c {z}_1} \left(\left|\vec{
{r}}_{1i}\right|^2- \left|\vec{ {r}}_{2i}\right|^2\right)\right]
\cr &&\times \exp\left[\frac{i \mathrm{m}^2 \omega} {2 c {z}_1}
\left(\left|\vec{ {r}}_{1i}\right|^2- \left|\vec{
{r}}_{2i}\right|^2\right)\right] \cr && \times  \int d \vec{ {u}}
~d \vec{ {v}}~ {\mathcal{G}}\left(0,\vec{ {u}},\vec{ {v}}\right)
\exp\left[i \mathrm{m} \left(\vec{ {r}}_{1i}\cdot\vec{ {u}}- \vec{
{r}}_{2i}\cdot\vec{ {v}}\right) \right]~, \label{propimf}
\end{eqnarray}
that can be rewritten as the analogous of Eq. (\ref{cpact5}):

\begin{eqnarray}
{G}(  {z}_i,\vec{  {r}}_{1i},\vec{  {r}}_{2i}) &=& {\mathrm{m}^2}
\exp\left[\frac{i \mathrm{m} \omega} {2 c {z}_1} \left(\left|\vec{
{r}}_{1i}\right|^2- \left|\vec{ {r}}_{2i}\right|^2\right)\right]
\cr && \times \exp\left[\frac{i \mathrm{m}^2\omega} {2 c {z}_1}
\left(\left|\vec{ {r}}_{1i}\right|^2- \left|\vec{
{r}}_{2i}\right|^2\right)\right] {G}\left( 0,{-\mathrm{m}\vec{
{r}}_{1i}},{-\mathrm{m}\vec{ {r}}_{2i}}\right)~.\cr &&
\label{propimf2}
\end{eqnarray}

\subsubsection{\label{subsub:incl2} Effect of aperture size}

Effects of aperture size can be included in strict analogy with
Section \ref{subsub:incl1}. Similarly to Section
\ref{subsub:incl1} the following mnemonic rule can be applied to
include the effects of the pupil in Eq. (\ref{crsfin}) or Eq.
(\ref{propimf2}). First, divide Eq. (\ref{crsfin}) or Eq.
(\ref{propimf2}) by the first phase factor, corresponding to the
phase factor outside the integral sign in Eq. (\ref{crspecprop3}).
Second, convolve twice with ${{\mathcal{P}}}$ and
${{\mathcal{P}}}^*$, ${{\mathcal{P}}}$ having already been defined
in Eq. (\ref{PFT}). Third, put the phase factor back. We will
denote with $G_P$ the cross-spectral density including the effects
due to the presence of the pupil. On the focal plane, in analogy
with Eq. (\ref{EFP}), one obtains

\begin{eqnarray}
G_P(z_f, \vec{r}_{1f}, \vec{r}_{2f}) &=& \exp\left[\frac{i \omega}
{2 c{f}} \left(\left|\vec{  {r}}_{1f}\right|^2- \left|\vec{
{r}}_{2f}\right|^2\right)\right]\cr &&\times\int d\vec{u}~
d\vec{v}~ \mathcal{P}\left(\frac{\omega \vec{r}_{1f}}{c
f}-\vec{u}\right) \mathcal{P}^*\left(\frac{\omega \vec{r}_{2f}}{c
f}-\vec{v}\right)\cr &&\times\exp\left[\frac{i c f} {2 \omega}
\left(\left|\vec{v}\right|^2- \left|\vec{u}\right|^2\right)\right]
G\left(z_f, \frac{c f \vec{u}}{\omega}, \frac{c f
\vec{v}}{\omega}\right) \label{GPF}
\end{eqnarray}
while on the image plane, in analogy with Eq. (\ref{EIP}), one has

\begin{eqnarray}
G_P(z_i, \vec{r}_{1i}, \vec{r}_{2i}) &=& \exp\left[\frac{i
\mathrm{m} \omega} {2 c {z}_1} \left(\left|\vec{
{r}}_{1i}\right|^2- \left|\vec{
{r}}_{2i}\right|^2\right)\right]\cr &&\times \int d\vec{u}~
d\vec{v} ~\mathcal{P}\left(\frac{\omega \mathrm{m}\vec{r}_{1i}}{c
z_1}-\vec{u}\right)  \mathcal{P}^*\left(\frac{\omega \mathrm{m}
\vec{r}_{2i}}{c z_1}-\vec{v}\right)\cr &&\times \exp\left[\frac{i
 c z_1} {2 \omega \mathrm{m}} \left(\left|\vec{v}\right|^2-
\left|\vec{u}\right|^2\right)\right]G\left(z_i, \frac{ c z_1
\vec{u}}{\omega \mathrm{m}},\frac{c z_1 \vec{v}}{\omega
\mathrm{m}}\right)~. \label{GPI}
\end{eqnarray}
In the limit for large apertures, $\mathcal{P}$ and
$\mathcal{P}^*$ can be substituted by $\delta$-Dirac functions in
both Eq. (\ref{GPF}) and Eq. (\ref{GPI}). In this case, from Eq.
(\ref{GPF}) we recover $G_P(z_f,\vec{r}_{1f},\vec{r}_{2f}) =
G(z_f,\vec{r}_{1f},\vec{r}_{2f})$, that is given by Eq.
(\ref{crsfin}). From Eq. (\ref{GPI}) instead, we have
$G_P(z_i,\vec{r}_{1i},\vec{r}_{2i}) =
G(z_i,\vec{r}_{1i},\vec{r}_{2i})$, that is given by Eq.
(\ref{propimf2}).

Similarly to the case analyzed in Section \ref{sub:imcoh}, unless
particular conditions are met, the phase factor under the
integrals in Eq. (\ref{GPF}) and Eq. (\ref{GPI}) compensate only
partially the phase in Eq. (\ref{crsfin}) and Eq.
(\ref{propimf2}). In the following we will restrict our attention
to the image plane. In this case, complete phase compensation is
achieved when the lens is in the far field.

The far field limit of Eq. (\ref{propimf2})  can be obtained by
substitution of Eq. (\ref{maintrick}) in Eq.
(\ref{crspecpropim0}). After calculating the inverse
transformation of Eq. (\ref{trG}) one obtains:

\begin{eqnarray}
{G}(  {z}_i,\vec{  {r}}_{1i},\vec{  {r}}_{2i}) &=& {\mathrm{m}^2}
\exp\left[\frac{i \mathrm{m} \omega} {2 c {z}_1} \left(\left|\vec{
{r}}_{1i}\right|^2- \left|\vec{ {r}}_{2i}\right|^2\right)\right]
{G}\left( 0,{-\mathrm{m}\vec{ {r}}_{1i}},{-\mathrm{m}\vec{
{r}}_{2i}}\right)~.\cr && \label{propimf2farf}
\end{eqnarray}
Eq. (\ref{propimf2farf}) can also be obtained directly from Eq.
(\ref{propimf2}) neglecting the second phase factor on the right
hand side, that is possible when

\begin{equation} \frac{\mathrm{m}^2\omega} {2 c {z}_1}
\left(\left|\vec{ {r}}_{1i}\right|^2- \left|\vec{
{r}}_{2i}\right|^2\right)\ll 1 \label{phasesmallfar2}
\end{equation}
for any pair of points $(\vec{r}_{1i},\vec{r}_{2i})$ on the image
pattern. When the coherence length on the image plane is much
smaller than the characteristic size of the image, condition
(\ref{phasesmallfar2}) constitutes, similarly to condition
(\ref{phasesmall}) before (that holds in the case of coherent
light), the requirement to be satisfied for the lens to be in the
far zone. When condition (\ref{phasesmallfar2}) is satisfied, Eq.
(\ref{propimf2farf}) holds instead of Eq. (\ref{propimf2}). As
explained in Section \ref{sub:imcoh}, in this paper we will study
pupil effects only under this assumption.

Explicit substitution of Eq. (\ref{propimf2farf}) in Eq.
(\ref{GPI}) yields, the following far field limit expression:

\begin{eqnarray}
G_P(z_i, \vec{r}_{1i}, \vec{r}_{2i}) &=& \exp\left[\frac{i
\mathrm{m} \omega} {2 c {z}_1} \left(\left|\vec{
{r}}_{1i}\right|^2- \left|\vec{ {r}}_{2i}\right|^2\right)\right]
\int d\vec{u}~ d\vec{v} ~\mathcal{P}\left(\frac{\omega
\mathrm{m}\vec{r}_{1i}}{c z_1}-\vec{u}\right)\cr &&\times
\mathcal{P}^*\left(\frac{\omega \mathrm{m} \vec{r}_{2i}}{c
z_1}-\vec{v}\right){G}\left( 0,-\frac{c z_1 \vec{u}}{\omega
},-\frac{c z_1 \vec{v}}{\omega }\right)~, \label{GPIfar}
\end{eqnarray}
that is the analogous of Eq. (\ref{EIPfar}) in Section
\ref{subsub:incl1}.

\section{\label{sec:fila} Image formation with a perfectly coherent undulator source}

In order to give an explicit expression for the cross-spectral
density at position $z_2$ we need to know an explicit expression
for the cross-spectral density ${G}( {z}_1,\vec{ {r}'}_1, \vec{
{r}'}_2)$ \textit{immediately in front} of the lens. This was the
subject of our previous work \cite{OURS}. In this and in the
following Sections we will present the image formation problem and
its solution with the help of that reference. In particular, in
the present Section \ref{sec:fila} we will begin with the simpler
deterministic case of zero electron beam emittance. One may think
of a filament electron beam or equivalently, as concerns the
cross-spectral density, of a single electron. We will start
neglecting the presence of the pupil. At the end we will
generalize our results to account for it. In the following parts
of this paper we will then generalize the results obtained in the
present Section \ref{sec:fila} to include emittance effects, thus
taking full advantage of the Statistical Optics formulation.

\begin{figure}
\begin{center}
\includegraphics*[width=110mm]{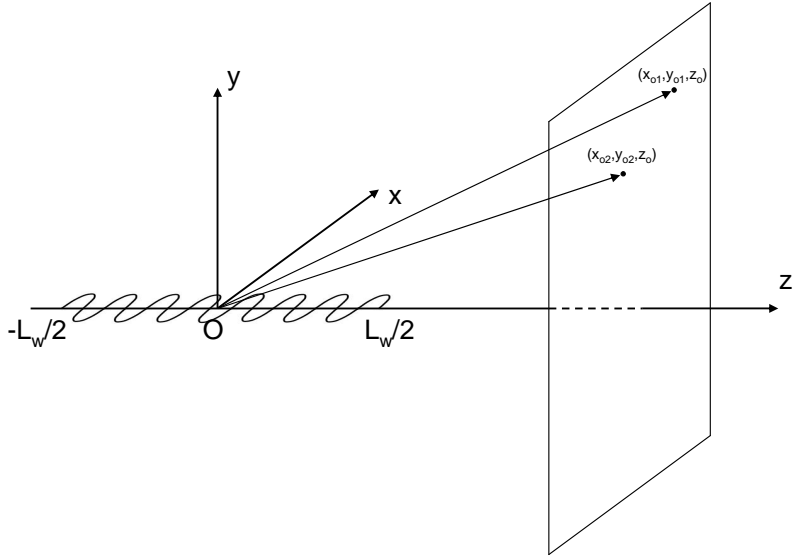}
\caption{\label{geo} Illustration of the undulator geometry and of
the observation plane after the undulator. }
\end{center}
\end{figure}
Our starting point is an expression, derived in reference
\cite{OURS}, for the complex envelope $\tilde{E}_{s\bot}$ of the
Fourier transform of the electric field produced by a single
electron moving through a planar undulator at any distance from
the exit of the undulator. That expression accounts for a given
offset and deflection angle of the particle trajectory with
respect to the undulator axis. Referring to Fig. \ref{geo} we
found

\begin{eqnarray}
\tilde{E}_{s\bot}&=& -\frac{K \omega e  A_{JJ}}{c^2  \gamma}
\int_{-L_w/2}^{L_w/2} d z' \frac{1}{{z}_o-{z}'} \cr &&\times\exp
\left\{i \left[\left(C+\frac{\omega \left|\vec{\eta}\right|^2}{2
c}\right){z}' + \frac{\omega\left(\vec{r}_{\bot
o}-\vec{l}-\vec{\eta}z' \right)^2 }{2 c(z_o-z')}\right] \right\} .
\label{undunormfin}
\end{eqnarray}
Here $K$ is the undulator parameter, $L_w$, as before, is the
undulator length, $(-e)$ is the electron charge and $\gamma$ is
the relativistic Lorentz factor. Moreover

\begin{equation}
A_{JJ} = J_0\left( \frac{K^2}{4+2K^2}\right) - J_1\left(
\frac{K^2}{4+2K^2}\right) ~,\label{AJJdef}
\end{equation}
$J_n$ being the Bessel function of the first kind of order $n$.
Also,

\begin{equation}
\omega_o = \frac{4\pi c \gamma^2}{\lambda_w \left(1+K^2/2\right)}~
\label{omegazerodef}
\end{equation}
is the fundamental frequency of the undulator, $\lambda_w$ being
the undulator period. Finally,

\begin{equation}
C = \frac{2 \pi}{\lambda_w}
\frac{\omega-\omega_o}{\omega_o}\label{Cnon}
\end{equation}
is the detuning parameter, which accounts for small deviations in
frequency from resonance.

Eq. (\ref{undunormfin}) is valid for frequencies about the
fundamental harmonic $\omega_o$. This means that we are
considering a large number of undulator periods $N_w \gg 1$ and
that we are looking at frequencies near the fundamental at angles
within the main lobe of the directivity diagram of the radiation.
In this situation one can neglect the vertical $y$-polarization
component of the field with an accuracy $(4\pi N_w)^{-1}$. This
constitutes a great simplification of the problem. At any position
of the observer, we may consider the temporal Fourier transform of
the electric field as a complex scalar quantity corresponding to
the surviving $x$-polarization component of the original vector
quantity. Moreover it should be noted that, in deriving Eq.
(\ref{undunormfin}), we assumed that no influence of focusing is
present inside the undulator. $\vec{\eta}$ and $\vec{l}$ are to be
understood as deflection angles and offset of the electron at the
position $z=0$.

\begin{figure}
\begin{center}
\includegraphics*[width=110mm]{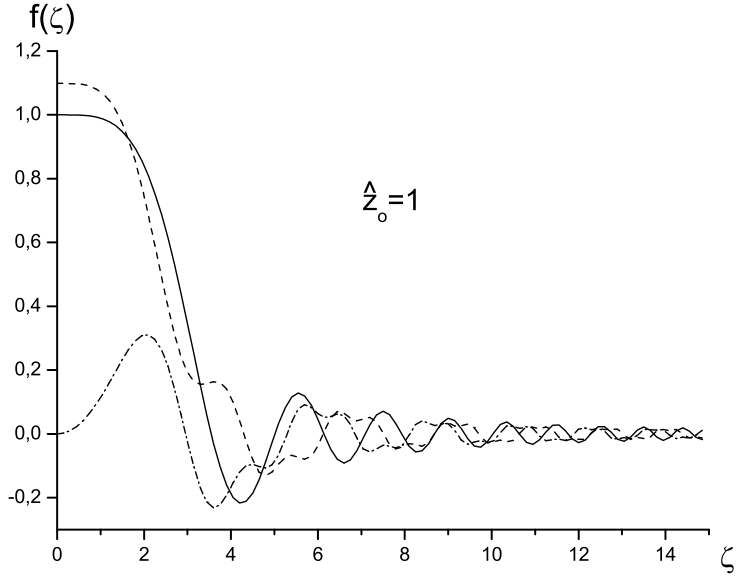}
\caption{\label{S1} Comparison between
$f(\zeta)=\mathrm{sinc}(\zeta^2/4)$ (solid line), the real (dashed
line) and the imaginary (dash-dotted line) parts of
$f(\zeta)=\hat{z}_o S(0,\hat{z}_o,\zeta^2)$ at $\hat{z}_o = 1$.}
\end{center}
\end{figure}

\begin{figure}
\begin{center}
\includegraphics*[width=110mm]{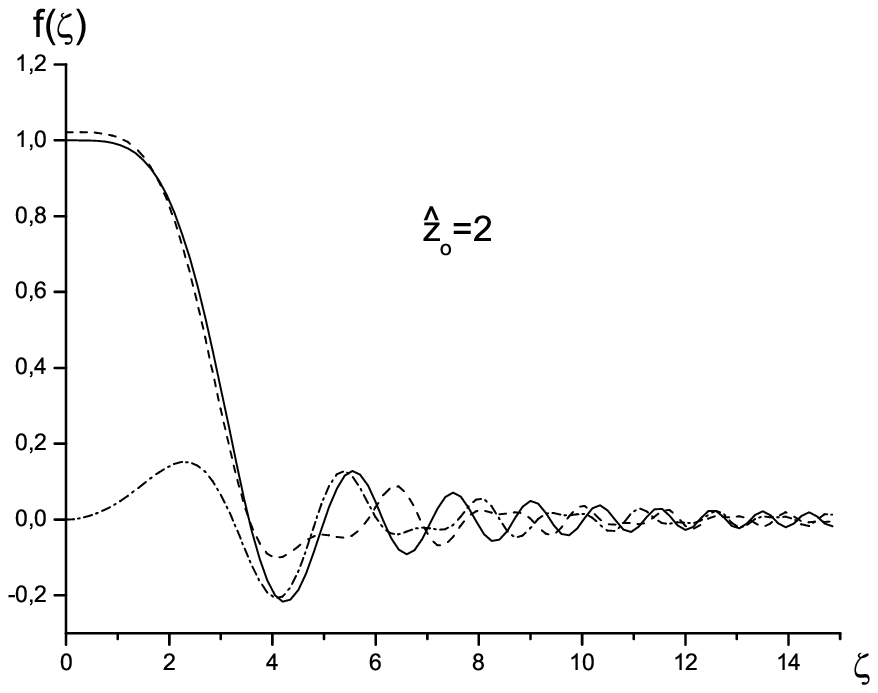}
\caption{\label{S2} Comparison between
$f(\zeta)=\mathrm{sinc}(\zeta^2/4)$ (solid line), the real (dashed
line) and the imaginary (dash-dotted line) parts of
$f(\zeta)=\hat{z}_o S(0,\hat{z}_o,\zeta^2)$ at $\hat{z}_o = 2$.}
\end{center}
\end{figure}

\begin{figure}
\begin{center}
\includegraphics*[width=110mm]{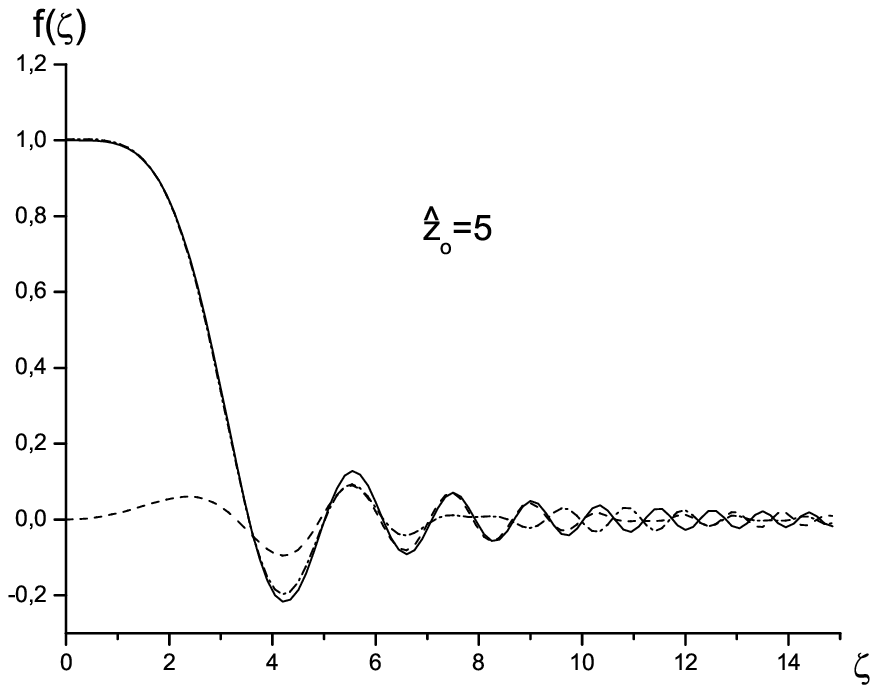}
\caption{\label{S5} Comparison between
$f(\zeta)=\mathrm{sinc}(\zeta^2/4)$ (solid line), the real (dashed
line) and the imaginary (dash-dotted line) parts of
$f(\zeta)=\hat{z}_o S(0,\hat{z}_o,\zeta^2)$ at $\hat{z}_o = 5$.}
\end{center}
\end{figure}

\begin{figure}
\begin{center}
\includegraphics*[width=110mm]{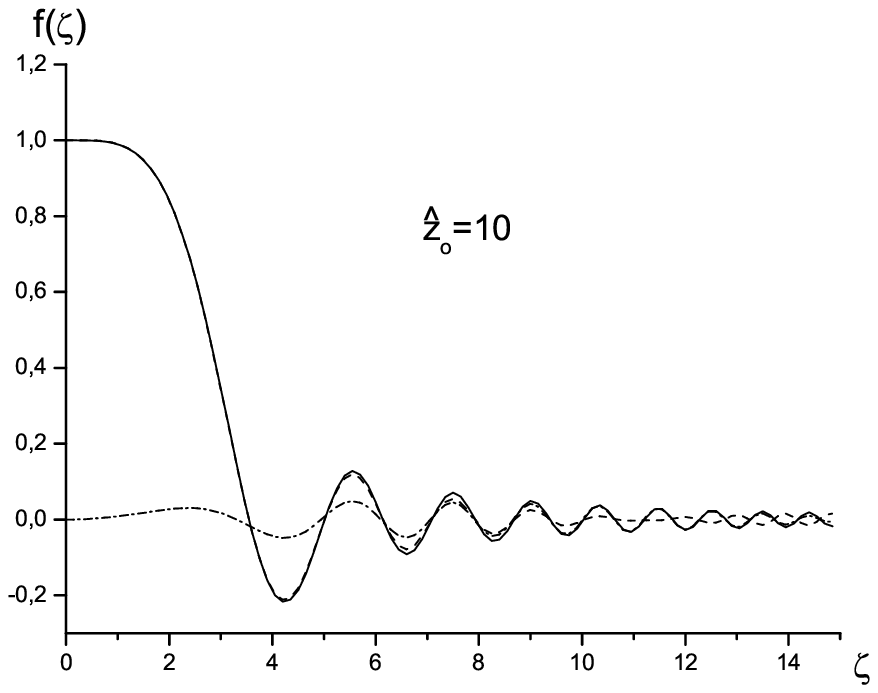}
\caption{\label{S10} Comparison between
$f(\zeta)=\mathrm{sinc}(\zeta^2/4)$ (solid line), the real (dashed
line) and the imaginary (dash-dotted line) parts of
$f(\zeta)=\hat{z}_o S(0,\hat{z}_o,\zeta^2)$ at $\hat{z}_o = 10$.}
\end{center}
\end{figure}

Let us introduce normalized units \footnote{The relation between
$\hat{E}_{s\bot}$ and $\tilde{E}_{s\bot}$ in Eq. (\ref{Cnorm})
differs from the analogous one in Eq. (25) of reference
\cite{OURS} for a factor $\hat{z}_o = z_o/L_w$. The reason for
this discrepancy is related to the different subjects treated. In
\cite{OURS} we considered only the free space case, while in this
paper we extend our considerations to an optical element, thus
introducing another privileged longitudinal position (the lens
position) other that the observation plane. The definition of
$\hat{E}_{s\bot}$ in \cite{OURS} is no more a convenient one here
and would lead to artificial complications in the following parts
of this paper. Therefore it has been slightly modified as in Eq.
(\ref{Cnorm}). This leads to slight changes (related with the
factor $\hat{z}_o$) in some of the following equations when
compared to the analogous quantities in \cite{OURS}.}

\begin{eqnarray} && \hat{E}_{s\bot} = -\frac{c^2  \gamma
\tilde{E}_{s\bot}}{K \omega e  A_{JJ}} ~,\cr && \vec{\hat{\eta}}
=\vec{{\eta}}\sqrt{\frac{\omega L_w}{c}}  ~,\cr&& \hat{C} = L_w C
= 2 \pi N_w \frac
{\omega-\omega_o}{\omega_o}~,\cr&&\vec{\hat{r}}_{\bot o}
={\vec{r}}_{\bot o} \sqrt{\frac{\omega}{{L_w
c}}}~,\cr&&\vec{\hat{l}} ={\vec{l}}\sqrt{\frac{\omega}{{L_w
c}}}~,\cr &&\hat{z}=\frac{z}{L_w}~.\label{Cnorm}
\end{eqnarray}
As shown in Appendix B of \cite{OURS}, after some algebraic
manipulation, Eq. (\ref{undunormfin}) can be rewritten in
normalized units as

\begin{eqnarray}
\hat{E}_{s\bot} =    \int_{-1/2}^{1/2} \frac{ d\hat{z}'
}{\hat{z}_o-\hat{z}'} \exp \left\{i \left[\Phi_U +\hat{C} \hat{z}'
+ \frac{\hat{z}_o \hat{z}'}{2 (\hat{z}_o-\hat{z}')}
\left(\vec{\hat{\theta}}- \frac{\vec{\hat{l}}}{\hat{z}_o}-
\vec{\hat{\eta}}\right)^2 \right] \right\}~,
\label{undunormfinult}
\end{eqnarray}
where

\begin{eqnarray}
\vec{\hat{\theta}}= \frac{{\vec{\hat{r}}}_{\bot
o}}{\hat{z}_o}~\label{Cnorm2}
\end{eqnarray}
represents the observation angle and $\Phi_U$ is given by

\begin{equation}
\Phi_U =
\left(\vec{\hat{\theta}}-\frac{\vec{\hat{l}}}{\hat{z}_o}\right)^2
 \frac{\hat{z}_o}{2} ~.\label{phisnorm}
\end{equation}
Eq. (\ref{undunormfinult}) is of the form

\begin{equation}
\hat{E}_{s\bot}\left(\hat{C},\hat{z}_o,\vec{\hat{\theta}}-
\frac{\vec{\hat{l}}}{\hat{z}_o}- \vec{\hat{\eta}}\right) =
\exp{(i\Phi_U)} S\left[\hat{C},\hat{z}_o,\left(\vec{\hat{\theta}}-
\frac{\vec{\hat{l}}}{\hat{z}_o}-
\vec{\hat{\eta}}\right)^2\right]~. \label{Esum}
\end{equation}
It is possible to show that the expression for the function
$S(\cdot)$ reduces to a $\mathrm{sinc(\cdot)}$ function as
$\hat{z}_o \gg 1$~. In this limiting case, the expression for the
electric field from a single particle, Eq. (\ref{undunormfinult}),
is simplified to

\begin{eqnarray}
\hat{E}_{s\bot} =  \exp{(i\Phi_U)} \int_{-1/2}^{1/2}
\frac{d\hat{z}'}{\hat{z}_o} \exp \left\{i \hat{z}' \left[\hat{C} +
\frac{1}{2}\left(\vec{\hat{{\theta}}}-\frac{\vec{\hat{l}_x}}{\hat{z}_o}-
\vec{\hat{\eta}}\right)^2 \right]\right\} ~. \label{undunormfin2}
\end{eqnarray}
Eq. (\ref{undunormfin2}) can be integrated analytically giving

\begin{equation}
\hat{E}_{s\bot}= \exp{(i \Phi_U)} \frac{1}{\hat{z}_o}
\mathrm{sinc}\left(\frac{\hat{C}}{2}+\frac{\zeta^2}{4}\right) ~,
\label{endangle}
\end{equation}
where

\begin{equation}
\zeta =
\vec{\hat{{\theta}}}-\frac{\vec{\hat{l}}}{\hat{z}_o}-\vec{\hat{\eta}}~
\label{zeta}
\end{equation}
and where the $\mathrm{sinc}$ function has been defined as

\begin{equation}
\mathrm{sinc}(x) = \frac{\sin(x)}{x}~. \label{sincdef}
\end{equation}
For simplicity, in this paper we will restrict our attention to
the case $\hat{C}=0$. In the particular case $\hat{C} =0$, the
function $S$ can be represented in terms of the exponential
integral function $\mathrm{Ei}(\cdot)$ as

\begin{equation}
S\left(0,\hat{z}_o,\zeta^2\right) = \exp (-i \hat{z}_o \zeta^2/2)
\left[ \mathrm{Ei} \left(\frac{i \hat{z}_o^2 \zeta^2
}{-1+2\hat{z}_o}\right)- \mathrm{Ei} \left(\frac{i \hat{z}_o^2
\zeta^2 }{1+2\hat{z}_o}\right) \right] ~.\label{SfuncEi}
\end{equation}
It is interesting to study the behavior of the $S$ function as the
distance from the undulator center $\hat{z}_o$ increases. This
gives an idea of how good the asymptotic approximation of the $S$
function for $\hat{z}_o \gg 1$ (that is a $\mathrm{sinc}$
function) is. A comparison between $\mathrm{sinc}(\zeta^2/4)$ and
the real and imaginary parts of $\hat{z}_o S(0,\hat{z}_o,\zeta^2)$
for $\hat{z}_o = 1$, $\hat{z}_o = 2$, $\hat{z}_o = 5$ and
$\hat{z}_o = 10$ is given respectively in Fig. \ref{S1}, Fig.
\ref{S2}, Fig. \ref{S5} and Fig. \ref{S10}.

When the electron beam has zero emittance we are dealing with a
perfectly coherent wavefront. The evolution of the radiation
wavefront through our optical system
can be obtained with the help of Eq. (\ref{Esum}). 
In the following  we will study such evolution assuming  $\hat{C}
=0$, $\vec{\hat{l}}=0$ and $\vec{\hat{\eta}}=0$. These assumptions
mean that the radiation frequency is perfectly tuned to the
fundamental frequency of the undulator and that the electron beam
is moving on the $z$ axis. In this case Eq. (\ref{endangle})
describes, in the far field region, a spherical wave with the
source in the center of the undulator. This remark allows one to
consider the undulator center as a privileged point. In other
words, the phase factor in Eq. (\ref{endangle}) represents, in
paraxial approximation, the phase difference (characterizing a
spherical wave) between the point
$(\hat{x}_o,\hat{y}_o,\hat{z}_o)$ and the point $(0,0,\hat{z}_o)$
on the observation plane.

\subsection{\label{subsub:negl3} Large non-limiting aperture}

\begin{figure}
\begin{center}
\includegraphics*[width=110mm]{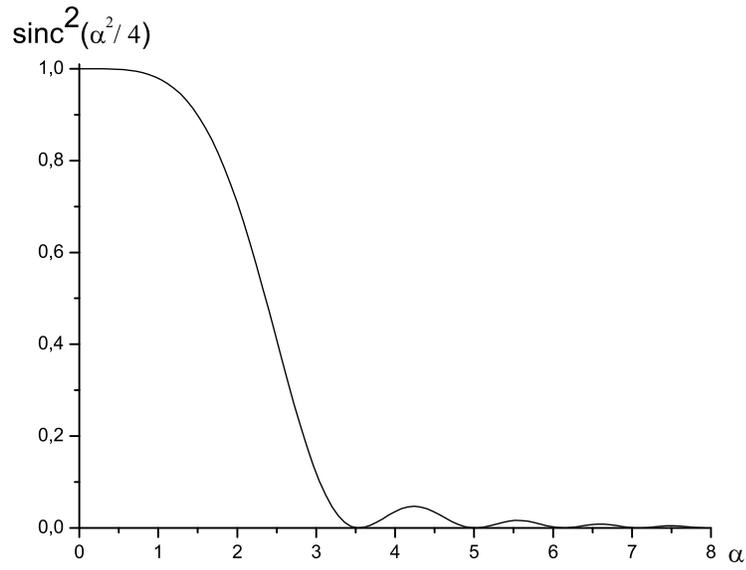}
\caption{\label{focindet} Universal function
$\mathrm{sinc}^2(\alpha^2/4)$ used to calculate (according to Eq.
(\ref{Icpact31})) the focal intensity of a single electron at the
fundamental harmonic at perfect resonance.}
\end{center}
\end{figure}

We will first study the case when the pupil function can be
neglected. Let us introduce a normalized version of the spatial
Fourier transform of the field, analogous to Eq. (\ref{ftfield}),
that is

\begin{eqnarray}
\hat{\mathrm{F}}\left(\hat{z},\vec{\hat{u}}\right) = \int
d\vec{\hat{r'}} \hat{E}\left(\hat{z},\vec{\hat{r'}}\right)
\exp\left[i
\vec{\hat{r'}}\cdot\vec{\hat{u}}\right]~.\label{FTspatF}
\end{eqnarray}
The spatial Fourier transform
$\mathrm{\hat{F}}\left(\hat{z},\vec{\hat{u}}\right)$ can be
calculated directly from Eq. (\ref{undunormfinult}) (compare also
with Eq. (184) of reference \cite{OURS}) and gives

\begin{eqnarray}
\mathrm{\hat{F}}\left(\hat{z},\vec{\hat{u}}\right) = -{2 \pi i }
~\mathrm{sinc}\left(\frac{\left|\vec{\hat{u}}\right|^2}{4}\right)
\exp{\left[ -\frac{i \left|\vec{\hat{u}}\right|^2 \hat{z}}{2
}\right]}~. \label{fieldpropback3b}
\end{eqnarray}
Eq. (\ref{cpact1}) and Eq. (\ref{cpact2}) can  be respectively
rewritten in normalized units as

\begin{eqnarray}
{ \hat{E}}_\bot( \hat{z}_f,\vec{ \hat{r}}_f) &=& \frac{i }{2 \pi
\hat{f}}  \exp{\left[ \frac{i  | \vec{\hat{r}}_f |^2}{2  \hat{f}}
\right]}
\mathrm{\hat{F}}\left(\hat{z}_1,-\frac{\vec{\hat{r}}_f}{\hat{f}}\right)
\label{cpact11}
\end{eqnarray}
and

\begin{eqnarray}
{ \hat{E}}_\bot( \hat{z}_i,\vec{ {r}}_i) &=& \frac{\mathrm{m}}{4
\pi^2} \exp{\left[ \frac{i \mathrm{m} | \vec{\hat{r}}_i |^2}{2
\hat{z}_1} \right]}\int d \vec{\hat{u}}~
\mathrm{\hat{F}}\left(\hat{z}_1,\vec{\hat{u}}\right)\exp{\left[\frac{i
\hat{z}_1 }{2}
\left(-\frac{\mathrm{m}\vec{\hat{r}}_i}{\hat{z}_1}-\vec{\hat{u}}\right)^2
\right]}~,\cr && \label{cpact21}
\end{eqnarray}
where $\hat{f}=f/L_w$. Substitution of Eq. (\ref{fieldpropback3b})
in Eq. (\ref{cpact11}) and in Eq. (\ref{cpact21}) yields results
respectively for the focal and the image plane. In the focal plane
we have

\begin{eqnarray}
{ \hat{E}}_\bot( \hat{z}_f,\vec{ \hat{r}}_f) &=& \frac{1}{\hat{f}}
\exp{\left[ \frac{i  | \vec{\hat{r}}_f |^2}{2 \hat{f}}
\right]}\exp{\left[-\frac{i \hat{z}_1 | \vec{\hat{r}}_f |^2}{2
\hat{f}^2} \right]} \mathrm{sinc}\left(\frac{\left|\vec{\hat{
r}}_f\right|^2}{4 \hat{f}^2}\right)~.\label{cpact31}
\end{eqnarray}
Note that the relative distribution  of intensity in the focus
reproduces the angular distribution of intensity in the far field,
that is

\begin{eqnarray}
{ \hat{I}}\left({\hat{r}_f}\right) &=&
\mathrm{sinc}^2\left(\frac{\left|\vec{\hat{ r}}_f\right|^2}{4
\hat{f}^2}\right)~.\label{Icpact31}
\end{eqnarray}
A plot of the universal function $\mathrm{sinc}^2(\alpha^2/4)$ is
given in Fig. \ref{focindet}. Note that all expressions pertaining
undulator radiation from a single electron are azimuthal symmetric
but do not admit factorization in the product of functions
separately depending on the $x$ and the $y$ coordinates. As we
have shown in \cite{OURS}, the absence of factorization leads to
an influence of the presence of the horizontal emittance on the
coherent properties of undulator radiation in the vertical
direction. As a result, even in the case of zero vertical
emittance one cannot have perfect coherence in the vertical
direction.

\begin{figure}
\begin{center}
\includegraphics*[width=110mm]{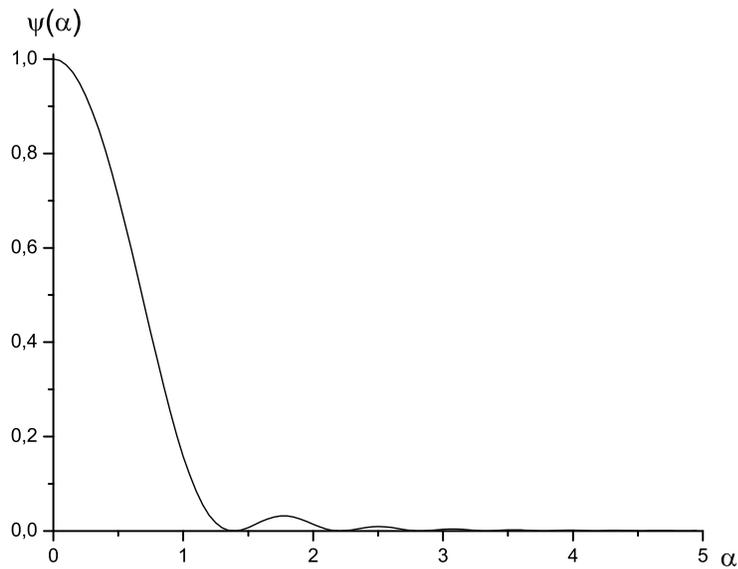}
\caption{\label{imindet} Universal function $\Psi(\alpha)$ used to
calculate (according to Eq. (\ref{cpact61})) the intensity
distribution on the image plane from a single electron at the
fundamental harmonic at perfect resonance.}
\end{center}
\end{figure}
For the image plane we obtain:

\begin{eqnarray}
{ \hat{E}_\bot}( \hat{z}_i,\vec{\hat{r}}_i) &=& -\frac{ i
\mathrm{m}}{2\pi} \exp{\left[ \frac{i \mathrm{m} | \vec{\hat{r}}_i
|^2}{2 \hat{z}_1} \right]}\exp{\left[ \frac{i \mathrm{m}^2 |
\vec{\hat{r}}_i |^2}{2 \hat{z}_1}\right]} \cr &&\times \int d
\vec{\hat{u}}~ \mathrm{sinc}\left(\frac{\left|\vec{\hat{
u}}\right|^2}{4}\right) \exp{\left[{i \mathrm{m}
\vec{\hat{r}}_i}\cdot\vec{\hat{u}} \right]}~.\cr &&
\label{cpact41}
\end{eqnarray}
As already seen in Eq. (190) of \cite{OURS} (Appendix C), the
Fourier transform in Eq. (\ref{cpact41}) can be calculated in
terms of a Fourier-Bessel transform:

\begin{eqnarray}
\int d \vec{\hat{u}}~ \mathrm{sinc}\left(\frac{\left|\vec{\hat{
u}}\right|^2}{4}\right) \exp{\left[{i \mathrm{m} }
{\vec{\hat{r}}_i}\cdot\vec{\hat{u}} \right]} &=& 2 \pi
\int_0^{\infty} d u~ u J_o\left( {\mathrm{m} |\vec{\hat{r}}_i}| u
\right) \mathrm{sinc}\left(\frac{u^2}{4}\right)  \cr & = &  2\pi
\left[\pi - 2\mathrm{Si}\left({\mathrm{m}^2 |\vec{\hat{r}}_i|^2}
\right)\right]~.\label{cpact51}
\end{eqnarray}
Eq. (\ref{cpact41}) can now be written as

\begin{eqnarray}
{ \hat{E}_\bot}( \hat{z}_i,\vec{\hat{r}}_i) &=& -{i}\mathrm{m}
\exp{\left[ \frac{i  \mathrm{m} |\vec{\hat{r}}_i |^2}{2 \hat{z}_1}
\right]} \exp{\left[ \frac{i \mathrm{m}^2 |\vec{\hat{r}}_i |^2}{
2\hat{z}_1}\right]}\cdot \left[\pi -
2\mathrm{Si}\left(\mathrm{m}^2 |\vec{\hat{r}}_i|^2
\right)\right]~. \label{cpact61a}
\end{eqnarray}
It is convenient to introduce the following universal function
normalized to unity:

\begin{eqnarray}
\Psi(\alpha) = \frac{1}{\pi^2}\left[\pi -
2\mathrm{Si}\left(\alpha^2 \right)\right]^2 \label{psiuni}
\end{eqnarray}
The relative intensity on the image plane is related to the
universal function $\Psi$ through the scaling factor $\mathrm{m}$:

\begin{eqnarray}
{\hat{I}}\left(|\vec{\hat{r}_i}|\right) &=& \Psi\left({\mathrm{m}
|\vec{\hat{r}}_i}| \right)~. \label{cpact61}
\end{eqnarray}
A plot of the universal function $\Psi(\alpha)$ is given in Fig.
\ref{imindet}.

Comparison of Eq. (\ref{cpact61a}) with Eq. (\ref{cpact5}) allows
one to conclude that the equivalent source for a single electron
moving on the $z$ axis is a source characterized by a plane
wavefront and an intensity distribution related to the universal
function $\Psi$, that can be written as

\begin{eqnarray}
{ \hat{E}_\bot}( 0,\vec{\hat{r}}) &=& -{i} \left[\pi -
2\mathrm{Si}\left(|\vec{\hat{r}}|^2 \right)\right]~.
\label{virsof}
\end{eqnarray}
By means of an inverse Fourier transformation it follows that

\begin{eqnarray}
\mathrm{\hat{F}}\left(0,\vec{\hat{u}}\right) = -{2 \pi i }
~\mathrm{sinc}\left(\frac{\left|\vec{\hat{u}}\right|^2}{4}\right)
~, \label{fieldpropback3bbis}
\end{eqnarray}
in agreement with Eq. (\ref{fieldpropback3b}). Comparison of Eq.
(\ref{virsof}) with a normalized version of Eq.
(\ref{fieldpropback3tris}) show that the phase of the field of the
virtual source is shifted of a quantity $-\pi/2$ with respect to
the spherical wave in the far zone. Such phase shift is the
analogous of the Guoy phase shift in laser physics. A single
electron produces a laser-like radiation beam that has a (virtual)
waste much larger than the radiation wavelength located in the
center of the undulator.

The intensity distribution from the most elementary undulator
source, i.e. the radiation from a single electron (or,
equivalently, from an electron beam with zero emittance) was just
described analytically. Such analytical description, Eq.
(\ref{cpact61}), can immediately be applied in situations of
practical relevance. In \cite{YAB1,YAB2} a characterization of the
vertical emittance in Spring-8 is reported. It is based on the
measurement of the X-Ray beam coherence length in the far zone.
The experiment was performed at the beamline BL29XU. Based on the
assumption of validity of the van Cittert-Zernike theorem, it was
found that the rms electron beam size at the undulator center
(corresponding to the minimal value of the beta function) was $s_y
\simeq 4.5 ~\mu$m, and that the coupling factor between horizontal
and vertical emittance was down to the value $\chi \simeq 0.12
\%$, which corresponds to an extremely small vertical emittance
$\epsilon_y = 3.6$ pm$\cdot$rad. A resolution limit of this method
was also discussed, based on numerical calculations of the
radiation size from a single electron $s_p \simeq 1.6 ~\mu$m at
$E_p = 14.41 $ keV for the $4.5$ m long undulator used in the
experiment. The resolution limit of the measurement of $s_y$ was
estimated to be about $1~ \mu$m.

Based on Eq. (\ref{cpact61}), we can determine the virtual source
size of undulator radiation from a single electron. Let us
consider the case when the single electron is emitting photons at
the fundamental harmonic with energy $E = 14.41$ keV. The angular
frequency of light oscillations is given, in this case, by $\omega
= 2.2 \cdot 10^{19}$ Hz. For an undulator length $L_w = 4.5$ m,
the normalization factor for the transverse size introduced in Eq.
(\ref{Cnorm}), $(L_w c/\omega)^{1/2}$, is about $8~\mu$m. From
Fig. \ref{imindet} obtain the dimensionless Half Width Half
Maximum (HWHM) radiation size from a single electron (i.e. the
HWHM width of the intensity distribution at the virtual source,
located at the center of the undulator). This HWHM dimensionless
value is about $0.7$. It follows that the HWHM value of the
radiation spot size from a single electron is about $0.7 \cdot (c
L_w/\omega)^{1/2} \simeq 6~ \mu$m. Therefore, the rms value $s_p
\simeq 1.6~ \mu$m  in \cite{YAB1,YAB2}, which was calculated
numerically \cite{SPECT}, is an underestimation of the correct
value.

Note that the HWHM radiation spot size from a \textit{single
electron} is larger than the \textit{rms electron beam size} $s_y
\simeq 4.5 ~ \mu$m found by means of coherence measurements. One
concludes that the uncertainty due to finite resolution is larger
than the measured electron beam size. This suggests that the
method used in \cite{YAB1,YAB2} may be inconsistent. Such
inconsistency may be traced to the fact that authors of
\cite{YAB1,YAB2} assume the validity of the van Cittert-Zernike
theorem in the vertical direction. If one assumes their result of
a vertical emittance ${\epsilon}_y \simeq 0.3 \lambda/(2 \pi)$, it
follows \textit{a posteriori} that the van Cittert-Zernike theorem
could not have been applied in first instance (in this experiment
the value of the beta function was ${\beta} \simeq L_w$). Hence
the inconsistency of the method follows. Analysis of experimental
results should have been based, instead, on the study of
transverse coherence for non-homogeneous undulator sources in free
space made in \cite{OURS}.

\subsection{\label{subsub:incl3} Effect of aperture size}

The effects due to the presence of the pupil can be included in
the treatment by means of a normalized version of Eq. (\ref{EFP})
and Eq. (\ref{EIP}) on the focal plane

\begin{eqnarray}
\hat{E}_P(\hat{z}_f,\vec{\hat{r}}_f) &=& \exp\left[\frac{i
|\vec{\hat{r}}_f|^2}{2  \hat{f}} \right] \cr &&\times\int d
\vec{\hat{u}}~ \hat{\mathcal{P}}\left(\frac{ \vec{\hat{r}}_f}{
\hat{f}}-\vec{\hat{u}}\right)\cdot \exp{\left[- \frac{i  \hat{f} |
\vec{\hat{u}} |^2}{2 } \right]}\hat{E}_\bot\left(\hat{z}_f,
\hat{f} \vec{\hat{u}}\right) ~,\label{EFPn}
\end{eqnarray}
and on the image plane

\begin{eqnarray}
\hat{E}_P(\hat{z}_i,\vec{\hat{r}}_i) &=& \exp\left[\frac{i
\mathrm{m} |\vec{\hat{r}}_i|^2}{2 \hat{z}_1} \right] \cr
&&\times\int d \vec{\hat{u}}~
\hat{\mathcal{P}}\left(\frac{\mathrm{m}
\vec{\hat{r}}_i}{\hat{z}_1}-\vec{\hat{u}}\right)\cdot\exp{\left[-
\frac{i \hat{z}_1 | \vec{\hat{u}} |^2}{2  \mathrm{m}} \right]}
\hat{E}_\bot \left(z_i,\frac{ \hat{z}_1 \vec{\hat{u}}}{
\mathrm{m}}\right) ~, \label{EIPn}
\end{eqnarray}
where

\begin{eqnarray}
\hat{\mathcal{P}}(\vec{\hat{u}}) &=& \int d\vec{\hat{r'}}
P(\vec{\hat{r'}}) \exp{\left[-i \vec{\hat{r'}}\cdot
\vec{\hat{u}}\right]}~. \label{PFThat}
\end{eqnarray}
Eq. (\ref{EIPn}) is valid independently of the position of the
lens. However, as explained in Section \ref{sub:imcoh} and in
Section \ref{sub:impc} we will limit ourselves to the case when
the lens is in the far zone. From Eq. (\ref{cpact61}) we see that
the characteristic size of the source is of order unity, because
$\Psi$ is a universal function. As a result, the far field zone
for a single particle is defined by the condition

\begin{eqnarray}
\hat{z}_1 \gg 1  ~.\label{farfnorm}
\end{eqnarray}
By substitution of  Eq. (\ref{fieldpropback3tris}) in Eq.
(\ref{fieldpropexp}), followed by use of the lens equation Eq.
(\ref{lensagain}) and normalization, one obtains an expression for
the field valid in the case the lens is placed in the far zone.
Such expression is equivalent, in that limit, to Eq. (\ref{EIPn}).
If the pupil function is not set to unity, we have

\begin{eqnarray}
{ \hat{E}}_P( \hat{z}_i,\vec{ \hat{r}_i}) &=& -\frac{
\mathrm{m}}{4 \pi^2  \hat{z}_1^2}  \exp{\left[ \frac{i \mathrm{m}
| \vec{\hat{r}}_i |^2 }{2  \hat{z}_1} \right]}\cr && \times\int d
\vec{ \hat{r}'}~ { \hat{F}} \left(0,-\frac{\vec{
\hat{r}'}}{\hat{z}_1}\right) P\left(\vec{\hat{r}'}\right)
\exp{\left[- \frac{i  \mathrm{m}
(\vec{\hat{r}}_i\cdot\vec{\hat{r}'})}{ \hat{z}_1}\right]}~.\cr &&
\label{fieldpropexpimaff}
\end{eqnarray}
Eq. (\ref{fieldpropexpimaff}) can also be obtained by substitution
of Eq. (\ref{cpact21}) in Eq. (\ref{EIPn}) followed by application
of the convolution theorem.

In the far field case Eq. (\ref{EIPn}) (or Eq.
(\ref{fieldpropexpimaff})) can be more easily used to calculate
the effects of the pupil on the intensity. A natural example to
study is the case of a lens with azimuthal symmetry and no
aberrations. Consider a pupil of radius $a$. After introduction of
the normalized pupil radius:

\begin{equation}
\hat{a} = a \sqrt{\frac{\omega}{L_w c}} ~\label{arad}
\end{equation}
we set the pupil function:

\begin{eqnarray}
P(\vec{\hat{r}}) &=& \left \{
\begin{tabular}{c}
1 ~~~~~~~if~$|\vec{\hat{r}}|<\hat{a}$\\
0 ~~~~otherwise~.
\end{tabular}
\right. \label{Pcirc}
\end{eqnarray}
Using the Fourier-Bessel transform one obtains

\begin{eqnarray}
\hat{\mathcal{P}}(\vec{\hat{u}}) &=& \frac{2 \pi
\hat{a}}{|\vec{\hat{u}}|} J_1\left(\hat{a}
|\vec{\hat{u}}|\right)~.\label{FBP}
\end{eqnarray}
Substitution of Eq. (\ref{cpact61a}) in Eq. (\ref{EIPn}) and use
of the far zone assumption leads to

\begin{eqnarray}
\hat{E}_P(\hat{z}_i,\vec{\hat{r}}_i) &=&-2 \pi i \mathrm{m}
\hat{a} \exp\left[\frac{i \mathrm{m} |\vec{\hat{r}}_i|^2}{2
\hat{z}_1} \right] \cr &&\times\int d \vec{\hat{u}}~
\left|\frac{\mathrm{m}
\vec{\hat{r}}_i}{\hat{z}_1}-\vec{\hat{u}}\right|^{-1}
J_1\left(\hat{a}\left|\frac{\mathrm{m}
\vec{\hat{r}}_i}{\hat{z}_1}-\vec{\hat{u}}\right|\right)\left[\pi -
2\mathrm{Si}\left(\hat{z}_1^2 |\vec{\hat{u}}|^2 \right)\right]
~.\cr && \label{EIPn2}
\end{eqnarray}
Eq. (\ref{EIPn2}) is, essentially, the convolution product of the
Fourier transform of two known functions with circular symmetry.
Therefore, recalling Eq. (\ref{cpact51}) and Eq. (\ref{Pcirc}),
Eq. (\ref{EIPn2}) can be written in terms of the following
Fourier-Bessel transform:

\begin{eqnarray}
\hat{E}_P(\hat{z}_i,\vec{\hat{r}}_i) &=&-{i}
 \exp\left[\frac{i \mathrm{m} |\vec{\hat{r}}_i|^2}{2
\hat{z}_1} \right] \int_0^{\hat{a}/\hat{z}_1} d \hat{u}~
\hat{u}J_0\left({\mathrm{m} |\vec{\hat{r}}_i|\hat{u}}\right)
\mathrm{sinc}\left(\frac{\hat{u}^2}{4}\right)  ~. \label{EIPn3}
\end{eqnarray}
\begin{figure}
\begin{center}
\includegraphics*[width=140mm]{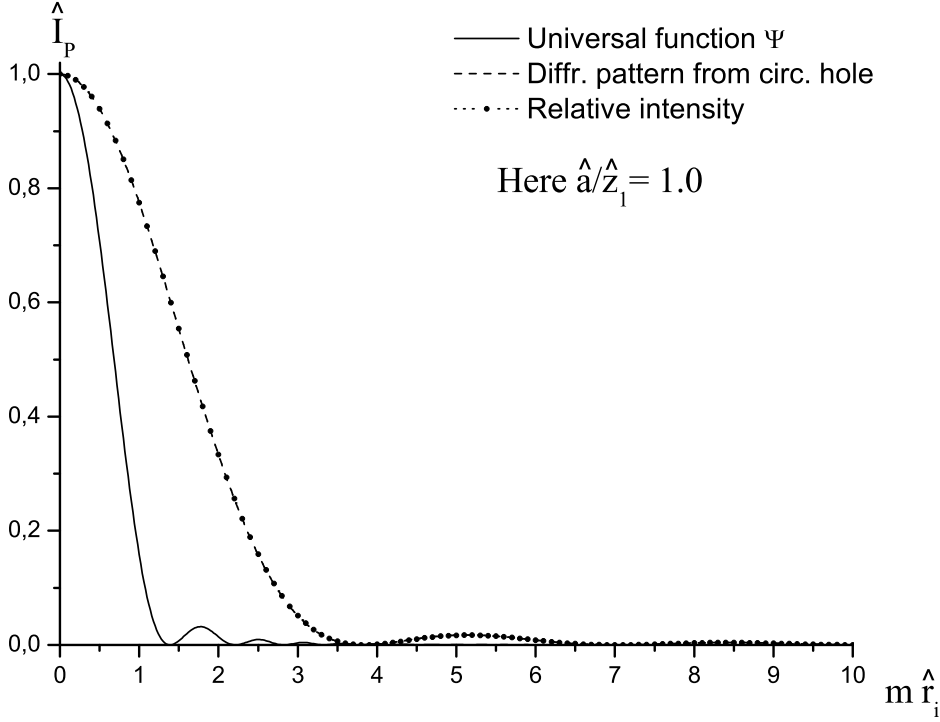}
\caption{\label{IP1}  Comparison between the relative intensity
for a single electron at the image plane $\hat{I}_P$, Eq.
(\ref{EIPn4}), the universal function $\Psi$, Eq. (\ref{cpact61}),
and the (Airy) diffraction pattern from a circular hole, Eq.
(\ref{EIPn5}), as a function of $\mathrm{m} \hat{r}_i$. Here
$\hat{a}/\hat{z}_1 = 1$.}
\end{center}
\end{figure}
\begin{figure}
\begin{center}
\includegraphics*[width=140mm]{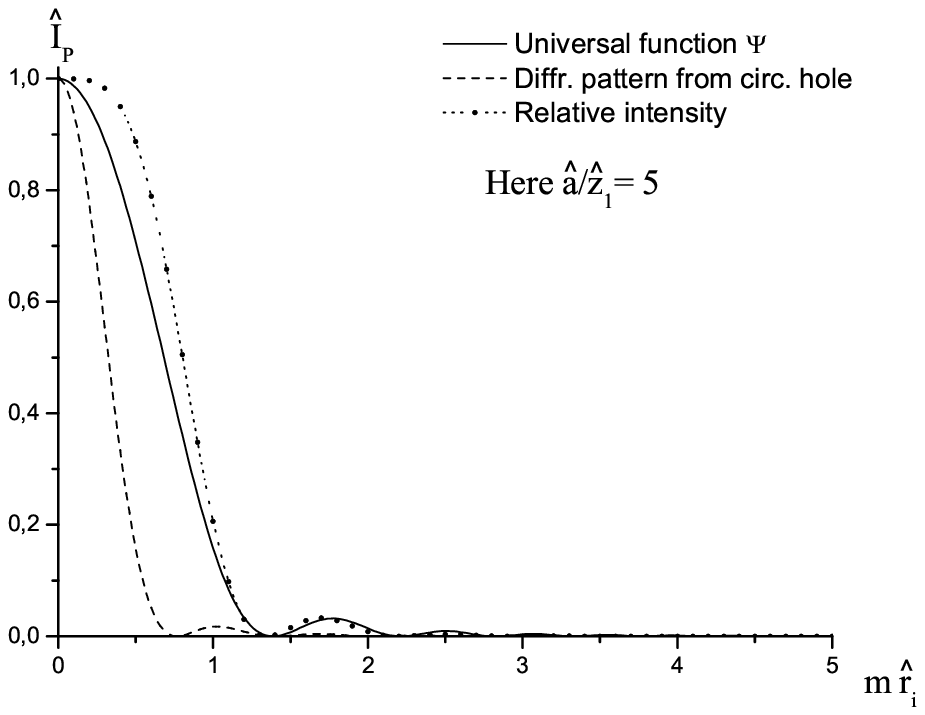}
\caption{\label{IP5} Comparison between the relative intensity for
a single electron at the image plane $\hat{I}_P$, Eq.
(\ref{EIPn4}), the universal function $\Psi$, Eq. (\ref{cpact61}),
and the (Airy) diffraction pattern from a circular hole, Eq.
(\ref{EIPn5}), as a function of $\mathrm{m} \hat{r}_i$. Here
$\hat{a}/\hat{z}_1 = 5$.}
\end{center}
\end{figure}
\begin{figure}
\begin{center}
\includegraphics*[width=140mm]{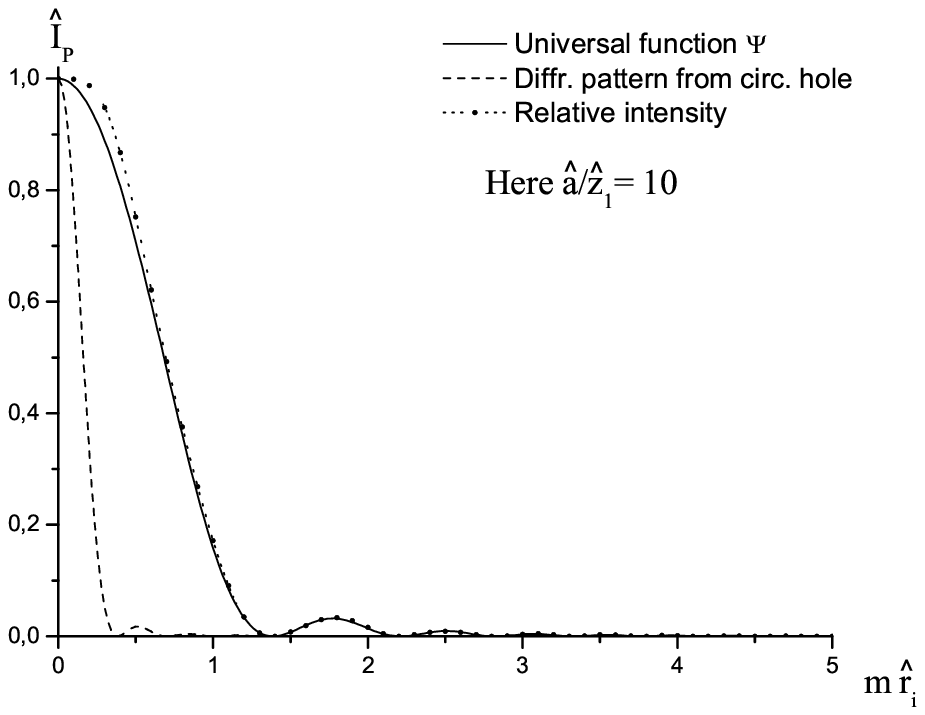}
\caption{\label{IP10}  Comparison between the relative intensity
for a single electron at the image plane $\hat{I}_P$, Eq.
(\ref{EIPn4}), the universal function $\Psi$, Eq. (\ref{cpact61}),
and the (Airy) diffraction pattern from a circular hole, Eq.
(\ref{EIPn5}), as a function of $\mathrm{m} \hat{r}_i$. Here
$\hat{a}/\hat{z}_1 = 10$.}
\end{center}
\end{figure}

Eq. (\ref{EIPn3}) can also be directly obtained using the
Fourier-Bessel integration formula and Eq.
(\ref{fieldpropexpimaff}). Eq. (\ref{EIPn3}) corresponds to a
relative intensity

\begin{eqnarray}
{\hat{I}_P}\left(|\vec{\hat{r}_i}|\right) &=&\frac{1}{4}
\left[{\mathrm{Si}\left(\frac{\hat{a}^2}{4
\hat{z}_1^2}\right)}\right]^{-2} \left| \int_0^{\hat{a}/\hat{z}_1}
d \hat{u}~ \hat{u}J_0\left({\mathrm{m}
|\vec{\hat{r}}_i|\hat{u}}\right)
\mathrm{sinc}\left(\frac{\hat{u}^2}{4}\right) \right|^2 ~.
\label{EIPn4}
\end{eqnarray}
In the limit  $\hat{a}/\hat{z}_1\gg1$ Eq. (\ref{EIPn3}) gives back
Eq. (\ref{cpact61a}) without the second phase factor, and Eq.
(\ref{EIPn4}) gives back Eq. (\ref{cpact61}) as it should be.

When $\hat{a}/\hat{z}_1\ll1$ the $\mathrm{sinc}(\cdot)$ drops out
of the integral in Eq. (\ref{EIPn4}) giving the diffraction
pattern from a circular hole:

\begin{eqnarray}
{\hat{I}_P}\left(|\vec{\hat{r}_i}|\right) &=& \frac{4
\hat{z}_1^4}{\hat{a}^4}\left| \int_0^{\hat{a}/\hat{z}_1} d
\hat{u}~ \hat{u}J_0\left({\mathrm{m}
|\vec{\hat{r}}_i|\hat{u}}\right) \right|^2 = \frac{4
\hat{z}_1^2}{\hat{a}^2 \mathrm{m}^2 |\vec{\hat{r}}_i|^2 }
J_1^2\left(\frac{\hat{a}\mathrm{m}\hat{r}_i}{\hat{z}_1}\right)~.
\label{EIPn5}
\end{eqnarray}
Further, analysis of Fig. \ref{IP1} actually shows that Eq.
(\ref{EIPn5}) retains its validity also for $\hat{a}/\hat{z}_1
\lesssim 1$.

Comparisons between the relative intensity $\hat{I}_P$, Eq.
(\ref{EIPn4}), the universal function $\Psi$, Eq. (\ref{cpact61}),
and the diffraction pattern from a circular hole, Eq.
(\ref{EIPn5}), are plotted as a function of $\mathrm{m} \hat{r}_i$
in Fig. (\ref{IP1}) for $\hat{a}/\hat{z}_1 = 1$, in Fig.
(\ref{IP5}) for $\hat{a}/\hat{z}_1 = 5$ and in Fig. (\ref{IP10})
for $\hat{a}/\hat{z}_1 = 10$.

\section{\label{sec:emit} Image formation with partially coherent undulator source}

\subsection{\label{subsub:negl4} Coherence properties of undulator source in the presence of electron beam emittance}

In the last Section we dealt with the image formation problem in
the case of a filament beam, i.e. when the electron beam emittance
is zero. In this Section we will generalize the previous results
to the case when the electron beam has finite emittance. In this
situation, methods from Statistical Optics must be applied in
order to solve the image formation problem. As discussed before,
the cross-spectral density of an undulator source must first be
calculated at the lens position and subsequently propagated
through the lens and the forthcoming optical beamline up to the
experimental plane.

In \cite{OURS} we proposed a method, based on Eq.
(\ref{undunormfinult}), to calculate the cross-spectral density
from undulator sources at any position in free space after the
undulator. Let us follow \cite{OURS} and use Eq.
(\ref{undunormfinult}) to calculate the cross-spectral density.
The cross-spectral density $G$ is given, in dimensional units and
as a function of dimensional variables, by Eq. (\ref{coore}).
Since the field in Eq. (\ref{undunormfinult}) is given in
normalized units and as a function of normalized variables
$\hat{z}_o$, $\vec{\hat{\theta}}_{x,y}$ and $\hat{C}$, it is
convenient to introduce a version of $G$ defined by means of the
field in normalized units

\begin{eqnarray}
\hat{G}(\hat{z}_o,\vec{\hat{\theta}}_1,\vec{\hat{\theta}}_2,\hat{C})
 &=& \Bigg \langle \hat{E}_{s\bot}\left(\hat{C},\hat{z}_o,\vec{\hat{\theta}}_1-
\frac{\vec{\hat{l}}}{\hat{z}_o}- \vec{\hat{\eta}}\right)\cr
&&\times
\hat{E}_{s\bot}^*\left(\hat{C},\hat{z}_o,\vec{\hat{\theta}}_2-
\frac{\vec{\hat{l}}}{\hat{z}_o}- \vec{\hat{\eta}}\right)
\Bigg\rangle_{\vec{\hat{\eta}},\vec{\hat{l}}} ~.\label{Gnormdef1}
\end{eqnarray}
Transformation of  $G$ in Eq. (\ref{coore}) to $\hat{G}$ (and
viceversa) can be  performed shifting from dimensional to
normalized variables and multiplying $G$ by an inessential factor

\begin{equation}
\hat{G} =  \left(\frac{c^2  \gamma}{K \omega e  A_{JJ}}\right)^2 G
~.\label{inetransf}
\end{equation}
Substituting Eq. (\ref{Esum}) in Eq. (\ref{Gnormdef1}) we obtain

\begin{eqnarray}
\hat{G}(\hat{z}_o,\vec{\hat{\theta}}_1,\vec{\hat{\theta}}_2,\hat{C})
 &=& \Bigg \langle S\left[\hat{C},\hat{z}_o,\left(\vec{\hat{\theta}}_1-
\frac{\vec{\hat{l}}}{\hat{z}_o}- \vec{\hat{\eta}}\right)^2\right]
S^*\left[\hat{C},\hat{z}_o,\left(\vec{\hat{\theta}}_2-
\frac{\vec{\hat{l}}}{\hat{z}_o}- \vec{\hat{\eta}}\right)^2\right]
\cr && \times
\exp\left\{i\left[\left(\vec{\hat{\theta}}_{1}-\frac{\vec{\hat{l}}}{\hat{z}_o}\right)^2
-
\left(\vec{\hat{\theta}}_{2}-\frac{\vec{\hat{l}}}{\hat{z}_o}\right)^2
\right]
 \frac{\hat{z}_o}{2}  \right\}
\Bigg\rangle_{\vec{\hat{\eta}},\vec{\hat{l}}}~. \label{Gzlarge}
\end{eqnarray}
Expanding the exponent in the exponential factor in the right hand
side of Eq. (\ref{Gzlarge}), one can see that terms in
$\hat{l}_{x,y}^2$ cancel out. Terms in $\hat{\theta}_{x,y}^2$
contribute for a common factor, and only linear terms in
$\hat{l}_{x,y}$ remain inside the ensemble average sign.
Substitution of the ensemble average with integration over the
beam distribution function leads to

\begin{eqnarray}
\hat{G}(\hat{z}_o,\vec{\hat{\theta}}_1,\vec{\hat{\theta}}_2,\hat{C})
&=&
\exp{\left[i\left(\vec{\hat{\theta}}_1^2-\vec{\hat{\theta}}_2^2\right)\frac{\hat{z}_o}
{2}\right]}
 \int d \vec{\hat{\eta}} d
\vec{\hat{l}}~
F_{\vec{\hat{\eta}},\vec{\hat{l}}}\left(\vec{\hat{\eta}},\vec{\hat{l}}\right)
\cr && \times \exp\left[i
(\vec{\hat{\theta}}_{2}-\vec{\hat{\theta}_{1}})\cdot \vec{
\hat{l}} \right]
S\left[\hat{C},\hat{z}_o,\left(\vec{\hat{\theta}}_1-
\frac{\vec{\hat{l}}}{\hat{z}_o}-
\vec{\hat{\eta}}\right)^2\right]\cr &&\times
S^*\left[\hat{C},\hat{z}_o,\left(\vec{\hat{\theta}}_2-
\frac{\vec{\hat{l}}}{\hat{z}_o}- \vec{\hat{\eta}}\right)^2\right]
~.~ \label{Gzlarge2}
\end{eqnarray}
Here integrals $d \vec{\hat{\eta}}$ and in $d \vec{\hat{l}}$ are
to be intended as integrals over the entire plane spanned by the
$\vec{\hat{\eta}}$ and $\vec{\hat{l}}$ vectors. Eq.
(\ref{Gzlarge2}) is very general and can be used as a starting
point for computer simulations.

We assume that the distribution in the horizontal and vertical
planes are not correlated, so that
$F_{\vec{\hat{\eta}},\vec{\hat{l}}} = F_{\hat{\eta}_x,\hat{l}_x}
F_{\hat{\eta}_y,\hat{l}_y}$. If the transverse phase-space is
specified at the virtual-source position $\hat{z}_o = 0$
corresponding to the minimal values of the beta functions, we can
write $F_{\hat{\eta}_x,\hat{l}_x}= F_{\hat{\eta}_x} F_{\hat{l}_x}$
and $F_{\hat{\eta}_y,\hat{l}_y}= F_{\hat{\eta}_y} F_{\hat{l}_y}$
with

\begin{eqnarray}
F_{\hat{\eta}_x}(\hat{\eta}_x) = \frac{1}{\sqrt{2\pi \hat{D}_x}}
\exp{\left(-\frac{\hat{\eta}_x^2}{2 \hat{D}_x}\right)}~,&&\cr
F_{\hat{\eta}_y}(\hat{\eta}_y)  = \frac{1}{\sqrt{2\pi \hat{D}_y}}
\exp{\left(-\frac{\hat{\eta}_y^2}{2 \hat{D}_y}\right)}~,&&\cr
F_{\hat{l}_x}(\hat{l}_x) =\frac{1}{\sqrt{2\pi\hat{N}_x} }
\exp{\left(-\frac{\hat{l}_x^2}{2 \hat{N}_x}\right)}~, && \cr
F_{\hat{l}_y}(\hat{l}_y)=\frac{1}{\sqrt{2\pi\hat{N}_y} }
\exp{\left(-\frac{\hat{l}_y^2}{2 \hat{N}_y}\right)}~.\cr &&
\label{distr}
\end{eqnarray}
From Eq. (\ref{Cnorm}) and Eq. (\ref{Cnorm2}) it is possible to
see that

\begin{equation}
\hat{D}_{x,y} = \sigma_{x',y'}^2 {\frac{\omega L_w}{c}}\label{sig}
\end{equation}
\begin{equation}
\hat{N}_{x,y} = \sigma^2_{x,y} \frac{\omega}{c L_w}\label{enne}
\end{equation}
where $\sigma_{x,y}$ and $\sigma_{x',y'}$ are the rms transverse
bunch dimension and angular spread. Parameters $\hat{N}_{x,y}$
will be indicated as the beam diffraction parameters and are
analogous to Fresnel numbers. They correspond to the normalized
square of the electron beam sizes. $\hat{D}_{x,y}$ represent the
normalized square of the electron beam divergences instead. It is
also convenient to introduce the square of the apparent angular
size of the electron beam at the observer point position
$\hat{z}_o$, that is

\begin{equation}
\hat{A}_{x,y} = \frac{\hat{N}_{x,y}}{\hat{z}_o^2} ~.\label{adef}
\end{equation}
Substitution of relations (\ref{distr}) in Eq. (\ref{Gzlarge2})
yields\footnote{In Eq. (\ref{Gzlarge3}), for notational simplicity
we substituted the proper notation
$\hat{G}(\hat{z}_o,\vec{\hat{\theta}}_1,\vec{\hat{\theta}}_2,
\hat{C})$ with the simplified dependence
$\hat{G}(\hat{z}_o,\vec{\hat{\theta}}_1,\vec{\hat{\theta}}_2)$.
This is justified because we will be treating the case $\hat{C}=0$
only. Consistently, also $S {[\hat{z}_o,
(\vec{\hat{\theta}}-\vec{\hat{l}}/ {\hat{z}_o}-\vec{\hat{\eta}})^2
]}$ is to be understood as a shortcut notation for
$S[\hat{C},\hat{z}_o,(\vec{\hat{\theta}}-
\vec{\hat{l}}/\hat{z}_o-\vec{\hat{\eta}})^2]$ calculated at
$\hat{C}=0$.} at perfect resonance ($\hat{C} = 0$):

\begin{eqnarray}
\hat{G}(\hat{z}_o,\vec{\hat{\theta}}_1,\vec{\hat{\theta}}_2)
&=&\frac{\exp{\left[i\left(\vec{\hat{\theta}}_1^2-
\vec{\hat{\theta}}_2^2\right){\hat{z}_o}/{2}\right]}}{4\pi^2
\sqrt{\hat{D}_x\hat{D}_y\hat{N}_x\hat{N}_y}}
\int_{-\infty}^{\infty} d \hat{\eta}_{x }
\exp{\left(-\frac{\hat{\eta}_x^2}{2 \hat{D}_x}\right)}\cr &&\times
\int_{-\infty}^{\infty} d \hat{\eta}_{y }
\exp{\left(-\frac{\hat{\eta}_y^2}{2 \hat{D}_y}\right)}
\int_{-\infty}^{\infty} d \hat{l}_{x }
\exp{\left(-\frac{\hat{l}_x^2}{2 \hat{N}_x}\right)}\cr &&\times
\int_{-\infty}^{\infty} d \hat{l}_{y
}\exp{\left(-\frac{\hat{l}_y^2}{2 \hat{N}_y}\right)} \exp\left[i
(\vec{\hat{\theta}}_{2}-\vec{\hat{\theta}_{1}})\cdot \vec{
\hat{l}} \right] \cr && \times
S\left[\hat{z}_o,\left(\vec{\hat{\theta}}_1-
\frac{\vec{\hat{l}}}{\hat{z}_o}-
\vec{\hat{\eta}}\right)^2\right]S^*\left[\hat{z}_o,\left(\vec{\hat{\theta}}_2-
\frac{\vec{\hat{l}}}{\hat{z}_o}- \vec{\hat{\eta}}\right)^2\right]
 ~.\label{Gzlarge3}
\end{eqnarray}
Let us now introduce\footnote{Note that the definition of $\Delta
\hat{\theta}_{x}$ and $\Delta \hat{\theta}_{y}$ differ for a
factor 2 and a sign with respect to notations in
\cite{GOOD,GOOD2}.}

\begin{equation}
\Delta \hat{\theta}_{x} =
\frac{\hat{\theta}_{x1}-\hat{\theta}_{x2}}{2}~,~~ \bar{\theta}_x =
\frac{\hat{\theta}_{x1}+\hat{\theta}_{x2}}{2} \label{deltabarx}
\end{equation}
and

\begin{equation}
\Delta \hat{\theta}_{y} =
\frac{\hat{\theta}_{y1}-\hat{\theta}_{y2}}{2}~,~~ \bar{\theta}_y =
\frac{\hat{\theta}_{y1}+\hat{\theta}_{y2}}{2}~. \label{deltabary}
\end{equation}
With this variables redefinition we obtain

\begin{equation}
\hat{G}=\hat{G}(\hat{z}_o,\bar{\theta}_x,\bar{\theta}_y,\Delta
\hat{\theta}_x,\Delta \hat{\theta}_y) \label{Gshort}
\end{equation}
and, explicitly,

\begin{eqnarray}
\hat{G} &=&\frac{\exp{\left[i 2 \hat{z}_o\left(\bar{\theta}_x
\Delta \hat{\theta}_x+\bar{\theta}_y \Delta \hat{\theta}_y
\right)\right]}}{4\pi^2
\sqrt{\hat{D}_x\hat{D}_y\hat{N}_x\hat{N}_y}}
\int_{-\infty}^{\infty} d \hat{\eta}_{x }
\exp{\left(-\frac{\hat{\eta}_x^2}{2 \hat{D}_x}\right)}\cr &&\times
\int_{-\infty}^{\infty} d \hat{\eta}_{y }
\exp{\left(-\frac{\hat{\eta}_y^2}{2 \hat{D}_y}\right)}
\int_{-\infty}^{\infty} d \hat{l}_{x }
\exp{\left(-\frac{\hat{l}_x^2}{2 \hat{N}_x}\right)}\cr &&\times
\int_{-\infty}^{\infty} d \hat{l}_{y
}\exp{\left(-\frac{\hat{l}_y^2}{2 \hat{N}_y}\right)} \exp\left[-2
i \left(\Delta \hat{\theta}_x \hat{l}_x+\Delta \hat{\theta}_y
\hat{l}_y\right) \right] \cr && \times
S~\left[\hat{z}_o,\left({\bar{\theta}}_x+\Delta {\hat{\theta}}_x-
\frac{{\hat{l}}_x}{\hat{z}_o}-
{\hat{\eta}}_x\right)^2+\left({\bar{\theta}}_y+\Delta
{\hat{\theta}}_y- \frac{{\hat{l}}_y}{\hat{z}_o}-
{\hat{\eta}}_y\right)^2\right]\cr && \times
S^*\left[\hat{z}_o,\left({\bar{\theta}}_x-\Delta {\hat{\theta}}_x-
\frac{{\hat{l}}_x}{\hat{z}_o}-
{\hat{\eta}}_x\right)^2+\left({\bar{\theta}}_y-\Delta
{\hat{\theta}}_y- \frac{{\hat{l}}_y}{\hat{z}_o}-
{\hat{\eta}}_y\right)^2\right]
 ~.\label{Gzlarge5}
\end{eqnarray}
A double change of variables $\hat{\eta}_{x,y} \longrightarrow
\hat{\eta}_{x,y} + \bar{\theta}_{x,y}$ followed by the
substitution $\hat{l}_{x,y}/\hat{z}_o \longrightarrow
\hat{\phi}_{x,y} - \hat{\eta}_{x,y}$ and by analytical calculation
of the integrals in $d \hat{\eta}_{x,y}$ leads to

\begin{eqnarray}
\hat{G} &=&\frac{1}{4 \pi^2\sqrt{\hat{A}_x \hat{D}_x \hat{A}_y
\hat{D}_y }}\cr && \times {\exp{\left[i 2
\bar{\theta}_x\hat{z}_o\Delta \hat{\theta}_x \right]}}
\exp{\left[-\frac {\bar{\theta}_x^2 + 4 \hat{A}_x \hat{z}_o^2
\Delta \hat{\theta}_x^2 \hat{D}_x + 4 i \hat{A}_x \bar{\theta}_x
\hat{z}_o \Delta \hat{\theta}_x }{2(\hat{A}_x+\hat{D}_x)}\right]}
\cr&&\times {\exp{\left[i 2\bar{\theta}_y\hat{z}_o\Delta
\hat{\theta}_y \right]}}\exp{\left[ - \frac {\bar{\theta}_y^2 + 4
\hat{A}_y \hat{z}_o^2 \Delta \hat{\theta}_y^2 \hat{D}_y + 4 i
\hat{A}_y \bar{\theta}_y \hat{z}_o \Delta \hat{\theta}_y
}{2(\hat{A}_y+\hat{D}_y)}\right] } \cr&& \times
\int_{-\infty}^{\infty} d \hat{\phi}_x \int_{-\infty}^{\infty} d
\hat{\phi}_y
\exp{\left[-\frac{\hat{\phi}_x^2+2\hat{\phi}_x\left(\bar{\theta}_x+2i
\hat{A}_x \hat{z}_o \Delta \hat{\theta}_x \right)}{2
(\hat{A}_x+\hat{D}_x)}\right]} \cr && \times
\exp{\left[-\frac{\hat{\phi}_y^2+2\hat{\phi}_y\left(\bar{\theta}_y+2i
\hat{A}_y \hat{z}_o \Delta \hat{\theta}_y \right)}{2
(\hat{A}_y+\hat{D}_y)}\right]}\cr && \times
S^*{\left[\hat{z}_o,(\hat{\phi}_x-\Delta
\hat{\theta}_x)^2+(\hat{\phi}_y-\Delta
\hat{\theta}_y)^2\right]}\cr &&\times
S~{\left[\hat{z}_o,(\hat{\phi}_x+\Delta
\hat{\theta}_x)^2+(\hat{\phi}_y+\Delta
\hat{\theta}_y)^2\right]}~.\label{G2D}
\end{eqnarray}
Eq. (\ref{G2D}) is a valid expression for the cross-spectral
density in free space \textit{after} the undulator device (i.e.
for $\hat{z}_o > 1/2$) and can be used together with equations
from (\ref{crspecprop3}) to (\ref{propimf2}).

Let us now introduce the dimensionless version of Eq. (\ref{trG})
with the help of $\vec{\bar{r}} = \hat{z}_o \vec{\bar{\theta}}$
and $\Delta \vec{\hat{r}} = \hat{z}_o \Delta \vec{\hat{\theta}}$
\footnote{A short digression about Eq. (\ref{trGbis}) is due here.
As the reader may have noticed, $\hat{\mathcal{G}}$ coincides with
the Fourier transform, done with respect to $\vec{\bar{r'}}$, of
the Wigner distribution $\hat{\Phi}(\hat{z},\vec{\bar{r}},
\vec{\bar{u}}) = \int d \Delta\vec{ \hat{r}'}~ \hat{G}(
\hat{z},\vec{ \bar{r}}, \Delta\vec{\hat {r}'}) \exp [2i (\vec{
\bar{u}}\cdot\Delta\vec{ \hat{r}'} )]$. The knowledge of the
Wigner distribution is mathematically equivalent to the knowledge
of $\hat{\mathcal{G}}$ or $\hat{G}$. A formalism based on the
Wigner distribution may be thus developed, which is mathematically
equivalent to the one developed here. In the case of
quasi-homogeneous sources, the Wigner distribution amounts to Eq.
(\ref{phsp}), that is the phase space distribution.
Interpretations of such a function as a sort of generalized phase
space distribution in more generic cases for non-homogeneous
sources have been proposed. However, there is no practical
advantage in considering such an approach in our case. Moreover,
the Wigner distribution is a quantity that cannot be directly
measured. Therefore, we prefer to use a formalism based on the
cross-spectral density which is a physically measurable quantity.
The cross-spectral density may be directly measured by means of
Young's double pinhole interferometer, whereas the Wigner function
is a mathematical transformation of the cross-spectral density.}:

\begin{eqnarray}
\hat{\mathcal{G}}\left(\hat{z},\vec{\bar{u}}, \Delta
\vec{\hat{u}}\right) &=& \int d \vec{ \bar{r}'}~d \Delta\vec{
\hat{r}'}~ \hat{G}\left( \hat{z},\vec{ \bar{r}'}, \Delta\vec{\hat
{r}'}\right)  \exp \left[2i \left(\vec{ \bar{u}}\cdot\Delta\vec{
\hat{r}'} +\Delta\vec{\hat{u}}\cdot\vec{ \bar{r'}}\right)\right]~.
\label{trGbis}
\end{eqnarray}
Its inverse is given by

\begin{eqnarray}
\hat{G}\left( \hat{z},\vec{ \bar{r}}, \Delta\vec{\hat {r}}\right)
&=& \frac{1}{(2\pi)^4} \int d \vec{ \bar{u}'}~d \Delta\vec{
\hat{u}'}\cr && \times
\hat{\mathcal{G}}\left(\hat{z},\vec{\bar{u}'}, \Delta
\vec{\hat{u}'}\right)  \exp \left[-2i \left(\vec{
\bar{u}'}\cdot\Delta\vec{ \hat{r}} +\Delta\vec{\hat{u}'}\cdot\vec{
\bar{r}}\right)\right]~. \label{trGtris}
\end{eqnarray}
Similarly as before, we consider coordinates $ \vec{\bar{\theta}}
= \vec{\bar{r}}/\hat{z}_o$ and $\Delta \vec{\hat{\theta}} =
\Delta\vec{\hat{r}}/\hat{z}_o$ in the limit  $\hat{z}_o
\longrightarrow \infty$ but for finite ratios $
\vec{\bar{\theta}}$ and $\Delta \vec{\hat{\theta}}$.  The
dimensionless version of Eq. (\ref{maintrick}) then reads:

\begin{eqnarray}
\hat{{G}}\left(\hat{z}_o,\vec{\bar{\theta}}, \Delta
\vec{\hat{\theta}}\right) &=& \frac{1}{4\pi^2 \hat{z}_o^2}
\exp\left[2{i \hat{z}_o } \vec{
\bar{\theta}}\cdot\Delta\vec{\hat{\theta}}\right]\hat{\mathcal{G}}\left(0,-\vec{\bar{\theta}},
-\Delta \vec{\hat{\theta}}\right)~. \label{maintrick2}
\end{eqnarray}
This result will be widely used in what follows. Moreover an
analogous of Eq. (\ref{factprop}) is:

\begin{eqnarray}
\hat{\mathcal{G}} \left(\hat{z},\vec{\bar{\theta}}, \Delta
\vec{\hat{\theta}}\right) &=&
\hat{\mathcal{G}}\left(0,\vec{\bar{\theta}}, \Delta
\vec{\hat{\theta}}\right) \exp\left[-i 2 \hat{z}
\vec{\bar{\theta}}\cdot \Delta \vec{\hat{\theta}}\right]~.
\label{ftGprop}
\end{eqnarray}
This result means that, aside for a phase factor, the spatial
Fourier transform of the cross-correlation function,
$\hat{\mathcal{G}}$, does not depend on $\hat{z}$, as it follows
from the analogous property of the Fourier transform of the
electric field discussed in Section \ref{sec:elem}.

Before proceeding, let us  introduce the spectral degree of
coherence $g$, which can be presented as a function of $\bar{r}$
and $\Delta \hat{r}$:

\begin{equation}
g\left(\vec{\bar{r}},\Delta \vec{\hat{r}}\right) =
\frac{\hat{G}\left(\vec{\bar{r}},\Delta
\vec{\hat{r}}\right)}{\left\langle \left
|\hat{E}_{s\bot}\left(\vec{\bar{r}}+\Delta
\vec{\hat{r}}\right)\right|^2\right \rangle ^{1/2} \left \langle
\left|\hat{E}_{s\bot}\left(\vec{\bar{r}}-\Delta
\vec{\hat{r}}\right)\right|^2\right\rangle^{1/2}} ~.
\label{normfine}
\end{equation}
\begin{figure}
\begin{center}
\includegraphics*[width=150mm]{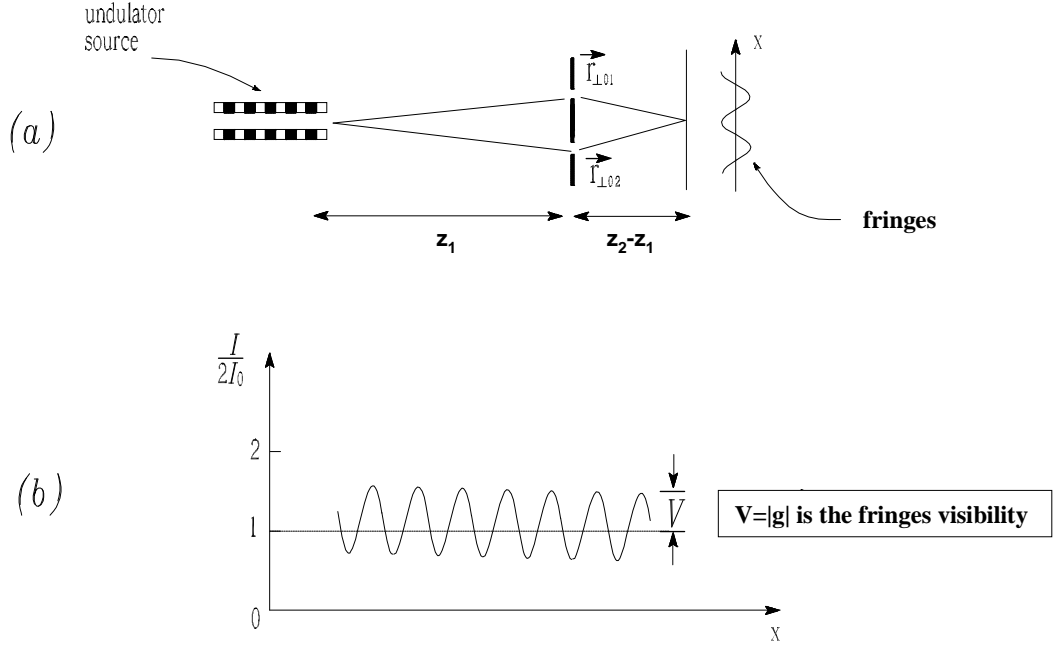}
\caption{\label{ed12c} Measurement of the cross-spectral density
of an undulator source. (a) Young's double-pinhole interferometer
demonstrating the coherence properties of undulator radiation.
Radiation must be spectrally filtered by a monochromator or
detector (not shown in figure). (b) In the quasi-homogeneous case
the fringe visibility $\mathrm{V}$ of the resultant interference
pattern is equal to the absolute value of the spectral degree of
coherence: $\mathrm{V} = |g|$.}
\end{center}
\end{figure}
With reference to Fig. \ref{ed12c}, the modulus of the spectral
degree of coherence, $|g|$, mathematically describes the fringe
visibility of the interference pattern from a Young's
double-pinhole interferometric measure. The phase of the spectral
degree of coherence is related, instead, to the position of the
fringes. The cross-spectral density gives amplitude \textit{and}
position of the fringes. In general, the process may not be
quasi-homogeneous. In this case, the result of Young's experiment
varies with $\vec{\bar{r}}$. In this case, the relation between
the visibility $V$ of the fringes and $g(\vec{\bar{r}},\Delta
\vec{\hat{r}})$ reads

\begin{eqnarray}
V = 2 \frac{ \left<\left|\hat{E}_{s\bot}\left(\vec{\bar{r}}+\Delta
\vec{\hat{r}}\right)\right|^2\right>^{1/2}
\left<\left|\hat{E}_{s\bot}\left(\vec{\bar{r}}-\Delta
\vec{\hat{r}}\right)\right|^2\right>^{1/2}}{
\left<\left|\hat{E}_{s\bot}\left(\vec{\bar{r}}+\Delta
\vec{\hat{r}}\right)\right|^2\right>+
\left<\left|\hat{E}_{s\bot}\left(\vec{\bar{r}}-\Delta
\vec{\hat{r}}\right)\right|^2\right>}\left|g\left(\vec{\bar{r}},\Delta
\vec{\hat{r}}\right)\right|~. \label{Vfrin}
\end{eqnarray}
In the quasi-homogeneous limit $V \longrightarrow \left
|{g}\left(\Delta \vec{\hat{r}}\right)\right|$.

\subsection{\label{subsub:lar} Large non-limiting aperture}

As explained before we start neglecting, at first, the effects
from a finite pupil dimension, assuming a perfect lens with no
aberrations. The imaging problem for an ideal lens is solved once
we find the cross-spectral density of the equivalent virtual
source for the undulator source. On the focal plane we can write
Eq. (\ref{crsfin}) in normalized units as

\begin{eqnarray}
\hat{G}( \hat{z}_f,\vec{ \bar{r}}_f,\Delta\vec{\hat{r}}_f) &=&
\frac{1}{4\pi^2 \hat{f}^2}  \exp\left[\frac{2i}{\hat{f}} \vec{
\bar{r}}_f\cdot\Delta\vec{\hat{r}}_f \right] \cr && \times
\exp\left[-\frac{2i\hat{z}_1}{\hat{f}^2} \vec{
\bar{r}}_f\cdot\Delta\vec{\hat{r}}_f \right]
\hat{\mathcal{G}}\left(0,-\frac{\vec{ \bar{r}}_{f}}{\hat{f}},
-\frac{\Delta\vec{ \hat{r}}_{f}}{\hat{f}}\right)~,\label{fundfoc}
\end{eqnarray}
while on the image plane, Eq. (\ref{propimf2}) in normalized units
reads

\begin{eqnarray}
\hat{G}( \hat{z}_i,\vec{ \bar{r}}_i,\Delta\vec{\hat{r}}_i)  &=&
{\mathrm{m}} \exp\left[\frac{2i \mathrm{m}}{\hat{z}_1}
\vec{\bar{r}}_i \cdot \Delta \vec{\hat{r}}_i \right]\cr &&\times
\exp\left[\frac{2i \mathrm{m}^2}{\hat{z}_1} \vec{\bar{r}}_i \cdot
\Delta \vec{\hat{r}}_i \right]\hat{G}\left( 0,{-\mathrm{m}\vec{
\bar{r}}_i},{-\mathrm{m} \Delta\vec{\hat{r}}_i}\right)~.
\label{fundima}
\end{eqnarray}
In all cases considered in this paper the position $\hat{z}=0$ is
well within the radiation formation length of the undulator.
Therefore the cross-spectral density $\hat{G}$, calculated at
$\hat{z}=0$ has no direct physical meaning, and must be considered
as a quantity characterizing the virtual source only. From the
definitions of virtual source and cross-spectral density follows
that the virtual source produces not only the same field but also
the same cross-spectral density of the real undulator source, at
any distance from the exit of the undulator.

In the present study case of a radiation spot size smaller than
the area of the lens and of a lens with no aberrations, Eq.
(\ref{G2D}) and Eq. (\ref{ftGprop}), together with Eq.
(\ref{fundfoc}) and Eq. (\ref{fundima}), solve the problem of
characterizing the cross-spectral density on the focal plane (with
the help of Eq. (\ref{fundfoc})) and on the image plane (with the
help of Eq. (\ref{fundima})). The situation of a radiation spot
size smaller than the area of the lens is practically achievable
for Synchrotron Radiation due to its high directionality. In this
case, vignetting effects are not present. However, even in this
case, in order to use Eq. (\ref{fundfoc}) and Eq. (\ref{fundima})
one must further assume that aberrations can be neglected.

\subsection{\label{subsub:incl4} Effect of aperture size}

Accounting for the presence of the pupil, in analogy with Eq.
(\ref{crspecprop3}) one has the following normalized expression
for the cross-spectral density on any observation plane at
position $\hat{z}_2$ along the beamline behind the lens:

\begin{eqnarray}
\hat{G}( \hat{z}_2,\vec{ \bar{r}},\Delta\vec{\hat{r}}) &=&
\frac{1}{4\pi^2 ( \hat{z}_2- \hat{z}_1)^2} \exp\left[\frac{2i
\vec{ \bar{r}}\cdot\Delta\vec{\hat{r}}
 } {  \hat{z}_2- \hat{z}_1} \right]\cr && \times \int d
\vec{ \bar{r'}} ~d \Delta\vec{\hat{r'}}~ \Bigg\{\hat{G}\left(
\hat{z}_1,\vec{ \bar{r'}},\Delta\vec{\hat{r'}}\right) P\left(\vec{
\bar{r'}}+\Delta\vec{\hat{r'}}\right) P^*\left(\vec{
\bar{r'}}-\Delta\vec{\hat{r'}}\right) \cr && \times \exp\left[2i
\left(-\frac{1}{ \hat{f}}+ \frac{1}{\hat{z}_2- \hat{z}_1}
\right)\vec{ \bar{r'}}\cdot\Delta\vec{\hat{r'}}\right]\Bigg\} \cr
&& \times \exp \left[-\frac{2 i }{ \hat {z}_2-\hat
{z}_1}\left(\vec{ \bar{r}}\cdot \Delta\vec{\hat{r'}}+\vec{
\bar{r'}}\cdot \Delta\vec{\hat{r}}\right)\right]~.
\label{crspecprop3bis}
\end{eqnarray}
Let us consider, more specifically, the focal and the image plane.
Results can be obtained directly from Eq. (\ref{crspecprop3bis}).
Alternatively, in analogy with Eq. (\ref{GPF}) and Eq.
(\ref{GPI}), one can use our previous results, Eq. (\ref{fundfoc})
and Eq. (\ref{fundima}),   divide them by the first phase factor,
convolve them twice with ${\hat{\mathcal{P}}}$ and
${\hat{\mathcal{P}}}^*$ and, finally, put the phase factor back. A
normalized version of the cross-spectral density $\hat{G}_P$
including pupil effects at the focal is then found and can be
written as

\begin{eqnarray}
\hat{G}_P(\hat{z}_f, \vec{\bar{r}}_f, \Delta \vec{\hat{r}}_f) &=&
\exp\left[\frac{2i}{\hat{f}}\vec{\bar{r}}_f\cdot\Delta\vec{\hat{r}}_f\right]\cr
&&\times \int d\vec{\bar{u}} ~d \Delta \vec{\hat{u}}~
\exp\left[-{2i
\hat{f}}\vec{\bar{u}}\cdot\Delta\vec{\hat{u}}\right]\hat{G}(\hat{z}_f,
\hat{f}\vec{\bar{u}},\hat{f} \Delta \vec{\hat{u}})\cr &&\times
\hat{\mathcal{P}}\left[\frac{\vec{\bar{r}}_f+ \Delta
\vec{\hat{r}}_f}{\hat{f}}-\vec{\bar{u}}- \Delta
\vec{\hat{u}}\right]
\hat{\mathcal{P}}^*\left[\frac{\vec{\bar{r}}_f- \Delta
\vec{\hat{r}}_f}{\hat{f}}-\vec{\bar{u}}+ \Delta
\vec{\hat{u}}\right] ~, \cr &&\label{GPFn}
\end{eqnarray}
while on the image plane one obtains

\begin{eqnarray}
\hat{G}_P(\hat{z}_i, \vec{\bar{r}}_i, \Delta \vec{\hat{r}}_i)&=&
\exp\left[\frac{2i\mathrm{m}}{\hat{z}_1}\vec{\bar{r}}_i\cdot\Delta\vec{\hat{r}}_i\right]
\int d\vec{\bar{u}} ~d \Delta \vec{\hat{u}}~ \exp\left[-\frac{2i
\hat{z}_1
}{\mathrm{m}}\vec{\bar{u}}\cdot\Delta\vec{\hat{u}}\right] \cr &&
\times \hat{G}\left(\hat{z}_i,
\frac{\hat{z}_1}{\mathrm{m}}\vec{\bar{u}},\frac{\hat{z}_1}{\mathrm{m}}
\Delta \vec{\hat{u}}\right)\hat{\mathcal{P}}\left[\frac{
\mathrm{m}}{\hat{z}_1}\left(\vec{\bar{r}}_i+ \Delta
\vec{\hat{r}}_i\right)-\vec{\bar{u}}- \Delta
\vec{\hat{u}}\right]\cr && \times \hat{\mathcal{P}}^*\left[\frac{
\mathrm{m}}{\hat{z}_1}\left(\vec{\bar{r}}_i- \Delta
\vec{\hat{r}}_i\right)-\vec{\bar{u}}+ \Delta \vec{\hat{u}}\right]
~.\label{GPIn}
\end{eqnarray}
As said before, we will treat particular situations in the image
plane when the lens is in the far field. Using coordinates
$\vec{\bar{r}}_i$ and $\Delta \vec{\hat{r}}_i$ the analogous of
condition (\ref{phasesmallfar2}) reads

\begin{equation}
\frac{2 \mathrm{m}^2}{\hat{z}_1} \vec{\bar{r}}_i \cdot \Delta
\vec{\hat{r}}_i \ll 1 \label{farlens}
\end{equation}
for any pair of points on the image pattern.

Explicit substitution of Eq. (\ref{fundima}) in Eq. (\ref{GPIn})
yields the following far field limit expression, which accounts
for condition (\ref{farlens}):

\begin{eqnarray}
\hat{G}_P(\hat{z}_i, \vec{\bar{r}}_i, \Delta \vec{\hat{r}}_i)&=&
{\mathrm{m}}\exp\left[\frac{2i\mathrm{m}}{\hat{z}_1}\vec{\bar{r}}_i\cdot\Delta\vec{\hat{r}}_i\right]
\int d\vec{\bar{u}} ~d \Delta \vec{\hat{u}}~ \hat{G}\left(
0,-{\hat{z}_1}\vec{\bar{u}},{- {\hat{z}_1} \Delta
\vec{\hat{u}}}\right) \cr && \times\hat{\mathcal{P}}~\left[\frac{
\mathrm{m}}{\hat{z}_1}\left(\vec{\bar{r}}_i+ \Delta
\vec{\hat{r}}_i\right)-\vec{\bar{u}}- \Delta
\vec{\hat{u}}\right]\cr &&\times \hat{\mathcal{P}}^*\left[\frac{
\mathrm{m}}{\hat{z}_1}\left(\vec{\bar{r}}_i- \Delta
\vec{\hat{r}}_i\right)-\vec{\bar{u}}+ \Delta \vec{\hat{u}}\right]
~. \label{GPInfarfl}
\end{eqnarray}
Eq. (\ref{GPInfarfl}) is in analogy with Eq. (\ref{GPIfar}) and
Eq. (\ref{EIPfar}).

\section{\label{sec:qhso} Imaging of quasi-homogeneous Gaussian undulator sources by a lens with large non-limiting aperture}

In this Section we specialize our discussion to the particular
case of quasi-homogeneous Gaussian undulator sources, assuming a
lens with large non-limiting aperture and no aberrations. A
Statistical Optics treatment is not the only one possible in this
particular study case. A Geometrical Optics approach can also be
applied, practically consisting in ray-tracing techniques. In this
Section we will consider the image formation problem from a
Statistical Optics viewpoint. In Section \ref{sub:imfoge} an
analysis in terms of Geometrical Optics will be given, and
agreement between these two methods will be demonstrated.

From this point on, we will systematically ignore unimportant
pre-factors appearing in the expressions for the cross-spectral
density. Moreover we will assume $\hat{N}_x \gg 1$ and $\hat{D}_x
\gg 1$ which is a reasonable approximation for third generation
light sources in the X-ray region. We will show that this
assumption leads to a major simplification: namely, horizontal and
vertical coordinates turn out to be factorized in the expression
for the cross-spectral density in free space. As a consequence,
Eq. (\ref{fundfoc}) and (\ref{fundima}) can also be factorized in
the product of a factor depending on the horizontal coordinates
and a factor depending on the vertical coordinates. These separate
factors will be obtained from Eq. (\ref{fundfoc}) and
(\ref{fundima}) substituting all vector quantities with scalar
quantities (horizontal or vertical components).



\subsection{Evolution of the cross-spectral density function in
free space}

Eq. (\ref{G2D}) is a valid expression for the cross-spectral
density in free space at perfect resonance, calculated under the
only assumptions that the system is ultra-relativistic (and,
therefore, the paraxial approximation can be applied) and that the
insertion device is characterized by a large number of undulator
periods. In this case the resonance approximation is enforced. Eq.
(\ref{G2D}) is quite generic and, with respect to first-principle
calculations, it involves the computation of a two-dimensional
integral, whereas the most generic calculations would require a
total of six integrations, two over the undulator length and four
over the electron beam transverse phase space (assuming that the
cross-correlation terms between different electrons is neglected).
From a computational viewpoint, the advantage of reducing the
number of integration is obvious and it can be appreciated even
more after the cross-spectral density is propagated through an
optical system with limiting apertures, which naturally increases
the dimensions of the integration to be performed.

When $\hat{N}_x \gg 1$ and $\hat{D}_x \gg 1$ the cross-spectral in
free space, Eq. (\ref{G2D}), can be written as the product of
factors separately depending on the $x$ and on the $y$ coordinate,
as has been shown in \cite{OURS}.  In fact, analyzing the
exponential factor outside the integral sign in Eq. (\ref{G2D}) it
is possible to see that the maximum value of $\Delta
\hat{\theta}_x^2$ is of order $(\hat{A}_x + \hat{D}_x)/(\hat{A}_x
\hat{D}_x \hat{z}_o^2) \ll 1$, where we remember $\hat{A}_x =
\hat{N}_x^2/\hat{z}_o^2$. As a result, $\Delta \hat{\theta}_x$ can
be neglected inside the $S$ functions in Eq. (\ref{G2D}).
Moreover, since $\hat{D}_x \gg 1$ one can also neglect the
exponential factor in $\hat{\phi}_x^2+2\hat{\phi}_x
\bar{\theta}_x$ inside the integral. This leads to

\begin{eqnarray}
\hat{G} &=&~~~ {\exp{\left[i 2 \bar{\theta}_x\hat{z}_o\Delta
\hat{\theta}_x \right]}} \exp{\left[-\frac {\bar{\theta}_x^2 + 4
\hat{A}_x \hat{z}_o^2 \Delta \hat{\theta}_x^2 \hat{D}_x + 4 i
\hat{A}_x \bar{\theta}_x \hat{z}_o \Delta \hat{\theta}_x
}{2(\hat{A}_x+\hat{D}_x)}\right]}\cr&&\times\exp{\left[i
2\bar{\theta}_y\hat{z}_o\Delta \hat{\theta}_y \right]} \exp{\left[
- \frac {\bar{\theta}_y^2 + 4 \hat{A}_y \hat{z}_o^2 \Delta
\hat{\theta}_y^2 \hat{D}_y + 4 i \hat{A}_y \bar{\theta}_y
\hat{z}_o \Delta \hat{\theta}_y }{2(\hat{A}_y+\hat{D}_y)}\right] }
\cr&& \times \int_{-\infty}^{\infty} d \hat{\phi}_x  \exp{\left[i
\hat{\phi}_x\frac{
 2  \hat{A}_x \hat{z}_o \Delta \hat{\theta}_x
}{\hat{A}_x+\hat{D}_x}\right]} \cr &&\times\int_{-\infty}^{\infty}
d \hat{\phi}_y
\exp{\left[-\frac{\hat{\phi}_y^2+2\hat{\phi}_y\left(\bar{\theta}_y+2i
\hat{A}_y \hat{z}_o \Delta \hat{\theta}_y \right)}{2
(\hat{A}_y+\hat{D}_y)}\right]}\cr&& \times
S^*{\left[\hat{z}_o,\hat{\phi}_x^2+(\hat{\phi}_y-\Delta
\hat{\theta}_y)^2\right]}
S{\left[\hat{z}_o,\hat{\phi}_x^2+(\hat{\phi}_y+\Delta
\hat{\theta}_y)^2\right]}~.\label{G2Dnewsimplif}
\end{eqnarray}
Following the same reasoning in \cite{OURS}, we can also neglect
the phase factor in $\hat{\phi}_x$ under the integral in
$d\hat{\phi}_x$ in Eq. (\ref{G2Dnewsimplif}). As a result, when
$\hat{N}_x \gg 1$ and $\hat{D}_x \gg 1$ horizontal and vertical
coordinates are factorized and we obtain the following equation
for $\hat{G}$:

\begin{equation}
\hat{G}(\hat{z}_o,\bar{\theta}_x,\bar{\theta}_y,\Delta
\hat{\theta}_x,\Delta \hat{\theta}_y) =
\hat{G}_x(\hat{z}_o,\bar{\theta}_x,\Delta \hat{\theta}_x)~
\hat{G}_y(\hat{z}_o,\bar{\theta}_y,\Delta
\hat{\theta}_y)~,\label{sepa}
\end{equation}
where

\begin{eqnarray}
\hat{G}_x ={\exp{\left[i 2 \bar{\theta}_x\hat{z}_o\Delta
\hat{\theta}_x \right]}}  \exp{\left[-\frac {\bar{\theta}_x^2 + 4
\hat{A}_x \hat{z}_o^2 \Delta \hat{\theta}_x^2 \hat{D}_x + 4 i
\hat{A}_x \bar{\theta}_x \hat{z}_o \Delta \hat{\theta}_x
}{2(\hat{A}_x+\hat{D}_x)}\right]}~\label{G2Dnewsimplif2x}
\end{eqnarray}
and

\begin{eqnarray}
\hat{G}_y &=&{\exp{\left[i 2\bar{\theta}_y\hat{z}_o\Delta
\hat{\theta}_y \right]}}  \exp{\left[ - \frac {\bar{\theta}_y^2 +
4 \hat{A}_y \hat{z}_o^2 \Delta \hat{\theta}_y^2 \hat{D}_y + 4 i
\hat{A}_y \bar{\theta}_y \hat{z}_o \Delta \hat{\theta}_y
}{2(\hat{A}_y+\hat{D}_y)}\right] } \cr && \times
\int_{-\infty}^{\infty} d \hat{\phi}_y
\exp{\left[-\frac{\hat{\phi}_y^2+2\hat{\phi}_y\left(\bar{\theta}_y+2i
\hat{A}_y \hat{z}_o \Delta \hat{\theta}_y \right)}{2
(\hat{A}_y+\hat{D}_y)}\right]}\cr && \times
\int_{-\infty}^{\infty} d \hat{\phi}_x
S^*{\left[\hat{z}_o,\hat{\phi}_x^2+(\hat{\phi}_y-\Delta
\hat{\theta}_y)^2\right]}
S{\left[\hat{z}_o,\hat{\phi}_x^2+(\hat{\phi}_y+\Delta
\hat{\theta}_y)^2\right]}~.\label{G2Dnewsimplif2y}
\end{eqnarray}
To begin our investigation of quasi-homogeneous sources we will
consider the limit $\hat{N} \gg 1$ and $\hat{D} \gg 1$, when the
photon-beam phase space in a certain (horizontal or both
horizontal and vertical) direction is an exact replica of the
electron-beam phase space. Calculations can be performed in one
dimension, suppressing indexes $x$ and $y$. In the case of Second
Generation light sources the results that we are going to derive
constitute a realistic description of the radiation
characteristics in both horizontal and vertical directions. Then,
Eq. (\ref{G2Dnewsimplif2y}) coincides with Eq.
(\ref{G2Dnewsimplif2x}).


Let us present  Eq. (\ref{G2Dnewsimplif2x}), i.e. the asymptotic
expression for the cross-spectral density in the limit  $\hat{N}
\gg1 $ and $\hat{D} \gg 1$, in terms of coordinates $\bar{r}
=\hat{z} \bar{\theta}$ and $\Delta \hat{r} = \hat{z} \Delta
\hat{\theta}$. We have

\begin{eqnarray}
\hat{G}(\hat{z},\bar{r},\Delta \hat{r}) &=&
\exp\left[-\frac{\bar{r}^{2}}{2(\hat{A}+\hat{D}) \hat{z}^2}\right]
\exp\left[2 i \frac{\bar{r}\Delta \hat{r}}{\hat{z}}\right]\cr
&&\times \exp\left[- 2 i \frac{\hat{A}\bar{r}\Delta
\hat{r}}{\hat{z} (\hat{A}+\hat{D})}\right]
\exp\left[-2\frac{\hat{A}\hat{D} (\Delta
\hat{r})^{2}}{(\hat{A}+\hat{D}) }\right] ~,\label{crossx1}
\end{eqnarray}
where

\begin{equation}
\hat{A} = \frac{\hat{N}}{\hat{z}^2} ~.\label{a2set}
\end{equation}
Note that here, depending on the situation, $r$ may assume the
meaning of either variable $x$ or $y$.

In the far field limit, when $\hat{A} \ll \hat{D}$ one obtains the
following limiting expression of Eq. (\ref{crossx1}):

\begin{eqnarray}
\hat{G}(\hat{z},\bar{r},\Delta \hat{r}) &=& \exp\left[2 i
\frac{\bar{r}\Delta \hat{r}}{\hat{z}}\right]
\exp\left[-\frac{\bar{r}^{2}}{2 \hat{D} \hat{z}^2}\right]
\exp\left[-2{\hat{A} (\Delta \hat{r})^{2}}\right] ~.
\label{crossx1vcz}
\end{eqnarray}
With the help of  Eq. (\ref{crossx1vcz}) and using Eq.
(\ref{maintrick2}) one can find the expression for
$\hat{\mathcal{G}}$ and $\hat{{G}}$ at the virtual source position
in the center of the undulator. In the case under study
($\hat{N}\gg1 $ and $\hat{D}\gg 1$) the virtual source is a
Gaussian quasi-homogeneous source. Aside for unessential
multiplication constants we have

\begin{eqnarray}
\hat{\mathcal{G}}\left(0,{ \bar{\theta}}, {\Delta{
\hat{\theta}}}\right)&=& \exp\left[-{2 \hat{N}
\Delta\hat{\theta}^{2}}\right] \exp\left[-\frac{\bar{\theta}^2}{2{
\hat{D}}}\right] ~.\label{crossx1lb}
\end{eqnarray}
Therefore, using Eq. (\ref{trGtris}) we also obtain

\begin{eqnarray}
\hat{G}(0,\bar{r},\Delta \hat{r})&=&
\exp\left[-\frac{\bar{r}^{2}}{2\hat{N}}\right] \exp\left[-2{
\hat{D} (\Delta \hat{r})^{2} }\right] ~.\label{crossx1l}
\end{eqnarray}
From Eq. (\ref{crossx1l}) we conclude that the intensity
distribution of the virtual source is a replica of the electron
beam density distribution at the position of minimal beta function
of the undulator (i.e. at the undulator center). Moreover, in this
particular study case, if the position of the minimal beta
function does not coincide with the undulator center, the virtual
source corresponding to the description in Eq. (\ref{crossx1l}) is
simply translated, and is always located at the position where the
beta function of the electron beam is minimal.

It should be noted that the far field limit $\hat{A} \ll \hat{D}$
corresponds with the applicability region of the van
Cittert-Zernike theorem. In virtue of the van Cittert-Zernike
theorem the modulus of the spectral degree of coherence in the far
field, i.e. $\exp[-2{\hat{A} (\Delta \hat{r})^{2}}]$ from Eq.
(\ref{crossx1vcz}), forms a Fourier pair with the intensity
distribution of the virtual source, i.e.
$\exp[-{\bar{r}^{2}}/{2\hat{N}}]$ from Eq. (\ref{crossx1l}).  In
particular one concludes that the rms width of the virtual source
is $\sqrt{\hat{N}}$, as it can be seen directly from Eq.
(\ref{crossx1l}). In our study case for $\hat{N} \gg 1$ and
$\hat{D} \gg 1$, such a relation between the rms width of the
spectral degree of coherence in the far field and the rms
dimension of the virtual source is also a relation between the rms
width of the cross-spectral density function in the far field and
the rms dimension of the electron beam at the plane of minimal
beta function in the center of the undulator. In dimensional units
one can write the value $\sigma_c$ of the rms width of the
spectral degree of coherence ${g}(\Delta \vec{\hat{r}})$ in the
far field as

\begin{equation}
\sigma_c = \frac{\lambda z}{2\pi \sigma} ~,\label{eqdimboh}
\end{equation}
$\sigma$ being, as usual, the rms dimension of the electron beam.
These few last remarks help to clarify what is the size of the
source in the van Cittert-Zernike theorem, that is far from being
a trivial question. For instance, assume that the van
Cittert-Zernike theorem can be applied. Then, the rms electron
beam size can be recovered from the measurement of the transverse
coherence length. In this regard, in \cite{YAB1} Section V, one
may find a statement according to which the rms electron beam size
"is only the average value along the undulator" because "the beta
function has a large variation along the undulator". However, as
we have seen before, the concept of virtual source does not
require a small variation of the beta function. In the most
general case, any variation of the beta function does not affect
the virtual source size and, in our case of quasi-homogeneous
Gaussian source, the virtual source size is also the transverse
size of the electron beam at the position where the beta function
is minimal. Another example dealing with the same issue is given
in reference \cite{PFEI}. This paper (as well as reference
\cite{YAB1}) reports experimental results. However, authors of
\cite{PFEI} observe a disagreement between the electron beam rms
size reconstructed from the van Cittert-Zernike theorem and beam
diagnostics result of about a factor $2$. They ascribe this
variation to the variation of the electron beam size along the
undulator. In footnote [25] of reference \cite{PFEI}, one may
read: "The precise shape and width of the x-ray intensity
distribution in the source plane are directly connected to the
properties of the electron beam. It would not be surprising if the
limited depth of focus of the parabolically shaped electron beta
function in the undulator translates into a virtually enlarged
x-ray source size.". At first glance it looks like if the
Synchrotron Radiation source has a finite longitudinal dimension.
However, based on the previous discussion we conclude that the
virtual source size is equal to the electron beam size at the
point where the beta functions have their minimum and that it is
not affected by variations of the beta function along the
undulator. As a result, one should not observe any virtually
enlarged X-ray source size because of this reason.

\subsection{Evolution of the cross-spectral density function behind the lens}

If we neglect the effect of the pupil in the particular case under
examination ($\hat{N} \gg 1$ and $\hat{D} \gg 1$), it is possible
to find an analytical expression for the cross-spectral density
for any observation plane, and not only for the focal or the image
plane. This is due to the fact that, in this particular case, the
virtual source is gaussian. In the most general case instead, one
has to make use of Eq. (\ref{crspecprop3bis}).

As usual we will neglect, at first, the effect of the pupil
function and group all the phase terms in $\Delta \hat{r}'
\bar{r'}$ in Eq. (\ref{crspecprop3bis}) with the help of the
definition

\begin{equation}
\hat{Q} =  \frac{1}{\hat{z}_1}-\frac {   \hat{A}
 }{\hat{z}_1(\hat{A}+\hat{D})}
-\frac{1}{\hat{f}}+\frac{1}{\hat{z}_2-\hat{z}_1}~, \label{Qfact}
\end{equation}
where

\begin{equation}
\hat{A} = \frac{\hat{N}}{\hat{z}_1^2} ~.\label{a2setbbis}
\end{equation}
With this in mind, after substitution of Eq. (\ref{crossx1})
calculated at $\hat{z}=\hat{z}_1$ in Eq. (\ref{crspecprop3bis}),
one obtains

\begin{eqnarray}
\hat{G}(\hat{z}_2,\bar{r},\Delta \hat{r}) &=&
\exp\left[2i\frac{\bar{r} \Delta \hat{r} } {\hat{z}_2-\hat{z}_1}
\right] \exp\left[-\frac{ 2(\hat{A}+\hat{D}) \hat{z}_1^2(\Delta
\hat{r})^2 }{(\hat{z}_2-\hat{z}_1)^2}\right] \cr &&
\times\exp\left\{- \frac{(\hat{A}+\hat{D}) \left[\bar{r}+ 2 i
\hat{Q} (\hat{A}+\hat{D})\hat{z}_1^2 \Delta \hat{r}
\right]^2}{\left[2 \hat{A}\hat{D}+2 (\hat{A}+\hat{D})^2 \hat{Q}^2
\hat{z}_1^2
\right](\hat{z}_2-\hat{z}_1)^2}\right\}~.\label{finxexpl2}
\end{eqnarray}
which corresponds to a relative intensity

\begin{eqnarray}
\hat{I}(\hat{z}_2,\bar{r}) &=& \exp\left\{-
\frac{(\hat{A}+\hat{D}) \bar{r}^2}{\left[2 \hat{A}\hat{D}+2
(\hat{A}+\hat{D})^2 Q^2 \hat{z}_1^2
\right](\hat{z}_2-\hat{z}_1)^2}\right\}~.\label{intxexpl2}
\end{eqnarray}
and to a modulus of the spectral degree of transverse coherence

\begin{eqnarray}
\left|{g}(\hat{z}_2,\bar{r},\Delta \hat{r})\right| &=&
\exp\left\{-\frac{ 2 \hat{A}\hat{D} (\hat{A}+\hat{D})
\hat{z}_1^2(\Delta \hat{r})^2 }{\left[ \hat{A}\hat{D}+
(\hat{A}+\hat{D})^2 \hat{Q}^2 \hat{z}_1^2
\right](\hat{z}_2-\hat{z}_1)^2}\right\} ~.\label{specxexpl2}
\end{eqnarray}
Letting $\hat{Q} = - \hat{A}/[\hat{z}_1(\hat{A}+\hat{D})]$ the
reader can specialize the results to the case of the image plane.
For $\hat{Q} = \hat{1}/\hat{z}_1 -
\hat{A}/[\hat{z}_1(\hat{A}+\hat{D})]$ one gets the results for the
focal plane. Also, the intensity and the modulus of the spectral
degree of coherence on the image plane can be obtained from those
on the focal plane exchanging $\hat{A}$ with $\hat{D}$. This
symmetry can be explained in terms of Fourier transforms. Phase
factors aside, the cross-spectral density on the image plane is
equal to the cross-spectral density on the object plane. The
cross-spectral density on the focal plane instead, is equal (phase
factors aside) to the Fourier transform of the cross-spectral
density on the object plane.

As we have seen, in the case for a Gaussian electron beam with
$\hat{N} \gg 1$, $\hat{D} \gg 1$ and for a perfect lens with
non-limiting aperture and no aberrations, the Gaussian
approximation for the cross-spectral density at the virtual source
in Eq. (\ref{crossx1l}) can be used, and the cross-spectral
density in free space at any position $\hat{z}$ can be calculated
with the help of Eq. (\ref{crossx1}). Then, Eq.
(\ref{crspecprop3bis}) can be simplified to recover both the
intensity and the modulus of the spectral degree of coherence (Eq.
(\ref{intxexpl2}) and Eq. (\ref{specxexpl2}) respectively), which
are Gaussian functions for any value of $\hat{z}_2$ and
$\hat{z}_1$. Even for quasi-homogeneous sources though, there are
a number of examples when it is difficult to obtain analytical
results from Eq. (\ref{crspecprop3bis}) for any value of
$\hat{z}_2$. Nevertheless it is possible to calculate the
cross-spectral density at the image plane and at the focal plane
(for any value of $\hat{z}_1$)  with the help of Eq.
(\ref{fundfoc}) and Eq. (\ref{fundima}). This can be done relying
on the calculation of the cross-spectral density at the
virtual-source position (and its Fourier transform), which allows
further use of Eq. (\ref{fundfoc}) and Eq. (\ref{fundima}). We
will first use Eq. (\ref{fundfoc}) and Eq. (\ref{fundima}) to deal
with the case that we just discussed when $\hat{N} \gg 1$ and
$\hat{D} \gg 1$. This is not a simple repetition of already known
results, because the particular way of reasoning used for the
focal and the image plane, through Eq. (\ref{fundfoc}) and Eq.
(\ref{fundima}), will be widely used in the following parts of
this paper too. The Statistical Optics method conjugated to
Fourier Optics results allows us to predict, by manipulations of
Eq. (\ref{crossx1}), the cross-spectral density (and, therefore,
the intensity and the absolute value of the spectral degree of
coherence) on the focal and on the image plane. In order to use
Eq. (\ref{fundfoc}) and Eq. (\ref{fundima}) we must take advantage
of the expressions for $\hat{\mathcal{G}}$ and $\hat{{G}}$ at the
virtual source position, Eq. (\ref{crossx1lb}) and Eq.
(\ref{crossx1l}) respectively.

With the help of Eq. (\ref{fundfoc}) and Eq. (\ref{crossx1lb}), on
the focal plane we obtain

\begin{eqnarray}
\hat{G}( \hat{z}_f,\vec{ \bar{r}}_f,\Delta\vec{\hat{r}}_f)  &=&
\exp\left[\frac{i 2(\hat{f}-\hat{z}_1)}{\hat{f}^2} \bar{r}_f
\Delta \hat{r}_f\right] \exp\left[-\frac{2 \hat{N}
\Delta\hat{r}_f^{2}}{\hat{f}^2}\right]
\exp\left[-\frac{\bar{r}_f^2}{2{ \hat{D} \hat{f}^2}}\right]~.\cr
&& \label{focalcrspden}
\end{eqnarray}
For \textit{any} value of $\hat{z}_1$ we have a relative intensity
on the focal plane given by

\begin{equation}
\hat{I} =\exp\left[-\frac{\bar{r}_f^2}{2{ \hat{D}
\hat{f}^2}}\right]~,\label{Infrel}
\end{equation}
while the modulus of the spectral degree of coherence (again, for
any position $\hat{z}_1$ of the lens\footnote{Note that the
\textit{modulus} of the spectral degree of coherence in Eq.
(\ref{spectrf}) is independent of $\bar{r}_f$. However, the
spectral degree of coherence depends on $\bar{r}_f$ through a
phase factor. This situation corresponds, according to a
definition given by us in \cite{OURS}, to a \textit{weakly
quasi-homogeneous wavefront}. This remark is valid for many
expressions of the modulus of the cross-spectral density given in
this work.}) is

\begin{eqnarray}
\left|{g}\left(\hat{z}_f, \bar{r}_f, {\Delta{
\hat{r}}_{f}}\right)\right|&=& \exp\left[-\frac{2 \hat{N}
\Delta\hat{r}_f^{2}}{\hat{f}^2}\right] ~.\label{spectrf}
\end{eqnarray}
These results are intuitively sound. In Section \ref{sec:elem} we
explained that we expect to find, on the focal plane, the spatial
Fourier transform of the wavefront on the object plane (except for
a phase and a proportionality factor). Therefore, it is intuitive
that the intensity on the focal plane must depend on the electron
beam divergence only and that the modulus of the spectral degree
of coherence must depend on the electron beam size only. In fact,
the exchange of roles of $\hat{N}$ and $\hat{D}$ passing from the
virtual source plane to the  the focal plane is related to the
operation of Fourier transform. Also note that the Fourier
transform of the field depends on $\hat{z}_1$ through a phase
factor only, and free space basically acts as a Fourier transform
itself (see Section \ref{sec:elem}): what we find on the focal
plane in terms of intensity and modulus of the spectral degree of
coherence we must also find in the far field after propagation in
free space. The reader may check that, after substitution
$\bar{r}_f/\hat{f} \longrightarrow \bar{\theta}$ and $\Delta
\hat{r}_f/\hat{f} \longrightarrow \Delta \hat{\theta}$, Eq.
(\ref{Infrel}) and Eq. (\ref{spectrf}) can be found from Eq. (54)
and Eq. (61) in \cite{OURS} in the limit $\hat{z}_o \gg 1$, which
describe propagation in free space, as it should be.

A similar simplified reasoning can be applied for the
cross-spectral density on the image plane. With the help of  Eq.
(\ref{fundima}) and  Eq. (\ref{crossx1l}), describing the
cross-spectral density of the virtual source we obtain:

\begin{eqnarray}
\hat{G}( \hat{z}_i,{ \bar{r}}_i,\Delta{\hat{r}}_i)  &=&
\exp\left[\frac{2i \mathrm{m}(\mathrm{m}+1){
\bar{r}}_i\Delta{\hat{r}}_i }{\hat{z}_1} \right]
\exp\left[-\frac{\mathrm{m}^2 \bar{r}^{2}_i}{2\hat{N}}\right] \cr
&& \times \exp\left[-2{ {\hat{D}}\mathrm{m}^2 (\Delta
\hat{r}_i)^{2} }\right]~. \label{crspecpropimbist}
\end{eqnarray}
The relative intensity on the image plane is given by

\begin{equation}
\hat{I} =\exp\left[-\frac{\mathrm{m}^2 \bar{r}^{2}_i}{2
\hat{N}}\right]~,\label{Infrelim}
\end{equation}
while the modulus of the spectral degree of coherence is

\begin{eqnarray}
\left|{g}\left(\hat{z}_i,\bar{r}_i,\Delta
\hat{r}_{i}\right)\right|&=& \exp\left[-2{ \mathrm{m}^2 \hat{D}}
(\Delta \hat{r}_i)^{2} \right] ~.\label{spectr}
\end{eqnarray}
These results are very natural. By definition of image plane, when
we image an object with an ideal lens with a large non-limiting
pupil aperture, we obtain a magnified version of the object
(virtual, in this case).

We remarked before that the intensity and the modulus of the
spectral degree of coherence on the image plane can be obtained
from that on the focal plane exchanging $\hat{A}$ with $\hat{D}$.
This symmetry though is not evident from the expressions in Eq.
(\ref{Infrel}), Eq. (\ref{Infrelim}), Eq. (\ref{spectrf}) and Eq.
(\ref{spectr}): to display it one has to express these equations
in terms of $\hat{z}_2$ and $\hat{z}_1$.

\section{\label{sec:nong} Imaging of quasi-homogeneous
non-Gaussian undulator sources by a lens with large non-limiting
aperture}

In the previous Section \ref{sec:qhso} we treated the case for
$\hat{N} \gg 1$ and $\hat{D} \gg 1$.  In the present Section
\ref{sec:nong} we will deal with other quasi-homogeneous cases,
always assuming $\hat{N}_x \gg 1$ and $\hat{D}_x \gg 1$. The
quasi-homogeneous situations that remain to be treated under this
assumption are for either $\hat{N}_y \gg 1$ \textit{or} $\hat{D}_y
\gg 1$. In fact, the situation for both $\hat{N}_y \gg 1$
\textit{and} $\hat{D}_y \gg 1$ is automatically included in
Section \ref{sec:qhso}. Moreover, in all quasi-homogeneous cases,
the cross-spectral density in the horizontal direction obeys Eq.
(\ref{finxexpl2}). Therefore, we will focus our attention on the
cross-spectral density in the vertical direction only.

\subsection{Source with non-Gaussian angular distribution in the vertical direction}

Let us use of our Statistical Optics approach to solve a somewhat
complicated image formation problem. After assuming separability
of the horizontal and vertical directions (${\hat{N}_x\gg1}$,
${\hat{D}_x\gg 1 }$) we suppose that the electron beam has a
vertical transverse size much larger than the diffraction size,
$\hat{N}_y \gg 1$, and a finite divergence $\hat{D}_y>0$. As
usual, we will first neglect the influence of the pupil function.
The difference with respect to the case treated in the previous
Section \ref{sec:qhso} is that Eq. (\ref{crspecprop3bis}) cannot
be explicitly calculated for any value of $\hat{z}_1$ and
$\hat{z}_2$. However, as said before, the Statistical Optics
method conjugated to Fourier Optics results allows us to predict,
for any value of $\hat{z}_1$, the cross-spectral density on the
focal ($\hat{z}_2 = \hat{z}_f$) and on the image ($\hat{z}_2 =
\hat{z}_i$) plane by means of Eq. (\ref{fundfoc}) and Eq.
(\ref{fundima}). In order to use these equations we must first
calculate $\hat{{G}}$ and $\hat{\mathcal{G}}$ at the virtual
source at position $\hat{z}_o = 0$. This can be done taking the
limit $\hat{z}_o^2 \gg \hat{N}/\hat{D}$ of Eq.
(\ref{G2Dnewsimplif2y}), i.e. calculating the far zone limit of
Eq. (\ref{G2Dnewsimplif2y}), and using Eq. (\ref{maintrick2}).
First, under the assumption $\hat{N}_y \gg 1$ we can neglect
$\Delta \hat{\theta}_y$ in the $S$ functions in Eq.
(\ref{G2Dnewsimplif2y}), thus obtaining the vertical
cross-spectral density function in the far field limit

\begin{eqnarray}
\hat{G} &=& {\exp{\left[i 2\bar{\theta}_y\hat{z}_o\Delta
\hat{\theta}_y \right]}} \exp{\left[ - {2 \hat{N}_y \Delta
\hat{\theta}_y^2 }\right] } \int_{-\infty}^{\infty} d \hat{\phi}_y
\exp{\left[-\frac{(\hat{\phi}_y+\bar{\theta}_y)^2}
{2\hat{D}_y}\right]}\hat{I}_S(\hat{\phi}_y)~, \cr &&
\label{G2Dmeno1}
\end{eqnarray}
where

\begin{figure}
\begin{center}
\includegraphics*[width=140mm]{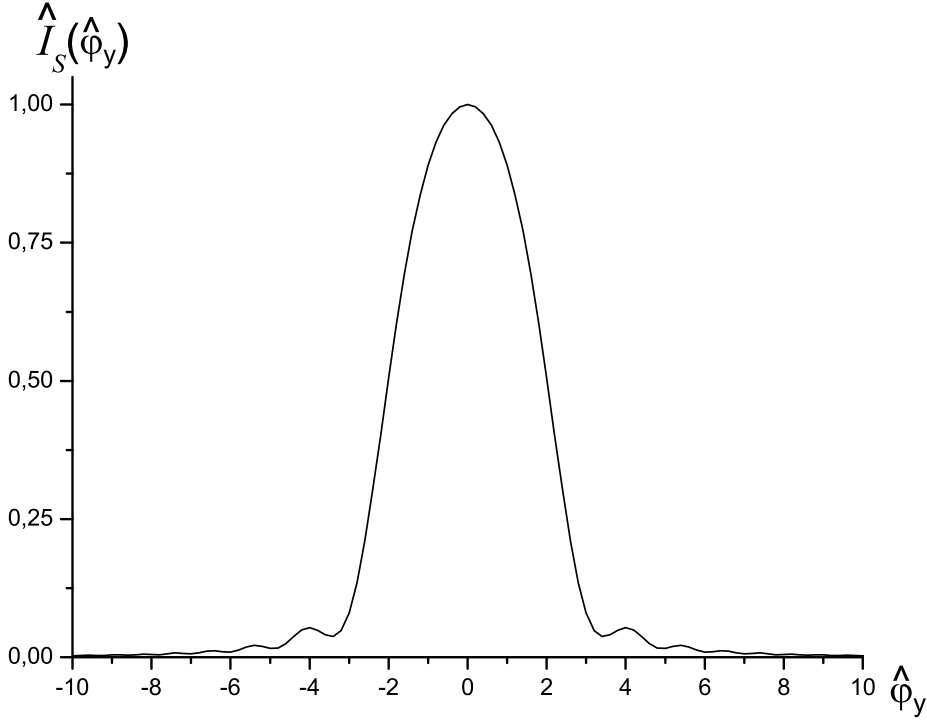}
\caption{\label{ISnorm} The universal function $\hat{I}_S$, used
to calculate the focal intensity of a quasi-homogeneous source at
$\hat{N}_x \gg 1$, $\hat{D}_x \gg 1$ and $\hat{N}_y\gg 1$.}
\end{center}
\end{figure}
\begin{equation}
{\hat{I}_S}(\hat{\phi}_y) =\frac{3}{8 \sqrt{\pi}}
\int_{-\infty}^{\infty} d \hat{\phi}_x ~ \mathrm{sinc}^2
\left[\left({\hat{\phi}_x^2+\hat{\phi}_y^2}\right)/4\right]
~\label{bingo2}
\end{equation}
is  a universal function related to undulator radiation. A plot of
$\hat{I}_S$ is given in Fig. \ref{ISnorm}. Eq. (\ref{G2Dmeno1}),
substituted into Eq. (\ref{maintrick2}), gives the Fourier
transform of the cross-spectral density at $\hat{z}_o=0$, i.e. at
the virtual-source position:

\begin{eqnarray}
\hat{\mathcal{G}}\left(0,{ \bar{u}}, {\Delta{ \hat{u}}}\right) &=&
\exp{\left[ - {2 \hat{N}_y \Delta \hat{u}^2 }\right] }
\int_{-\infty}^{\infty} d \hat{\phi}_y
\exp{\left[-\frac{(\hat{\phi}_y+\bar{u})^2}
{2\hat{D}_y}\right]}\hat{I}_S(\hat{\phi}_y)~. \cr &&
\label{G2Dmeno2}
\end{eqnarray}
Inverse transforming Eq. (\ref{G2Dmeno2}) according to the
definition in Eq. (\ref{trGtris}), we obtain the cross spectral
density at the virtual source position

\begin{eqnarray}
\hat{G}(0,\bar{y},\Delta \hat{y}) = \exp\left[-\frac{\bar{y}^2}{2
\hat{N}_y} \right] \exp\left[-2 \hat{D}_y \Delta \hat{y}^2
\right]\gamma(\Delta \hat{y})~,\label{Ngrande}
\end{eqnarray}
where function $\gamma(\Delta \hat{y})$ is an inverse Fourier
transform, normalized to unity, of  $\hat{I}_S$, defined as from
Eq. (93) in \cite{OURS}:

\begin{eqnarray}
\gamma(\Delta \hat{y}) = \frac{1}{2 \pi^2} \int_{-\infty}^{\infty}
d \hat{\phi}_y \exp{\left[i\left(- 2 \Delta \hat{y}
\right)\hat{\phi}_y\right]}
{\hat{I}_S}(\hat{\phi}_y)~.\label{G2D3lastlastno2}
\end{eqnarray}
It has been shown in \cite{OURS} that $\gamma$ can be expressed in
terms of the sine integral function $\mathrm{Si}(\cdot)$ and of
the cosine integral function $\mathrm{Ci}(\cdot)$. One has

\begin{eqnarray}
\gamma(\Delta \hat{y}) =\frac{2}{\pi} \left[\frac{\pi}{2}+ 2\Delta
\hat{y}^2 \mathrm{Ci}\left(2\Delta \hat{y}^2\right)-
\sin\left(2\Delta \hat{y}^2\right)- \mathrm{Si}\left(2\Delta
\hat{y}^2\right) \right]~.\label{G2D3lastlastno5}
\end{eqnarray}
This means that $\gamma$ is a real function. Moreover, in Eq.
(\ref{Ngrande}), $\Delta \hat{y}$ and $\bar{y}$ are separated and,
since $\hat{N}_y \gg 1$, the typical correlation length is much
smaller than the radiation spot, independently of the value of
$\hat{D}_y$. This shows that Eq. (\ref{Ngrande}) models a
quasi-homogeneous source. From Eq. (\ref{fundfoc}) and Eq.
(\ref{G2Dmeno2}) we obtain the cross-spectral density on the focal
plane

\begin{eqnarray}
\hat{G}( \hat{z}_f,{ \bar{y}}_f,\Delta{\hat{y}}_f) &=&
\exp\left[\frac{2i}{\hat{f}^2} \left(\hat{f}-\hat{z}_1\right){
\bar{y}}_f\Delta{\hat{y}}_f\right]\exp\left[- \frac{2\hat{N}_y
\Delta{\hat{y}}_f^2 }{ \hat{f}^2}\right] \cr && \times
\int_{-\infty}^{\infty} d\hat{\phi}_y \exp \left[-
\frac{\left(\bar{y}_f/\hat{f}+\hat{\phi}_y\right)^2}{2\hat{D}_y }
\right] \hat{I}_S(\hat{\phi}_y)~.\label{xdngrandefoc}
\end{eqnarray}
The relative intensity on the focal plane is therefore given by

\begin{eqnarray}
\hat{I}(\hat{z}_f,\bar{y}_f) &=& \int_{-\infty}^{\infty}
d\hat{\phi}_y \exp \left[-
\frac{\left(\bar{y}_f/\hat{f}+\hat{\phi}_y\right)^2}{2\hat{D}_y}
\right] \hat{I}_S(\hat{\phi}_y)\cr && \times \left\{
\int_{-\infty}^{\infty} d\hat{\phi}_y \exp \left[-
\frac{\hat{\phi}_y^2}{2\hat{D}_y} \right]
\hat{I}_S(\hat{\phi}_y)\right\}^{-1}~,\label{Ingrandefoc}
\end{eqnarray}
while the modulus of the spectral degree of coherence reads

\begin{eqnarray}
|{g}( \hat{z}_f,\bar{y}_f,\Delta \hat{y}_f)| &=&\exp\left[-
2\frac{\hat{N}_y \Delta{\hat{y}}_f^2
}{\hat{f}^2}\right]~.\label{spdngrandefoc}
\end{eqnarray}
For the image plane, Eq. (\ref{fundima}) and Eq. (\ref{Ngrande})
give the following cross-spectral density:

\begin{figure}
\begin{center}
\includegraphics*[width=140mm]{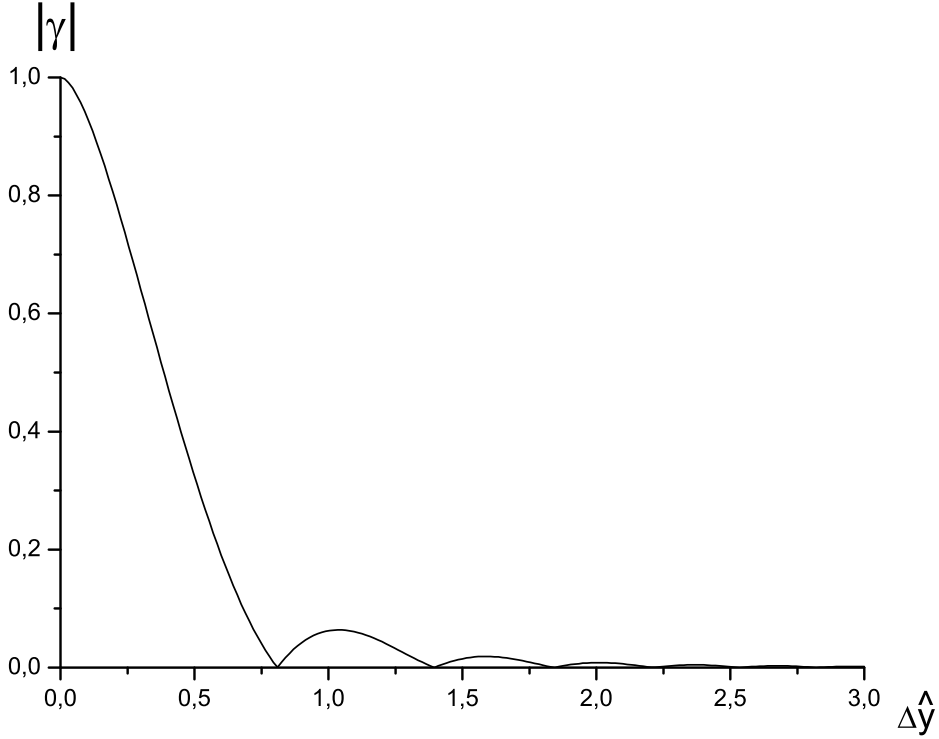}
\caption{\label{GAM} Absolute value of the universal function
$\gamma$, used to calculate, on the image plane, the spectral
degree of coherence of a quasi-homogeneous undulator source when
$\hat{N}_x \gg1 $, $\hat{D}_x \gg 1$ and $\hat{N}_y \gg 1$.}
\end{center}
\end{figure}
\begin{eqnarray}
\hat{G}( \hat{z}_i,{ \bar{y}}_i,\Delta{\hat{y}}_i)  &=&
\exp\left[\frac{i \mathrm{m}(\mathrm{m}+1){ \bar{y}}_i
\Delta{\hat{y}}_i }{2 \hat{z}_1} \right]\cr &&\times
\exp\left[-\frac{\mathrm{m}^2\bar{y}_i^2}{2 \hat{N}_y} \right]
\exp\left[-2 \hat{D}_y \mathrm{m}^2 \Delta \hat{y}_i^2
\right]\gamma(\mathrm{m}\Delta \hat{y}_i)~, \label{xdngandeima}
\end{eqnarray}
corresponding to a relative intensity on the image plane

\begin{eqnarray}
\hat{I}( \hat{z}_i,{ \bar{y}}_i)  &=&
\exp\left[-\frac{\mathrm{m}^2\bar{y}_i^2}{2 \hat{N}_y} \right]~.
\label{Ingandeima}
\end{eqnarray}
Eq. (\ref{Ingandeima}) is the (magnified) image of the electron
beam in the object plane $\hat{z} = 0$. The modulus of the
spectral degree of coherence is:

\begin{eqnarray}
|{g}( \hat{z}_i,\bar{y}_i, \Delta{\hat{y}}_i)|  &=& \exp\left[-2
\hat{D}_y \mathrm{m}^2 \Delta \hat{y}_i^2
\right]|\gamma(\mathrm{m}\Delta \hat{y}_i)|~. \label{spdngandeima}
\end{eqnarray}
A plot of $|\gamma(\Delta \hat{y})|$ is given in Fig. \ref{GAM}.

\subsection{Source with non-Gaussian intensity distribution in the vertical direction}

Let us now consider the case when $\hat{D}_y\gg 1$ and $\hat{N}_y$
assumes arbitrary values. In this case, Eq. (115) and Eq. (125) of
\cite{OURS} allow reconstruction of the cross-spectral density in
the far zone, that is

\begin{eqnarray}
\hat{G}= \exp{\left[2 i \bar{\theta}_y \hat{z}_o \Delta
\hat{\theta}_y \right]} \exp\left[-2 \hat{N}_y \Delta
\hat{\theta}_y^2 \right]
\exp\left[-\frac{\bar{\theta}_y^2}{2\hat{D}_y} \right]\beta(\Delta
\hat{\theta}_y)~,\label{Dgrande}
\end{eqnarray}
where the function $\beta(\Delta \hat{y})$ is defined in Eq. (113)
of \cite{OURS} and reads:

\begin{eqnarray}
\beta(\Delta\hat{\theta}_y) &=& \frac{1}{2\pi^2}
\int_{-\infty}^{\infty} d \hat{\phi}_y \int_{-\infty}^{\infty} d
\hat{\phi}_x \cr && \times
\mathrm{sinc}{\left[\frac{\hat{\phi}_x^2+(\hat{\phi}_y-\Delta
\hat{\theta}_y)^2}{4}\right]}
\mathrm{sinc}{\left[\frac{\hat{\phi}_x^2+(\hat{\phi}_y+\Delta
\hat{\theta}_y)^2}{4}\right]}~.\label{G2D2perDG}
\end{eqnarray}
Eq. (\ref{Dgrande}) may be obtained directly from Eq.
(\ref{G2Dnewsimplif2y}) in the limit $\hat{z}_o^2
\gg\hat{N}_y/\hat{D}_y$, i.e. in the far zone. Note that if
$\hat{N}_y /\hat{D}_y \ll 1$ ($\hat{N}_y \lesssim 1$ is our main
case of interest since we have already treated the case when both
$\hat{N}_y \gg 1$ and $\hat{D}_y \gg 1$) the far zone begins
already at the exit of the undulator, when $\hat{z}_1 \sim 1$ (see
also \cite{OURS} for details). Let us introduce, in analogy with
Eq. (\ref{G2D3lastlastno2}) the following (inverse) Fourier
transform of the function $\beta$:

\begin{eqnarray}
\hat{\mathcal{B}}(\bar{y}) = \frac{1}{\mathcal{K}}
\int_{-\infty}^{\infty} d \hat{\phi}_y \exp\left[i (-2\bar{y})
\hat{\phi}_y\right] \beta(\hat{\phi}_y) ~.\label{BB}
\end{eqnarray}
Here $\mathcal{K}$ is the normalization factor

\begin{eqnarray}
\mathcal{K} = \int_{-\infty}^{\infty} d \hat{\phi}_y
\beta(\hat{\phi}_y) \simeq 2.200~,\label{BB}
\end{eqnarray}
and has been calculated numerically.

Both $\beta(\Delta \hat{\theta}_y)$ and
$\hat{\mathcal{B}}(\bar{y})$ admit representations in terms of a
one-dimensional integral (note that the representation for
$\beta(\Delta \hat{\theta}_y)$ has been already introduced in
\cite{OURS}). In order to see this, let us first consider the
function:

\begin{eqnarray}
\tilde{f}(\Delta\hat{\theta}_x',\Delta\hat{\theta}_y') &=&
\frac{1}{2\pi^2} \int_{-\infty}^{\infty} d \hat{\phi}_y
\int_{-\infty}^{\infty} d \hat{\phi}_x \cr &&\times
\mathrm{sinc}{\left[\frac{(\hat{\phi}_x-\Delta
\hat{\theta}_x'/2)^2+(\hat{\phi}_y-\Delta
\hat{\theta}_y'/2)^2}{4}\right]} \cr && \times
\mathrm{sinc}{\left[\frac{(\hat{\phi}_x+\Delta
\hat{\theta}_x'/2)^2+(\hat{\phi}_y+\Delta
\hat{\theta}_y'/2)^2}{4}\right]}~.\label{G2D2perDFgenApp}
\end{eqnarray}
The function $\tilde{f}$  is  circularly symmetric. This can be
seen switching to polar coordinates:

\begin{eqnarray}
\hat{\phi}_x &=& \hat{r}_\phi \cos(\hat{\eta}_\phi) \cr
\hat{\phi}_y &=& \hat{r}_\phi \sin(\hat{\eta}_\phi) \label{pol1}
\end{eqnarray}
and

\begin{eqnarray}
\Delta \hat{\theta}_x'/2 &=& \hat{r}_\theta
\cos(\hat{\eta}_\theta) \cr \Delta \hat{\theta}_y'/2 &=&
\hat{r}_\theta \sin(\hat{\eta}_\theta)~. \label{pol1}
\end{eqnarray}
Then, Eq. (\ref{G2D2perDFgenApp}) can be rewritten as

\begin{eqnarray}
\tilde{f}(\hat{r}_\theta) &=& \frac{1}{2\pi^2} \int_{0}^{\infty} d
\hat{r}_\phi \int_{0}^{2 \pi} d \hat{\eta}_\phi\cr &&\times
\mathrm{sinc}{\left[\frac{\hat{r}_\phi^2 +  \hat{r}_\theta^2 -
2\hat{r}_\phi \hat{r}_\theta
\cos\left(\hat{\eta}_\phi-\hat{\eta}_\theta\right)}{4}\right]} \cr
&& \times \mathrm{sinc}{\left[\frac{\hat{r}_\phi^2 +
\hat{r}_\theta^2 + 2\hat{r}_\phi \hat{r}_\theta
\cos\left(\hat{\eta}_\phi-\hat{\eta}_\theta\right)}{4}\right]}~,\label{G2D2perDFgenApppolar}
\end{eqnarray}
which does not depend on $\hat{\eta}_\theta$, as can be seen
switching to the integration variable $\hat{\eta}' =
\hat{\eta}_\phi-\hat{\eta}_\theta$. The following relation
follows:

\begin{equation}
\beta(\Delta{\theta}_y) =
\tilde{f}(\Delta\hat{\theta}_x',\Delta\hat{\theta}_y')
\label{relbf}
\end{equation}
for any $(\Delta\hat{\theta}_x',\Delta\hat{\theta}_y')$ such that

\begin{equation}
\Delta\hat{\theta}_y =
\sqrt{(\Delta\hat{\theta}_x'/2)^2+(\Delta\hat{\theta}_y'/2)^2}~.
\label{relbfexpl}
\end{equation}
The function $\beta$ can be seen as a restriction of the function
$\tilde{f}$. The reason why $\tilde{f}$ has been introduced is
that it allows the use the autocorrelation theorem to obtain the
following relation:

\begin{eqnarray}
\int_{-\infty}^{\infty} d \Delta\hat{\theta}_x'
\int_{-\infty}^{\infty} d\Delta\hat{\theta}_y'  \exp{[i (\alpha_x
\Delta\hat{\theta}_x'+ \alpha_y \Delta\hat{\theta}_y')]}
\tilde{f}(\Delta\hat{\theta}_x',\Delta\hat{\theta}_y') = \cr
\frac{1}{2\pi^2}\left|\int_{-\infty}^{\infty} d \hat{\phi}_x
\int_{-\infty}^{\infty} d \hat{\phi}_y \exp{[i (\alpha_x
\hat{\phi}_x+ \alpha_y \hat{\phi}_y)]}
\mathrm{sinc}{\left[\frac{\hat{\phi}_x^2+\hat{\phi}_y^2}{4}\right]}
\right|^2~. \label{appliedapp}
\end{eqnarray}
The integral in the right hand side of Eq. (\ref{appliedapp}) has
been already calculated, in practice, in Eq. (\ref{cpact51}), and
it can be expressed in terms of the universal function $\Psi$
defined in Eq. (\ref{psiuni}). It follows that

\begin{eqnarray}
&&\int_{-\infty}^{\infty} d \Delta\hat{\theta}_x'
\int_{-\infty}^{\infty} d\Delta\hat{\theta}_y'  \exp{[i (\alpha_x
\Delta\hat{\theta}_x'+ \alpha_y \Delta\hat{\theta}_y')]}
\tilde{f}(\Delta\hat{\theta}_x',\Delta\hat{\theta}_y')= \cr &&
2\left[\pi-2\mathrm{Si}(\alpha_x^2+\alpha_y^2)\right]^2 = {2\pi^2}
\Psi\left(\sqrt{\alpha_x^2+\alpha_y^2}\right) ~.
\label{appliedapp3}
\end{eqnarray}
We can now inverse transform Eq. (\ref{appliedapp3}) using the
Fourier-Bessel formula, thus obtaining

\begin{eqnarray}
\tilde{f}(\Delta\hat{\theta}_x',\Delta\hat{\theta}_y')={\pi}
\int_{0}^{\infty} d\alpha ~\alpha J_o\left({\alpha}
{\sqrt{(\Delta\hat{\theta}_x')^2+(\Delta\hat{\theta}_y')^2}}\right)
\Psi(\alpha) ~.\label{fbessel}
\end{eqnarray}
Letting $\Delta \hat{\theta}_x' = 0$ and using Eq. (\ref{relbf})
and Eq. (\ref{relbfexpl}) we obtain the following representation
for $\beta$:

\begin{eqnarray}
\beta(\Delta \hat{\theta}_y)={\pi} \int_{0}^{\infty} d\alpha~
\alpha J_o\left(2{\alpha} \Delta\hat{\theta}_y\right) \Psi(\alpha)
~.\label{brepr}
\end{eqnarray}
Applying the definition of $\hat{\mathcal{B}}$ in Eq. (\ref{BB})
we obtain

\begin{eqnarray}
\hat{\mathcal{B}}(\bar{y})=\frac{\pi}{\mathcal{K}}
\int_{0}^{\infty} d\alpha ~ \alpha \Psi(\alpha)
\int_{-\infty}^{\infty} d \Delta \hat{\theta}_y \exp\left[i
(-2\bar{y}) \Delta \hat{\theta}_y \right] J_o\left(2{\alpha}
\Delta\hat{\theta}_y\right)  ~.\label{brepr2}
\end{eqnarray}

\begin{figure}
\begin{center}
\includegraphics*[width=140mm]{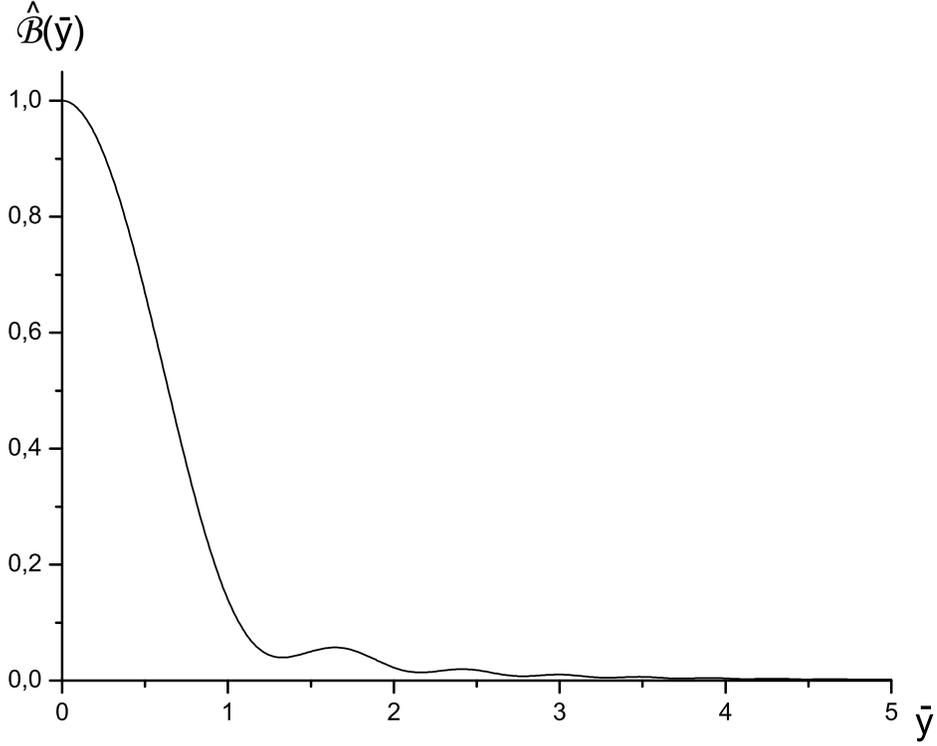}
\caption{\label{ftbetaplot} The universal function
$\hat{\mathcal{B}}$, used to calculate intensity on the image
plane of a quasi-homogeneous undulator source when $\hat{N}_x
\gg1$, $\hat{D}_x \gg1$ and $\hat{D}_y \gg 1$. }
\end{center}
\end{figure}
The Fourier integral in $\Delta \hat{\theta}_y$ can be performed
analytically (see \cite{GOOD}, Appendix A.3.), thus giving the
following representation of $\hat{\mathcal{B}}$:

\begin{eqnarray}
\hat{\mathcal{B}}(\bar{y})=\frac{\pi}{\mathcal{K}}
\int_{0}^{\infty} d\alpha ~
\frac{\mathrm{rect}\left[\bar{y}/(2\alpha)\right]}{ \left[1 -
\left(\bar{y}/\alpha\right)^2\right]^{1/2}}~
\Psi(\alpha)~,\label{ftbrepr}
\end{eqnarray}
where the function $\mathrm{rect}(x)$ is defined, following
\cite{GOOD}, to be unity for $|x|\leqslant 1/2$ and zero
otherwise. A plot of the universal $\hat{\mathcal{B}}$ is given in
Fig. \ref{ftbetaplot}.

Using Eq. (\ref{maintrick2}) and Eq. (\ref{Dgrande}) we obtain the
Fourier transform of the cross-spectral density at the
virtual-source position

\begin{eqnarray}
\hat{\mathcal{G}}\left(0,{ \bar{u}}, {\Delta{ \hat{u}}}\right) &=&
\exp\left[-2 \hat{N}_y \Delta \hat{u}^2 \right]
\exp\left[-\frac{\bar{u}^2}{2\hat{D}_y} \right]\beta(\Delta
\hat{u})~. \cr && \label{Ngr2}
\end{eqnarray}
Inverse transforming Eq. (\ref{Ngr2}) we can write the
cross-spectral density at the virtual-source position

\begin{eqnarray}
\hat{G}(0,\bar{y},\Delta \hat{y}) = \exp\left[-2 \hat{D}_y \Delta
\hat{y}^2 \right] \int_{-\infty}^{\infty} d \eta
\exp\left[-\frac{\left(\eta+\bar{y}\right)^2}{2 \hat{N}_y} \right]
\hat{\mathcal{B}}(\eta) ~.\label{Ngr3}
\end{eqnarray}
Then, from Eq. (\ref{fundfoc}) and Eq. (\ref{Ngr2}) we obtain the
cross-spectral density on the focal plane

\begin{eqnarray}
\hat{G}( \hat{z}_f,{ \bar{y}}_f,\Delta{\hat{y}}_f) &=&
\exp\left[\frac{2i}{\hat{f}^2} \left(\hat{f}-\hat{z}_1\right){
\bar{y}}_f\Delta{\hat{y}}_f\right]\exp\left[- \frac{2\hat{N}_y
\Delta{\hat{y}}_f^2 }{ \hat{f}^2}\right] \cr && \times
\exp\left[-\frac{\bar{y}_f^2}{2\hat{D}_y \hat{f}^2}
\right]\beta\left(\frac{\Delta
\hat{y}_f}{\hat{f}}\right).\label{xdngrandefoc}
\end{eqnarray}
The relative intensity on the focal plane is therefore given by

\begin{eqnarray}
\hat{I}(\hat{z}_f,\bar{y}_f)
&=&~\exp\left[-\frac{\bar{y}_f^2}{2\hat{D}_y \hat{f}^2}
\right],\label{Ingrandefoc}
\end{eqnarray}
while the modulus of the spectral degree of coherence reads

\begin{eqnarray}
|{g}( \hat{z}_f,\bar{y}_f, \Delta \hat{y}_f)| &=&\exp\left[-
\frac{2\hat{N}_y \Delta{\hat{y}}_f^2 }{ \hat{f}^2}\right]\left|
\beta\left(\frac{\Delta
\hat{y}_f}{\hat{f}}\right)\right|~.\label{spdngrandefoc}
\end{eqnarray}
\begin{figure}
\begin{center}
\includegraphics*[width=140mm]{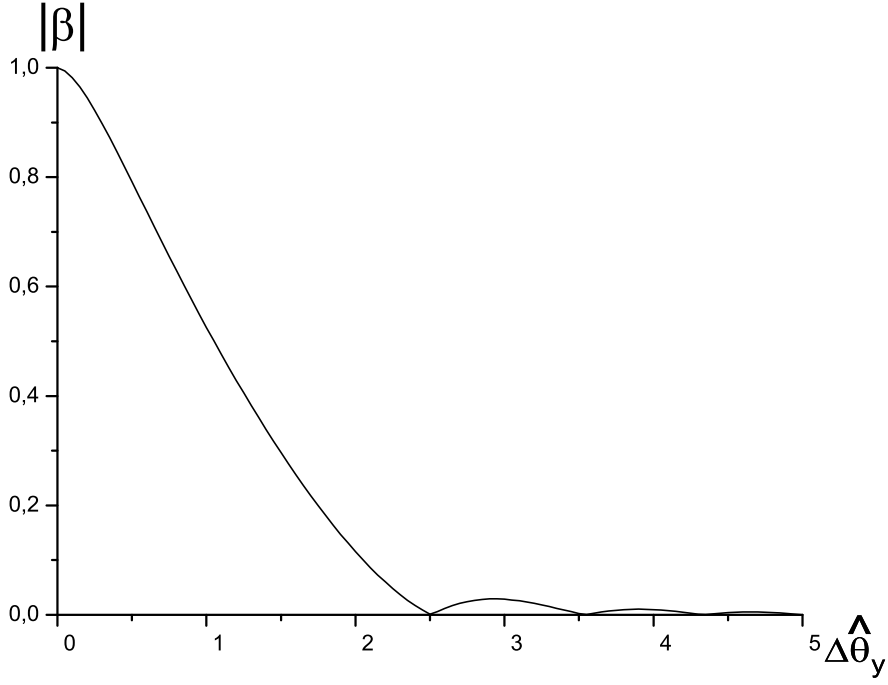}
\caption{\label{betaf} Absolute value of the universal function
$\beta$, used to calculate coherence on the focal plane of a
quasi-homogeneous undulator source when $\hat{N}_x \gg1$,
$\hat{D}_x \gg1$ and $\hat{D}_y \gg 1$. }
\end{center}
\end{figure}
A plot of $|\beta(\Delta \hat{\theta}_y)|$ is given in Fig.
\ref{betaf}. For the image plane, Eq. (\ref{fundima}) and Eq.
(\ref{Ngr3}) give the following cross-spectral density:

\begin{eqnarray}
\hat{G}( \hat{z}_i,{ \bar{y}}_i,\Delta{\hat{y}}_i)  &=&
\exp\left[\frac{i {\mathrm{m}(\mathrm{m}+1) \bar{y}}_i
\Delta{\hat{y}}_i }{2\hat{z}_1} \right]\cr &&\times  \exp\left[-2
\hat{D}_y \mathrm{m}^2\Delta \hat{y}_i^2 \right]
\int_{-\infty}^{\infty} d\eta
\exp\left[-\frac{\left(\eta+\mathrm{m} \bar{y}_i\right)^2}{2
\hat{N}_y} \right] \hat{\mathcal{B}}( \eta)~,\cr &&
\label{xdngandeimadoppio}
\end{eqnarray}
corresponding to a relative intensity on the image plane

\begin{eqnarray}
\hat{I}( \hat{z}_i,{ \bar{y}}_i)  &=& \int_{-\infty}^{\infty}
d\eta \exp\left[-\frac{\left(\eta+\mathrm{m} \bar{y}_i\right)^2}{2
\hat{N}_y} \right] \hat{\mathcal{B}}(
\eta)\Bigg/\left\{\int_{-\infty}^{\infty} d\eta
\exp\left[-\frac{\eta^2}{2 \hat{N}_y} \right] \hat{\mathcal{B}}(
\eta)\right\}~.\cr && \label{Ingandeima2}
\end{eqnarray}
The modulus of the spectral degree of coherence is

\begin{eqnarray}
|{g}( \hat{z}_i, \bar{y}_i, \Delta{\hat{y}}_i)|  &=& \exp\left[-2
\hat{D}_y \mathrm{m}^2\Delta \hat{y}_i^2 \right] ~.
\label{spdngandeima2}
\end{eqnarray}
Finally, in the particular case for $\hat{N}_y \ll 1$, Eq.
(\ref{Ingandeima2}) reduces to

\begin{eqnarray}
\hat{I}( \hat{z}_i,{ \bar{y}}_i)  &=&  \hat{\mathcal{B}}(
\mathrm{m} \bar{y}_i)~. \label{Ingandeima2redux}
\end{eqnarray}

\section{\label{sub:imfoge} Analysis of the image formation
mechanism for quasi-homogeneous undulator sources in terms of
Geometrical Optics}

In the Introduction we have stressed that the image formation
problem is twofold: one should be able to provide a
characterization of the virtual source as well as to track the
cross-spectral density of the source trough the optical beamline.

Let us first analyze the problem of source characterization. In
Section \ref{sec:intro} we have seen that, in the asymptotic limit
for a large electron beam emittance ${\epsilon}_{x,y} \gg
\lambda/(2\pi)$, Geometrical Optics may be used equally well as
Statistical Optics to fulfill this task. Here we will discuss more
in detail the relation between the Statistical Optics approach and
the Geometrical Optics approach with particular attention to the
applicability region of the latter.


Let us start with a remark, which applies not only to undulator
radiation sources but also to sources of other kind (e.g. bending
magnets). In the Introduction, in order to decide wether
Geometrical Optics or Wave Optics is applicable, we compared the
\textit{electron beam} emittance with the radiation wavelength.
This is acceptable in many cases when undulator radiation is
involved but not, for instance, when bending magnet radiation is
considered. In all generality one should separately compare the
\textit{photon beam} size and divergence with the radiation
diffraction size and diffraction angle, which are quantities
pertaining the single electron radiation. Let us fix a given
direction $x$ or $y$. The square of the diffraction angle is
defined by $(\sigma_d')^2 \sim \lambda/(2 \pi L_f)$, $L_f$ being
the formation length of the radiation at wavelength $\lambda$ as
defined in \cite{OUR1}. The diffraction size of the source is
given by $\sigma_d \sim \sigma'_d L_f$. In calculating the photon
beam size and divergence one should always include diffraction
effects. As a result, if $\sigma^2$ and $(\sigma')^{2}$ indicate
the square of the electron beam size and divergence, the
corresponding square of the photon beam size and divergence will
be respectively of order $\max[\sigma^2, \sigma_d^2]$ and
$\max[(\sigma')^{2}, (\sigma_d')^2]$. These quantities can be
rewritten in terms of the electron beam emittance  as
$\max[\epsilon \beta, \sigma_d^2]$ and $\max[\epsilon/\beta,
(\sigma'_d)^2]$, $\beta$ being the minimal beta function value,
defining the virtual source position for the radiator (undulator,
bending magnet, or other). Dividing these two quantities
respectively by $\sigma_d^2$ and $(\sigma'_d)^2$  give natural
values, normalized to unity, for the photon beam size $\max[2 \pi
\epsilon \beta/(L_f \lambda), 1]$ and divergence $\max[2 \pi
\epsilon L_f/ (\beta \lambda), 1]$. When the product between these
two quantities is much larger than unity one can use a Geometrical
Optics approach. In this case, this product represents the
normalized \textit{photon beam} emittance. When $\beta \sim L_f$,
as in many undulator cases, one may compare, for rough
estimations, the electron beam emittance and the radiation
wavelength as we have done before. However, in the case of a
bending magnet one may typically have $\beta$ of order $10$ m and
$L_f$ of order $10^{-3} \div 10^{-2} $ m. The ratio $\beta/L_f \gg
1$ now constitutes an extra large parameter of the problem. In
this case, even if the electron beam emittance is two order of
magnitude smaller than the wavelength, due to diffraction effects
one can still apply a Geometrical Optics approach, because $\max[2
\pi \epsilon \beta/(L_f \lambda), 1] \cdot \max[2 \pi \epsilon
L_f/ (\beta \lambda), 1] \gg 1$, i.e. the photon beam emittance is
much larger than the wavelength. As a result, dimensional analysis
suggests that bending magnet radiation may be treated exhaustively
in the framework of Geometrical Optics even for third generation
light sources.

As discussed above, when $\beta \sim L_f$ a large electron beam
emittance (compared with the radiation wavelength) is a necessary
and sufficient condition for the Geometrical Optic approach to
apply. In spite of that, when $\beta \gg L_f$ or $\beta \ll L_f$,
a large electron beam emittance is a sufficient, but not necessary
condition for the Geometrical Optic approach to be possibly used
for source characterization. Let us prove this statement with
undulator sources in mind\footnote{Note that, even though in the
case of undulator sources one often has $\beta \sim L_f$, there
are situations when $\beta \gg L_f$ or $\beta \ll L_f$ and when
the electron emittance is of order of the wavelength. However, in
the undulator case, very large values of the ratio $\beta/L_f$ of
order $10^3 \div 10^4$, typical of the bending magnet case, are
unrealistic.}. It is enough to prove that the wider class of
quasi-homogeneous sources, which includes situations when the
electron beam emittance is not larger than the wavelength, can be
described in terms of Geometrical Optics. 

Let us then consider the class of quasi-homogeneous virtual
sources for undulator devices. The cross-spectral density of the
virtual source (positioned at $z=0$, i.e. at the virtual source
plane) can be written as in Eq. (\ref{introh}), that we rewrite
here for convenience in terms of coordinates $\bar{r}_{x,y}$ and
$\Delta {\hat{r}_{x,y}}$:

\begin{eqnarray}
\hat{G}_o({\bar{r}_x},\bar{r}_y,\Delta {\hat{r}_x},\Delta
{\hat{r}_y}) = \hat{I}\left(\bar{r}_x,\bar{r}_y\right) {g}(\Delta
{\hat{r}_x},\Delta {\hat{r}_y}) ~.\label{introhdopo}
\end{eqnarray}
As usual,  the Fourier transform of Eq. (\ref{introhdopo}) with
respect to all variables will be indicated with

\begin{eqnarray}
\hat{\mathcal{G}}_o( {\bar{\theta}_x}, {\bar{\theta}_y},{\Delta
\hat{\theta}_x},\Delta \hat{\theta}_y) &=& \int_{-\infty}^{\infty}
d \Delta { \hat{r}_x'}\int_{-\infty}^{\infty}d \Delta {
\hat{r}_y'} \int_{-\infty}^{\infty} d {
\bar{r}_x'}\int_{-\infty}^{\infty}d
 { \bar{r}_y'} ~ \hat{G}_o({ \bar{r}_x'},{ \bar{r}_y'},
\Delta{\hat {r}'_x},\Delta{\hat {r}'_y}) \cr &&\times\exp [2i (
\bar{\theta}_x \Delta{ \hat{r}'_x}+\bar{\theta}_y \Delta{
\hat{r}'_y} )]\exp [2i ( \Delta \hat{\theta}_x {
\bar{r}'_x}+\Delta \hat{\theta}_y { \bar{r}'_y} )]~.\cr &&
\label{ftgdeffor}
\end{eqnarray}
The two quantities
$\hat{I}(\bar{r}_x,\bar{r}_y)=\hat{G}_o({\bar{r}_x},\bar{r}_y,0,0)$
and $\hat{\Gamma}({\bar{\theta}_x}, {\bar{\theta}_y})
=\hat{\mathcal{G}}_o( {\bar{\theta}_x}, {\bar{\theta}_y},0,0)$ are
always positive, because, by definition of $\hat{G}_o$, they are
ensemble averages of quantities under square modulus.

Let us now introduce the Fourier transform of Eq.
(\ref{introhdopo}) with respect to $\Delta {\hat{r}_{x,y}}$:

\begin{eqnarray}
\hat{\Phi}_o({\bar{r}_x},\bar{r}_y, {\bar{\theta}_x},
{\bar{\theta}_y}) &=& \int_{-\infty}^{\infty} d \Delta {
\hat{r}_x'}\int_{-\infty}^{\infty}d \Delta { \hat{r}_y'}~
\hat{G}_o({ \bar{r}_x},{ \bar{r}_y}, \Delta{\hat
{r}'_x},\Delta{\hat {r}'_y}) \cr &&\times\exp [2i ( \bar{\theta}_x
\Delta{ \hat{r}'_x}+\bar{\theta}_y \Delta{ \hat{r}'_y} )]~.
\label{wigdef}
\end{eqnarray}
Accounting for Eq. (\ref{introhdopo}), i.e. in the particular case
of a (virtual) quasi-homogeneous source, Eq. (\ref{wigdef}) can be
written as

\begin{eqnarray}
\hat{\Phi}_o({\bar{r}_x},\bar{r}_y, {\bar{\theta}_x},
{\bar{\theta}_y}) &=& \hat{I}\left(\bar{r}_x,\bar{r}_y\right)
\hat{\Gamma}({\bar{\theta}_x}, {\bar{\theta}_y})~,
\label{wigdefdef}
\end{eqnarray}
having recognized that $\hat{\Gamma}({\bar{\theta}_x},
{\bar{\theta}_y}) =\hat{\mathcal{G}}_o( {\bar{\theta}_x},
{\bar{\theta}_y},0,0)$ is the Fourier transform of the spectral
degree of coherence ${g}$. The distribution $\hat{\Phi}_o$, being
the product of two positive quantities, never assumes negative
values. Therefore it may always be interpreted as a phase space
distribution\footnote{It should be remarked that this result has
been obtained only on the ground of mathematical basis, i.e.
without ascribing to $\hat{I}$ and $\hat{\Gamma}$ any physical
meaning. In other words, we simply considered the cross-spectral
density $\hat{G}_o$ as the ensemble-averaged product of
$\hat{E}(\hat{\vec{r}}_1)$ and $\hat{E}^*(\hat{\vec{r}}_2)$
without ascribing to the function $\hat{E}$ any physical meaning.
Physically, as has been said in the Introduction, in the
quasi-homogeneous case $\hat{\Gamma}$ can be identified with the
radiant intensity of the virtual source (compare with Eq.
(\ref{phsp})). This follows from a statement similar to the van
Cittert-Zernike theorem for quasi-homogeneous sources (see
\cite{MAND}). Note that the intensity and the Fourier transform of
the spectral degree of coherence are obtained back from the phase
space distribution, Eq. (\ref{wigdefdef}), by integration over
coordinates $\bar{\theta}_{x,y}$ and $\bar{r}_{x,y}$
respectively.}. This analysis shows that quasi-homogeneous sources
can always be characterized in terms of Geometrical Optics. It
also shows that, in this particular case, the coordinates in the
phase space, $\bar{r}_{x,y}$ and $\bar{\theta}_{x,y}$, are
separable.

Eq. (\ref{wigdef}) is the definition of a Wigner distribution. In
the case of quasi-homogenous sources, as we have just seen, the
Wigner distribution is never negative and, therefore, can always
be interpreted as a phase space distribution.  In the case of non
quasi-homogeneous sources one may still define a Wigner
distribution using Eq. (\ref{wigdef}). The integral of the Wigner
function over its coordinates must still be finite\footnote{This
ensures that $\hat{\Phi}_o$ has finite integral over its
variables.  $\hat{\Phi}_o$ is the Fourier transform of a
correlation function (the cross-spectral density) of
electromagnetic fields. The fields being physical quantities can
carry only a finite amount of energy and they are limited in
spatial extent. As a result the cross-spectral density must be an
integrable function over its variables and so must be, by
definition, $\hat{\Phi}_o$. \label{foot7}}. However the Wigner
function itself is not always a positive function. As a
consequence it cannot always be interpreted as a phase space
distribution. On the one hand, quasi-homogeneity is a sufficient
condition for the Geometrical Optics approach to be possibly used
in the representation of the source. On the other hand though,
necessary and sufficient conditions for $\hat{\Phi}_o$ to be a
positive function are more difficult to find.

One may observe that Bochner's theorem\footnote{Bochner's theorem
"in its elementary form asserts that every non-negative definite
function of a broad class has a non-negative Fourier transform
and, conversely, that the Fourier transform of every non-negative
function of a broad class is non-negative definite. This class
includes functions which fall off sufficiently rapidly to infinity
to ensure that their Fourier transforms are continuous functions"
[cited from paragraph 1.4.2. of reference \cite{MAND}]. It should
hereby be stressed the difference in the mathematical language
between a positive function $f(\xi) \geqslant 0$ for every real
value $\xi$ and a \textit{non-negative definite} function. The
function $h$ is said to be non-negative definite when "for an
arbitrary set of $N$ real numbers $\xi_1$, $\xi_2$,...,$\xi_N$ and
$N$ arbitrary complex numbers $a_1$, $a_2$, ..., $a_N$,
$\sum_{i=1}^{N}\sum_{j=1}^{N} a_i^* a_j h(\xi_j-\xi_i) \geqslant
0$."[cited from paragraph 1.4.2. of reference \cite{MAND}]. Based
on the assumption of a quasi-homogeneous source authors of
\cite{MAND} use Bochner's theorem instead of our previous
discussion to demonstrate that the spectral degree of coherence
${g}$ in Eq. (\ref{introhdopo}) is necessarily non-negative
definite. In fact, since the intensity $\hat{I}$ is a positive
function,  the sign of $\sum_{i=1}^{N}\sum_{j=1}^{N} a_i^* a_j
{g}(\Delta \hat{r}_{x~ij},\Delta \hat{r}_{y~ij})$ is the same of
the sign of $\sum_{i=1}^{N}\sum_{j=1}^{N} a_i^* a_j
\hat{I}(\bar{r}_{x~ij},\bar{r}_{y~ij}){g}(\Delta
\hat{r}_{x~ij},\Delta \hat{r}_{y~ij})=\sum_{i=1}^{N}\sum_{j=1}^{N}
a_i^* a_j \left\langle E(\hat{r}_j) E^*(\hat{r}_i)\right\rangle =
\left\langle \left|\sum_{i=1}^{N} a_i
E(\hat{r}_i)\right|^2\right\rangle \geqslant 0$, \textit{quantum
erat demonstrandum}. This demonstration is more involved than
ours, even though it is based, as the ours, on the positivity of
the square modulus of quantities. The reason for this complexity
is that it uses a more general theorem, i.e. Bochner's theorem.
However, as we have seen, it is not necessary to invoke Bochner's
theorem in the quasi-homogeneous case. \label{foot6}} may be used
to investigate whether the Wigner function can be interpreted as a
phase space distribution in the case of non-homogeneous sources.
In particular, it is necessary and sufficient to look for
non-negative definite cross-spectral density functions. However,
in general, it is not trivial to investigate wether a function is
non-negative definite (see footnote \ref{foot6}) and therefore
this observation does not seem to constitute a simplification. We
will simply leave the search for necessary and sufficient
condition for $\hat{\Phi}_o$ to be a positive function as an open
question. We did not rule out, for undulator setups, the
possibility of having a positive Wigner distribution in the non
quasi-homogeneous case. At first glance it may look like such a
case brings advantages in the formulation of the imaging theory,
because the source can be described in terms of Geometrical
Optics. On the one hand, in the case the Wigner distribution is
positive,  the evolution of the radiation in free space can be
described by a ray-tracing approach, as the Wigner distribution
can be interpreted as a phase space distribution. On the other
hand though, such fact is almost irrelevant because it is not of
help when optical elements are considered. As we will see later
on, there are two particular conditions at the basis of a
simplified formulation of the imaging theory based on the
incoherent point spread function of the optical system. The first
is the separability of the cross-spectral density is the product
of a factor depending on $\vec{\bar{r}}$ and a factor depending on
$\Delta \vec{r}$. The second is a transverse dimension of the
source much larger than the transverse coherence length. As we
have already seen  these two conditions, together, define a
quasi-homogeneous source. Quasi-homogeneous sources are
necessarily characterized by a positive Wigner function. However,
the positivity of the Wigner function alone is not sufficient to
obtain a simplified formulation of the imaging theory in terms of
incoherent point-spread function. A similar remark holds for a
particular kind of sources often considered in literature also in
connection with undulator radiation (see \cite{COI1,COI2}). These
sources, characterized by a cross-spectral density $G =
\sqrt{I(r_1)}\sqrt{I(r_2)} g(r_1-r_2)$ are called Shell sources
(in particular, in \cite{COI1,COI2} Gaussian-Shell sources are
discussed which assume gaussian a profile for both $I$ and $g$).
They exhibit separability of the cross-spectral density, but are
not quasi-homogeneous because the transverse dimension of the
source fails to be much larger than the transverse coherence
length: a simplified formulation of the imaging theory does not
hold in this case either. Moreover, as it will be more extensively
discussed here below and in Section \ref{sec:supp}, the Shell
model (and, in particular, the Gaussian-Shell model) may be useful
for describing light sources other than undulator-based or for
educational purposes, but does not describe any practical
realization of an undulator source. 

We are now interested to find, in particular, equivalent
conditions for quasi-homogeneity in terms of the electron beam
sizes $\hat{N}_{x,y}$ and divergences $\hat{D}_{x,y}$ that apply
to our case of interest, i.e. third generation light sources.  In
order to do so we start deriving an expression $\hat{\mathcal{G}}$
for the Fourier transform of the cross-spectral density at the
virtual source position. This is given by calculating the limit of
Eq. (\ref{G2D}) for $\hat{z}_o \gg 1$ and taking advantage of Eq.
(\ref{maintrick2}). Aside for an inessential multiplicative
constant we obtain:

\begin{eqnarray}
\hat{\mathcal{G}}(0,\bar{\theta}_x,\bar{\theta}_y,\Delta
\hat{\theta}_x,\Delta \hat{\theta}_y) &&= \exp{\left[- 2\hat{N}_x
\Delta \hat{\theta}_x^2\right]} \exp{\left[- 2\hat{N}_y \Delta
\hat{\theta}_y^2\right]} \int_{-\infty}^{\infty} d \hat{\phi}_x
\int_{-\infty}^{\infty} d \hat{\phi}_y  \cr&& \times
\exp{\left[-\frac{\left(\hat{\phi}_x+\bar{\theta}_x\right)^2}{2
\hat{D}_x}\right]}
\exp{\left[-\frac{\left(\hat{\phi}_y+\bar{\theta}_y\right)^2}{2
\hat{D}_y}\right]}\cr && \times
\mathrm{sinc}{\left[\frac{(\hat{\phi}_x-\Delta
\hat{\theta}_x)^2+(\hat{\phi}_y-\Delta
\hat{\theta}_y)^2}{4}\right]} \cr &&\times
\mathrm{sinc}~{\left[\frac{(\hat{\phi}_x+\Delta
\hat{\theta}_x)^2+(\hat{\phi}_y+\Delta
\hat{\theta}_y)^2}{4}\right]}.\cr && \label{calGo}
\end{eqnarray}
We have said that the quasi-homogeneity of the virtual source is
equivalent to (i) separability of the cross-spectral density
$\hat{{G}}$ in the product of two factors respectively depending
on ${\bar{\theta}}_{x,y}$ and $\Delta {\hat{\theta}_{x,y}}$ and
(ii) a large characteristic scale of $\bar{\theta}_{x,y}$ with
respect to the characteristic scale of $\Delta
\hat{\theta}_{x,y}$. From condition (i) follows that the virtual
source is quasi-homogeneous \textit{only if} it is possible to
factorize the Fourier transform of the cross-spectral density,
$\hat{\mathcal{G}}$  in Eq. (\ref{calGo}), in the product of two
factors separately depending on ${\bar{\theta}_{x,y}}$ and $\Delta
{\hat{\theta}_{x,y}}$. Such factorization, for third generation
light sources, is equivalent to a particular choice of the region
of parameters for the electron beam: $\hat{N}_x \gg 1$, $\hat{D}_x
\gg 1$ and either (or both) $\hat{N}_y \gg 1$ and $\hat{D}_y \gg
1$\footnote{It should be remarked here, that these conditions
describe the totality of third generation quasi-homogeneous
sources. In fact, while a purely mathematical analysis indicates
that factorization of Eq. (\ref{calGo}) is equivalent to more
generic conditions ($\hat{N}_x \gg1$ and $\hat{N}_y \gg 1$, or
$\hat{D}_x \gg1$ and $\hat{D}_y \gg 1$), comparison with third
generation source parameters reduces such conditions to the
already mentioned ones.}. In this case, the second condition (ii)
is automatically verified as one can verify inspecting Eq.
(\ref{calGo}).

An intuitive picture in the real space is given by a (virtual)
quasi-homogeneous source with characteristic (normalized) square
sizes $\max(\hat{N}_{xy},1)$ and characteristic (normalized)
correlation length square of order $\min(1/\hat{D}_{x,y},1)$. As
already remarked before, in the quasi-homogeneous situation the
horizontal and the vertical directions can be treated separately,
because Eq. (\ref{calGo}) factorizes in the product of factors
separately depending on the horizontal and on the vertical
coordinates.  This corresponds to a large number of independently
radiating sources given by the product

\begin{equation}
M_{x,y}= \max(\hat{N}_{x,y},1)\max(\hat{D}_{x,y},1)~.
\label{modesappr} \end{equation}
The number $M_{x,y}$ is, in other words, an estimation of the
number of coherent modes in the horizontal and in the vertical
direction\footnote{This is in agreement with an intuitive picture
where the photon-beam phase space reproduces the electron-beam
phase space up to the limit imposed by the intrinsic diffraction
of undulator radiation. Imagine to start from a situation with
$\hat{N}_{x,y}\gg 1$ and $\hat{D}_{x,y} \gg 1$ and to "squeeze"
the electron-beam phase space by diminishing $\hat{N}_{x,y}$ and
$\hat{D}_{x,y}$. On the one hand the characteristic sizes of the
phase space of the electron beam are always of order
$\hat{N}_{x,y}$ and $\hat{D}_{x,y}$. On the other hand the
characteristic sizes of the phase space of the photon beam are of
order $\max(\hat{N}_{x,y},1)$ and $\max(\hat{D}_{x,y},1)$:
diffraction effects limit the "squeezing" of the phase space of
the photon beam. }. The number $M_{x,y}^{-1}$ is the accuracy of
Geometrical Optics results compared with Statistical Optics
results or, better, the accuracy of the quasi-homogeneous
assumption. It should be noted that, as $M_{x,y}$ approaches
unity, the accuracy of the quasi-homogeneous assumption becomes
worse and worse and $M_{x,y}$ cannot be taken anymore as a
meaningful estimation of the number of modes: it should be
replaced by a more accurate concept based on Statistical Optics.
To complete the previous statement we should add that $M_{x,y}$
completely loses the meaning of "number of modes" when Geometrical
Optics cannot be applied. For instance when both $\hat{N}_y$ and
$\hat{D}_y$ are of order unity (or smaller), one can state that
the Geometrical Optics approach fails in the vertical direction
because the phase space area is getting near to the uncertainty
limit. In this case it is not possible to ascribe the meaning of
"number of modes" to the number $M_y$ simply because the
Geometrical Optics approach in the vertical direction fails.
However, when $\hat{N}_y$ and $\hat{D}_y$ are of order unity (or
smaller), but both $\hat{N}_x \gg 1$ and $\hat{D}_x \gg 1$, the
cross-spectral density admits factorization in the horizontal and
in the vertical direction and the source in the horizontal
direction can be still described, independently, with the help of
Geometrical Optics.

Up to now we discussed about the roles of Geometrical and
Statistical optics in the characterization of the source only.
However, as already remarked, the specification of the source
constitutes only part of the solution of the imaging problem. One
has, in fact, to track information regarding the source through
the optical beamline up to the observation plane. Depending on the
situation Geometrical Optics may be used or not. For instance, a
quasi-homogeneous source may well be described in terms of a phase
space distribution, but if diffraction effects dominate the photon
beam transport to the observation plane, one cannot use
ray-tracing techniques to calculate the intensity profile at the
observation plane.  However, as we will see in the next Section,
if the virtual source is quasi-homogeneous, the intensity at the
observation plane can always be expressed as a convolution product
between the impulse response of the optical system and the
intensity which would be recovered at the observation plane in the
case of an ideal optical system (i.e. one with no aberration and
non-limiting pupil apertures). In this case, the entire line may
be studied with the help of ray-tracing programs if and only if
the impulse response of the system can be recovered by means of
Geometrical Optics techniques.

In Geometrical Optics, a Hamiltonian description of the optical
system holds so that interaction with optical media (i.e. the
system evolution) is conveniently modelled in terms of symplectic
transformations. A given symplectic transformation $\mathcal{S}$
acts on point $\tilde{\rho}_o=({\bar{r}}_{ox},
{\bar{\theta}}_{ox},{\bar{r}}_{oy}, {\bar{\theta}}_{oy})$ of the
phase space $\hat{\Phi}_o$ at $\hat{z}_o=0$ and maps it to a point
$\tilde{\rho}=({\bar{r}}_{x}, {\bar{\theta}}_{x},{\bar{r}}_{y},
{\bar{\theta}}_{y})$, of the phase space $\hat{\Phi}_{\hat{z}}$ at
$\hat{z}_o=\hat{z}$ according to

\begin{equation}
\tilde{\rho}=\mathcal{S}(\tilde{\rho}_o)~. \label{trsrh}
\end{equation}
The phase space distribution is therefore transformed according to

\begin{equation}
\hat{\Phi}_{\hat{z}}\left(\tilde{\rho}\right) =
\hat{\Phi}_o\left[\mathcal{S}^{-1}\left(\tilde{\rho}\right)\right]~.
\label{tranphsp}
\end{equation}
According to Liouville's theorem, the area of the phase space is
conserved during this process. In the particular case of linear
transformations, one can use a matrix formalism. $\mathrm{N}$
successive linear transformations are represented by $\mathrm{N}$
matrices $\mathrm{L}_1$ ... $\mathrm{L}_\mathrm{N}$ and the
resulting transformation is represented by $\mathrm{N}$ successive
matrix multiplications, which give the matrix $\mathrm{L} =
\mathrm{L}_\mathrm{N} \cdot ... \cdot \mathrm{L}_1$. The action of
$\mathrm{L}$ on an element of the phase space is then naturally
represented by multiplication. The variables $\tilde{\rho}$ in
phase space characterize a ray with a certain direction and offset
with respect to the optical axis. The task of calculating the
phase space distribution after a given number of optical elements
through

\begin{equation}
\hat{\Phi}_{\hat{z}}\left(\tilde{\rho}\right) =
\hat{\Phi}_o\left[\mathrm{L}^{-1}\cdot
\left(\tilde{\rho}\right)^{\mathrm{t}}\right]~, \label{tranphsp2}
\end{equation}
where $\mathrm{t}$ indicates transposition, or through the more
general Eq. (\ref{tranphsp}) can be solved by ray-tracing
programs. Once $\hat{\Phi}_{\hat{z}}$ is known, these codes
usually integrate it over the variable ${\bar{\theta}_x}$ and
${\bar{\theta}_y}$ to give the intensity distribution

\begin{equation}
\hat{I}\left(\hat{z},\bar{r}_x,\bar{r}_y\right) =
\int_{-\infty}^{\infty} d  {\bar{\theta}_x}
\int_{-\infty}^{\infty} d {\bar{\theta}_y}~
\hat{\Phi}_{\hat{z}}\left({\bar{r}}_{x},
{\bar{\theta}}_{x},{\bar{r}}_{y}, {\bar{\theta}}_{y}\right)
~.\label{intephsp}
\end{equation}
However, the same programs may also be used to calculate the
Fourier transform of the spectral degree of coherence through

\begin{equation}
\hat{\Gamma}\left(\hat{z},{\bar{\theta}_x},
{\bar{\theta}_y}\right) = \int_{-\infty}^{\infty} d {\bar{r}_x}
\int_{-\infty}^{\infty} d {\bar{r}_y}~
\hat{\Phi}_{\hat{z}}\left({\bar{r}}_{x},
{\bar{\theta}}_{x},{\bar{r}}_{y},{\bar{\theta}}_{y})\right)
~.\label{degcohphsp}
\end{equation}
In particular, in free space, Eq. (\ref{tranphsp2}) becomes

\begin{eqnarray}
\hat{\Phi}_{\hat{z}}\left({\bar{r}}_{x},
{\bar{\theta}}_{x},{\bar{r}}_{y}, {\bar{\theta}}_{y}\right)&=&
\hat{I} \left(0,{\bar{r}}_x-\hat{z}
{\bar{\theta}}_x,{\bar{r}}_y-\hat{z} {\bar{\theta}}_y\right)
\hat{\Gamma}\left(0,{\bar{\theta}}_x, {\bar{\theta}}_y\right)~,\cr
&& \label{free1}
\end{eqnarray}
while Eq. (\ref{intephsp}) and Eq. (\ref{degcohphsp}) reduce to
convolutions:

\begin{eqnarray}
\hat{I}\left(\hat{z},\bar{r}_x,\bar{r}_y\right) &=&
\int_{-\infty}^{\infty} d {\bar{\theta}_x} \int_{-\infty}^{\infty}
d {\bar{\theta}_y}~ \hat{I} \left(0,{\bar{r}}_x-\hat{z}
{\bar{\theta}}_x,{\bar{r}}_y-\hat{z} {\bar{\theta}}_y\right)
\hat{\Gamma}\left(0, {\bar{\theta}}_x, {\bar{\theta}}_y\right)
~\label{free2}
\end{eqnarray}
and

\begin{eqnarray}
\hat{\Gamma}\left(\hat{z},{\bar{\theta}_x},
{\bar{\theta}_y}\right) &=& \int_{-\infty}^{\infty} d {\bar{r}_x}
\int_{-\infty}^{\infty} d {\bar{r}_y}~ \hat{I}
\left(0,{\bar{r}}_x-\hat{z} {\bar{\theta}}_x,{\bar{r}}_y-\hat{z}
{\bar{\theta}}_y\right) \hat{\Gamma}\left(0, {\bar{\theta}}_x,
{\bar{\theta}}_y\right)\cr &=& \hat{\Gamma}\left(0,
{\bar{\theta}}_x, {\bar{\theta}}_y\right) ~.\label{free3}
\end{eqnarray}
Note that $\hat{\Gamma}$ calculated at $\hat{z}=0$ has direct
physical sense as the intensity distribution in the far zone, i.e.
the angular spectrum. Then, Eq. (\ref{free3}) tells that, at
arbitrary distance $\hat{z}$, the angular spectrum does not vary.

The intensity recovered at the image plane in the case of an ideal
optical system is a scaled copy of that at the virtual source,
regardless of the source. Generally, although as we will see
exceptions apply, such correspondence between the intensity of the
source and the observed intensity is only true in the case the
observation plane is the image plane. In the case of a
quasi-homogeneous virtual source, Geometrical Optics as well as
Statistical Optics techniques can be employed to recover the
intensity at the observation plane. Results from the Geometrical
Optics and from the Statistical Optics approach must then
coincide. Let us prove this fact considering the particular case
$\hat{N} \gg 1$ and $\hat{D} \gg 1$ in a given direction and
showing that we are able to recover Eq. (\ref{intxexpl2}) by means
of Geometrical Optics techniques, namely by means of the matrix
formalism employed in ray-tracing codes.

In this particular situation, the photon beam can be modelled as
if a Gaussian photon beam was present at $\hat{z}=0$ with the same
horizontal phase space of the electron beam. This is an
\textit{ansatz} on the virtual quasi-homogeneous source based on
the phase space picture described above since strictly speaking it
does not make sense to talk about a Gaussian photon beam inside
the undulator, i.e. within the radiation formation length. If,
however, this \textit{ansatz} is made, we can describe the optical
equivalent of the Twiss matrix at $\hat{z}=0$. Let us first
introduce the notion of normalized Twiss parameters as:

\begin{eqnarray}
\hat{\alpha}_T&=& \alpha_T~,\cr \hat{\beta}_T &=& L_w^{-1}
\beta_T~,\cr \hat{\gamma}_T &=& L_w \gamma_T~,\cr
\hat{\epsilon}~~&=& (\omega/c) \epsilon~, \label{adquaV}
\end{eqnarray}
where ${\alpha}_T$, ${\beta}_T$ and ${\gamma}_T$ are the Twiss
parameters and ${\epsilon}$ is the emittance pertaining the photon
beam \footnote{The Twiss parameters are the second moments of the
phase space distribution of the photons divided by the
emittance.}. In the case under study they are identical to the
analogous electron beam parameters. We have

\begin{equation}
\sigma_{|_{\hat{z}=0}} \equiv \hat{\epsilon} \left(
\begin{array}{cc}
\hat{\beta}_T(0) & -\hat{\alpha}_T(0) \\ -\hat{\alpha}_T(0) &
\hat{\gamma}_T(0)
\end{array}\right) = \left(
\begin{array}{cc}
\hat{N} & 0 \\ 0 & \hat{D}
\end{array}\right)~,
\label{sigma0}
\end{equation}
For this exemplification we will assume a non-limiting pupil
aperture. Then, the linear transformation mapping a phase-space
point in $\hat{z}=0$ to a phase-space point in $\hat{z}=\hat{z}_2$
is represented by the matrix $\mathrm{L}$. In our particular case
of interest we have

\begin{eqnarray}
\mathrm{L} &=& \left(
\begin{array}{cc}
1 & ~~\hat{z}_2-\hat{z}_1 \\ 0 & 1
\end{array}\right) \cdot
\left(
\begin{array}{cc}
1 & 0 \\ -1/\hat{f}~~ & 1
\end{array}\right) \cdot
\left(
\begin{array}{cc}
 1~~ & \hat{z}_1 \\ 0 ~~& 1
\end{array}\right)\cr &&= \left(
\begin{array}{cc}
-{\hat{z}_2}/{\hat{z}_1}+1 ~& ~0  \\ ~&~ \\
{\hat{z}_2}/({\hat{z}_1^2-\hat{z}_1\hat{z}_2})~ &~
{\hat{z}_1}/({\hat{z}_1-\hat{z}_2})
\end{array}\right)~.
\label{M}
\end{eqnarray}
As one can see from Eq. (\ref{M}), $L$ describes a free-space
flight followed by a focusing element and a second free-space
flight. A point $(\hat{l}_o,\hat{\eta}_{o})$ of the photon beam
phase space at $\hat{z}=0$ is transformed, at $\hat{z}=\hat{z}_2$
into

\begin{equation}
\left(
\begin{array}{c}
\hat{l}_1 \\ \hat{\eta}_{1}
\end{array}\right) = \mathrm{L} \left( \begin{array}{c}
\hat{l}_o \\ \hat{\eta}_{o}
\end{array}\right)~,
\label{transM}
\end{equation}
while the Twiss parameters for the photon beam at $\hat{z}
=\hat{z}_2$ are described by the matrix

\begin{eqnarray}
\sigma_{|_{\hat{z}=\hat{z}_2}} &=& \mathrm{L}\cdot
\sigma_{|_{\hat{z}=0}}\cdot \mathrm{L}^{\mathrm{t}} =
\hat{\epsilon} \left(
\begin{array}{cc}
\hat{\beta}_T(\hat{z}_2) & -\hat{\alpha}_T(\hat{z}_2) \\
-\hat{\alpha}_T(\hat{z}_2) & \hat{\gamma}_T(\hat{z}_2)
\end{array}\right)\label{sigmaz2}
\end{eqnarray}
with

\begin{eqnarray}
\hat{\epsilon}\hat{\alpha}_T(\hat{z}_2)&=&
\frac{1}{\left(\hat{A}+\hat{D}\right)}\Bigg\{\hat{z}_1\left[\hat{A}+\left(\hat{A}+\hat{D}\right)
Q \hat{z}_1 \right]\left[\hat{D}+\left(\hat{A}+\hat{D}\right) Q
\hat{z}_1 \right] \cr &&
-\left[\hat{A}\hat{D}+\left(\hat{A}+\hat{D}\right)^2 Q^2
\hat{z}_1^2 \right]\hat{z}_2\Bigg\}~,\label{alpha}
\end{eqnarray}
\begin{equation}
\hat{\epsilon}\hat{\beta}_T(\hat{z}_2) =
\frac{\left[\hat{A}\hat{D}+\left(\hat{A}+\hat{D}\right)^2 Q^2
\hat{z}_1^2
\right](\hat{z}_1-\hat{z}_2)^2}{\hat{A}+\hat{D}}\label{beta}
\end{equation}
and

\begin{eqnarray}
\hat{\epsilon}\hat{\gamma}_T(\hat{z}_2)&=&
\frac{\hat{A}\hat{D}}{\hat{A}+\hat{D}}+\left(\hat{A}+\hat{D}\right)Q^2
\hat{z}_1^2 +\frac{\left(\hat{A}+\hat{D}\right)\hat{z}_1^2}
{(\hat{z}_1-\hat{z}_2)^2} +\frac{2\left(\hat{A}+\hat{D}\right)
Q\hat{z}_1^2}{\hat{z}_1-\hat{z}_2}~. \cr && \label{gamma2}
\end{eqnarray}
It should be recalled that parameters $Q$ and $\hat{A}$ have been
defined in Eq. (\ref{Qfact}) and Eq. (\ref{a2setbbis}).

The photon phase space distribution at $\hat{z}=\hat{z}_2$ is
described by

\begin{equation}
f_{|_{\hat{z}=\hat{z}_2}}= \frac{1}{2\pi
\hat{\epsilon}}\exp\left[-\frac{\hat{\gamma}_T(\hat{z}_2)\hat{l}_1^2+
2\hat{\alpha}_T(\hat{z}_2)\hat{l}_1\hat{\eta}_{1} +
\hat{\beta}_T(\hat{z}_2)\hat{\eta}_{1}^2 }{2
\hat{\epsilon}}\right]~. \label{distrph}
\end{equation}
The relative intensity  is derived from Eq. (\ref{distrph})
integrating over the $\hat{\eta}_{1}$-coordinate, which gives

\begin{equation}
I(\hat{z}_2) = \exp{\left[-\frac{2 \hat{l}_1^2}{2
\hat{\epsilon}\hat{\beta}_T(\hat{z}_2)}\right]}~. \label{integeo}
\end{equation}
Finally, substitution of the expression for
$\hat{\epsilon}\hat{\beta}_T(\hat{z}_2)$ obtained in Eq.
(\ref{beta}), yields back Eq. (\ref{intxexpl2}), as it should be.
Similar conclusions may be obtained for the spectral degree of
coherence integrating over the $\hat{l}_1$-coordinate and inverse
Fourier transforming the result.


In spite of these results, we should stress again that Statistical
Optics is the only mean to deal with the stochastic nature of
Synchrotron Radiation in general. Only in particular cases
Synchrotron Radiation can be treated in terms of Geometrical
Optics. As we have just discussed, one of these cases is
constituted by second generation light sources, when
$\hat{N}_{x,y} \gg 1$ and $\hat{D}_{x,y} \gg 1$. Experiments in
this region of parameters can take advantage of ray-tracing code
techniques.

\begin{figure}
\begin{center}
\includegraphics*[width=110mm]{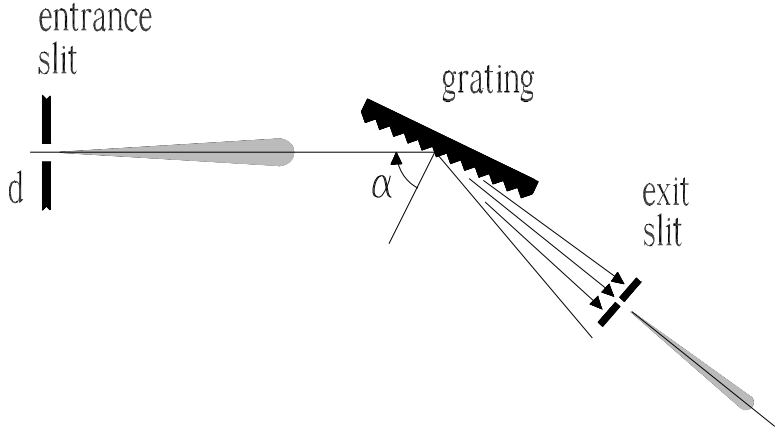}
\caption{\label{mono} Illustration of a grating monochromator. }
\end{center}
\end{figure}

To conclude this Section,  we would like to make a much stronger
statement: there are practical cases of interest when the
Statistical Optics approach must be used even for second
generation light sources. This should not sound too awkward since,
as we have stated before, the impulse response of an optical line
may not be treatable in terms of Geometrical Optics. Consider, for
instance, the setup illustrated in Fig. \ref{mono}.
This consists of an entrance slit, a grating and an exit slit,
that is a grating monochromator. The grating equation, which
describes how the monochromator works, relies on the principle of
interference applied to the light coming from adjacent grooves.
Such principle though, can only be applied when phase and
amplitude variations of the electromagnetic field are well defined
across the grating, that is when the field is perfectly
transversely coherent. If the transversely coherent spot of the
radiation is smaller than the grating, not all the grating is
taken advantage of, resulting in a decrease of resolution in
wavelength. To better explain this point, with reference to Fig.
\ref{mono}, let us indicate the width of the entrance slit with
$d$ and the angle of incidence of the incoming radiation with
$\alpha$. Moreover, let $D_g$ be the typical dimension of the
grating, $N_g$ the total groove number, $m$ the order of
diffraction and $z$ the distance between the entrance slit and the
grating. The maximal (relative) resolution which can be obtained
with a particular grating is given by $(m N_g)^{-1}$.
Qualitatively, to obtain such maximal resolution we must have a
transverse coherence area of at least the size $D_g \cos
(\alpha)$. If it is smaller, not all the grating is used. We now
need to transform this qualitative requirement into a quantitative
requirement.

In practical situations, the grating is placed in the far zone
with respect to the entrance slit. This is because the radiation
spot size at the grating should be at least of order $D_g
\cos(\alpha)$, which is much larger than the slit aperture $d$. If
we assume the slit uniformly illuminated, we can consider the slit
itself like a quasi-homogeneous source with rectangular profile.
Then, the van Cittert-Zernike theorem applies at the grating
position in the far zone. As a result, the modulus of the spectral
degree of coherence $|g|$ in the far zone is equal to the modulus
of the Fourier transform of the intensity profile at the slit,
which is a rectangular profile. The following expression for $|g|$
is found in the dispersion direction:

\begin{equation}
\left|g(\Delta r)\right| = \left|\mathrm{sinc}\left(\frac{\pi d
\Delta r}{\lambda z} \right)\right| ~.\label{spdcoh}
\end{equation}
A quantitative requirement for the coherence property of the
radiation at the grating can be given imposing that $|g|$ varies
within a fixed interval. For instance, one may require $0.8 <
\left|g(\Delta r)\right| < 1$. This requirement may be changed,
and is somewhat subjective. However it corresponds to a
quantification of the transverse coherence properties of the
radiation on the grating. In particular, if the criterium $0.8 <
\left|g(\Delta r)\right| < 1$ is chosen, the argument of the
$\mathrm{sinc}$ function in Eq. (\ref{spdcoh}) is allowed to vary
a certain range $[-X,X]$ with $X \simeq 1.13$. Moreover, the
maximal value for $\Delta r$ is $D_g \cos(\alpha)$, i.e. the
grating dimension. Putting all together we obtain the following
condition for the monochromator setup parameters:

\begin{equation}
\frac{\pi d D_g \cos(\alpha)}{\lambda z} = X \simeq 1.13~.
\label{spdcoh2}
\end{equation}
The same calculation may be repeated with a different choice for
the minimal allowed value of $\left|g(\Delta r)\right|$. This
would lead to a different value of $X$.

Our result is in agreement with the conclusion that one may draw
considering the following relation \cite{WEST}:

\begin{equation}
\frac{\Delta \lambda}{\lambda} = \left( \frac{d D_g \cos(\alpha)
}{\lambda z}\right) \frac{1}{  m N_g}~. \label{reso}
\end{equation}
Eq. (\ref{reso}) describes how the entrance slit width limits the
resolution in wavelength according to. The second factor on the
right hand side of Eq. (\ref{reso}), i.e. $(m N_g)^{-1}$ is,
again, the maximal relative resolution. This resolution can be
obtained by setting the first factor to unity. This yields a
result in parametrical agreement with Eq. (\ref{spdcoh2}). The
right parametric dependence in the condition for the maximal
resolution can also be obtained in another way. If the radiation
sent through the slit has, at the grating, a spot size equal to
$D_g \cos (\alpha)$, the condition for transverse coherence is
given by the space-angle product:

\begin{equation}
d\cdot\theta = \frac{d D_g \cos(\alpha) }{2z} \simeq \frac {
\lambda}{2\pi} ~.\label{spangle}
\end{equation}
Again, the qualitative estimation in Eq. (\ref{spangle}) is in
parametrical agreement with the quantitative calculation in Eq.
(\ref{spdcoh}).

The grating works with a resolution near to the theoretical limit
$(m N_g)^{-1}$ only with transversely coherent radiation. We may
say that the purpose of the entrance slit is to supply a
transversely coherent radiation spot at the grating in order to
allow the monochromator to work with a certain resolution. This
fact must hold for any light source, and in particular for second
generation light sources. The bottom line is that this
monochromator setup cannot be described in terms of Geometrical
Optics even in the case of a second generation light source:
transversely coherent radiation means that the image on the exit
slit is close to the diffraction limit. Therefore, in this case,
Geometrical Optics can only be used for approximate estimations,
while a correct treatment must involve the application of
Statistical Optics techniques.

\section{\label{par:ngdgb} Imaging of quasi-homogeneous undulator sources: effect of aperture size}

We will now consider, with the help of Eq. (\ref{GPIn}), the
effects of a pupil in the one-dimensional case (that can be
practically realized with the help of a slit aperture and a
cylindrical lens) when aberrations are not present.  First, in
Section \ref{sub:qhpup}, we will consider Gaussian
quasi-homogeneous sources. Then, in Section \ref{sub:emoco} we
will see that the arguments for Gaussian sources can be
generalized without modifying their substance with the help of
some notational change to treat the case of non-Gaussian
quasi-homogeneous sources as well.

\subsection{\label{sub:qhpup} Quasi-homogeneous Gaussian undulator sources}

In the case for $\hat{N}_x \gg 1$ and $\hat{D}_x \gg 1$ we can
treat the horizontal and the vertical direction separately. Then,
also the function $P$ and $\hat{\mathcal{P}}$ can be expressed as
the product of factors in the horizontal and in the vertical
direction. In particular we may consider the pupil function

\begin{eqnarray}
P({\hat{r}}) &=& \left \{
\begin{tabular}{c}
1 ~~~~~~if~$|{\hat{r}}|<\hat{a}$\\
0 ~~~~otherwise~
\end{tabular}
\right.~, \label{Pyfunc}
\end{eqnarray}
where $\hat{r}$ may represent either the variable $\hat{x}$ or the
variable $\hat{y}$. According to the definition in Eq.
(\ref{PFThat}) this gives:

\begin{equation}
\hat{\mathcal{P}}(\hat{u}) =  2 \hat{a} \mathrm{sinc}(\hat{a}
\hat{u}) \label{verpy}~.
\end{equation}
We can use Eq. (\ref{GPIn}) and Eq. (\ref{crspecpropimbist}) to
describe the case when the lens is in the far zone, that is when
condition (\ref{farlens}) is satisfied. From Eq. (\ref{crossx1l})
we can estimate the characteristic size of the source, that is of
order $\sqrt{\hat{N}}$, and of the correlation length at the
source, that is of order $1/\sqrt{\hat{D}}$. According to
condition (\ref{farlens}), the lens is in the far zone when
$\sqrt{\hat{N}}/\hat{z}_1 \ll \sqrt{\hat{D}}$, in agreement with
the far zone limit of Eq. (\ref{crossx1}), which is obtained for
$\hat{A} \ll \hat{D}$\footnote{Condition (\ref{farlens}) is
usually not discussed in textbooks describing thermal sources. In
fact, for perfectly incoherent thermal sources, the far zone is
defined by $\hat{z} \gg \hat{\sigma}_o$, $\sigma_o$ being the
source transverse size, i.e. when the paraxial approximation is
valid. Therefore, in this case, the pupil is always in the far
zone. }. In this limit, Eq. (\ref{GPIn}) and Eq.
(\ref{crspecpropimbist}) give

\begin{eqnarray} \hat{G}_P(\hat{z}_i,
{\bar{r}}_i, \Delta {\hat{r}}_i)&=&4 \hat{a}^2
\exp\left[\frac{2i\mathrm{m}}{\hat{z}_1}{\bar{r}}_i\Delta{\hat{r}}_i\right]
\cr && \times \int d{\bar{u}} ~d \Delta {\hat{u}}~
\exp\left[-\frac{\hat{z}_1^2 \bar{u}^{2}}{2\hat{N}}\right]
\exp\left[-2{ {\hat{D}}\hat{z}_1^2 (\Delta \hat{u})^{2}
}\right]\cr &&\times\mathrm{sinc}\left\{\hat{a}\left[\frac{
\mathrm{m}}{\hat{z}_1}\left({\bar{r}}_i+ \Delta
{\hat{r}}_i\right)-{\bar{u}}- \Delta {\hat{u}}\right]\right\}\cr
&& \times\mathrm{sinc}\left\{\hat{a}\left[\frac{
\mathrm{m}}{\hat{z}_1}\left({\bar{r}}_i- \Delta
{\hat{r}}_i\right)-{\bar{u}}+\Delta {\hat{u}}\right]\right\} ~.
\label{GPInb1}
\end{eqnarray}
According to the far field limit  of Eq. (\ref{crossx1}), the
quantity $\hat{D} \hat{z}_1^2$ is the square of the radiation spot
size on the pupil, while $\hat{z}_1^2/\hat{N}$ is essentially  the
square of the coherence length on the pupil. Two interesting
limiting cases of Eq. (\ref{GPInb1}) can be obtained comparing
these two characteristic scales with $\hat{a}^2$, that is the
square of the pupil size.

First, let us consider the case $ \hat{z}_1^2/\hat{N} \lesssim
\hat{a}^2 \ll \hat{D} \hat{z}_1^2 $. As we will demonstrate later
on, in all situations when the quasi-homogeneous assumption is
verified, the exponential function in $\Delta \hat{u}$ inside the
integral in Eq. (\ref{GPInb1}) behaves like a $\delta$-Dirac
distribution, and one obtains

\begin{eqnarray} \hat{G}_P(\hat{z}_i,
{\bar{r}}_i, \Delta {\hat{r}}_i)&=&4 \hat{a}^2
\exp\left[\frac{2i\mathrm{m}}{\hat{z}_1}{\bar{r}}_i\Delta{\hat{r}}_i\right]
\int_{-\infty}^{\infty}  d{\bar{u}}~ \exp\left[-\frac{\hat{z}_1^2
\bar{u}^{2}}{2\hat{N}}\right] \cr
&&\times\mathrm{sinc}\left\{\hat{a}\left[\frac{
\mathrm{m}}{\hat{z}_1}\left({\bar{r}}_i+ \Delta
{\hat{r}}_i\right)-{\bar{u}}\right]\right\}\cr &&
\times\mathrm{sinc}\left\{\hat{a}\left[\frac{
\mathrm{m}}{\hat{z}_1}\left({\bar{r}}_i- \Delta
{\hat{r}}_i\right)-{\bar{u}}\right]\right\} ~. \label{GPInb55}
\end{eqnarray}
This corresponds to a relative intensity

\begin{eqnarray} \hat{I}_P(\hat{z}_i,
{\bar{r}}_i)&=& \frac{1}{\mathcal{C}} \int_{-\infty}^{\infty}
d{\bar{u}}~ \exp\left[-\frac{\hat{z}_1^2
\bar{u}^{2}}{2\hat{N}}\right] \left|
\mathrm{sinc}\left[\hat{a}\left(\frac{
{\bar{r}}_i}{\hat{d}_i}-{\bar{u}}\right)\right]\right|^2~,
\label{GPInb55in}
\end{eqnarray}
where the normalization constant $\mathcal{C}$ is given by

\begin{eqnarray}
\mathcal{C} &=& \int_{-\infty}^{\infty} d{\bar{u}}~
\exp\left[-\frac{\hat{z}_1^2 \bar{u}^{2}}{2\hat{N}}\right] \left|
\mathrm{sinc}\left(\hat{a}{\bar{u}}\right)\right|^2 ~.
\label{mathcalb}
\end{eqnarray}
Eq. (\ref{GPInb55in}) expresses the image as a convolution of the
geometrical image with the slit diffraction pattern (in two
dimensions this would be the Airy pattern). It is valid for values
$\hat{z}_1^2/\hat{N} \lesssim \hat{a}^2\ll \hat{D} \hat{z}_1^2$
and also for $\hat{z}_1^2/\hat{N}  \ll \hat{a}^2\ll \hat{D}
\hat{z}_1^2$: the difference between these two cases is that in
the first the pupil influence is significant, while in the second
it is not. Inspection of Eq. (\ref{GPInb55in}) and use of Eq.
(\ref{magn2}) yields the ratio between the size of the diffraction
pattern and the geometrical image:
$\left(|\mathrm{M}|\sqrt{\hat{N}}\hat{a}/\hat{d}_i\right)^{-1} =
\left(\hat{a} \sqrt{\hat{N}}/\hat{z}_1\right)^{-1}$. When
$\hat{z}_1^2/\hat{N} \lesssim \hat{a}^2$ such ratio is comparable
with unity, i.e. the diffraction pattern significantly influences
the image formation process. When, $\hat{z}_1^2/\hat{N} \ll
\hat{a}^2$, this ratio is much smaller than unity. As a result,
the pupil influence is not significant and the image is given by
the geometrical image. The ratio between the size of the
diffraction pattern and the geometrical image, $\hat{z}_1/\hat{a}
\sqrt{\hat{N}}$, gives the resolution of the image due to
diffraction effects. It is also interesting to note that in the
limiting case when $\hat{z}_1^2/\hat{N}  \ll \hat{a}^2\ll \hat{D}
\hat{z}_1^2$, Eq. (\ref{GPInb55}) presents the asymptotic behavior

\begin{eqnarray} \hat{G}_P(\hat{z}_i,
{\bar{r}}_i, \Delta {\hat{r}}_i)&=&4 \hat{a}^2
\exp\left[\frac{2i\mathrm{m}}{\hat{z}_1}{\bar{r}}_i\Delta{\hat{r}}_i\right]
\exp\left[-\frac{\mathrm{m}^2\bar{r}_i^{2}}{2\hat{N}}\right] \cr
&&\times \int_{-\infty}^{\infty}  d{\bar{u}}~
\mathrm{sinc}\left\{\hat{a}\left[\frac{ \mathrm{m}}{\hat{z}_1}
\Delta {\hat{r}}_i-{\bar{u}}\right]\right\}
\mathrm{sinc}\left\{\hat{a}\left[\frac{ \mathrm{m}}{\hat{z}_1}
\Delta {\hat{r}}_i+{\bar{u}}\right]\right\} ~.\cr &&
\label{GPInb55limit}
\end{eqnarray}
The convolution theorem yields the following expression for the
spectral degree of coherence:

\begin{eqnarray} {g}_P(\hat{z}_i,
 \Delta {\hat{r}}_i)&=&\frac{1}{\mathcal{D}}
\int_{-\infty}^{\infty}  d{\hat{r}'} \left|P(\hat{r}')\right|^2
\exp\left[-2i\frac{\hat{r}'\Delta \hat{r}_i}{\hat{d}_i}\right] ~,
\label{GPInb55limit2}
\end{eqnarray}
the normalization factor $\mathcal{D}$ being given by

\begin{eqnarray} {\mathcal{D}} =
\int_{-\infty}^{\infty}  d{\hat{r}'} \left|P(\hat{r}')\right|^2 ~.
\label{matcald}
\end{eqnarray}

After the substitution $\bar{u} \longrightarrow \bar{u}'
\sqrt{N}/\hat{z}_1$ we may rewrite  Eq. (\ref{GPInb55in}) as a
function of $\xi = \hat{a} \mathrm{m} \bar{r}_i/\hat{z}_1$ with
the help of the only parameter $p = \hat{a} \sqrt{N}/\hat{z}_1$,
that is easier to plot

\begin{figure}
\begin{center}
\includegraphics*[width=140mm]{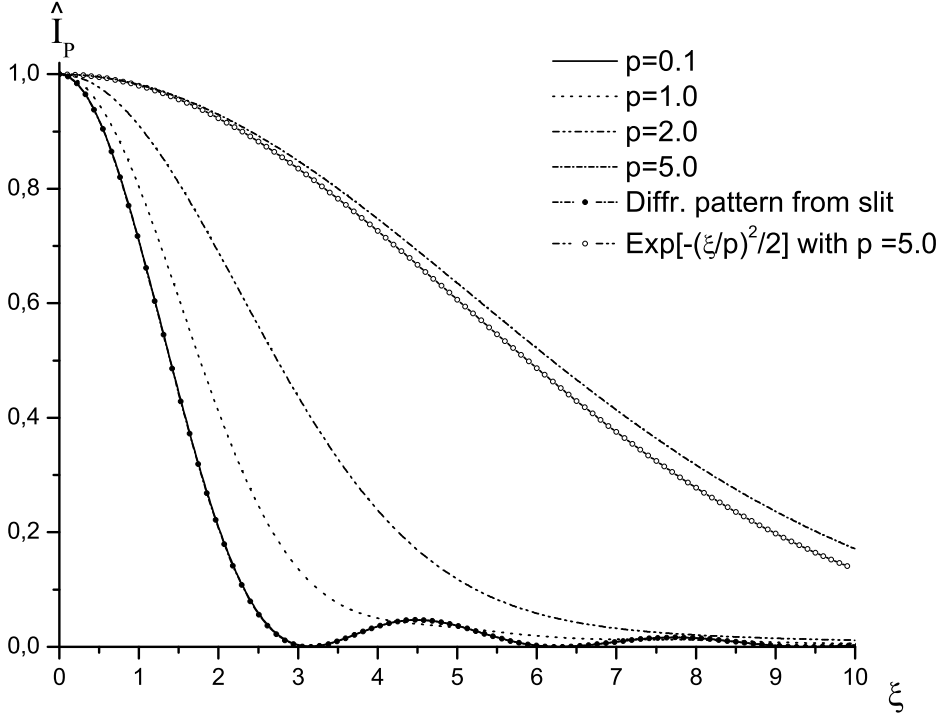}
\caption{\label{ipqh} Image intensity for a quasi-homogeneous
source, $\hat{I}_P$, as a function of $\xi = \hat{a} \mathrm{m}
\bar{r}_i /\hat{z}_1$, calculated with Eq. (\ref{GPInb55in2b2}),
for different values of the parameter $p = \hat{a}
\sqrt{\hat{N}}/\hat{z}_1$. The plot illustrates the
one-dimensional image formation problem (slit aperture,
cylindrical lens). }
\end{center}
\end{figure}
\begin{eqnarray} \hat{I}_P(\xi)&=& \frac{1}{\mathcal{C}} \int_{-\infty}^{\infty} d{\bar{u}'}~
\exp\left[-\frac{\bar{u}'^{2}}{2}\right] \left|
\mathrm{sinc}\left(\xi-p \bar{u}\right)\right|^2~,
\label{GPInb55in2b2}
\end{eqnarray}
where $\mathcal{C}$ can explicitly be calculated as:

\begin{eqnarray}
\mathcal{C}= \sqrt{\frac{\pi}{2}}\frac{1}{p^2}
\left\{-1+\exp\left[-2 p^2\right]\right\}+\frac{\pi}{p}
~\mathrm{erf}\left[\sqrt{2} p\right]~. \label{mathcalb2}
\end{eqnarray}
The function $\mathrm{erf}(\cdot)$ in Eq. (\ref{mathcalb2})
indicates the error function. Note that the variable $\xi =
\hat{a} \mathrm{m} \bar{r}_i/\hat{z}_1$ may also be written as
$\xi = (\hat{a}/\hat{d}_i)\bar{r}_i$, where $\hat{d}_i =
\hat{z}_i-\hat{z}_1$ and $|\mathrm{M}| = \hat{d}_i/\hat{z}_1 =
\mathrm{m}^{-1}$. It is interesting to remark that the ratio
$\hat{d}_i/\hat{a}$ is the dimensionless characteristic size of
the Fresnel zone, i.e. $\lambda d_i/(2 \pi a)$. $\hat{I}_P$ is
plotted in Fig. \ref{ipqh} for several values of the parameter
$p$.  In Fig. \ref{ipqh} we also plot the asymptotic behaviors of
$\hat{I}_p$ for small values of $p$, i.e. $\hat{I}_p =
\mathrm{sinc}^2(\xi)$, that is the diffraction pattern from a
slit, and for large values of $p$, i.e. $\exp[-(\xi/p)^2/2]$.

\begin{figure}
\begin{center}
\includegraphics*[width=140mm]{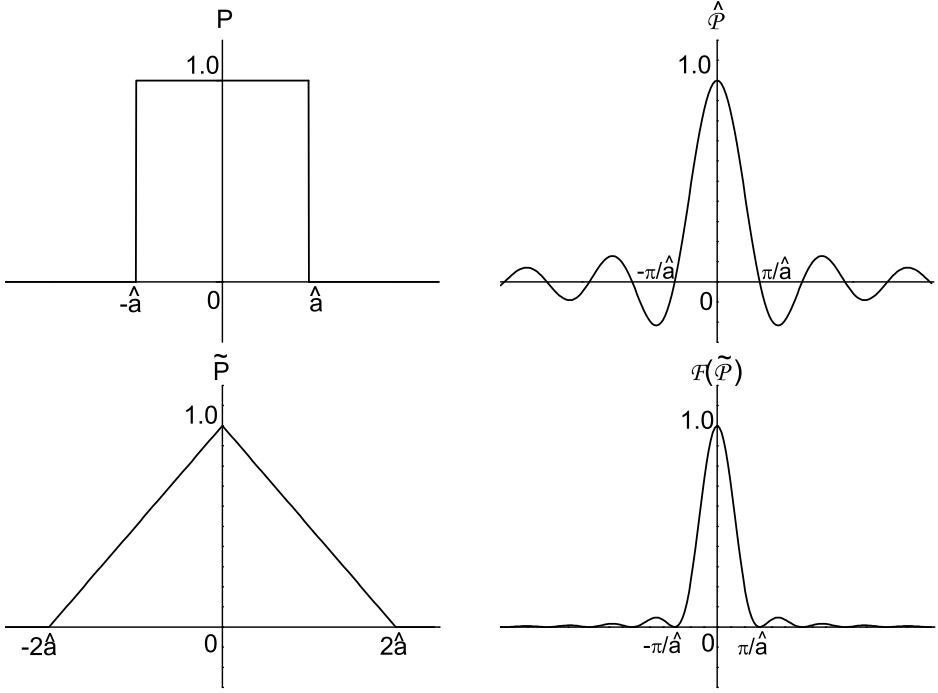}
\caption{\label{pupils} Upper plots:  profile of the pupil
function $P$ (slit aperture), together with $\hat{\mathcal{P}}$.
Lower plots: autocorrelation function of the pupil, $\tilde{P}$,
together with $\mathcal{F}(\tilde{\mathcal{P}})$. }
\end{center}
\end{figure}
On the one hand, the integral in $d \bar{u}$ in Eq.
(\ref{GPInb55in}) is a convolution. On the other hand, aside for
numerical factors, the Fourier transform of
$|\hat{\mathcal{P}}(\hat{u})|^2=|2\hat{a} \mathrm{sinc}(\hat{a}
\hat{u})|^2 $ can be given in terms of the triangular function
$\mathrm{tri(\cdot)}$, defined as\footnote{It should be noted that
$\chi = 2 \Delta \hat{r}$. The reason why we introduced the new
variable $\chi$ is to keep a certain homogeneity of notation when
comparing with reference \cite{GOOD}. Since our definition of
$\Delta \hat{r}$ differs for a factor $2$ with respect to that in
\cite{GOOD}, it is somewhat convenient to introduce $\chi = 2
\Delta \hat{r}$.}

\begin{eqnarray}
\mathrm{tri}\left(\frac{\chi}{2\hat{a}}\right) &=& \left \{
\begin{tabular}{c}
$1+\chi/(2 \hat{a})$ ~~~~~~~if~$-2 \hat{a}<\chi<0$\\
$1-\chi/(2 \hat{a})$ ~~~~~~~if~~~$0<\chi<2 \hat{a}$\\
$0$ ~~~~~~~~~~~~otherwise~.
\end{tabular}
\right. \label{Tr}
\end{eqnarray}
By means of the convolution theorem, Eq. (\ref{GPInb55in}) can be
written as

\begin{eqnarray} \hat{I}_P(\hat{z}_i,
{\bar{r}}_i)&=& \frac{1}{\mathcal{N}}\int_{-\infty}^{\infty} d
\chi\exp\left[-\frac{\hat{N} \chi^{2}}{2 \hat{z}_1^2}\right]
\mathrm{tri}\left(\frac{\chi}{2\hat{a}}\right) \exp\left[-i
\chi\frac{ \mathrm{m}}{\hat{z}_1}{\bar{r}}_i \right]~.
\label{GPInb55inbis}
\end{eqnarray}
The normalization factor $\mathcal{N}$ is given by

\begin{eqnarray}
{\mathcal{N}} &=& \int_{-\infty}^{\infty} d
\chi\exp\left[-\frac{\hat{N} \chi^{2}}{2 \hat{z}_1^2}\right]
\mathrm{tri}\left(\frac{\chi}{2\hat{a}}\right) = \cr &&
\frac{\hat{z}_1}{2\hat{a} \hat{N}}
\left\{\left(1+\hat{a}^2\right)\left[-1+\exp\left(-2\frac{\hat{a}^2
\hat{N}}{\hat{z}_1^2} \right)\right]\hat{z}_1+2 \hat{a} \sqrt{2\pi
\hat{N}} ~\mathrm{erf}\left[\frac{\sqrt{2
\hat{N}}~\hat{a}}{\hat{z}_1}\right] \right\}~.\cr &&
\label{GPInb55inbisnormalN}
\end{eqnarray}
Eq. (\ref{GPInb55inbis}) can be written as an analogous of Eq.
(5.7-10) of \cite{GOOD}, that sometimes goes under the name of
Shell's theorem. Let us define the autocorrelation function of the
pupil as

\begin{eqnarray}
\tilde{P}(\chi) = \int_{-\infty}^{\infty} d u P\left(u+
\frac{\chi}{2}\right) P^*\left(u-
\frac{\chi}{2}\right)~.\label{corrP}
\end{eqnarray}
%
In Fig. \ref{pupils} we plot the profile of the pupil function
$P$, together with $\hat{\mathcal{P}}$ and the autocorrelation
function of the pupil $\tilde{P}$, together with its Fourier
transform $\mathcal{F}({\tilde{P}})$. With the help of Eq.
(\ref{corrP}), Eq. (\ref{GPInb55inbis}) can be written as

\begin{eqnarray} \hat{I}_P(\hat{z}_i,
{\bar{r}}_i)&=&\frac{1}{\mathcal{N}}\int_{-\infty}^{\infty} d \chi
\exp\left[-\frac{\hat{N} \chi^{2}}{2 \hat{z}_1^2}\right]
\tilde{P}(\chi) \exp\left[-i \chi\frac{
\mathrm{m}}{\hat{z}_1}{\bar{r}}_i \right]~, \label{GPInb55inbis2}
\end{eqnarray}
that is Eq. (5.7-10) of reference \cite{GOOD}.

A second limiting case of interest is found when $\hat{a}^2 \ll
\hat{z}_1^2/\hat{N}\ll \hat{D} \hat{z}_1^2$. The pupil is
coherently and uniformly illuminated. In this case both the
exponential functions inside the integral in Eq. (\ref{GPInb1})
behave like $\delta$-Dirac functions yielding

\begin{eqnarray} \hat{G}_P(\hat{z}_i,
{\bar{r}}_i, \Delta {\hat{r}}_i)&=&4 \hat{a}^2
\exp\left[\frac{2i\mathrm{m}}{\hat{z}_1}{\bar{r}}_i\Delta{\hat{r}}_i\right]
\mathrm{sinc}\left\{\hat{a}\left[\frac{
\mathrm{m}}{\hat{z}_1}\left({\bar{r}}_i+ \Delta
{\hat{r}}_i\right)\right]\right\}\cr &&
\times\mathrm{sinc}\left\{\hat{a}\left[\frac{
\mathrm{m}}{\hat{z}_1}\left({\bar{r}}_i- \Delta
{\hat{r}}_i\right)\right]\right\} ~. \label{GPInb22}
\end{eqnarray}
corresponding to a relative intensity

\begin{eqnarray} \hat{I}_P(\hat{z}_i,
{\bar{r}}_i)&=& \mathrm{sinc}^2\left[\frac{\hat{a}
\mathrm{m}}{\hat{z}_1}{\bar{r}}_i\right]=\frac{1}{4 \hat{a}^2}
\left|\hat{\mathcal{P}}\left(\frac{
\mathrm{m}}{\hat{z}_1}{\bar{r}}_i\right)\right|^2~.
\label{GPInb23}
\end{eqnarray}
Also, using the definition of the spectral degree of coherence
given in Eq. (\ref{normfine}) one sees that $|g|=1$, i.e. the
pupil is coherently illuminated. Eq. (\ref{GPInb23}) is the
analogous of Eq. (5.7-14) in \cite{GOOD}.

Results obtained here deal both with the cross-spectral density
$\hat{G}_P$ and the relative intensity $\hat{I}_P$ in the presence
of the pupil. We classified these results comparing the square of
the extent of the pupil $\hat{a}^2$ with the square of the
radiation spot size $\hat{D} \hat{z}_1^2$ and of the coherence
length $\hat{z}_1^2/\hat{N}$ at the pupil location.

Let us now consider the intensity $\hat{I}_P$ only. In the case of
quasi-homogeneous sources, there is a more general method taken
from the theory of linear systems to account for the pupil
presence. This method can be extended to account for aberrations
as well, and allows a more compact treatment of the pupil effects
because it does not depend on how the extent of the pupil scales
with respect to the coherence length and to the radiation spot
size. It is based on the concept of line spread function. A linear
time-invariant system in two dimensions can be characterized by
the knowledge of the point spread function (or impulse response)
$h(x,y)$. Given a certain input $f(x_1,y_1)$, the output at any
point $(x,y)$ is given by the convolution of the point-spread
function $h$ and the input $f$. The point spread function $h(x,y)$
is the response to a $\delta$-Dirac signal at position $(0,0)$,
i.e. $\delta(x_1,y_1)$. The line spread $l(x)$ is, instead, the
response obtained from a line input $\delta(x_1)$, which is
independent of $y_1$, and can be calculated by integrating the
point spread function $h(x,y)$ with respect to the $y$-variable
(see, for instance, \cite{DAIN} Sec 6.2.).

In the quasi-homogeneous case, the intensity $\hat{I}_P$ can be
written as a convolution between a suitable line spread function
$l$\footnote{ Here we are treating a two-dimensional system, but
we are considering the case when we have separability properties
for both the source and the pupil. For the source this means that
$\hat{N}_x \gg 1$ and $\hat{D}_x \gg 1$, while for the pupil it
means that the pupil is rectangular. This case is practically
realized with the help of slit apertures and cylindrical optics.
The line spread function $l$ is, then, the proper tool to
consider. If one wants to consider the situation when no
separability property for the pupil is present, one should take
advantage of an approach based on the point spread function.} and
the intensity $\hat{I}$ which does not account for the pupil
presence, that is

\begin{equation}
\hat{I}_P(\hat{z}_i, \bar{r}_i) = [\hat{I} \ast l](\hat{z}_i,
\bar{r}_i) \label{conprod}
\end{equation}
The line spread function $l$ acts as the passport of the imaging
system, and depends on the properties of the lens only. In this
case we are only considering the effect of a finite pupil
dimension, i.e. we are accounting for diffraction effects from the
pupil. More in general, $l$ may depend on lens apodization or
aberrations too.

In the case under study here, the line spread function of the
system is given by

\begin{eqnarray}
l(\bar{r}_i) =
\mathrm{sinc}^2\left(\frac{\hat{a}}{\hat{d}_i}\bar{r}_i\right)~.
\label{lsfun}
\end{eqnarray}
For instance, it is straightforward to see that substitution of
the (magnified) input signal $\hat{I}=\exp[-
\bar{u}^2/(2|\mathrm{M}|^2\hat{N})]$ with the input signal
$\hat{I} = \delta(\bar{u}/|\mathrm{M}|)$, that is a line input, in
Eq. (\ref{GPInb55in}) gives back Eq. (\ref{lsfun}).

It should be emphasized that the resolution due to diffraction
effects is of order $\hat{z}_1/(\hat{a} \sqrt{\hat{N}})$. In all
cases when this resolution is better than (i.e.
$\hat{z}_1/(\hat{a} \sqrt{\hat{N}})$ is smaller than) the
resolution of the ideal image (which does not account for the
pupil presence)  the pupil does not play any role and, with the
accuracy of the calculation of the ideal intensity (see Section
\ref{sec:supp}), the $l$ function in Eq. (\ref{lsfun}) cannot be
distinguished from a $\delta$-Dirac. In our study case, the ideal
intensity $\hat{I}=\exp[-\hat{z}_1^2 \bar{u}^2/(2\hat{N})]$ is
calculated with an accuracy which is much worse than the
quasi-homogenous accuracy (see Section \ref{sec:supp} for details)
and is of order $\max(1/\sqrt{\hat{D}},1/\sqrt{\hat{N}})$. In
general, it is important to compare the resolution due to
diffraction effects with the accuracy of the calculation of the
ideal intensity. For example, on the basis of such comparison, one
may conclude that the case for $\hat{z}_1^2/\hat{N}\ll \hat{a}^2
\lesssim \hat{D} \hat{z}_1^2$ can be calculated assuming that the
line spread function in Eq. (\ref{lsfun}) is a $\delta$-Dirac. In
fact $\hat{z}_1/(\hat{a} \sqrt{\hat{N}}) \sim
1/\sqrt{\hat{N}\hat{D}}$ since $\hat{a}^2 \lesssim \hat{D}
\hat{z}_1^2$ and therefore $\hat{z}_1/(\hat{a} \sqrt{\hat{N}}) \ll
\max(1/\sqrt{\hat{D}},1/\sqrt{\hat{N}})$. This kind of reasoning
can be used to treat any quasi-homogeneous case, and the $l$
function can be modified to include aberrations and apodization
effects as well.  In the most general case, accounting for the
effects of the pupil in one dimension, Eq. (\ref{GPIn}) yields the
relative intensity

\begin{eqnarray}
\hat{I}_P(\hat{z}_i, {\bar{r}}_i)&=& \frac{1}{\mathcal{S}}\int
d{\bar{u}} ~d \Delta {\hat{u}}~
\exp\left[-\frac{2i\hat{z}_1}{\mathrm{m}}{\bar{u}}\cdot\Delta{\hat{u}}\right]
\hat{G}\left(\hat{z}_i,
\frac{\hat{z}_1}{\mathrm{m}}{\bar{u}},\frac{\hat{z}_1}{\mathrm{m}}
\Delta {\hat{u}}\right) \cr &\times&\hat{\mathcal{P}}\left[\frac{
\mathrm{m}}{\hat{z}_1}{\bar{r}}_i-{\bar{u}}- \Delta
{\hat{u}}\right]\hat{\mathcal{P}}^*\left[\frac{
\mathrm{m}}{\hat{z}_1}{\bar{r}}_i-{\bar{u}}+ \Delta
{\hat{u}}\right]~ , \label{IPIn}
\end{eqnarray}
where the normalization factor $\mathcal{S}$ is given by

\begin{eqnarray}
{\mathcal{S}} &=& \int d{\bar{u}} ~d \Delta {\hat{u}}~
\exp\left[-\frac{2i\hat{z}_1}{\mathrm{m}}{\bar{u}}\cdot\Delta{\hat{u}}\right]
\hat{G}\left(\hat{z}_i,
\frac{\hat{z}_1}{\mathrm{m}}{\bar{u}},\frac{\hat{z}_1}{\mathrm{m}}
\Delta {\hat{u}}\right) \cr &\times&\hat{\mathcal{P}}\left[
-{\bar{u}}- \Delta {\hat{u}}\right]\hat{\mathcal{P}}^*\left[
-{\bar{u}}+ \Delta {\hat{u}}\right] ~. \label{IPInmathcalS}
\end{eqnarray}
In the quasi-homogeneous case (including cases when the source is
not Gaussian, see Section \ref{sec:nong}),  if the lens is in the
far field, one can write

\begin{equation}
\hat{G}\left(\hat{z}_i,
\frac{\hat{z}_1}{\mathrm{m}}{\bar{u}},\frac{\hat{z}_1}{\mathrm{m}}
\Delta {\hat{u}}\right) = \hat{I}\left(\hat{z}_i,
\frac{\hat{z}_1}{\mathrm{m}}{\bar{u}}\right) {g}\left(\hat{z}_i,
\frac{\hat{z}_1}{\mathrm{m}} \Delta
{\hat{u}}\right)\exp\left[\frac{2i\hat{z}_1}{\mathrm{m}}{\bar{u}}\cdot\Delta{\hat{u}}\right]~.
\label{Gbreak}
\end{equation}
Further on, within the accuracy of the quasi-homogeneous
approximation, the spectral degree of coherence ${g}$ behaves like
a Dirac $\delta$-function in the calculation of both the intensity
and the cross-spectral density. In fact, the accuracy of the
incoherent impulse response $|\hat{\mathcal{P}}|^2$ is also the
accuracy of the quasi-homogenous assumption, and this is the
accuracy with which we can substitute ${g}$ with a Dirac
$\delta$-function on the image plane. As a result, in analogy with
Eq. (\ref{GPInb55}) one has

\begin{eqnarray}
\hat{G}_P(\hat{z}_i, \bar{r}_i,\Delta \hat{r}_i) &=&
\int_{-\infty}^{\infty} d\bar{u}~ \hat{I}\left(\hat{z}_i,
\frac{\hat{z}_1}{\mathrm{m}}{\bar{u}}\right)\cr
&&\times\hat{\mathcal{P}}\left[\frac{\mathrm{m}}
{\hat{z}_1}\left(\bar{r}_i+\Delta \hat{r}_i
\right)-\bar{u}\right]\hat{\mathcal{P}}^*\left[\frac{\mathrm{m}}
{\hat{z}_1}\left(\bar{r}_i-\Delta \hat{r}_i \right)-\bar{u}\right]
~,\label{IPIn2biss}
\end{eqnarray}
while the relative intensity can be written as

\begin{eqnarray}
\hat{I}_P(\hat{z}_i, \bar{r}_i) =\frac{1}{\mathcal{D}}
\int_{-\infty}^{\infty} d\bar{u}
\left|\hat{\mathcal{P}}\left(\frac{\mathrm{m}}
{\hat{z}_1}\bar{r}_i-\bar{u}\right)\right|^2
\hat{I}\left(\hat{z}_i,
\frac{\hat{z}_1}{\mathrm{m}}{\bar{u}}\right)~,\label{IPIn2}
\end{eqnarray}
$\mathcal{D}$ being defined in Eq. (\ref{matcald}). The line input
response is obtained by setting $ \hat{I} (\hat{z}_i,
\hat{z}_1{\bar{u}}/\mathrm{m})$ to a Dirac $\delta$-function, thus
obtaining the line spread function:

\begin{eqnarray}
l({\bar{r}}_i)&=& \left| \hat{\mathcal{P}}\left[\frac{
\mathrm{m}}{\hat{z}_1}{\bar{r}}_i\right]\right|^2 =
\mathcal{F}^{-1}(\tilde{P})~. \label{lgene}
\end{eqnarray}
The line spread function is therefore the inverse Fourier
transform of the autocorrelation function of the pupil. The
autocorrelation function of the pupil, i.e. the Fourier transform
of the line spread function, is also known as the Optical Transfer
Function (OTF). Other relevant quantities introduced in linear
system theory are the phase of the Optical Transfer Function and
its modulus, which is known as the Modulation Transfer Function
(MTF) \cite{DAIN}.

Eq. (\ref{IPIn2}) and, consequently, the line spread function
approach, constitutes a universal description of the intensity in
all quasi-homogeneous cases. In literature the line spread
function is used to describe perfectly incoherent sources only.
Note that, in general, radiation produced by an electron beam in
an undulator is similar to an incoherent sum of many independent
laser-like beams. Yet, it cannot be considered as an incoherent
sum of point sources because, as we have seen in Section
\ref{sec:fila}, a single electron cannot be considered as a
point-like radiation source. Radiation produced by a single
electron is similar to a laser beam. If no influence of focusing
is present in the undulator, this laser-like beam has a waist
located in the center of the undulator. At the waist the radiation
wavefront is plane and the radiation spot size is much larger than
the wavelength. We extended the use of the line spread function
approach to the realm of quasi-homogeneous sources.

In the case of third generation light sources the line spread
function method can almost always be applied in the horizontal
direction. However, it fails in the vertical direction, where
third generation light sources are seldom quasi-homogeneous. If
the source is not quasi-homogeneous, the cross-spectral density
cannot be factorized in the product of the intensity and of the
spectral degree of coherence, or the coherence length is not short
(compared with the size of the source). As a result, the
incoherent line spread function $l$ cannot be used to describe the
system. The function $\hat{\mathcal{P}}$ is known as the coherent
line spread function and must be used in its place. In fact, when
the source starts to exhibit a high degree of transverse coherence
(i.e. in the non quasi-homogeneous case), the coherent line spread
function, $\hat{\mathcal{P}}$ acts on the field at the image plane
analogously to the way the incoherent line spread function acts on
the intensity at the image plane. To see this, it is sufficient to
inspect Eq. (\ref{EIPn}).

It is interesting to compare this viewpoint with what can be found
in literature. For instance, in \cite{ATT2}, where a condenser
system is discussed, one may read: "The intrinsic divergence of
the extreme ultraviolet (EUV) undulator considered here is
$\theta_\mathrm{cen} = 80 \mu$rad, which is larger than the
beamline acceptance $\theta_\mathrm{accept}$ of $48 \mu$rad.
Therefore it is evident that the incoherent source approximation
holds here and the term incoherence source is used accordingly in
this paper". In the following Sections of their work, authors of
\cite{ATT2} use a point-spread function approach to account for
aberration effects: in their paper, the intensity at the source is
used, instead of the cross-spectral density, in order to evaluate
both the intensity and the degree of coherence at the image plane.
Such an approach is justified in the passage above, where they
state that the source is incoherent.

The statement in \cite{ATT2} about the incoherence of the source
is a misconception. According to such statement, \textit{perfectly
coherent} undulator radiation produced by an electron beam with
zero emittance should exhibit \textit{incoherent} properties when
the radiation divergence is larger than the acceptance of the
optical system. In contrast with the assertion made in
\cite{ATT2},  the coherence properties of the source are
independent of the beamline elements which follow. In order to
discuss about the coherence properties of the source one has to
refer to the radiation field at the virtual plane location only.
In particular, the fact that the source is coherent (or not) does
not depend on how the beam acceptance angle scales with the
intrinsic (single particle) divergence of the undulator radiation.
Our conclusion is that the only parameters which describe wether a
source is quasi-homogeneous or not are (in the vertical direction)
$\hat{N}_y$ and $\hat{D}_y$. If the source is quasi-homogeneous, a
point-spread function approach can be used. If not, the more
general results described in Section \ref{sub:hqhvn} should be
considered. In particular, in the case of \cite{ATT2}, the
vertical rms dimension of the source is $\sigma_y=16 ~\mu$m and
the radiation wavelength is $\lambda = 13.4~ n$m, while the
undulator (see \cite{ATT1}) is composed of $55$ periods, each one
$8$ cm long. This means $L_w = 4.4$ m. Moreover, the vertical
emittance at ALS is $\epsilon_y \simeq 0.1 n$m, while the vertical
beta function for beamline $12$ is $\beta_y =4.2 $ m $\simeq L_w$.
As a result both $\hat{N}_y \sim 0.1$ and $\hat{D}_y \sim 0.1$ and
the source is non-homogeneous. We conclude that, in this case,
approximations like Eq. (\ref{IPIn2biss}) or Eq. (\ref{IPIn2})
cannot be used. Eq. (\ref{ultimasp}) in Section \ref{sec:nonh}
should be considered instead.

As a final remark we should stress that, even in cases when the
virtual source is quasi-homogeneous, one should verify the
assumption that the lens is in the far zone, before applying a
point spread function formalism. In contrast to this, note that in
the usual framework of Statistical Optics, the radiant intensity
from thermal sources is distributed over a solid angle of order $2
\pi$, and optical elements can always be considered in the far
zone.

\subsection{\label{sub:emoco} Quasi-homogeneous non-Gaussian undulator sources}

In the present Section \ref{sub:emoco} we will extend results
obtained in Section \ref{sec:nong}. Results obtained in the
previous Section \ref{sub:qhpup} apply for a quasi-homogeneous
Gaussian undulator source only.  In particular, under the
assumptions $\hat{N}_x\gg 1$ and $\hat{D}_x \gg 1$ the
cross-spectral density can be factorized in a horizontal and in a
vertical contribution, and results in Section \ref{sub:qhpup} can
be applied in the horizontal direction. Note that, if $\hat{N}_y
\gg 1$ and $\hat{D}_y \gg 1$, one has, automatically, $\hat{D}_x
\gg 1$ and $\hat{D}_x \gg 1$ and the same results in Section
\ref{sub:qhpup} can be separately applied in both the horizontal
and the vertical directions. Here, with the help of Eq.
(\ref{GPIn}), we will include the effects of a pupil in the
one-dimensional case when the source is still quasi-homogeneous,
but non-Gaussian. In particular, we will still assume $\hat{N}_x
\gg1 $ and $\hat{D}_x \gg 1$ and concentrate our attention on the
vertical direction. First we will study the case when
$\hat{N}_y\gg 1$ and $\hat{D}_y$ is arbitrary and, then, the case
when $\hat{N}_y$ is arbitrary and $\hat{D}_y \gg 1$. We will see
that the reasoning applied in the case of quasi-homogeneous
Gaussian sources also holds in the case for quasi-homogeneous
non-Gaussian sources as it relies on the separability of the
cross-spectral density only. As a result we will present practical
examples of how, with minor substitutions, we can extend our
analysis of the pupil effects to the case of non-Gaussian sources.

\subsubsection{Source with non-Gaussian angular distribution in
the vertical direction}

Let us start considering the case when $\hat{D}_y$ is arbitrary
and $\hat{N}_y \gg 1$. The pupil function and $\hat{\mathcal{P}}$
are given by Eq. (\ref{Pyfunc}) and Eq. (\ref{verpy}). The
$r$-direction should be now substituted with the $y$-direction.

We can use Eq. (\ref{GPIn}) and an asymptotic expression of Eq.
(\ref{xdngandeima}) to describe the case when the lens is in the
far zone, that is when condition (\ref{farlens}) is satisfied.
From Eq. (\ref{Ngrande}) we can estimate the typical size of the
source that is of order $\sqrt{\hat{N}_y}$, and of the correlation
length at the source, that is of order
$\min[1/\sqrt{\hat{D}_y},1]$. According to Eq. (\ref{farlens}),
the lens is in the far zone when $\sqrt{\hat{N}_y}/\hat{z}_1 \ll
\max[{ \sqrt{\hat{D}_y},1}]$. In this limit, Eq. (\ref{GPIn}) and
Eq. (\ref{xdngandeima}) give

\begin{eqnarray} \hat{G}_P(\hat{z}_i,
{\bar{y}}_i, \Delta {\hat{y}}_i)&=&4 \hat{a}^2
\exp\left[\frac{2i\mathrm{m}}{\hat{z}_1}{\bar{y}}_i\Delta{\hat{y}}_i\right]
\cr && \times \int d{\bar{u}} ~d \Delta {\hat{u}}~
\exp\left[-\frac{\hat{z}_1^2 \bar{u}^{2}}{2\hat{N}_y}\right]
\exp\left[-2{ {\hat{D}_y}\hat{z}_1^2 (\Delta \hat{u})^{2} }\right]
\gamma(\hat{z}_1 \Delta \hat{u}) \cr
&&\times\mathrm{sinc}\left\{\hat{a}\left[\frac{
\mathrm{m}}{\hat{z}_1}\left({\bar{y}}_i+ \Delta
{\hat{y}}_i\right)-{\bar{u}}- \Delta {\hat{u}}\right]\right\}\cr
&& \times\mathrm{sinc}\left\{\hat{a}\left[\frac{
\mathrm{m}}{\hat{z}_1}\left({\bar{y}}_i- \Delta
{\hat{y}}_i\right)-{\bar{u}}+\Delta {\hat{u}}\right]\right\} ~.
\label{GPInb1b}
\end{eqnarray}
From Eq. (\ref{G2Dmeno1}), one sees that $\hat{z}_1^2
\max[\hat{D}_y,1]$ is of the order of the square of the radiation
spot size on the pupil, while $\hat{z}_1^2/\hat{N}_y$ is of the
order of the square of the coherence length on the pupil. Note
that the limiting expression obtained from Eq. (\ref{GPInb1b}) for
$\hat{D}_y \gg 1$ is Eq. (\ref{GPInb1}). As before, two
interesting limiting cases of Eq. (\ref{GPInb1}) can be obtained
comparing these two scales with $\hat{a}^2$, that is the square of
the pupil size.

First, let us consider the case $ \hat{z}_1^2/ \hat{N}_y \lesssim
\hat{a}^2 \ll \hat{z}_1^2 \max[1,\hat{D}_y]$. As we have already
discussed, in all situations when the quasi-homogeneous assumption
is verified, the exponential function in $\Delta \hat{u}$ inside
the integral in Eq. (\ref{GPInb1b}) behaves like a $\delta$-Dirac
distribution. As in Eq. (\ref{GPInb55}) one obtains

\begin{eqnarray} \hat{G}_P(\hat{z}_i,
{\bar{y}}_i, \Delta {\hat{y}}_i)&=&4 \hat{a}^2
\exp\left[\frac{2i\mathrm{m}}{\hat{z}_1}{\bar{y}}_i\Delta{\hat{y}}_i\right]
\int_{-\infty}^{\infty}  d{\bar{u}}~ \exp\left[-\frac{\hat{z}_1^2
\bar{u}^{2}}{2\hat{N}}\right] \cr
&&\times\mathrm{sinc}\left\{\hat{a}\left[\frac{
\mathrm{m}}{\hat{z}_1}\left({\bar{y}}_i+ \Delta
{\hat{y}}_i\right)-{\bar{u}}\right]\right\} \cr && \times
\mathrm{sinc}\left\{\hat{a}\left[\frac{
\mathrm{m}}{\hat{z}_1}\left({\bar{y}}_i- \Delta
{\hat{y}}_i\right)-{\bar{u}}\right]\right\} ~. \label{GPInb55b}
\end{eqnarray}
Setting $\Delta \hat{y}_i=0$ one yields the intensity

\begin{eqnarray} \hat{I}_P(\hat{z}_i,
{\bar{y}}_i)&=& \frac{1}{\mathcal{C}} \int_{-\infty}^{\infty}
d{\bar{u}}~ \exp\left[-\frac{\hat{z}_1^2
\bar{u}^{2}}{2\hat{N}}\right]
\left|\mathrm{sinc}\left[\hat{a}\left(\frac{
\mathrm{m}}{\hat{z}_1}{\bar{y}}_i-{\bar{u}}\right)
\right]\right|^2~, \label{Intenb55b}
\end{eqnarray}
where the normalization constant $\mathcal{C}$ has already been
defined in Eq. (\ref{mathcalb}). The result in Eq.
(\ref{Intenb55b}) is equivalent to the intensity already given in
Eq. (\ref{GPInb55in}). It can also be written as in Eq.
(\ref{GPInb55inbis}), that is

\begin{eqnarray} \hat{I}_P(\hat{z}_i,
{\bar{y}}_i)&=& \frac{1}{\mathcal{N}} \int_{-\infty}^{\infty} d
\chi \exp\left[-\frac{\hat{N} \chi^{2}}{2 \hat{z}_1^2}\right]
\mathrm{tri}\left(\frac{\chi}{2\hat{a}}\right) \exp\left[-i
\chi\frac{ \mathrm{m}}{\hat{z}_1}{\bar{y}}_i \right]~,
\label{GPInb55inbis2bbb}
\end{eqnarray}
where $\mathcal{N}$ is defined in Eq. (\ref{GPInb55inbisnormalN}).
After introduction of $\tilde{P}$ as in Eq. (\ref{corrP}), Eq.
(\ref{GPInb55inbis2bbb}) can be rewritten as Eq.
(\ref{GPInb55inbis2}),

\begin{eqnarray} \hat{I}_P(\hat{z}_i,
{\bar{y}}_i)&=& \frac{1}{\mathcal{N}} \int_{-\infty}^{\infty} d
\chi \exp\left[-\frac{\hat{N} \chi^{2}}{2 \hat{z}_1^2}\right]
\tilde{P}(\chi) \exp\left[-i \chi\frac{
\mathrm{m}}{\hat{z}_1}{\bar{y}}_i \right]~. \label{GPInb55inbis2b}
\end{eqnarray}
In this case, the expression for the intensity is the same as in
the case $\hat{D}_y \gg 1$. Also note that the results obtained
for the case $ \hat{z}_1^2/ \hat{N}_y \lesssim \hat{a}^2 \ll
\hat{z}_1^2 \max[1,\hat{D}_y]$ are valid in the limit $
\hat{z}_1^2/ \hat{N}_y \ll \hat{a}^2 \ll \hat{z}_1^2
\max[1,\hat{D}_y]$ as well. In this case Eq. (\ref{Intenb55b}) can
be simplified to

\begin{eqnarray} \hat{I}_P(\hat{z}_i,
{\bar{y}}_i)&=& \exp\left[-\frac{\mathrm{m}^2
\bar{y}_i^{2}}{2\hat{N}}\right]~. \label{Intenb55bbb}
\end{eqnarray}
The second limiting case that we can mention here for comparison
with what has been done in the Gaussian case is when $\hat{a}^2
\ll \hat{z}_1^2/\hat{N}\ll \max[\hat{D}_y,1] \hat{z}_1^2$ the
pupil is coherently and uniformly illuminated. In this case one
recovers the same results in Eq. (\ref{GPInb22}) and Eq.
(\ref{GPInb23}).

To sum up, we obtain, in all situations,  the same intensity as in
the case $\hat{D}_y \gg 1$ .

\subsubsection{Source with non-Gaussian intensity distribution in
the vertical direction}

Let us now study the case when $\hat{N}_y$ is arbitrary and
$\hat{D}_y \gg 1$.

We can use Eq. (\ref{GPIn}) and an asymptotic expression of Eq.
(\ref{xdngandeimadoppio}) to describe the case when the lens is in
the far zone, that is when condition (\ref{farlens}) is satisfied.
From Eq. (\ref{Ngr3}) we can estimate the typical size of the
source that is of order $\max[\sqrt{\hat{N}_y},1]$, and of the
correlation length at the source, that is of order
$1/\sqrt{\hat{D}_y}$. According to Eq. (\ref{farlens}), the lens
is in the far zone when $\max[\sqrt{\hat{N}_y},1]/\hat{z}_1 \ll
\sqrt{\hat{D}_y}$. In this limit, Eq. (\ref{GPIn}) and Eq.
(\ref{xdngandeimadoppio}) give

\begin{eqnarray} \hat{G}_P(\hat{z}_i,
{\bar{y}}_i, \Delta {\hat{y}}_i)&=&4 \hat{a}^2
\exp\left[\frac{2i\mathrm{m}}{\hat{z}_1}{\bar{y}}_i\Delta{\hat{y}}_i\right]
\int d{\bar{u}} ~d \Delta {\hat{u}}~ \cr && \times
\int_{-\infty}^{\infty} d\hat{\phi}_y
\exp\left[-\frac{\left(\hat{\phi}_y+\hat{z}_1 \bar{u}
\right)^2}{2\hat{N}_y}\right] \hat{\mathcal{B}}(\hat{\phi}_y)
\exp\left[-2{ {\hat{D}_y}\hat{z}_1^2 (\Delta \hat{u})^{2} }\right]
\cr &&\times\mathrm{sinc}\left\{\hat{a}\left[\frac{
\mathrm{m}}{\hat{z}_1}\left({\bar{y}}_i+ \Delta
{\hat{y}}_i\right)-{\bar{u}}- \Delta {\hat{u}}\right]\right\}\cr
&& \times\mathrm{sinc}\left\{\hat{a}\left[\frac{
\mathrm{m}}{\hat{z}_1}\left({\bar{y}}_i- \Delta
{\hat{y}}_i\right)-{\bar{u}}+\Delta {\hat{u}}\right]\right\} ~.
\label{GPInb1bc}
\end{eqnarray}
According to Eq. (\ref{Dgrande}), the quantity $\hat{z}_1^2
\hat{D}_y$ is of the order of the square of the radiation spot
size on the pupil, while $\hat{z}_1^2/\max[\hat{N}_y,1]$ is of the
order of the square of the coherence length on the pupil. Note
that the limiting expression obtained from Eq. (\ref{GPInb1bc})
for $\hat{D}_y \gg 1$ is Eq. (\ref{GPInb1}). We will study again
two limiting cases of Eq. (\ref{GPInb1bc}), which can be obtained
comparing these two scales with $\hat{a}^2$, that is the square of
the pupil size.

First, let us consider the case $ \hat{z}_1^2 /\max[\hat{N}_y,1]
\lesssim \hat{a}^2 \ll \hat{z}_1^2 \hat{D}$. As before, because of
the quasi-homogeneous assumption is verified, the exponential
function in $\Delta \hat{u}$ inside the integral in Eq.
(\ref{GPInb1bc}) behaves like a $\delta$-Dirac distribution. One
obtains

\begin{eqnarray} \hat{G}_P(\hat{z}_i,
{\bar{y}}_i, \Delta {\hat{y}}_i)&=&4 \hat{a}^2
\exp\left[\frac{2i\mathrm{m}}{\hat{z}_1}{\bar{y}}_i\Delta{\hat{y}}_i\right]\cr
&&\times \int_{-\infty}^{\infty}  d{\bar{u}}~
\int_{-\infty}^{\infty} d\hat{\phi}_y
\exp\left[-\frac{\left(\hat{\phi}_y+\hat{z}_1 \bar{u}
\right)^2}{2\hat{N}_y}\right] \hat{\mathcal{B}}(\hat{\phi}_y)\cr
&&\times\mathrm{sinc}\left\{\hat{a}\left[\frac{
\mathrm{m}}{\hat{z}_1}\left({\bar{y}}_i+ \Delta
{\hat{y}}_i\right)-{\bar{u}}\right]\right\} \cr &&\times
\mathrm{sinc}\left\{\hat{a}\left[\frac{
\mathrm{m}}{\hat{z}_1}\left({\bar{y}}_i- \Delta
{\hat{y}}_i\right)-{\bar{u}}\right]\right\} ~. \label{GPInb55bc}
\end{eqnarray}
This corresponds to the intensity

\begin{eqnarray} \hat{I}_P(\hat{z}_i,
\bar{y}_i)&=& \frac{1}{\tilde{\mathcal{C}}}
\int_{-\infty}^{\infty} d{\bar{u}}~ \int_{-\infty}^{\infty}
d\hat{\phi}_y \exp\left[-\frac{\left(\hat{\phi}_y+\hat{z}_1
\bar{u} \right)^2}{2\hat{N}_y}\right]
\hat{\mathcal{B}}(\hat{\phi}_y)\cr && \times
\left|\mathrm{sinc}\left[\hat{a}\left(\frac{
\mathrm{m}}{\hat{z}_1}{\bar{y}}_i-{\bar{u}}\right)\right]
\right|^2 ~, \cr && \label{GPInb55bc10}
\end{eqnarray}
where

\begin{eqnarray}
{\tilde{\mathcal{C}}} &=& \int_{-\infty}^{\infty} d{\bar{u}}~
\int_{-\infty}^{\infty} d\hat{\phi}_y
\exp\left[-\frac{\left(\hat{\phi}_y+\hat{z}_1 \bar{u}
\right)^2}{2\hat{N}_y}\right] \hat{\mathcal{B}}(\hat{\phi}_y)
\left|\mathrm{sinc}\left(\hat{a}{\bar{u}}\right) \right|^2 ~. \cr
&& \label{Gtildemathcalc}
\end{eqnarray}
In analogy with Eq. (\ref{GPInb55inbis2}), Eq. (\ref{GPInb55bc10})
can also be written as:

\begin{eqnarray} \hat{I}_P(\hat{z}_i,
{\bar{y}}_i)&=&
\frac{1}{\tilde{\mathcal{N}}}\int_{-\infty}^{\infty} d \chi\left\{
\exp\left[-\frac{\hat{N}_y \chi^2}{2\hat{z}_1^2}\right]
\beta\left(\frac{ \chi}{2\hat{z}_1}\right) \right\}
\tilde{P}(\chi) \exp\left[-i \chi\frac{
\mathrm{m}}{\hat{z}_1}{\bar{y}}_i \right]~,\cr &&
\label{GPInb55inbis2bc}
\end{eqnarray}
where the normalization factor ${\tilde{\mathcal{N}}}$ is defined
as

\begin{eqnarray}
{\tilde{\mathcal{N}}}&=&\int_{-\infty}^{\infty} d \chi\left\{
\exp\left[-\frac{\hat{N}_y \chi^2}{2\hat{z}_1^2}\right]
\beta\left(\frac{ \chi}{2\hat{z}_1}\right) \right\}
\tilde{P}(\chi) ~.\cr && \label{Gtildemathcaln}
\end{eqnarray}
Note that results obtained in the case $ \hat{z}_1^2
/\max[\hat{N}_y,1] \lesssim \hat{a}^2 \ll \hat{z}_1^2 \hat{D}$ are
also valid in the asymptote for $ \hat{z}_1^2 /\max[\hat{N}_y,1]
\ll \hat{a}^2 \ll \hat{z}_1^2 \hat{D}$. In this case, Eq.
(\ref{GPInb55bc10}) is simplified to

\begin{eqnarray}
\hat{I}_P(\hat{z}_i,
\bar{y}_i)&=& \frac{1}{\tilde{\mathcal{S}}}\int_{-\infty}^{\infty}
d\hat{\phi}_y \exp\left[-\frac{\left(\hat{\phi}_y+\mathrm{m}
\bar{y}_i \right)^2}{2\hat{N}_y}\right]
\hat{\mathcal{B}}(\hat{\phi}_y) ~, \label{GPInb55bc10bbb}
\end{eqnarray}
where

\begin{eqnarray}
{\tilde{\mathcal{S}}} &=& \int_{-\infty}^{\infty} d\hat{\phi}_y
\exp\left[-\frac{\hat{\phi}_y^2}{2\hat{N}_y}\right]
\hat{\mathcal{B}}(\hat{\phi}_y) ~. \label{tildemathcals}
\end{eqnarray}
The second limiting case that we will mention here for comparison
with what has been done in the Gaussian case is for $\hat{a}^2 \ll
\hat{z}_1^2/\max({\hat{N}_y,1})\ll \hat{D}_y \hat{z}_1^2$ the
pupil is coherently and uniformly illuminated. In this case one
recover the same results in Eq. (\ref{GPInb22}) and Eq.
(\ref{GPInb23}).

However, in general, the expression for the intensity is different
from that for $\hat{N}_y \gg 1$ .

\section{\label{sec:abe} Aberrations and imaging of quasi-homogeneous sources}

Up to now we have discussed cases when no aberrations are present.
Although there are widespread treatments of Aberration Theory in
literature, we will introduce our own here, so that this work is
self-consistent. In particular, we will focus on the
one-dimensional case, which has not been treated widely in books
and monographies, aside for some exception (see \cite{ONEI}). In
the present Section \ref{sec:abe} we will assume that the virtual
undulator source is quasi-homogeneous. Although we will begin
introducing the Optical Transfer Function (OTF) for the system, we
will mainly be concerned with the line spread function of the
system. In addition to that we will discuss  the case of severe
aberrations, presenting new analytical results for this asymptote
and comparison with numerical calculations. Finally, we will
intensively discuss  defocusing aberrations and present general
analytical results for this case too. Our particular consideration
of the defocusing case is justified by the fact that this is a
privileged kind of aberration in the framework of Fourier Optics.
In fact it shows a quadratic dependence of the phase error along
the pupil aperture, that is quite a natural dependence in the
Fourier Optics approach. Quadratic phase factors, which can be
interpreted as defocusing aberrations, are present even in the
case of a pinhole camera setup (that will be treated in the next
Section \ref{sec:cam}) and of imaging from ideal lenses in an
arbitrary plane of interest behind the lens (as it will be seen in
Section \ref{sec:focim} and \ref{sec:imany}).

\subsection{Aberrations and the effect of aperture size}

\begin{figure}
\begin{center}
\includegraphics*[width=140mm]{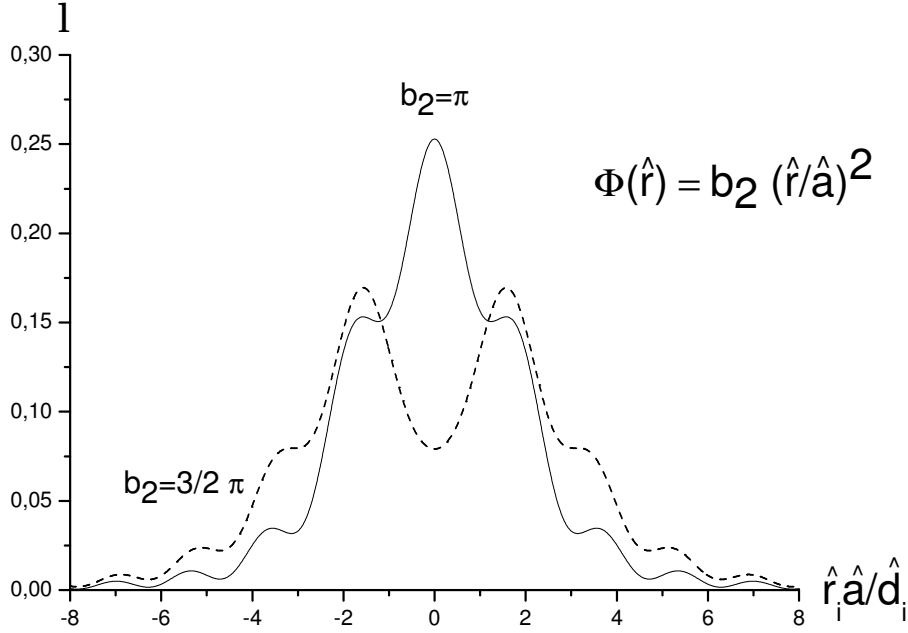}
\caption{The line spread $l$ function in the presence of
defocusing aberration\label{ftadefo}. }
\end{center}
\end{figure}
\begin{figure}
\begin{center}
\includegraphics*[width=140mm]{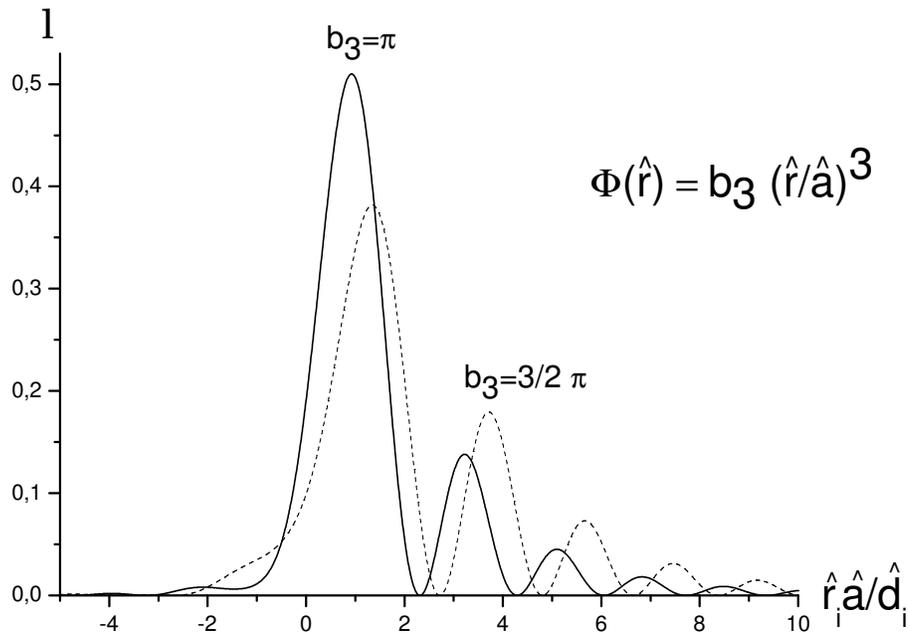}
\caption{The line spread $l$ function in the presence of coma
aberration\label{ftcoma}. }
\end{center}
\end{figure}
\begin{figure}
\begin{center}
\includegraphics*[width=140mm]{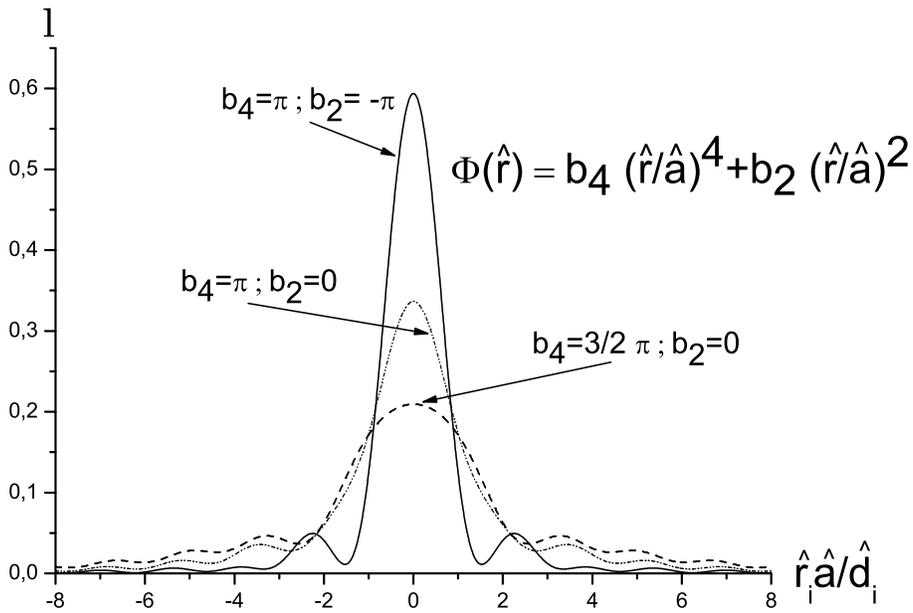}
\caption{The line spread $l$ function in the presence of both
spherical and defocusing aberrations\label{ftsphe}. }
\end{center}
\end{figure}
\begin{figure}
\begin{center}
\includegraphics*[width=140mm]{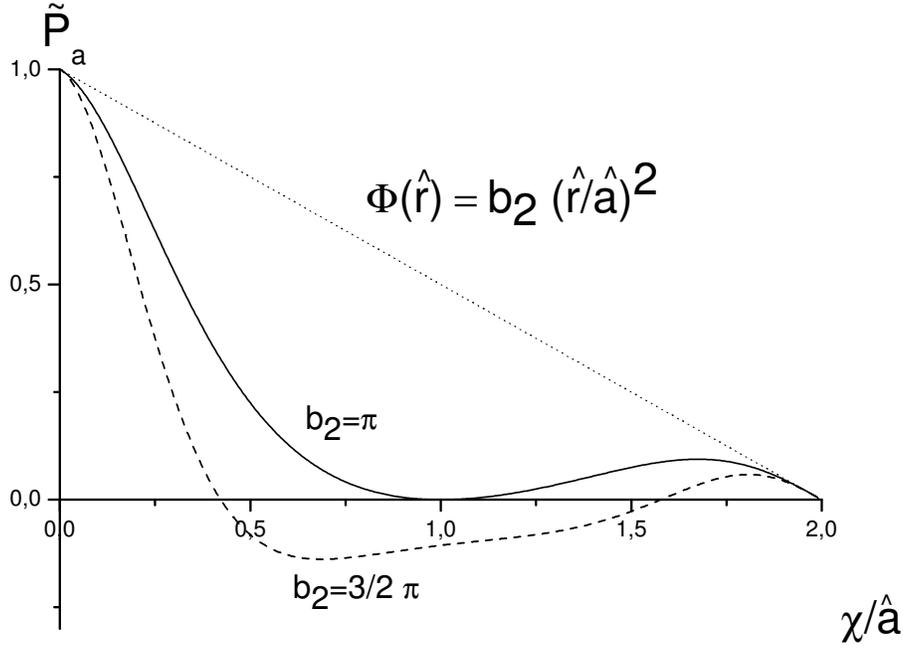}
\caption{The  transfer function in the presence of defocusing
aberration\label{expl1}. }
\end{center}
\end{figure}
\begin{figure}
\begin{center}
\includegraphics*[width=140mm]{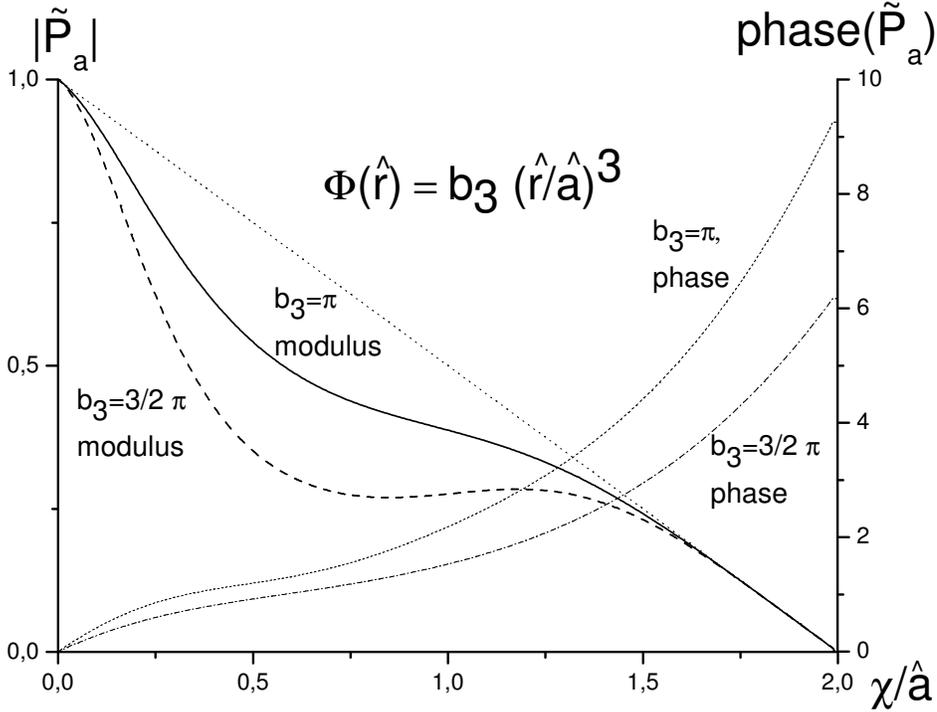}
\caption{\label{expl2} The  transfer function in the presence of
coma aberration. }
\end{center}
\end{figure}
\begin{figure}
\begin{center}
\includegraphics*[width=140mm]{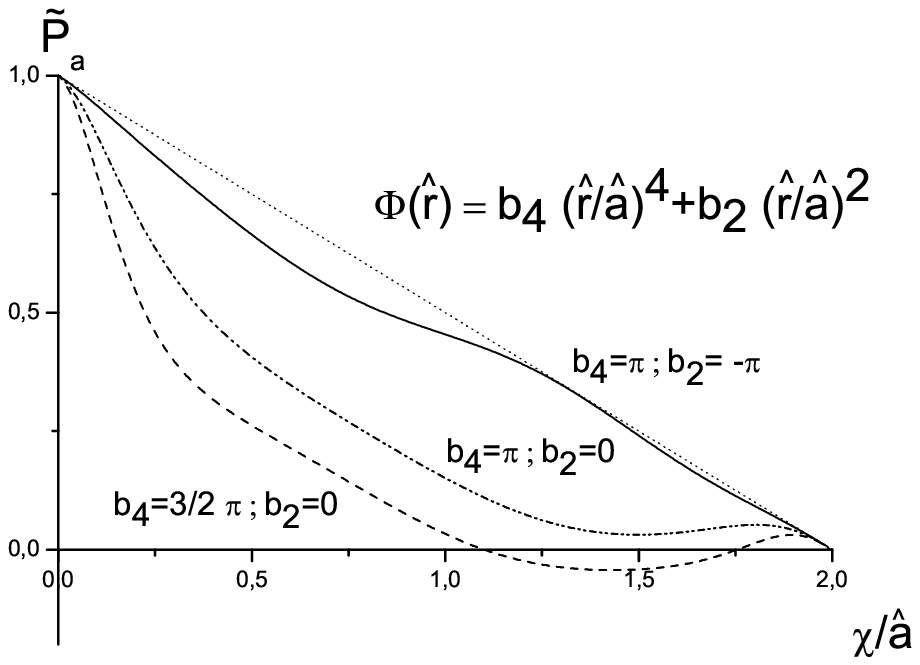}
\caption{\label{expl3} The  transfer function in the presence of
both spherical and defocusing aberrations. }
\end{center}
\end{figure}
In general, aberrations may be accounted for substituting the
pupil function $P$, the coherent line spread function
$\hat{\mathcal{P}}$, the autocorrelation function $\tilde{P}$ and
the Fourier transform of the autocorrelation function
$\mathcal{F}(\tilde{P})$ with new functions, respectively
$P_a$,$\hat{\mathcal{P}}_a$, $\tilde{P}_a$ and
$\mathcal{F}(\tilde{P}_a)$, that account for aberrations. Since
from the very beginning of the present Section \ref{sec:abe} we
considered only the one-dimensional case, we will consider
one-dimensional aberration theory only. Generalization is
possible, although calculations would become more cumbersome.

Mathematically, the presence of aberrations in one-dimension
modifies the pupil function by means of a phase error
$\Phi(\hat{r})$ leading to

\begin{eqnarray}
P_a(\hat{r}) = P(\hat{r}) \exp\left[i
\Phi\left(\hat{r}\right)\right]~. \label{Pa}
\end{eqnarray}
It is customary (see, for instance, \cite{ONEI}) to consider phase
errors of the form:

\begin{equation}
\Phi\left(\hat{r}\right)= {b}_n
\left(\frac{\hat{r}}{\hat{a}}\right)^n ~,\label{aberration}
\end{equation}
$n$ being an integer number, and $b_n$ being the maximum phase
error at the edge of the aperture.  The value $n=0$ corresponds to
a constant phase error and has no effects, as it cancels out when
one calculates the pupil autocorrelation function. The value $n=1$
contributes to the autocorrelation function for a phase term
linearly varying with the position. Its only effect is to shift
the image position. We will not deal with the cases $n=0$ and
$n=1$. We will focus instead on the values $n=2$ corresponding to
defocusing, $n=3$ corresponding to coma and $n=4$ corresponding to
spherical aberrations.

As we have seen in Section \ref{par:ngdgb}, the presence of the
pupil can be dealt with independently of how the pupil extension
$\hat{a}$ scales with respect to the radiation spot size and
coherence length at the pupil location. Every quasi-homogeneous
imaging system can  be described by means of a line spread
function that depends on the physical characteristics of the pupil
only. As has been shown in Section \ref{par:ngdgb}, the knowledge
of the line spread function $l$ and of the intensity at the image
plane without the pupil influence allows one to reconstruct the
actual image by means of a convolution operation. In the present
case of interest, the line spread function must be calculated
accounting for the presence of aberrations. With in mind the
purpose of calculating the line spread function in presence of
aberrations, and with the only assumption of a quasi-homogeneous
source, we start considering Eq. (\ref{GPInfarfl}). The
quasi-homogeneous assumption allows to represent the
cross-spectral density at the virtual source plane as
$\hat{G}(0,-\hat{z}_1 {\bar{u}},-z_1 \Delta {\hat{u}}) =
\hat{I}(0,-\hat{z}_1 {\bar{u}}) {g}(0,-\hat{z}_1 \Delta
{\hat{u}})$. By definition of quasi-homogeneity, and with the
accuracy of the quasi-homogeneous assumption
$(\max[1,\hat{N}_y]\cdot \max[1,\hat{D}_y])^{-1/2}$, the spectral
degree of coherence ${g}$ plays the role of a Dirac
$\delta$-function in the calculation of the intensity. Such
calculation begins from Eq. (\ref{GPInfarfl}). Accounting for the
pupil influence, the following expression for the intensity at the
image plane is therefore found:

\begin{eqnarray}
\hat{I}_P(\hat{z}_i, \bar{r}_i) =
\frac{1}{\mathcal{D}}\int_{-\infty}^{\infty} d\bar{u}
\left|\hat{\mathcal{P}}_a\left(\frac{\mathrm{m}}
{\hat{z}_1}\bar{r}_i-\bar{u}\right)\right|^2 \hat{I}(0,-\hat{z}_1
{\bar{u}})~,\label{Ipu1}
\end{eqnarray}
where we recall that $\mathcal{D}$ is defined in Eq.
(\ref{matcald}). Note that Eq. (\ref{Ipu1}) is obtained from Eq.
(\ref{GPInfarfl}) under the only assumption of quasi-homogeneity.
Also, when $|\hat{\mathcal{P}}_a|^2(\cdot) \longrightarrow
\delta(\cdot)$ we obtain back an ideal lens and
$\hat{I}_P(\hat{z}_i, \bar{r}_i)=\hat{I}(0,-\mathrm{m}
\bar{r}_i)$.

By definition, the line spread function can be obtained by letting
$\hat{I}(0,\hat{r}) \longrightarrow \delta(\hat{r})$ in Eq.
(\ref{Ipu1}). As a result we have

\begin{eqnarray}
l(\bar{r}_i) = \left|\hat{\mathcal{P}}_a\left(\frac{\mathrm{m}}
{\hat{z}_1}\bar{r}_i\right)\right|^2=\int_{-\infty}^{\infty} d\chi
~ \tilde{P}_a(\chi) \exp\left[-i \frac{\chi}{\hat{d}_i}
\bar{r}_i\right] ~.\label{Ipu11}
\end{eqnarray}
The proof of the last equality in Eq. (\ref{Ipu11}) is based on
the autocorrelation theorem, which states that if the
(two-dimensional) Fourier transform of a function $w(x,y)$ with
respect to variables $\alpha_x$ and $\alpha_y$ is indicated by
$\bar{w}(\alpha_x,\alpha_y)$, then the Fourier transform of the
two-dimensional autocorrelation function of $w(x,y)$ with respect
to the same variables $\alpha_x$ and $\alpha_y$ is given by
$|\bar{w}(\alpha_x,\alpha_y)|^2$. In formulas, after definition of
the autocorrelation function

\begin{equation}
\mathcal{A}[w](x,y) = \int_{-\infty}^{\infty} d\eta
\int_{-\infty}^{\infty} d\xi w(\eta+x,\xi+y)
w^*(\eta,\xi)~,\label{autoapp}
\end{equation}
which is equivalent to

\begin{equation}
\mathcal{A}[w](x,y) = \int_{-\infty}^{\infty} d\eta
\int_{-\infty}^{\infty} d\xi w(\eta+x/2,\xi+y/2)
w^*(\eta-x/2,\xi-y/2)~,\label{auto2app}
\end{equation}
the autocorrelation theorem states that

\begin{equation}
\int_{-\infty}^{\infty} dx \int_{-\infty}^{\infty} dy  \exp{[i
(\alpha_x x+ \alpha_y y)]} \mathcal{A}[w](x,y) =
|\bar{w}(\alpha_x,\alpha_y)|^2~. \label{fourapp}
\end{equation}
Eq. (\ref{Ipu11}) could have been written down immediately just
recalling that the line spread function is the Fourier transform
of the pupil autocorrelation function.

It is convenient to make the change of variable $\chi
\longrightarrow \Delta x \hat{a}$. In fact, because of the
definition of $b_n$ in Eq. (\ref{aberration}), only the ratio
$\Delta \hat{r}/\hat{a}$, or $\chi/\hat{a} = 2 \Delta
\hat{r}/\hat{a}$ is important. Aside for an unimportant factor
$\hat{a}$, the line spread function becomes

\begin{eqnarray}
l(\bar{r}_i) = \int_{-\infty}^{\infty} d\Delta{x} ~
\tilde{P}_a(\hat{a} \Delta x) \exp\left[-i
\frac{\hat{a}}{\hat{d}_i} \Delta x \bar{r}_i\right] ~.\label{Ipu2}
\end{eqnarray}
The autocorrelation function of the pupil $\tilde{P}_a$, which is
the Fourier transform of $l$, is known under the name of Optical
Transfer Function (OTF) of the system. The Optical Transfer
Function and the line spread function are obviously equivalent and
their knowledge solves the problem of accounting for aberrations.
The Optical Transfer Function can be written explicitly modifying
Eq. (\ref{corrP}) to include a phase error in the expression for
the pupil function, that is

\begin{eqnarray}
\tilde{P}_a(\hat{a} \Delta x) &=& \int d \bar{x}
~P\left[\hat{a}\left(\bar{x} + \frac{\Delta x}{2}\right)\right]
P\left[\hat{a}\left(\bar{x} - \frac{\Delta x}{2}\right)\right] \cr
&& \times \exp\left\{i \Phi\left[\hat{a}\left(\bar{x}
+\frac{\Delta x}{2}\right)\right]-i \Phi\left[\hat{a}
\left(\bar{x} - \frac{\Delta x}{2}\right) \right]\right\}~.\cr
&&\label{corrP2}
\end{eqnarray}
Accounting for the definition of the phase error $\Phi$ in Eq.
(\ref{aberration}) one obtains from Eq. (\ref{corrP2}) the
following expression for the Optical Transfer Function:

\begin{eqnarray}
\tilde{P}_a(\hat{a} \Delta x) &=& \int_{-\infty}^{\infty} d
\bar{x} ~P\left[\hat{a}\left(\bar{x} + \frac{\Delta
x}{2}\right)\right] P\left[\hat{a}\left(\bar{x} - \frac{\Delta
x}{2}\right)\right] \cr && \times \exp\left\{i b_n
\left[\left(\bar{x} + \frac{\Delta x}{2}\right)^n-\left(\bar{x} -
\frac{\Delta x}{2}\right)^n\right] \right\}~.\label{corrPabe}
\end{eqnarray}
Here $P$ is the pupil function with no aberrations or apodizations
and only an aberration term of order $n$ has been considered. If
$\bar{x}$ is outside the interval $[-1,1]$, at least one of the
$P$ functions in the integral gives zero value. As a result we may
substitute the integration limits in Eq. (\ref{corrPabe}) to
obtain the following expression for the Optical Transfer Function:

\begin{eqnarray}
\tilde{P}_a(\hat{a} \Delta x) &=& \int_{-1}^{1} d \bar{x}
~P\left[\hat{a}\left(\bar{x} + \frac{\Delta x}{2}\right)\right]
P\left[\hat{a}\left(\bar{x} - \frac{\Delta x}{2}\right)\right] \cr
&& \times \exp\left\{i b_n \left[\left(\bar{x} + \frac{\Delta
x}{2}\right)^n-\left(\bar{x}- \frac{\Delta x}{2}\right)^n\right]
\right\}~.\label{corrPabe2}
\end{eqnarray}
With the help of Eq. (\ref{aberration})  it is possible to
directly calculate how different aberrations modify the expression
for the autocorrelation function and its Fourier transform. These
computations can be carried out by means of numerical techniques
for any value of $n$ and $b_n$, in analogy with what has been done
in \cite{ONEI}.

For completeness we will now calculate the Optical Transfer
Function in several situations, which may also be found in
\cite{ONEI}. In addition, we will also present the line spread
function typical of these situation, which cannot be easily found
in textbooks. Other kind of aberrations can be treated in the same
fashion, and pupils with different shape may be selected. We are
interested in the case when the influence of the phase error is
comparable with the influence of diffraction effects on the pupil,
i.e. when $|b_n| \sim 1$. Here we will consider several cases for
defocusing ($n=2$), coma ($n=3$), spherical aberrations ($n=4$)
and a combination of defocusing and spherical aberrations as well.
This last situation is \textit{per se} interesting, because it
illustrates how it is possible to improve the quality of a lens
with spherical aberration by further introducing a defocusing
aberration (in the case under study, i.e. for $|b_n| \sim 1$). In
Fig. \ref{ftadefo}, Fig. \ref{ftcoma} and Fig. \ref{ftsphe} we
plot the line spread functions describing these aberration cases.
As an aside it is worth to anticipate here that Fig. \ref{ftadefo}
is strictly related with the resolution of a pinhole camera setup.
This will be demonstrated in  the next Section \ref{sec:cam}. As
has been already said, the knowledge of the line spread function
is completely equivalent to the knowledge of the transfer function
(OTF) of the system.  In Fig. \ref{expl1}, Fig. \ref{expl2}, and
Fig. \ref{expl3} we plot the transfer functions relative to the
same cases  treated in Fig. \ref{ftadefo}, Fig. \ref{ftcoma} and
Fig. \ref{ftsphe}.

In closing, it is interesting to deal with the limit when
$\hat{z}_1^2/\hat{N} \ll \hat{a}^2 \ll \hat{D} \hat{z}_1^2$. In
this case the coherence length at the pupil is much smaller than
the characteristic pupil size, or, equivalently,
$P_a(\hat{r}_1)P_a^*(\hat{r}_2) \simeq |P_a(\hat{r}_1)|^2 \simeq
|P_a(\hat{r}_2)|^2$, as presented in Eq. (7.2-15b) of \cite{GOOD}.
In order to retain this last simplification when aberrations are
present, we must require that the phase of the pupil function
$P_a$ (that is now a complex object) is not appreciably different
when $\hat{r}_1$ and $\hat{r}_2$ are separated by a distance of
order of the coherence length or smaller. This is equivalent to
the requirement that the characteristic scale of the lens
imperfections is much larger than the coherence length.
Mathematically, this means that $|{b}_n| \lesssim 1$. Under this
assumption, aberrations cannot affect the cross-spectral density
in the limit  $\hat{z}_1^2/\hat{N} \ll
\hat{a}^2 \ll \hat{D} \hat{z}_1^2$. 
This fact is known from a long time and, as reported in
\cite{GOOD}, it was first discovered by Zernike \cite{ZERN}. The
same limiting case can be presented in the line spread function
formalism. One should recall that the resolution due to
diffraction effects is of order $\hat{z}_1/(\hat{a}
\sqrt{\hat{N}})$, as has already been seen in the previous Section
\ref{par:ngdgb}. This means that in the limit
$\hat{z}_1^2/\hat{N} \ll \hat{a}^2 \ll \hat{D} \hat{z}_1^2$ it
makes sense to account for diffraction effects from the pupil,
because in this case the resolution due to diffraction effects is
worse than that related with the quasi-homogeneous approximation
($\hat{z}_1/(\hat{a} \sqrt{\hat{N}}) \gg
1/\sqrt{\hat{N}\hat{D}}$). However, one may choose to worsen the
resolution of the calculations from $1/\sqrt{\hat{N}\hat{D}}$ to
the resolution due to diffraction effects, $\hat{z}_1/(\hat{a}
\sqrt{\hat{N}}) \ll 1$. This is equivalent to neglect diffraction
effects. In this case, since $|b_n| \lesssim 1$, the
autocorrelation function of the pupil can be substituted with
unity or, equivalently, the line spread function $l$ plays the
role of a Dirac $\delta$-function in the calculation of the
intensity. Therefore, aberrations cannot affect the intensity
distribution at the image plane. Moreover, the expression for the
spectral degree of coherence Eq. (\ref{GPInb55limit2}), remains
valid for $|{b}_n| \lesssim 1$ because, as already discussed,
$P_a(\hat{r}_1)P_a^*(\hat{r}_2) \simeq |P_a(\hat{r}_1)|^2 \simeq
|P_a(\hat{r}_2)|^2$. One concludes that in this limit, and with
resolution $\hat{z}_1/(\hat{a} \sqrt{\hat{N}}) \ll 1$, aberrations
cannot affect the coherence properties on the image plane.

\subsection{\label{sub:abgeo} Severe aberrations}

It is now interesting to discuss an analytical treatment valid in
the case for $|b_n| \gg 1$, which exploits the simplifications
arising from the large parameter $|b_n|$. Under this constraint,
aberrations will be considered severe.

\subsubsection{\label{subsub:phyo} Physical Optics prediction of the line spread function}

Under the approximation $|b_n| \gg 1$, it is possible to present
an analytical calculation for the line spread function $l$ which
characterizes the imaging system in the case of quasi-homogeneous
sources. Then, once the line spread function is known, one obtains
the intensity at the image plane by convolving the line spread
function and the intensity from an ideal system (i.e. without
accounting for the pupil influence).

Let us now focus on that term in the phase factor of Eq.
(\ref{corrPabe2}) which is linear in $\Delta x$, i.e. on $n b_n
\Delta x \bar{x}^{n-1}$. We will assume that the integrand
contributes to the integral for all values of $\bar{x}$ inside the
interval $[-1,1]$, otherwise the autocorrelation function would be
suppressed, as the effective integration range would be smaller
than $[-1,1]$. Then, a typical scale of the autocorrelation
function is obtained in terms of $\Delta x$ by imposing $n |b_n|
\Delta x \bar{x}^{n-1} \sim 1$. In fact, as $n |b_n| \Delta x
\bar{x}^{n-1}
> 1$ the integrand starts to exhibit fast oscillatory
behavior, thus suppressing the integral. Thus, the characteristic
scale $\Delta x_{typ} \sim 1/(n |b_n|)$ is found. Since we assumed
$|b_n| \gg 1$, we can state that, with accuracy $1/|b_n|$, the
functions $P$ inside the integrand can be substituted with unity
and the nonlinear phase factors in $\Delta x^k$ with $k=2,3 ...$
can be neglected, at least for reasonable orders of $n$, as they
would give rise to typical scales of order $1/|b_n|^{1/k} \gg
1/|b_n|$. As a result we obtain the following major
simplification:

\begin{eqnarray}
\tilde{P}_a(\hat{a} \Delta x) &=& \int_{-1}^{1} d \bar{x}
\exp\left[i n b_n \Delta x \bar{x}^{n-1}
\right]~.\label{corrPabe3}
\end{eqnarray}
Eq. (\ref{corrPabe3}) can be integrated analytically for all
values of $n$ (therefore including defocusing, coma, spherical or
higher order aberrations). After definition of

\begin{eqnarray}
\mathcal{T}(n, b_n, \Delta x) &=& \frac{2}{(n-1)\left[-i n b_n
x\right]^{\frac{1}{n-1}}}\cr && \times\left[ (n-1)
\Gamma\left(0,\frac{n}{n-1}\right)-\Gamma\left(\frac{1}{n-1},-i n
b_n \Delta x \right) \right]~,\label{Tfunc}
\end{eqnarray}
$\Gamma(s,z)$ being the incomplete Euler gamma function

\begin{equation}
\Gamma(s,z) = \int_z^{\infty} dt ~t^{s-1} \exp[-t]~, \label{euler}
\end{equation}
we have the following result:

\begin{equation}
\tilde{P}_a(\hat{a} \Delta x) = \mathrm{Re}\left[\mathcal{T}(n,
b_n, \Delta x)\right] + \Pi(n) \cdot
\mathrm{Im}\left[\mathcal{T}(n, b_n, \Delta x)\right]
~,\label{panalyt}
\end{equation}
where $\Pi(n)$ is the parity of $n$, i.e. $\Pi(n)=0$ if $n$ is
even and $\Pi(n)=1$ if $n$ is odd. Eq. (\ref{panalyt}) is valid
for any value of $n>1$.

\begin{figure}
\begin{center}
\includegraphics*[width=140mm]{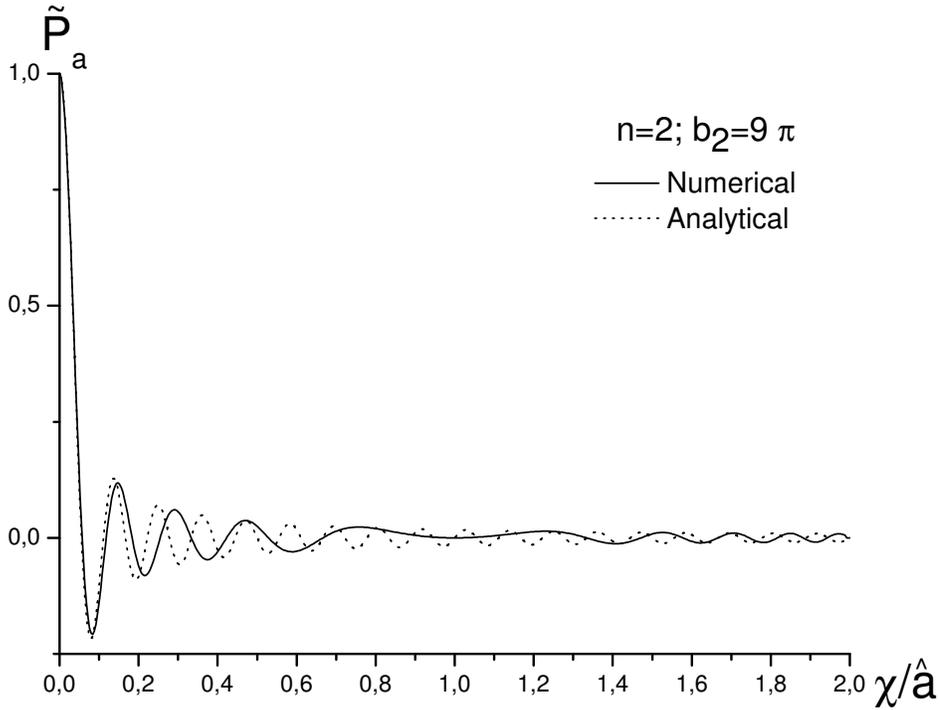}
\caption{\label{bl1} The transfer function in the presence of
defocusing aberration ($n=2$, $b_2 = 9 \pi$). Numerical techniques
have been used to calculate the exact autocorrelation function
that is compared to the analytical evaluation of the
autocorrelation function in the severe aberration asymptote.}
\end{center}
\end{figure}

\begin{figure}
\begin{center}
\includegraphics*[width=140mm]{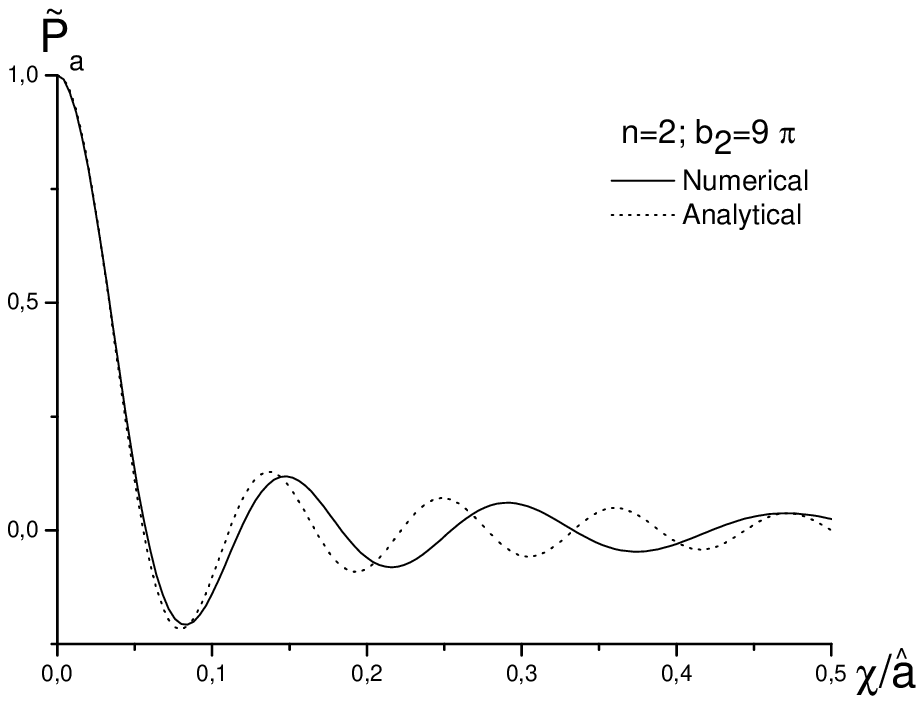}
\caption{\label{bl2} An enlarged version of Fig. \ref{bl1}.
Numerical techniques have been used to calculate the exact
autocorrelation function that is compared to the analytical
evaluation of the autocorrelation function in the severe
aberration asymptote.}
\end{center}
\end{figure}

\begin{figure}
\begin{center}
\includegraphics*[width=140mm]{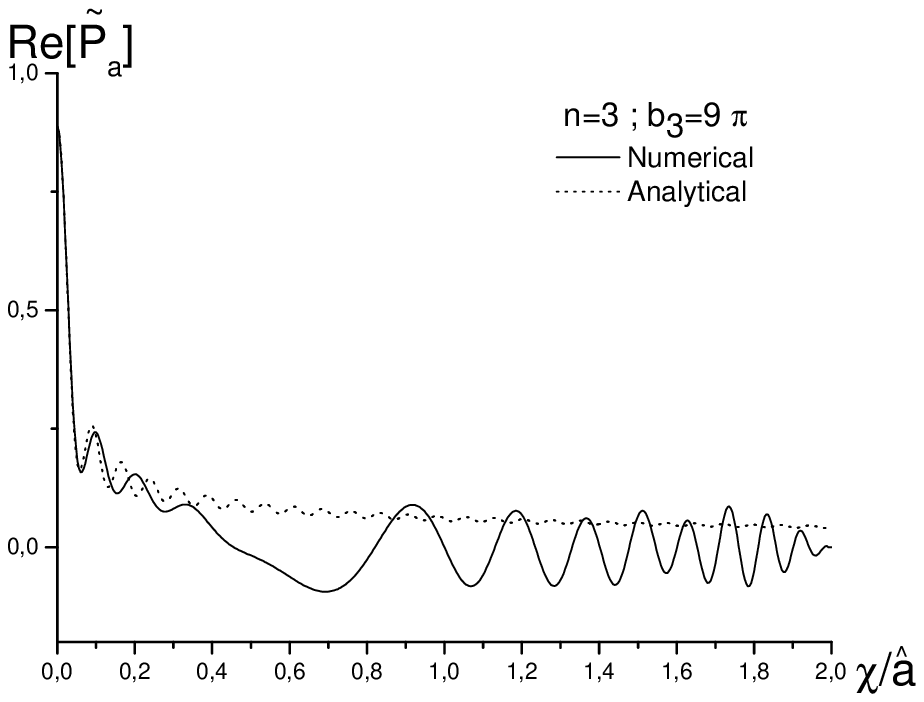}
\caption{\label{bl3} The real part of the transfer function in the
presence of coma aberration ($n=3$, $b_3 = 9\pi$).  Numerical
techniques have been use to calculate the exact autocorrelation
function that is compared to the analytical evaluation of the
autocorrelation function in the severe aberration asymptote.}
\end{center}
\end{figure}

\begin{figure}
\begin{center}
\includegraphics*[width=140mm]{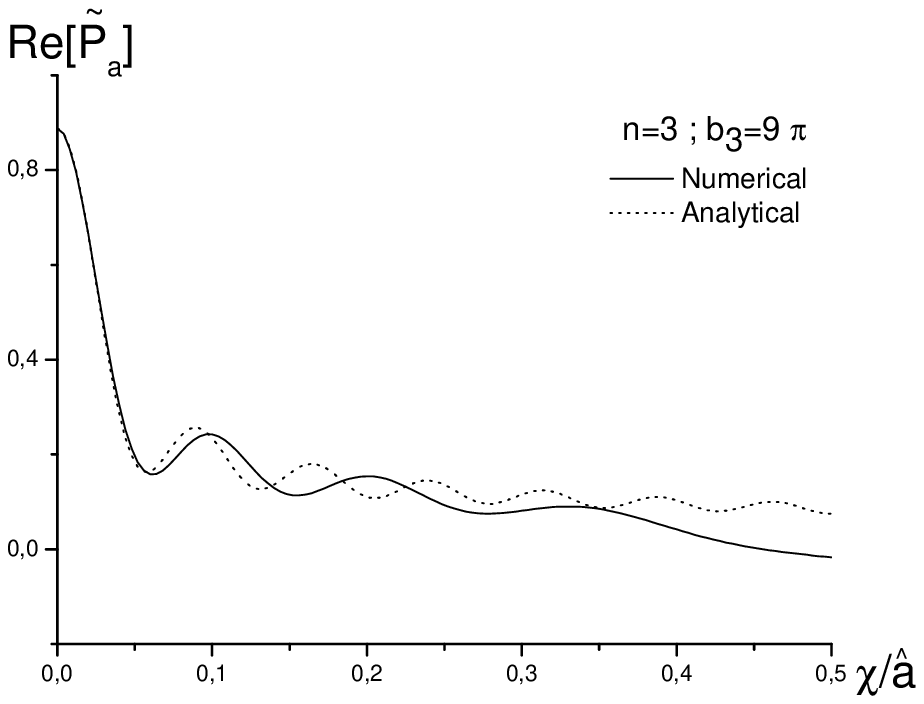}
\caption{\label{bl4} An enlarged version of Fig. \ref{bl3}.
Numerical techniques have been use to calculate the exact
autocorrelation function that is compared to the analytical
evaluation of the autocorrelation function in the severe
aberration asymptote.}
\end{center}
\end{figure}

\begin{figure}
\begin{center}
\includegraphics*[width=140mm]{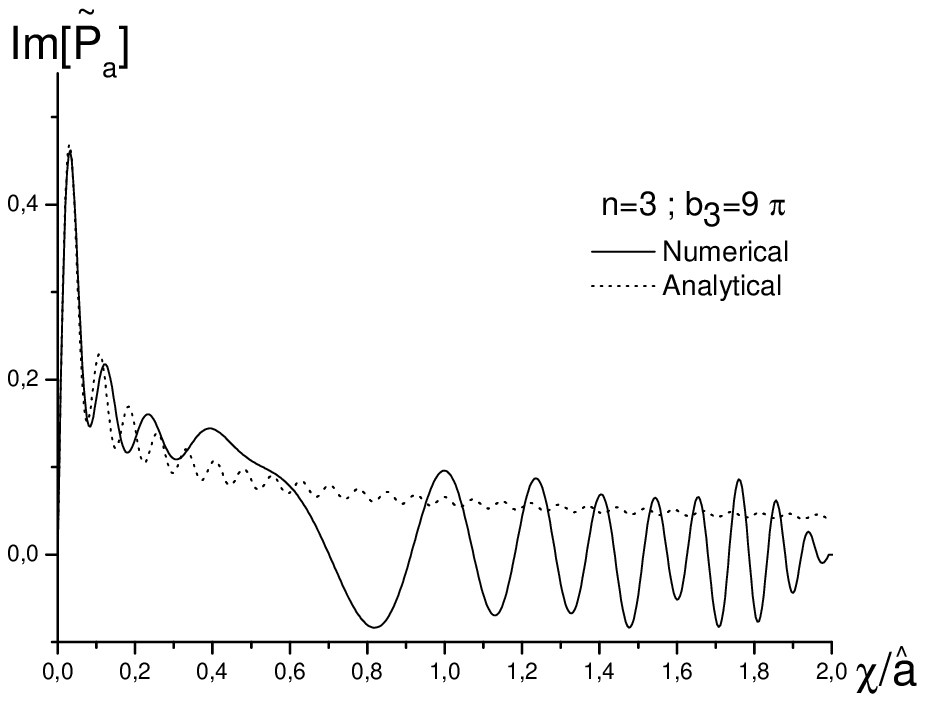}
\caption{\label{bl5} The imaginary part of the transfer function
in the presence of coma aberration ($n=3$, $b_3 = 9\pi$).
Numerical techniques have been use to calculate the exact
autocorrelation function that is compared to the analytical
evaluation of the autocorrelation function in the severe
aberration asymptote.}
\end{center}
\end{figure}

\begin{figure}
\begin{center}
\includegraphics*[width=140mm]{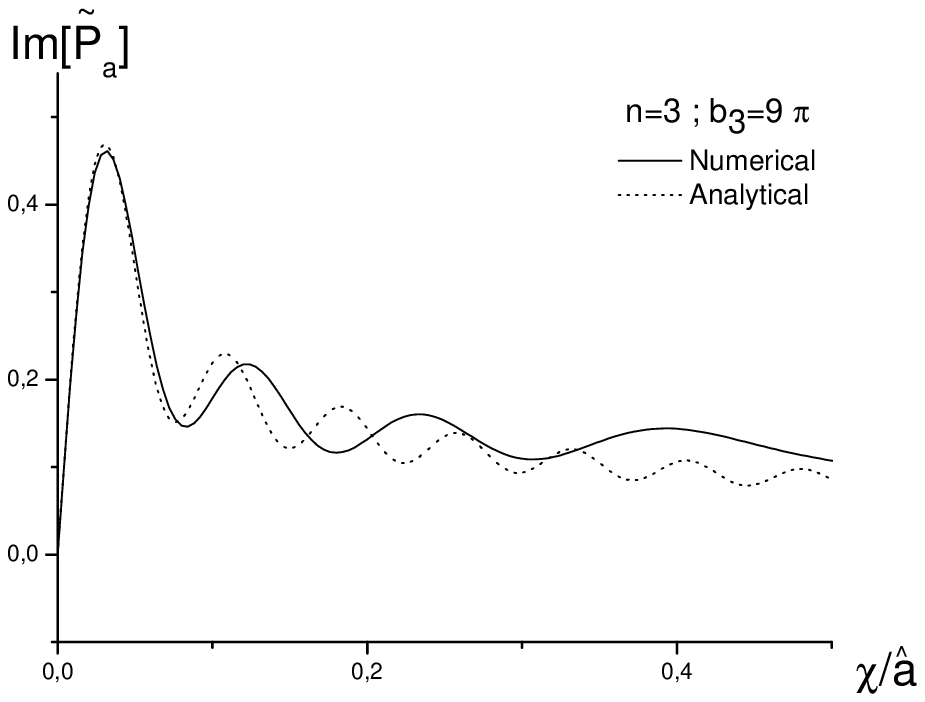}
\caption{\label{bl6} An enlarged version of Fig. \ref{bl5}.
Numerical techniques have been use to calculate the exact
autocorrelation function that is compared to the analytical
evaluation of the autocorrelation function in the severe
aberration asymptote.}
\end{center}
\end{figure}

It is interesting to compare the shape of the autocorrelation
function obtained in the limiting case $|b_n| \gg 1$ with that
obtained with numerical calculations which do not exploit the
simplification based on the large value of the parameter $|b_n|$.
They rely, instead, on the exact formula for the autocorrelation
function, Eq. (\ref{corrPabe2}) . This gives a visual idea of the
accuracy of the asymptotic. Fixing $b_n = 9 \pi$ we plot the
autocorrelation function for defocusing aberrations, with $n=2$,
in Fig. \ref{bl1} and Fig. \ref{bl2}. For coma aberrations, with
$n=3$, we plot the real part of the autocorrelation function in
Fig. \ref{bl3} and Fig. \ref{bl4}, while  the imaginary part  is
plotted in Fig. \ref{bl5} and Fig. \ref{bl6}. The function
$\mathcal{T}$ should be truncated as $\Delta x>2$.

Explicit substitution of Eq. (\ref{corrPabe3}) in Eq. (\ref{Ipu2})
gives

\begin{eqnarray}
l(\bar{r}_i) = \int_{-\infty}^{\infty} d\Delta{x} ~ \int_{-1}^{1}
d \bar{x} \exp\left[i n b_n \Delta x \bar{x}^{n-1}  \right]
\exp\left[-i \frac{\hat{a}}{\hat{d}_i} \Delta x \bar{r}_i\right]
~.\label{Ipu3}
\end{eqnarray}
Finally, exchange of the integration order and calculation of the
integral in $d \Delta{x}$ yields the following expression for the
line spread function, provided that $|b_n| \gg 1$:

\begin{eqnarray}
l(\bar{r}_i) &=&  \int_{-1}^{1} d \bar{x}~ \delta\left(n b_n
\bar{x}^{n-1}  - \frac{\hat{a}}{\hat{d}_i} \bar{r}_i\right)
~.\label{Ipu4}
\end{eqnarray}
Eq. (\ref{Ipu4}) may be explicitly evaluated with the help of the
new integration variable $y = n b_n \bar{x}^{n-1}$. Care must be
taken in separating the cases when $n$ is even and when $n$ is
odd.

When $n$ is even we obtain

\begin{eqnarray}
l(\bar{r}_i) &=& \frac{1}{2(n-1)} \int_{-n b_n}^{n b_n} d y~
\delta\left(y  - \frac{\hat{a}}{\hat{d}_i} \bar{r}_i\right)
\left(\frac{\mathrm{abs}(y)}{n b_n}\right)^{-\frac{n-2}{n-1}} \cr
&=& \frac{1}{2(n-1)}
\mathrm{rect}\left(\frac{\hat{a}{\bar{r}_i}}{2\hat{d}_i n
b_n}\right)\cdot
\left(\frac{\hat{a}~{\mathrm{abs}(\bar{r}_i)}}{\hat{d}_i n
b_n}\right)^{-\frac{n-2}{n-1}} ~.\label{Ipu5}
\end{eqnarray}
Here the function $\mathrm{rect}(x)$ is defined, as before,
following \cite{GOOD}, and is equal to unity for $|x|\leqslant
1/2$ and zero otherwise.  When $n$ is odd we have

\begin{eqnarray}
l(\bar{r}_i) &=& \frac{1}{n-1} \int_{0}^{n b_n} d y~ \delta\left(y
- \frac{\hat{a}}{\hat{d}_i} \bar{r}_i\right) \left(\frac{y}{n
b_n}\right)^{-\frac{n-2}{n-1}}\cr & =& \frac{1}{n-1}
\mathrm{rect}\left(\frac{\hat{a}{\bar{r}_i}}{\hat{d}_i n
b_n}-\frac{1}{2}\right)\cdot
\left(\frac{\hat{a}{\bar{r}_i}}{\hat{d}_i n
b_n}\right)^{-\frac{n-2}{n-1}} ~.\label{Ipu6}
\end{eqnarray}
Introduction of the new variable $\bar{r}'' = \hat{a}
\bar{r}_i/(\hat{d}_i n b_n)$ allows to write Eq. (\ref{Ipu5}) and
Eq. (\ref{Ipu6}) in a more compact way,

\begin{eqnarray}
l(\bar{r}'') &=& \frac{1}{2(n-1)}
\mathrm{rect}\left(\frac{\bar{r}''}{2}\right)\cdot
(\mathrm{abs}(\bar{r}''))^{-\frac{n-2}{n-1}} ~\label{Ipu7}
\end{eqnarray}
when $n$ is even, and

\begin{eqnarray}
l(\bar{r}'') &=&  \frac{1}{n-1}
\mathrm{rect}\left({\bar{r}''}-\frac{1}{2}\right)\cdot
\left({\bar{r}''}\right)^{-\frac{n-2}{n-1}} ~\label{Ipu8}
\end{eqnarray}
when $n$ is odd. Note that we have normalized Eq. (\ref{Ipu7}) and
Eq. (\ref{Ipu8})  in such a way that integration of $l$ in
$d\bar{r}''$ over the real field gives unity.

\begin{figure}
\begin{center}
\includegraphics*[width=140mm]{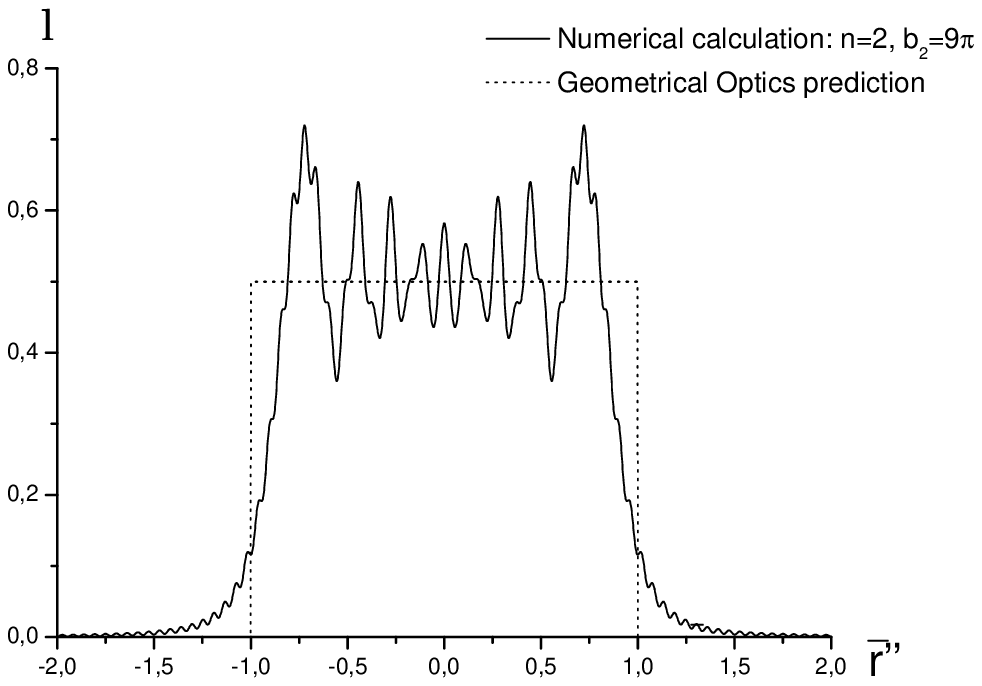}
\caption{The line spread function versus the reduced coordinate on
the image plane $\bar{r}'' = \hat{a} \bar{r}_i /(\hat{d}_i n b_n)$
in the case of severe defocusing aberration($n=2$, $b_2 = 9 \pi$)
and comparison with the geometrical optics prediction.
\label{abel2} }
\end{center}
\end{figure}

\begin{figure}
\begin{center}
\includegraphics*[width=140mm]{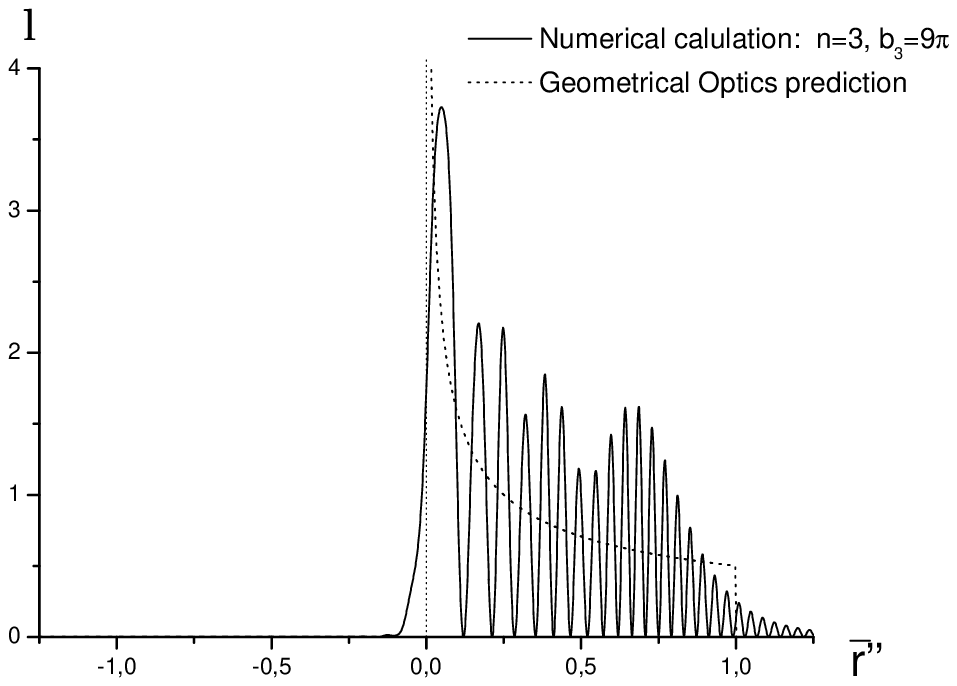}
\caption{The line spread function versus the reduced coordinate on
the image plane $\bar{r}'' = \hat{a} \bar{r}_i /(\hat{d}_i n b_n)$
in the case of severe coma aberration ($n=3$, $b_3 = 9 \pi$) and
comparison with the geometrical optics prediction. \label{abel3} }
\end{center}
\end{figure}

\begin{figure}
\begin{center}
\includegraphics*[width=140mm]{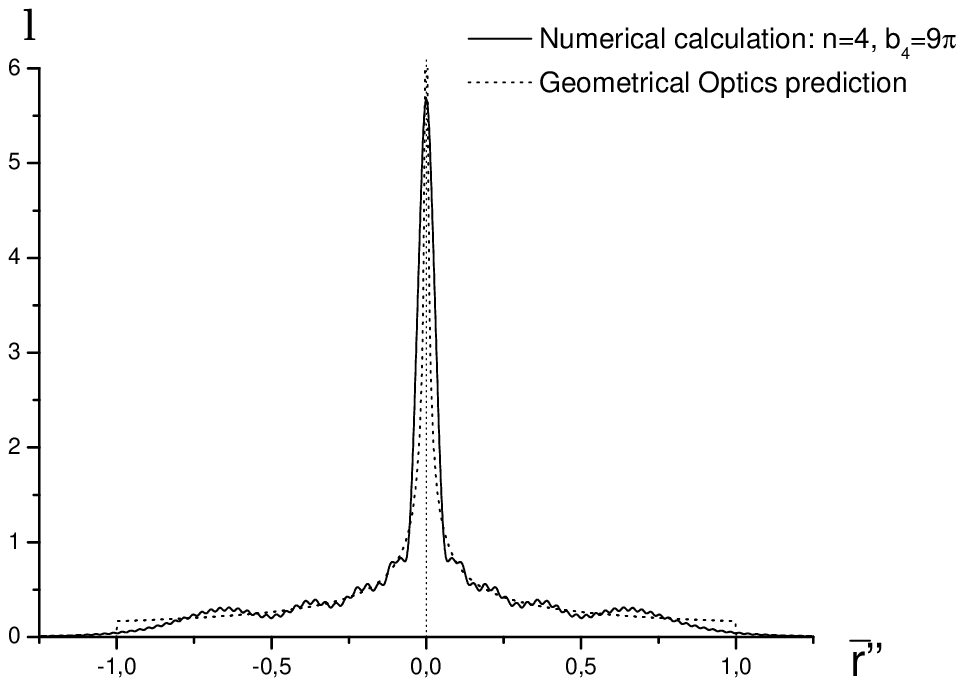}
\caption{The line spread function versus the reduced coordinate on
the image plane $\bar{r}'' = \hat{a} \bar{r}_i /(\hat{d}_i n b_n)$
in the case of severe spherical aberration ($n=4$, $b_4 = 9 \pi$)
and comparison with the geometrical optics
prediction.\label{abel4} }
\end{center}
\end{figure}

As we will see later on, Eq. (\ref{Ipu7}) and Eq. (\ref{Ipu8}) can
be found with the help of Geometrical Optics alone. We will refer
to such derivation as the Geometrical Optics prediction of the
line spread function. We plotted Eq. (\ref{Ipu7}) or Eq.
(\ref{Ipu8}) as a function of $\bar{r}''$ for different
aberrations. In Fig. \ref{abel2} we plotted the case of defocusing
aberration , in Fig. \ref{abel3} the case of coma and in Fig.
\ref{abel4} the case of spherical aberrations. Also, in these
figures, comparison with numerical calculations is shown for the
severe aberration cases $b_2 = 9\pi$,  $b_3 = 9\pi$ and $b_4 =
9\pi$. Note that Eq. (\ref{Ipu7}) is symmetric in $\bar{r}''$
(when $n$ is even $l$ is symmetric), while Eq. (\ref{Ipu8}) is not
(when $n$ is odd $l$ is not symmetric). This is consistent with
the fact that the Optical transfer function is real in the case
$n$ is even, while it has a non-zero imaginary part in the case
$n$ is odd (see, for instance, Fig. \ref{expl1}, Fig. \ref{expl2}
and Fig. \ref{expl3} and, later on, Fig. \ref{bl1}, Fig. \ref{bl3}
and Fig. \ref{bl5}). The same behavior is also found in Fig.
\ref{ftadefo}, Fig. \ref{ftcoma} and Fig. \ref{ftsphe}.
Furthermore it should be noted that the line spread functions in
Fig. \ref{abel3} and Fig. \ref{abel4} are not convergent for
values of $\bar{r}''$ near zero. However, the meaning of the line
spread function is that of the imaged intensity from a line input,
but a line input is not physical, and must be represented in terms
of a generalized function, a $\delta$-Dirac function. It is not
surprising that such an object may lead to a singular result.


\subsubsection{\label{subsub:phge} Physical Optics and Geometrical Optics}

In the previous Section \ref{subsub:phyo} we presented a
calculation of the line spread function $l$ for the case $|b_n|
\gg 1$ based on Physical Optics. This gave us Eq. (\ref{Ipu4}) for
the line spread function or, equivalently, Eq. (\ref{Ipu5}) and
Eq. (\ref{Ipu6}), that are an explicit evaluation of Eq.
(\ref{Ipu4}) in the case $n$ is even or odd. We have seen that the
analytical treatment in Section \ref{subsub:phyo} follows
mathematically from a major simplification of Eq.
(\ref{corrPabe2}), arising from the large parameter $|b_n| \gg 1$.
Physically, this condition means that effects of diffraction from
the pupil can be neglected. Consider the expression for the pupil
autocorrelation function, Eq. (\ref{corrP2}). On the one hand,
when $\bar{x} \sim 1/b_n$ the phase term due to aberration effects
in Eq. (\ref{corrP2}) becomes comparable to unity, thus leading to
oscillatory behavior of the integrand. On the other hand, the
pupil finite aperture limits the integration in Eq. (\ref{corrP2})
for $\bar{x} \sim 1$. Therefore, $|b_n| \gg 1$ diffraction effects
can be neglected, while they start to become important when $|b_n|
\lesssim 1$. It follows that it must be possible to obtain results
in Section \ref{subsub:phyo}, which have been derived with the
help of physical Optics, with the help of Geometrical Optics only.
In the next Section \ref{subsub:geop} we discuss how this can be
done.

Before doing that it is worth to discuss the relationship between
Geometrical and Physical Optics, which may otherwise be
misleading. We should make clear that when we discuss about
Geometrical or Physical Optics we are talking about possible ways
of calculating the line spread function of the system $l$.

In Section \ref{par:ngdgb} we have introduced the concept of line
spread function $l$ and we have demonstrated that, under the
assumption that the virtual source is quasi-homogeneous, $l$
constitutes a sort of passport for a given lens. It relates the
intensity from any quasi-homogeneous source imaged with a perfect
lens and the intensity obtained by using a particular non-ideal
lens. The intensity from a specific optical system can be
recovered as a convolution of the intensity obtained in the case
of an ideal optical system and the line spread function. The
intensity from an ideal system at the image plane is, by
definition, the intensity at the virtual source. Therefore, when
the quasi-homogeneous approximation is applicable, it is always
possible to break the imaging problem into two separate problems.
First, specify the intensity distribution of the source. Second,
specify the optical system through the $l$ function.

When diffraction effects are negligible with respect to aberration
effects, i.e. when $|b_n| \gg 1$, the line spread function of the
lens can be calculated by both Physical Optics considerations (as
in Section \ref{subsub:phyo}) and Geometrical Optics
considerations, as in the next Section \ref{subsub:geop}. In this
case, a finite aperture size does not influence the calculation of
the line spread function. It is responsible for the quantity of
the total energy transmitted only. Once the source characteristics
are specified, one may use a ray-tracing code to get the image
intensity from a non-ideal system or, equivalently, one may
calculate the $l$ function and convolve with a scaled version of
the intensity on the virtual source. When diffraction effects are
not negligible anymore, $l$ can be evaluated with the help of
Physical Optics considerations only. In this case, use of
ray-tracing codes to solve the imaging problem makes no sense.
Yet, the quasi-homogeneous approximation allows one to use a line
spread function approach. Convolution of $l$ with a scaled version
of the intensity on the virtual source solves the imaging problem.
From this viewpoint the quasi-homogeneity of the source is an
\textit{a priori} condition with respect to the possibility of
applying Geometrical Optics for the solution of imaging problems
from a non-ideal setup.

A final word of caution should be spent regarding notations
historically used to represent aberrations (see, for instance,
\cite{SHAN}). On the one hand, the knowledge of the phase error
$\Phi(\hat{r})=b_n (\hat{r}/\hat{a})^n$ is equivalent to the
knowledge of the surfaces of equal phase, i.e. of the wavefronts
of the electromagnetic field. In this case one usually talks about
"wave aberration". On the other hand, the knowledge of the
derivative $d \Phi(\hat{r})/d\hat{r}$ (or, in more dimensions, of
the gradient $\vec{\nabla} \Phi(\vec{\hat{r}})$) is equivalent to
the knowledge of the vector field orthogonal to the wavefronts of
the electromagnetic field. In the case diffraction effects are not
present, one may identify $d \Phi(\hat{r})/d\hat{r}$  with the
extra angular displacement of a ray and recover the deviation of
the transverse coordinate of a ray on the image plane. In this
case, usually, one talks about "Geometrical aberration"
\cite{SHAN}\footnote{In reference \cite{BORN} the term "Rays
aberration" is used in place of "Geometrical aberration"}. It
should be clear though, that the presentations in terms of "Wave
aberration" and "Geometrical aberration" are completely equivalent
from a mathematical viewpoint, regardless the value assumed by the
parameter $|b_n|$ with respect to unity. Therefore, one needs to
clearly distinguish between the language used (Geometrical or Wave
aberration) and the possibility of applying Geometrical Optics to
calculate the line spread function. When $|b_n| \lesssim 1$ this
is not possible, and one has to rely on Wave Optics predictions
for the line spread function $l$ only. When $|b_n| \gg 1$ one may
rely both on Wave Optics predictions (see Section
\ref{subsub:phyo}) or Geometrical Optics predictions (see Section
\ref{subsub:geop}), and the two predictions must coincide.

\subsubsection{\label{subsub:geop} Geometrical Optics prediction of the line spread function}

As has been already said in Section \ref{subsub:phge}, when
diffraction effects are negligible with respect to aberration
effects, i.e. when $|b_n| \gg 1$, the line spread function of the
lens can be calculated by both Physical Optics considerations (as
in Section \ref{subsub:phyo}) and Geometrical Optics
considerations. In this case, a finite aperture size does not
influence the calculation of the line spread function, and is
responsible for the quantity of the total energy transmitted only.

In the Geometrical Optics limit, Maxwell equations can be replaced
by the simpler Eikonal equation, which should be solved for
surfaces of equal phase. In our one-dimensional case of study,
these surfaces are indeed lines on the plane $\hat{r}-\hat{z}$ and
may be indicated with the family $\phi_p(\hat{r},\hat{z})$, where
$p$ identifies a particular value of the phase. Once the functions
$\phi_p$ are known, one can recover the usual ray-tracing
techniques remembering that rays are, at any point, normal to
surfaces with equal phase. The following ray equation holds:

\begin{equation}
\vec{s}(\hat{r},\hat{z}) = \vec{\nabla}\phi_p(\hat{r},\hat{z}) ~,
\label{rayeq}
\end{equation}
where $\vec{s}$ indicates a vector field tangent to the rays
expressed in normalized units. If the lens is ideal one recovers,
at the image plane, the intensity profile of the source reversed
and magnified. In particular, a line input would be mapped to a
line on the image plane. However, aberrations modify the surfaces
of equal phase in Eq. (\ref{rayeq}), because a phase error is to
be added to $\phi_p~$. As a result, when calculating $\vec{s}$ by
means of Eq. (\ref{rayeq}), one obtains an extra angular
displacement in normalized units for each ray dependent on the
transverse position of the ray, that is

\begin{figure}
\begin{center}
\includegraphics*[width=90mm]{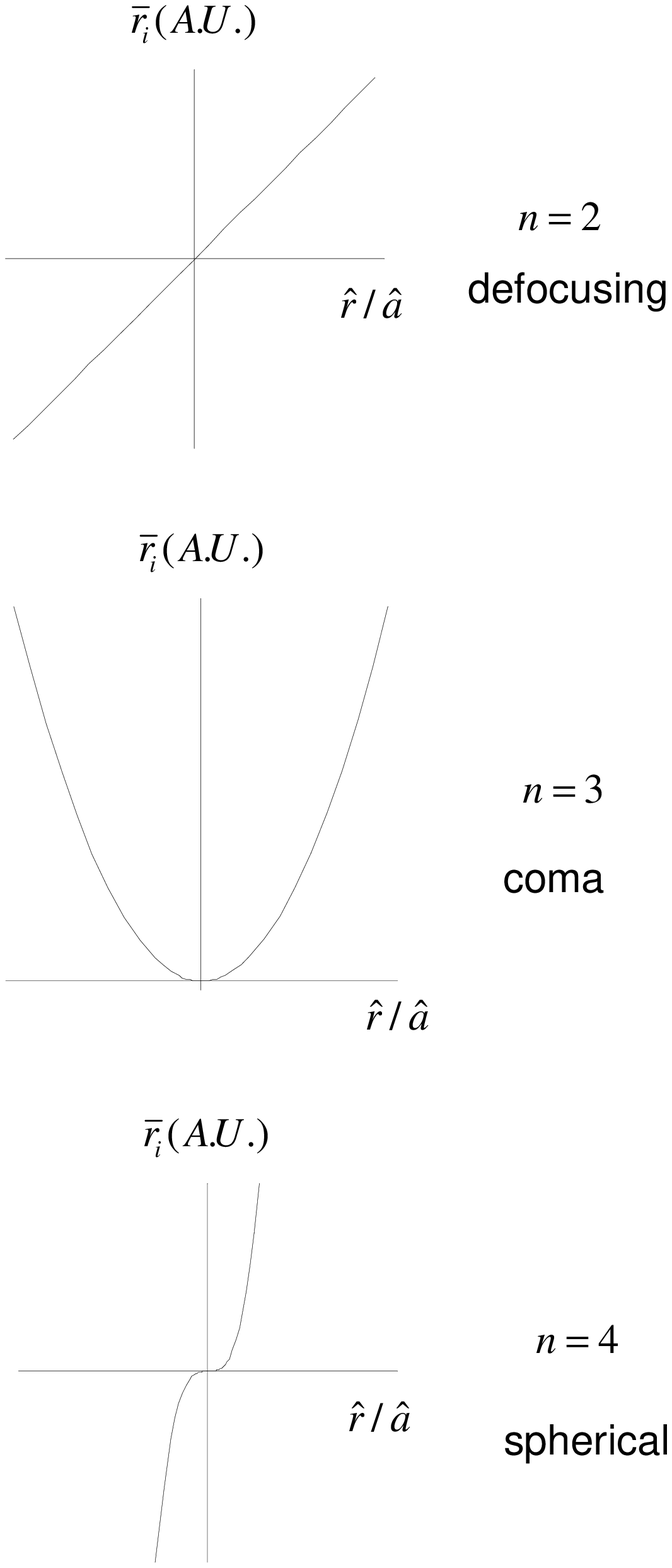}
\caption{\label{goabe} Amount by which a geometrically traced ray
departs from a desired location in the image formed by an optical
system. The ordinate for each curve is the height at which the ray
intersects the image plane. The abscissa is the coordinate of the
ray at the pupil plane. $n=2$ represents defocusing, $n=3$
represents coma, and $n=4$ represents spherical aberration cases
according to Eq. (\ref{transveim}). }
\end{center}
\end{figure}
\begin{equation}
\alpha(\hat{r}) = \frac{d \Phi(\hat{r})}{d\hat{r}} =
\frac{n}{\hat{a}} b_n \left(\frac{\hat{r}}{\hat{a}}\right)^{n-1}~.
\label{xtrangle}
\end{equation}
It follows that the coordinate of the ray going through the lens
at transverse position $\hat{r}$ has a transverse position, at the
image plane, given by

\begin{equation}
\bar{r}_i(\hat{r})=\frac{n\hat{d}_i}{\hat{a}} b_n
\left(\frac{\hat{r}}{\hat{a}}\right)^{n-1}~. \label{transveim}
\end{equation}
The relation between the coordinate of a ray at the pupil and its
transverse position at the image plane for $n=2$ (defocusing),
$n=3$ (coma), and $n=4$ (spherical aberration) is plotted in Fig.
\ref{goabe}.

If the lens is hit by a finite number of rays, the output of the
system at the image plane is constituted by a finite sum of
Dirac-$\delta$ functions. Each of these can be represented in the
implicit form $\delta\left[\bar{r}_i-\frac{n\hat{d}_i}{\hat{a}}
b_n \left(\frac{\hat{r}}{\hat{a}}\right)^{n-1}\right]$. Therefore,
if the lens is homogeneously illuminated by an infinite number of
rays from the input line source, we obtain the line spread
function

\begin{equation}
l(\bar{r}_i) = \int_{-a}^{a} d \hat{r}
\delta\left[\bar{r}_i-\frac{n\hat{d}_i}{\hat{a}} b_n
\left(\frac{\hat{r}}{\hat{a}}\right)^{n-1}\right]~.\label{lc1}
\end{equation}
Finally, using the new integration variable $\bar{x} =
\hat{r}/\hat{a}$ and normalizing $l$ so that the integral of $l$
gives unity yields

\begin{equation}
l(\bar{r}_i) = \int_{-1}^{1} d \bar{x}
\delta\left(\bar{r}_i-\frac{n\hat{d}_i}{\hat{a}} b_n
\bar{x}^{n-1}\right)~,\label{lc1}
\end{equation}
that is equivalent to Eq. (\ref{Ipu4}). From a ray-tracing
viewpoint, the problem of calculating the line spread function
reduces to the problem of transforming a uniform distribution of
rays into a non-uniform distribution related to the non-linear
transformation Eq. (\ref{transveim}). When $|b_n| \lesssim 1$
instead, Geometrical Optics cannot be used to calculate the $l$
function, and the Eikonal approximation fails. As we have seen in
Section \ref{subsub:phyo} this is equivalent, in the language of
Physical Optics, to a situation when the simplification in Eq.
(\ref{corrPabe3}) does not hold.

\section{\label{sec:cam} Pinhole optics}

\begin{figure}
\begin{center}
\includegraphics*[width=150mm]{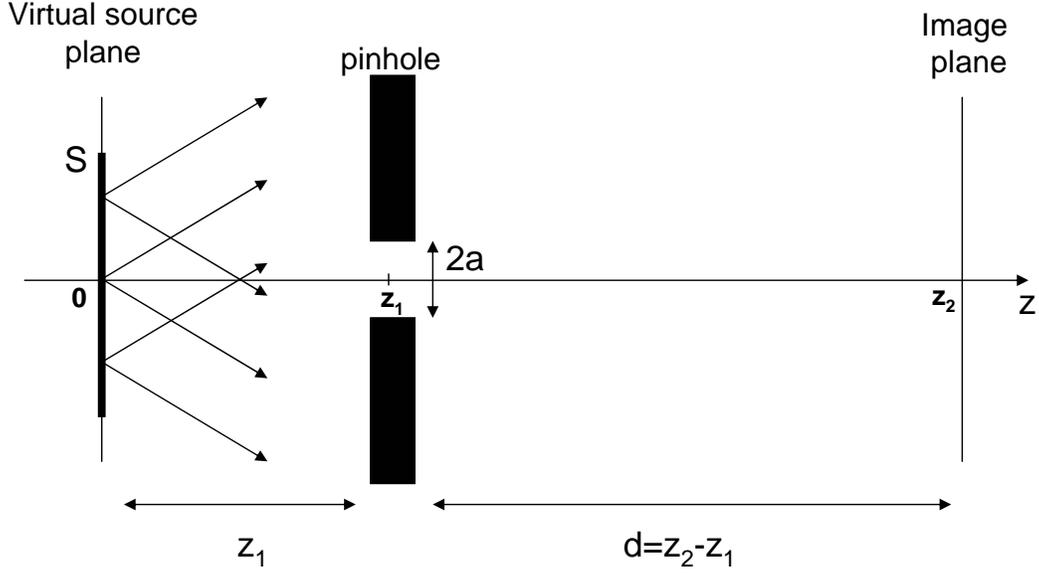}
\caption{\label{pinhgeo} Geometry for a pinhole camera setup. S
indicates the virtual source plane.}
\end{center}
\end{figure}

Taking advantage of Eq. (\ref{crspecprop3bis}), we will now study
the case when images of a virtual source can be obtained with the
help of a pinhole (i.e. a pupil without a lens), without further
lenses or mirrors. When a pinhole can be treated as an imaging
system, people refer to it as an (X-ray) pinhole camera. Here we
will consider the geometry in Fig. \ref{pinhgeo}. The study of
this relatively simple setup  will be helpful to reach a better
understanding of Section \ref{sec:focim}, dedicated  to imaging in
the focal pane, Section \ref{sec:imany}, where we will describe
imaging in any plane behind the lens, and Section \ref{sec:dof},
that will deal with the depth of focus of an imaging system.
Moreover, it will also suggestively show how a problem apparently
not related with the theory of aberrations (in the pinhole camera
setup there is not even a lens) can formally be treated like a
defocusing aberration problem. This is due to the appearance of a
quadratic phase factor in the equation for the intensity at the
image plane.

An X-ray pinhole camera has the same properties as the more
familiar visible light pinhole-camera. The main advantage of such
a lensless imaging system is that a pinhole is easier to fabricate
than a lens. Pinhole cameras can be combined with X-ray
Synchrotron Radiation sources and detectors for a number of
relatively specialized applications \cite{PIN1}.

The conditions under which the pinhole can be treated as an
imaging system are non-trivial, and are not always satisfied. With
the help of Eq. (\ref{crspecprop3bis}) we can investigate the
properties of the image in the limiting case for $\hat{f}
\longrightarrow \infty$, i.e. when there is no lens. We will
restrict ourselves to the one-dimensional case, thus simplifying
the vectorial notation in Eq. (\ref{crspecprop3bis}) to scalar
notation. The assumption of separability of the cross-spectral
density in the horizontal and in the vertical direction
($\hat{N}_x\gg1$ and $\hat{D}_x \gg 1$) suggests, in fact, to
discuss horizontal and vertical directions separately. In general,
one can see that the following conditions must be satisfied in
order to form an image of the source (in one dimension):

$~~~1.$ The pinhole must be in the far field. In this case, using
Eq. (\ref{maintrick2}), the cross-spectral density on the pupil
plane is given by

\begin{eqnarray}
\hat{{G}}\left(\hat{z}_1,{\bar{r}}, \Delta {\hat{r}}\right) &=&
\frac{1}{4\pi^2 \hat{z}_1^2} \exp\left[2i {
\bar{r}}\cdot\Delta{\hat{r}}/{ \hat{z}_1
}\right]\hat{\mathcal{G}}\left(0,-{\bar{r}}, -\Delta
{\hat{r}}\right)~. \label{maintrick2rev}
\end{eqnarray}

$~~~2.$ In the integrand of Eq. (\ref{crspecprop3bis}) two
specific phase factors appear. These are $\exp[2i
{\bar{r}'}\cdot\Delta{\hat{r}'}/{(\hat{z}_2 - \hat{z}_1 )}]$ and
$\exp[2i {\bar{r}'}\cdot\Delta{\hat{r}'}/{ \hat{z}_1 }]$, the
latter appearing through Eq. (\ref{maintrick2rev}). Both must be
negligible.

$~~~3.$ The pinhole size must be larger than the coherence length
on the pinhole plane, or the pupil functions $P$ in Eq.
(\ref{crspecprop3bis}) will modify the dependence of
$\hat{G}(\hat{z}_2)$ on ${\bar{r}}$ (i.e. the image will not have
a good resolution).

If conditions from 1. to  3. above are satisfied, the pinhole
camera works as an imaging system forming an inverted image of the
source. From Eq. (\ref{crspecprop3bis}) one may see that the image
is magnified of a quantity $|\mathrm{M}| = \hat{z}_2/\hat{z_1}$,
because of the same reason as the lens.

We will now give a physical interpretation of conditions 1.
through 3. stated above. We begin considering the limiting case
when $\hat{N} \gg 1$ and $\hat{D} \gg 1$, i.e. a Gaussian
quasi-homogeneous virtual source. Further on we will see up to
what extent this assumption can be relaxed. We assume a large
magnification constant $|\mathrm{M}| \simeq \hat{d}/\hat{z}_1 \gg
1$, where $\hat{d} = \hat{z}_2-\hat{z}_1$. Here this assumption
will be accepted for simplicity and relaxed, later on, to an
arbitrary value of $|\mathrm{M}|$ .

The second condition requires that two distinct phase factors in
Eq. (\ref{crspecprop3bis}) may be neglected. The assumption
$\hat{d} \gg \hat{z}_1$ leads to the single requirement ${\bar{r}
\Delta \hat{r}}/{\hat{z}_1} \ll 1$. On the one hand, $\bar{r}$ is
limited by the presence of the pinhole, i.e. we must impose
$\bar{r} \lesssim \hat{a}$. On the other hand, $\Delta \hat{r}$ is
limited by the coherence length at the pinhole, i.e. $\Delta
\hat{r} \lesssim \hat{z}_1 /\sqrt{\hat{N}}$, because otherwise the
cross-spectral density $\hat{G}$ in Eq. (\ref{crspecprop3bis})
drops to zero. As a result we obtain that the second condition
given above can be expressed in mathematical terms by

\begin{equation}
\hat{a} \ll \sqrt{\hat{N}}~. \label{seccond}
\end{equation}
It is possible to give a clear interpretation of condition
(\ref{seccond}) in terms of Geometrical Optics. In fact, on the
one hand, the minimal geometrical spot size from a line source is
given, at the image plane, by $|\mathrm{M}| \hat{a}$. On the other
hand, the size of the image is of order $|\mathrm{M}|
\sqrt{\hat{N}}$, by definition of magnification $|\mathrm{M}|$.
Then, in order to have a good resolution, we must require that the
image size of a point source be much smaller than the image size
of the object, i.e. $\sqrt{\hat{N}}/\hat{a} \gg 1$, that is
condition (\ref{seccond}).

The third condition given above requires that the pinhole size be
larger than the coherence length on the pinhole plane, otherwise
the pupil functions $P$ in Eq. (\ref{crspecprop3bis}) would modify
the dependence of $\hat{G}(\hat{z}_2)$ on $\vec{\bar{r}}$ (i.e.
the image would be influenced by the pupil). This can be
mathematically stated by requiring that

\begin{equation}
\hat{a} \gg \frac{\hat{z}_1}{\sqrt{\hat{N}}}~. \label{thicond}
\end{equation}
Condition (\ref{thicond}) has a natural explanation in terms of
diffraction theory. In fact, the diffraction spot due to the
presence of the pupil can be estimated as $\hat{d}/\hat{a}$. In
order to have a good resolution, we should impose that the
diffraction spot be much smaller than the image size of the
object, i.e. $|\mathrm{M}| \sqrt{\hat{N}}/(\hat{d}/\hat{a}) \gg
1$, that is condition (\ref{thicond}).

Finally, the first condition given above requires that the
cross-spectral density at the pinhole position be Eq.
(\ref{maintrick2rev}), i.e. the pinhole must be in the far zone
region. This fact can be alternatively stated by requiring that
the radiation spot size at the pupil be dominated by the angular
divergence $\hat{D}$, i.e. $\hat{z}_1  \sqrt{\hat{D}} \gg
\sqrt{\hat{N}}$. In the case $\hat{N} < \hat{D}$, one should
require in any case that $\hat{z}_1 \gg 1$.  This can be
mathematically expressed by the requirement

\begin{equation}
\hat{z}_1 \gg \max\left[\sqrt{\frac{\hat{N}}{\hat{D}}},1\right]~.
\label{fircond}
\end{equation}
Note that condition (\ref{fircond}) and the initial assumption of
a quasi-homogeneous source are equivalent to the requirement that
the pinhole be far enough for the van Cittert-Zernike theorem to
apply. In the case of a perfectly incoherent object (e.g. thermal
light), the source radiates over an angle $2\pi$. Then, the
validity of the van Cittert-Zernike theorem (i.e. the requirement
that the radiation spot size at the pupil be dominated by the
angular divergence) is equivalent to the condition that the
transverse dimension of the source be much smaller than the
distance between the source and the pupil. This is the same
condition for the paraxial approximation to be applicable. As a
result, condition (\ref{fircond}) is always considered satisfied
in usual treatments describing pinhole setups in the presence of
incoherent objects.

The three conditions (\ref{seccond}), (\ref{thicond}) and
(\ref{fircond}) can be summed up in the following:

\begin{equation}
\sqrt{\hat{N}} \gg \hat{a} \gg \frac{\hat{z}_1}{\sqrt{\hat{N}}}
\gg
\max\left[\frac{1}{\sqrt{\hat{D}}},\frac{1}{\sqrt{\hat{N}}}\right]~.
\label{megacond}
\end{equation}
Let us now discuss about the resolution of the pinhole camera.
Consistency of condition (\ref{megacond}) requires that

\begin{equation}
\hat{N} \gg \hat{z}_1~. \label{consis}
\end{equation}
Although condition (\ref{megacond}) requires that condition
(\ref{consis}) be satisfied, it does not pose any constraint on
the relative magnitude of $\hat{a}^2$ with respect to $\hat{z}_1$.
This observation suggests the presence of another characteristic
scale of the problem, $\hat{a}^2/\hat{z}_1$. This scale is linked
with the resolution of the system. When $\hat{a}^2/\hat{z}_1 \gg
1$ or $\hat{a}^2/\hat{z}_1 \ll 1$ we are in the presence of an
extra large or small parameter and, thus, we have two asymptotic
regimes. In order to systematically consider this issue we start
writing the expression for the intensity profile for the pinhole
camera image  with the help of Eq. (\ref{crspecprop3bis}) and Eq.
(\ref{maintrick2rev}), that is

\begin{eqnarray}
\hat{I}(\hat{z}_2,\bar{r}) &=& \frac{1}{\mathcal{Z}}\int
d\bar{r}'d\Delta\hat{r}'
\Bigg\{\hat{\mathcal{G}}(0,-\bar{r}',-\Delta\hat{r}')
\exp\left[i\left(\frac{2}{\hat{z}_1}+\frac{2}{\hat{d}}\right)\bar{r}'\Delta
\hat{r}' \right] \cr &&  \times
P(\bar{r}'+\Delta\hat{r}')P(\bar{r}'-\Delta \hat{r}')\Bigg\}
\exp\left[-\frac{2i}{\hat{d}}\bar{r}\Delta\hat{r}'\right]~.
\label{pineq}
\end{eqnarray}
As usual, the normalization factor $\mathcal{Z}$ is chosen in such
a way that $\hat{I}(\hat{z}_2,0)=1$. The phase in parenthesis
$\{...\}$ must be negligible under condition (\ref{seccond}), and
the influence of the pupil function $P$ must also be negligible
under condition (\ref{thicond}). Here we retain both these
factors, because we are interested in studying the resolution of
the pinhole camera, i.e. the accuracy of our calculations.

The phase contribution in $2/\hat{d}$ is always much smaller than
the phase contribution in $2/\hat{z}_1$, since we assumed
$|\mathrm{M}| \gg 1$. The integrand contributes to the integration
results only for those values of $\bar{r}'$ and $\Delta \hat{r}'$
for which the phase factor is not much larger than unity,
otherwise the integrand exhibits fast oscillatory behavior and it
effectively averages to zero. As a result we may neglect, with
some accuracy over the accuracy of the integral, the phase term in
$1/\hat{d}$, that will start oscillating for higher values of
$\bar{r}'$ and $\Delta \hat{r}'$.

Note that, under the accepted assumptions $\hat{N} \gg 1$ and
$\hat{D} \gg 1$, the expression for the Fourier transform of the
cross-spectral density at the virtual source position is given by

\begin{eqnarray}
\hat{\mathcal{G}}\left(0,{ \bar{r}'}, {\Delta{
\hat{r}'}}\right)&=& \exp\left[-\frac{2 \hat{N}
\Delta\hat{r}'^{2}}{\hat{z}_1^2}\right]
\exp\left[-\frac{\bar{r}'^2}{2{\hat{z}_1^2 \hat{D}}}\right]
~,\label{crspec0}
\end{eqnarray}
in agreement with Eq. (\ref{crossx1lb}).

Analysis of Eq. (\ref{crspec0}) shows that the exponential
function in $\bar{r}'$ exhibits a characteristic scale of order
$\hat{z}_1 \sqrt{\hat{D}}$. Conditions (\ref{seccond}) and
(\ref{fircond}) require $\hat{z}_1 \sqrt{\hat{D}} \gg
\sqrt{\hat{N}} \gg \hat{a}$. As a result, since $\hat{z}_1
\sqrt{\hat{D}}  \gg \hat{a}$, we can approximate the exponential
function in $\bar{r}'$ in Eq. (\ref{crspec0}) with unity to obtain
an equation still suitable for investigating the resolution of the
pinhole camera, that is

\begin{eqnarray}
\hat{I}(\hat{z}_2,\bar{r}) &=& \frac{1}{\mathcal{Z}}
\int_{-\infty}^{\infty} d\Delta\hat{r}' \exp\left[-\frac{2 \hat{N}
\Delta\hat{r}'^{2}}{\hat{z}_1^2}\right]
\exp\left[-\frac{2i}{\hat{d}}\bar{r}\Delta\hat{r}'\right]\cr &&
\times \Bigg\{\int_{-\infty}^{\infty} d \bar{r}'
\exp\left[i\left(\frac{2}{\hat{z}_1}\right)\bar{r}'\Delta \hat{r}'
\right] P(\bar{r}'+\Delta\hat{r}')P(\bar{r}'-\Delta
\hat{r}')\Bigg\} ~. \cr &&\label{pineq2}
\end{eqnarray}
The quantity in parenthesis $\{...\}$ in Eq. (\ref{pineq2}) is the
autocorrelation function of the pupil. It accounts for a phase
error, exactly as in the case of aberrations of the second order
(defocus). Condition (\ref{thicond}) states that the
characteristic scale of $\Delta \hat{r}'$ in Eq. (\ref{crspec0})
is small compared to $\hat{a}$, that is the characteristic scale
of the pupil function $P$. If also $\hat{a}^2/\hat{z}_1 \gg 1$ we
see that the characteristic scale of $P$ is large compared with
the scale imposed by the phase in parenthesis $\{...\}$ in Eq.
(\ref{pineq2}), because $\bar{r}' \Delta \hat{r}'/\hat{z}_1 \gg 1$
at $\Delta \hat{r}' \sim \hat{a}$ and $\bar{r}' \sim \hat{a}$.
Therefore, under the assumption $\hat{a}^2/\hat{z}_1 \gg 1$ we may
neglect the dependence of $P$ on $\Delta \hat{r}'$ in Eq.
(\ref{pineq2}) and obtain

\begin{eqnarray}
\hat{I}(\hat{z}_2,\bar{r}) &=& \frac{1}{\mathcal{Z}}
\int_{-\infty}^{\infty} d\Delta\hat{r}' \tilde{P}(\Delta\hat{r}')
\exp\left[-\frac{2 \hat{N} \Delta\hat{r}'^{2}}{\hat{z}_1^2}\right]
\exp\left[-\frac{2i}{\hat{d}}\bar{r}\Delta\hat{r}'\right] ~,
\label{pineq3}
\end{eqnarray}
where

\begin{eqnarray}
\tilde{P}(\Delta \hat{r}') = \int_{-\infty}^{\infty} d\bar{r}'
\left| P(\bar{r}') \right|^2
\exp\left[i\frac{2}{\hat{z}_1}\bar{r}'\Delta \hat{r}' \right]
\label{autocpin} ~,\end{eqnarray}
or, equivalently\footnote{We assume, as done before, that no
apodization is present.}:

\begin{eqnarray}
\tilde{P}(\Delta \hat{r}') = \int_{-\hat{a}}^{\hat{a}} d\bar{r}'
\exp\left[i\frac{2}{\hat{z}_1}\bar{r}'\Delta \hat{r}' \right]
\label{autocpin2} ~.\end{eqnarray}
Except for an unessential multiplicative factor $\hat{a}$, Eq.
(\ref{autocpin2}) is formally equivalent\footnote{It should be
noted that we switched back from notations $\bar{x}$ and $\Delta
x$, used in Section \ref{sec:abe}, to our usual notation
$\bar{r}'$ and $\Delta \hat{r}'$, with $\bar{r}' = \bar{x}$ and
$2\Delta \hat{r}' = \hat{a} \Delta x$. We remind that the reason
why we used notations $\bar{x}$ and $\Delta x$ in Section
\ref{sec:abe} was for the reader's convenience, as these notations
allow direct comparison with aberration theory developed in
standard textbooks. }  to Eq. (\ref{corrPabe3}) with $n=2$ and

\begin{eqnarray}
b_2 = \frac{\hat{a}^2}{2 \hat{z}_1} \gg 1~. \label{b2}
\end{eqnarray}
This means that we may study the problem of the resolution of the
pinhole camera as an aberration problem: in particular, a
defocusing aberration. The parameter range when
$\hat{a}^2/\hat{z}_1 \gg 1$ leads to equations similar to the case
of severe aberrations when $|b_n| \gg 1$, treated in Section
\ref{sub:abgeo}. The autocorrelation function of the pupil, that
is the Optical Transfer Function of the system, is then obtained
by integration of Eq. (\ref{autocpin2}) and reads:

\begin{eqnarray}
\tilde{P}(\Delta \hat{r}') =
\mathrm{sinc}\left(\frac{2\hat{a}}{\hat{z}_1} \Delta
\hat{r}'\right)~, \label{autocpin3}
\end{eqnarray}
where  Eq. (\ref{autocpin2}) has been used and an unessential
multiplicative factor $2\hat{a}$ has been neglected. Substitution
of Eq. (\ref{autocpin3}) in Eq. (\ref{pineq3}) yields

\begin{eqnarray}
\hat{I}(\hat{z}_2,\bar{r}) &=&
\frac{1}{\mathcal{Z}}\int_{-\infty}^{\infty} d\Delta\hat{r}'
\mathrm{sinc}\left(\frac{2 \hat{a}}{\hat{z}_1} \Delta
\hat{r}'\right) \exp\left[-\frac{2 \hat{N}
\Delta\hat{r}'^{2}}{\hat{z}_1^2}\right]
\exp\left[-\frac{2i}{\hat{d}}\bar{r}\Delta\hat{r}'\right].
\label{pineq4}
\end{eqnarray}
Eq. (\ref{pineq4}) allows an estimation of the resolution by
taking the ratio of the width of the sinc and of the exponential
function in $\Delta \hat{r}'$. On the one hand, the width of the
exponential function is of order $\hat{z}_1/(2\sqrt{\hat{N}})$. On
the other hand, the width of the sinc function is of order
$\hat{z}_1/(2 \hat{a})$. As a result, when $\hat{a}^2/\hat{z}_1
\gg 1$ the resolution of the camera is of order
$\hat{a}/\sqrt{\hat{N}}$. This procedure is justified by the fact
that Eq. (\ref{pineq4}) can be interpreted as a convolution of a
rectangular profile with a (new) Gaussian function, and that the
width of these two functions can be obtained by taking the inverse
widths of the sinc function and the Gaussian function in Eq.
(\ref{pineq}) and by multiplying them by the factor $\hat{d}$.

Let us now deal with the case when $\hat{a}^2/\hat{z}_1 \ll 1$.
Going back to Eq. (\ref{pineq2}),  we see that $\hat{a}$ is narrow
compared with the scale imposed by the phase in parenthesis
$\{...\}$ in Eq. (\ref{pineq2}), because $\bar{r}' \Delta
\hat{r}'/\hat{z}_1 \ll 1$ at $\Delta \hat{r}' \sim \hat{a}$ and
$\bar{r}' \sim \hat{a}$. Therefore we can neglect the phase factor
in parenthesis $\{...\}$, which corresponds to a case with no
aberrations. The autocorrelation function of the pupil can now be
written as a triangle function

\begin{eqnarray}
\tilde{P}(\Delta \hat{r}') = \mathrm{tri} \left(\frac{\Delta
\hat{r}'}{\hat{a}}\right)~, \label{autocpinal}
\end{eqnarray}
that should be substituted into Eq. (\ref{pineq3}) to give the
analogous of Eq. (\ref{pineq4}) in the limit $\hat{a}^2/\hat{z}_1
\ll 1$, that is

\begin{eqnarray}
\hat{I}(\hat{z}_2,\bar{r}) &=& \frac{1}{\mathcal{Z}}
\int_{-\infty}^{\infty} d\Delta\hat{r}' \mathrm{tri}
\left(\frac{\Delta \hat{r}'}{\hat{a}}\right) \exp\left[-\frac{2
\hat{N} \Delta\hat{r}'^{2}}{\hat{z}_1^2}\right]
\exp\left[-\frac{2i}{\hat{d}}\bar{r}\Delta\hat{r}'\right] ~. \cr
&&\label{pineq5}
\end{eqnarray}
Eq. (\ref{pineq5}) can also be interpreted as a convolution of a
$\mathrm{sinc}^2(\cdot)$ profile with a Gaussian profile.

Similarly to Eq. (\ref{pineq4}), Eq. (\ref{pineq5}) allows an
estimation of the resolution by taking the ratio of the width of
the triangular function and of the exponential function in $\Delta
\hat{r}'$ . As before, on the one hand the width of the
exponential function is of order $\hat{z}_1/(2\sqrt{\hat{N}})$. On
the other hand, the width of the triangular function is of order
$2 \hat{a}$. As a result, when $\hat{a}^2/\hat{z}_1 \ll 1$ the
resolution of the camera is of order $\hat{z}_1/(\hat{a}
\sqrt{\hat{N}})$.

When $\hat{a}^2/\hat{z}_1 \gg 1$ the resolution of the camera is
of order $\hat{a}/\sqrt{\hat{N}}$. One has better resolution as
the pupil becomes smaller and smaller, but the condition
$\hat{a}^2/\hat{z}_1 \gg 1$ becomes less and less satisfied. When
$\hat{a}^2/\hat{z}_1 \ll 1$ the resolution of the camera is of
order $\hat{z}_1/(\hat{a} \sqrt{\hat{N}})$. In this case one has
better resolution as the pupil becomes larger and larger, but the
condition $\hat{a}^2/\hat{z}_1 \ll 1$ becomes less and less
satisfied. As a result there must be an optimum for pupil
apertures of order $\hat{a}^2 \sim \hat{z}_1$. This optimum
depends on the object considered (in this discussion, for
instance, we assumed a Gaussian source) and on the definition of
the width of a function, that may vary depending on circumstances.
However, starting from Eq. (\ref{pineq2}), we may present an
expression for the Optical Transfer Function of the pinhole
camera, which is the quantity in parenthesis $\{...\}$. Such
quantity can be written as

\begin{eqnarray}
\tilde{P}_{\mathrm{pc}}(\Delta
\hat{r}',\hat{a},\hat{z}_1)=\int_{-\hat{a}}^{\hat{a}} d\bar{r}'
P(\bar{r}'+\Delta \hat{r}') P(\bar{r}'-\Delta \hat{r}')
\exp\left[i 2 \frac{\bar{r}'\Delta \hat{r}'}{\hat{z}_1} \right]~.
\label{accueq}
\end{eqnarray}
After introduction of $\xi=\bar{r}'/\hat{a}$ and
$\Omega=\hat{a}^2/\hat{z}_1$ we can write Eq. (\ref{accueq}) as

\begin{eqnarray}
\tilde{P}_{\mathrm{pc}}\left(\frac{\Delta
\hat{r}'}{\hat{a}},\Omega\right)=\int_{-1}^{1} d \xi
P\left(\xi+\frac{\Delta \hat{r}'}{\hat{a}}\right)
P\left(\xi-\frac{\Delta \hat{r}'}{\hat{a}}\right) \exp\left[i 2
\xi \Omega \frac{\Delta \hat{r}'}{\hat{a}} \right]~.
\label{accueq2}
\end{eqnarray}
Note that the limit $\Omega\longrightarrow \infty$ of Eq.
(\ref{accueq2}) is a $\mathrm{sinc}(\cdot)$ function, while the
limit $\Omega\longrightarrow 0$ is a $\mathrm{tri}(\cdot)$
function, as it should be. Eq. (\ref{accueq2}) enters in the
expression for the intensity as the term in $\{...\}$ in Eq.
(\ref{pineq2}), that can be reinterpreted as a convolution
product. Then, the Fourier transform of Eq. (\ref{accueq2}) with
respect to $\Delta\hat{r}'/\hat{a}$ is one of the factors in this
convolution product, and its width is related with the resolution
of the pinhole camera. This Fourier transform is the line spread
function for the pinhole camera. After introduction of $y =
\hat{a} \bar{r}_i/\hat{d}$ and $\eta = \Delta{\hat{r}'}/\hat{a}$
we can write the line spread function for the pinhole camera as

\begin{eqnarray}
l_\mathrm{pc}\left(\Omega,y\right)=\int_{-1}^{1} d\xi \left \{
\int_{-1}^{1} d \eta P\left(\xi+\eta\right) P\left(\xi-\eta\right)
\exp\left[i 2 \eta \left(\xi \Omega -  y\right)\right]\right\}~,
\label{accueq3}
\end{eqnarray}
where we accounted for the fact that the Fourier transform
integral in $d\xi$ is limited by the presence of the pinhole to
the range $[-1,1]$. On the one hand, the integral in parenthesis
$\{...\}$ in Eq. (\ref{accueq3}) is the Fourier transform of
$P(\xi+\eta) P(\xi-\eta)$ calculated with respect to $\eta$ as a
function of $2 (\xi \Omega + y)$. On the other hand, the function
$P(\xi+\eta) P(\xi-\eta)$ is a window function similar to the
pupil function $P$. It is equal to unity for values of
$|\eta|<1-|\xi|$, and it is zero elsewhere. Therefore, the
quantity in parenthesis $\{...\}$ in Eq. (\ref{accueq3}) can be
calculated analytically yielding

\begin{eqnarray}
l_\mathrm{pc}\left(\Omega,y\right)=\int_{-1}^{1} d\xi
\frac{\sin\left[2 (\xi \Omega + y)(1-|\xi|)\right]}{\xi \Omega +
y}~. \label{accueq4}
\end{eqnarray}
Note that if $\Omega \longrightarrow 0$ we have

\begin{eqnarray}
l_\mathrm{pc}\left(\Omega,y\right) \longrightarrow
\frac{1}{y}\int_{-1}^{1} d\xi {\sin\left[2 y (1-|\xi|)\right]} = 2
\mathrm{sinc}^2(y)~, \label{acculim1}
\end{eqnarray}
as it should be, while if $\Omega \longrightarrow \infty$, after
the change of variable $\xi \longrightarrow \xi' = \Omega \xi$ we
can rewrite Eq. (\ref{accueq4}) as

\begin{eqnarray}
l_\mathrm{pc}\left(\Omega,y\right) &\longrightarrow
&\frac{1}{\Omega} \int_{-\Omega}^{\Omega} d\xi' \frac{\sin\left[2
(\xi'  + y)\right]}{\xi'  + y} =\frac{1}{\Omega}
\mathrm{Si}[2(y+\xi)] \Bigg|_{-\Omega}^{\Omega} \cr
&=&\frac{1}{\Omega}\left\{
\mathrm{Si}[2(y+\Omega)]-\mathrm{Si}[2(y-\Omega)] \right\} ~,
\label{acculim2}
\end{eqnarray}
where $\mathrm{Si}(\cdot)$ indicates the sin integral function. It
can be seen that, as $\Omega\longrightarrow \infty$, the function
defined by Eq. (\ref{acculim2}) approximates more a more a
rectangular function which is constant for $-\Omega<y<\Omega$ and
equal to zero elsewhere, as it should be.

\begin{figure}
\begin{center}
\includegraphics*[width=140mm]{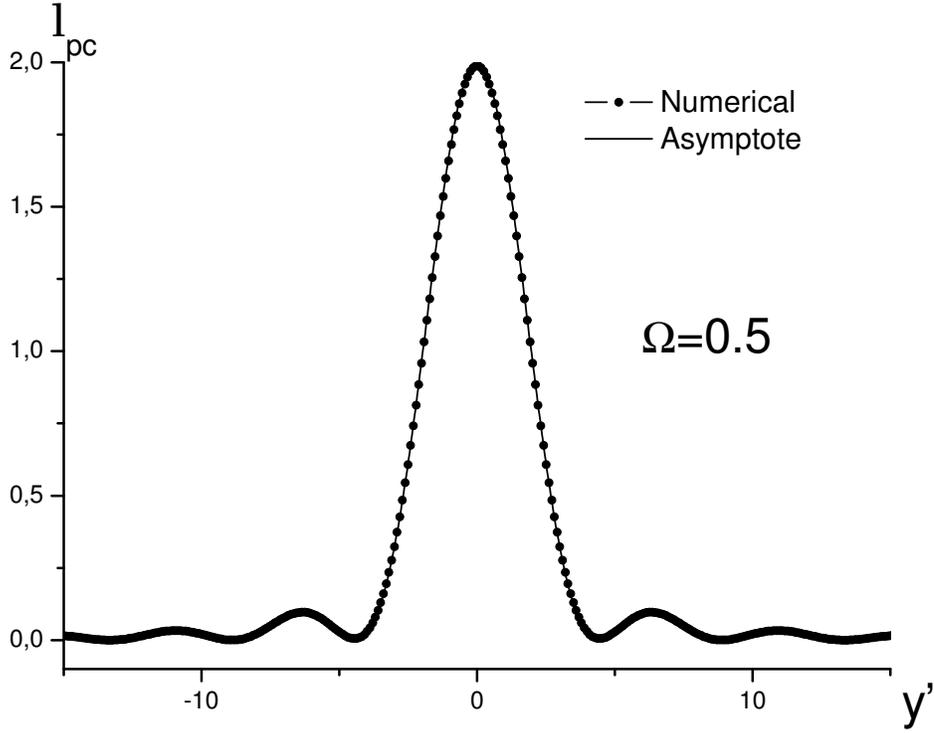}
\caption{\label{pinX05} The line spread function $l_\mathrm{pc}$
with $\Omega=0.5$. In this case the asymptote for $\Omega
\longrightarrow 0$, Eq. (\ref{acculim1}), is well-matched to the
numerical evaluations.}
\end{center}
\end{figure}
\begin{figure}
\begin{center}
\includegraphics*[width=140mm]{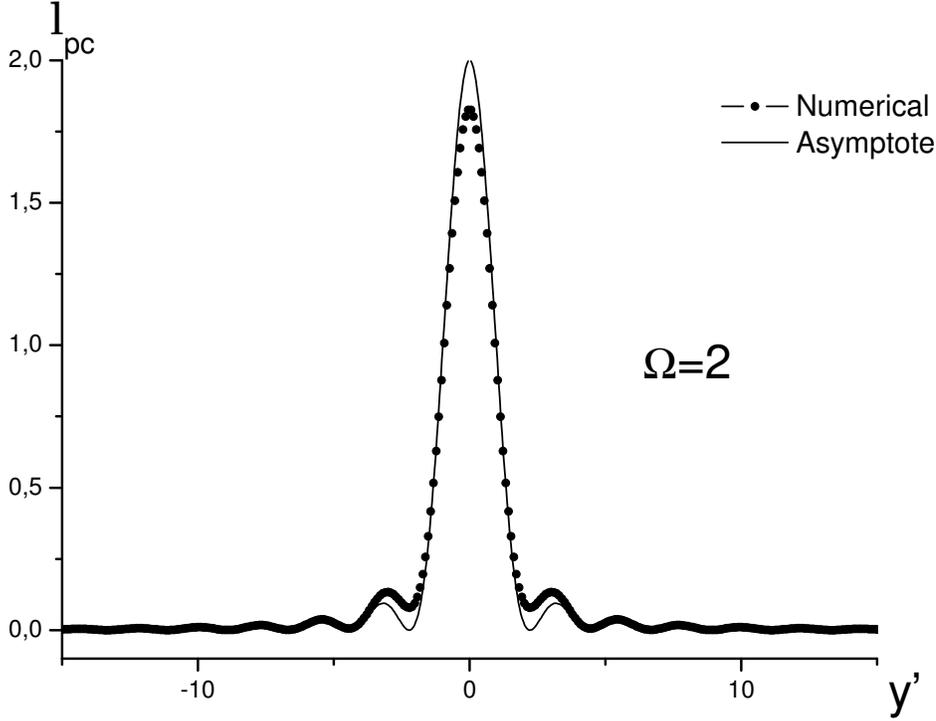}
\caption{\label{pinX2} The line spread function $l_\mathrm{pc}$
with $\Omega=2$. In this case the asymptote for $\Omega
\longrightarrow 0$, Eq. (\ref{acculim1}), starts to diverge from
numerical evaluations. The width of $l_\mathrm{pc}$ is close to
its minimum.}
\end{center}
\end{figure}
\begin{figure}
\begin{center}
\includegraphics*[width=140mm]{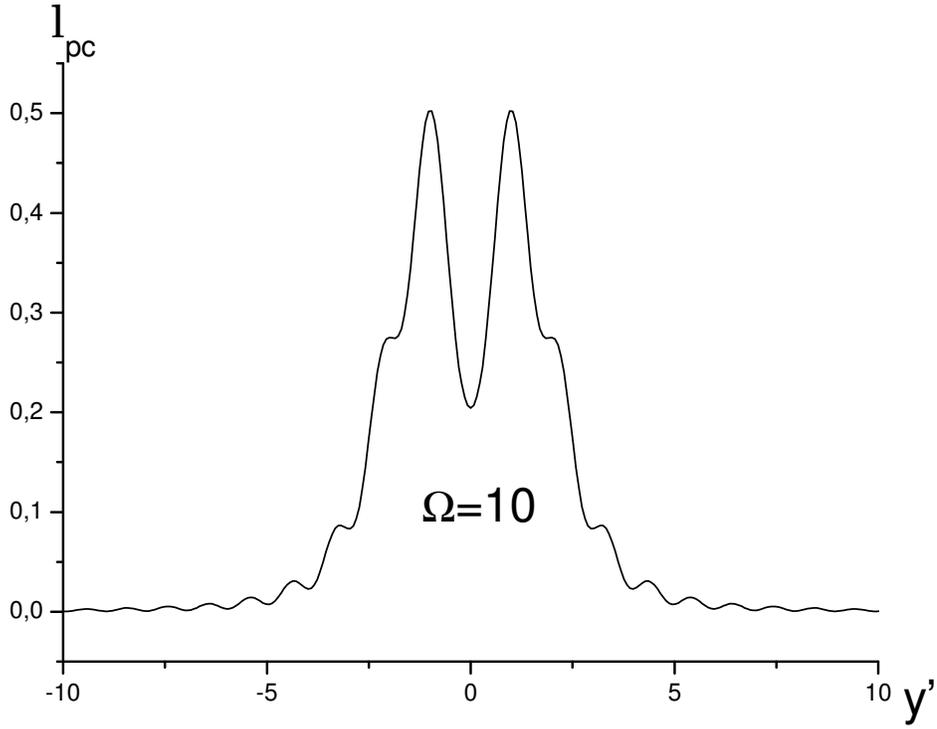}
\caption{\label{pinX10} The line spread function $l_\mathrm{pc}$
with $\Omega=10$. An intermediate situation between the two limits
for $\Omega \longrightarrow 0$ and $\Omega \longrightarrow
\infty$. }
\end{center}
\end{figure}
\begin{figure}
\begin{center}
\includegraphics*[width=140mm]{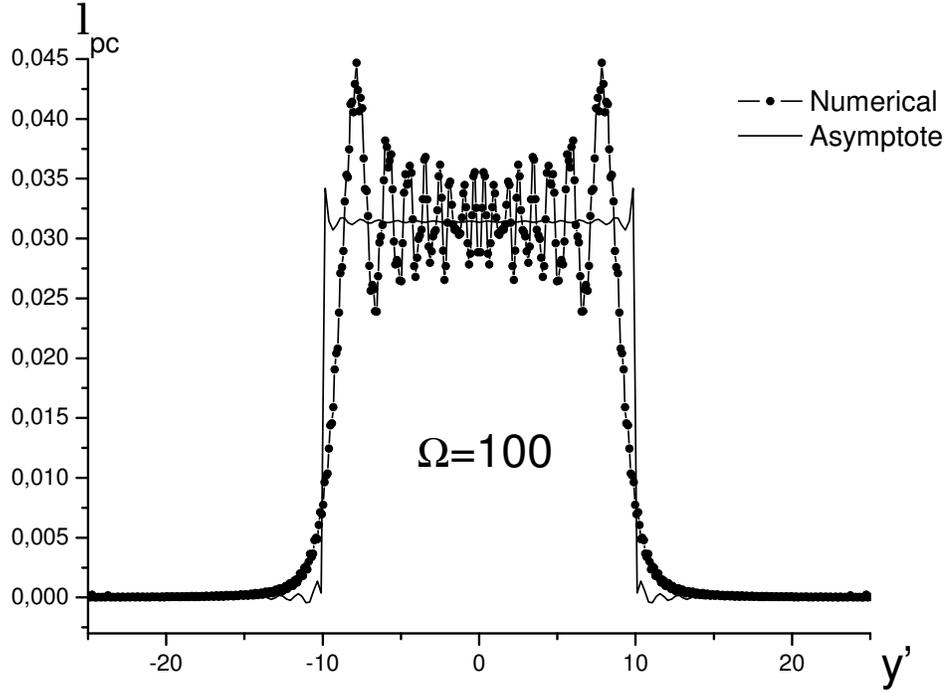}
\caption{\label{pinX100} The line spread function $l_\mathrm{pc}$
with $\Omega=100$. The asymptote in Eq. (\ref{acculim2}) starts to
match the numerical calculations. In the limit $\Omega
\longrightarrow \infty$ they both merge to a rectangular
function.}
\end{center}
\end{figure}
We may use a new variable $y'=y/\sqrt{\Omega}$ in Eq.
(\ref{accueq4}), thus obtaining:

\begin{eqnarray}
l_\mathrm{pc}\left(\Omega,y'\right)=\int_{-1}^{1} d\xi
\frac{\sin\left[2 (\xi \Omega + y'
\sqrt{\Omega})(1-|\xi|)\right]}{\xi \Omega + y' \sqrt{\Omega}}~.
\label{accueq5}
\end{eqnarray}
The reason for this is that, now, both the asymptotes Eq.
(\ref{acculim1}) for $\Omega\ll 1$ and Eq. (\ref{acculim2}) for
$\Omega\gg 1$ present a characteristic width $1/\sqrt{\Omega}$ and
$\sqrt{\Omega}$ respectively. In Fig. \ref{pinX05}, Fig.
\ref{pinX2}, Fig. \ref{pinX10} and Fig. \ref{pinX100} we present
various shapes of the $l_\mathrm{pc}$ function and its asymptotic
limit for different values of $\Omega$. Once a definition of the
width of the function $l_\mathrm{pc}$ is chosen, the optimal
operation point for the pinhole camera may be set requiring that
the width of $l_\mathrm{pc}$ be minimal. It should be remarked
that the definition of the width of $l_\mathrm{pc}$ is somewhat
subjective.

\begin{figure}
\begin{center}
\includegraphics*[width=140mm]{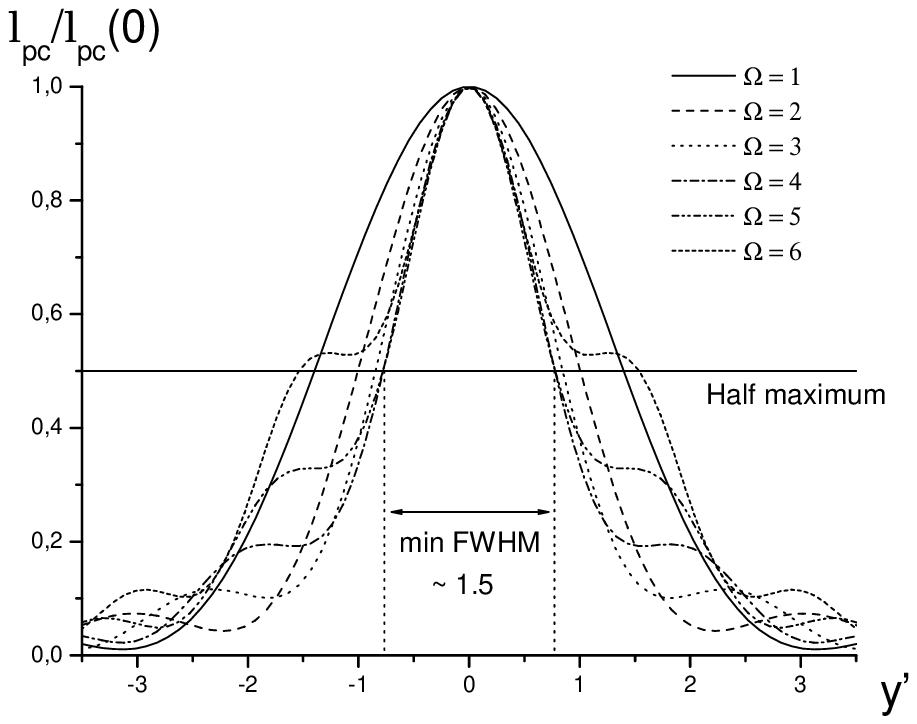}
\caption{Relative line spread function $l_{pc}/l_{pc}(0)$ for
different values of $\Omega$. \label{minfwhm} }
\end{center}
\end{figure}
One may, for instance, define the width of $l_\mathrm{pc}$ to be
the Full Width Half Maximum (FWHM). In Fig. \ref{minfwhm} we
present different plots for $l_\mathrm{pc}$ normalized to
$l_\mathrm{pc}(0)$ for different values of $\Omega = 1,2,3,4,5,6$.
The minimal value of the FWHM of the line spread function for the
pinhole camera happens to be located somewhere between $\Omega=4$
and $\Omega = 5$. It may be estimated to be about $1.5$. As a
result one concludes that the optimal aperture for the pinhole is
given by

\begin{equation}
\hat{a}^2 \simeq 4.5 \hat{z}_1 \label{best}
\end{equation}
or, in dimensional units,

\begin{equation}
{a}^2 \simeq 4.5 \frac{{z}_1 c}{\omega}~. \label{best2}
\end{equation}
We can estimate the resolution $\delta {r}$ of the pinhole camera
in dimensional units imposing $\delta y'$ to be equal to the  FWHM
of the line spread function, i.e. $1.5$, and taking advantage of
the definition of $y'$. Therefore we obtain

\begin{equation}
\delta {r} \simeq 1.5 \sqrt{\frac{c}{\omega {z}_1}} d~. \label{dr}
\end{equation}
Results in this Section have been obtained under the assumption
$|\mathrm{M}| \gg 1$. A generalization  can be presented
introducing

\begin{equation}
\hat{z}_{\mathrm{eff}} = \frac{\hat{z}_1
\hat{d}}{\hat{z}_1+\hat{d}} \label{zeff}
\end{equation}
and the analogous dimensional value $z_{\mathrm{eff}} =
\hat{z}_{\mathrm{eff}} L_w$ to be used in place of $\hat{z}_1$ and
$z_1$ in Eq. (\ref{best}), Eq. (\ref{best2}) and Eq. (\ref{dr}).
In dimensional units we have

\begin{equation}
{a}^2 \simeq 4.5 \frac{{z}_{\mathrm{eff}} c}{\omega}~
\label{best2s}
\end{equation}
and

\begin{equation}
\delta {r} \simeq 1.5 \sqrt{\frac{c}{\omega {z}_{\mathrm{eff}}}}
d~. \label{drs}
\end{equation}
Before proceeding we should discuss about the applicability of Eq.
(\ref{accueq5}). Up to now, in fact, we discussed about a Gaussian
quasi-homogeneous undulator source ($\hat{N} \gg 1$ and $\hat{D}
\gg 1$). Eq. (\ref{accueq5}) can be applied in a wider variety of
cases. In particular, Eq. (\ref{accueq5}) can be applied when (a)
the virtual source is quasi-homogeneous, (b) the pinhole camera is
installed in the far zone and  (c) $\hat{a}^2 \ll \max[1,\hat{D}]
\hat{z}_1^2$. When conditions (a), (b) and (c) are satisfied, the
line spread function for the pinhole camera, $l_\mathrm{pc}$ can
be calculated from Eq. (\ref{accueq5}). At the beginning of the
present Section \ref{sec:cam} we stated three conditions in order
to have an image of the source, besides the Gaussian
quasi-homogeneous assumption (that we relaxed, here, to the more
general case (a)). Condition 1. given at the beginning of the
present Section \ref{sec:cam} can be identified with (b).
Condition 1. and 2., together, are responsible for (c). The other
requirement was formulated in order to obtain a good quality image
of the source. However $l_\mathrm{pc}$ can be calculated
regardless of it and, in different situations, one may have a good
quality or a bad quality of the image, depending on how the width
of $l_\mathrm{pc}$ scales with the width of the ideal image. The
intensity profile at the observation plane located at $\hat{z}=
\hat{z}_2$ is given, in any case, by the convolution product
between the line spread function $l_\mathrm{pc}$ and the ideal
image. In particular, for an undulator source with Gaussian
intensity profile at the virtual plane such a product is given by
$l_\mathrm{pc} \ast \exp[-\bar{r}^2/(2 \mathrm{M}^2 \hat{N})]$,
where $|\mathrm{M}| = \hat{d}/\hat{z}_1$. When $\sqrt{\hat{N}} \gg
\hat{a} \gg \hat{z}_1/\sqrt{\hat{N}}$ a good quality image is
formed at the observation plane. In other situations instead, the
width of $l_\mathrm{pc}$ is of order of, or even larger than, the
width of the ideal image, and a bad quality image will be formed.
To sum up, in order to calculate the intensity distribution at the
observation plane located at position $\hat{z}=\hat{z}_2$ behind
the pinhole, we should first calculate $\Omega =
\hat{a}^2/\hat{z}_\mathrm{eff}$. Then use Eq. (\ref{accueq5}) to
calculate the line spread function $l_\mathrm{pc}$. Finally,
convolve with the ideal image from the source. The minimal width
of the line spread function, which correspond to the best accuracy
for the image, is for $\hat{a}^2/\hat{z}_\mathrm{eff} \simeq 4.5$.
Note that conditions (a), (b) and (c) above do not require the
source to be an undulator source. In particular, (a), (b) and (c)
are automatically satisfied for thermal sources, or perfectly
incoherent objects. As a result, our theory can be applied for
visible light imaging as well.

These remarks simplify our discussion and drastically widen the
applicability region of our results. In particular they can
describe bending magnet sources. For instance, at ESRF (European
Synchrotron Radiation Facility) a pinhole camera setup has been
developed for electron beam diagnostics \cite{PIN1}. Synchrotron
Radiation from a bending magnet is imaged through an X-ray pinhole
camera setup. Subsequent analysis of the pinhole camera image
allows one to retrieve the electron beam sizes. In the situation
studied in \cite{PIN1} the diffraction angle is of order of
$\gamma^{-1} \sim 10^{-4}$ rad. The system is reported to generate
an image of the source when the pinhole is moved at about $z_1=23$
m from the dipole magnet. At this distance, and at the critical
wavelength, the spot size due to divergence of the single particle
radiation is estimated to be about $\gamma^{-1} z_1 \simeq 2
\mathrm{mm} \gg \sigma_{x,y}$, $\sigma_{x,y}$ being the transverse
electron beam sizes. This guarantees that the pinhole is in the
far field, i.e. condition (b) is satisfied. The critical
wavelength is of order $\lambda \simeq 3 \cdot 10^{-11}$ m, since
one may read, in \cite{PIN1}, that "the contribution from
diffraction assumes a typical photon energy of $40$ keV". The
radiation diffraction size can be estimated to be of order
$\lambda \gamma/(2\pi) \simeq 10^{-1} \mu\mathrm{m} \ll
\sigma_{x,y}$, which demonstrates the quasi-homogeneity of the
source, i.e. condition (a). In reference \cite{PIN1} one can also
find $d_i = 11.5$ m. We obtain an optimal pinhole aperture of $a
\simeq 12 \mu$m. Since the pinhole aperture is much smaller than
the spot size due to divergence of the single particle radiation,
condition (c) is satisfied as well. As a result our theory can be
applied to the situation treated in \cite{PIN1}. Our results
should be compared with the actual choice in \cite{PIN1}. From
Fig. 2 in that reference we can conclude that and $2 a \simeq 50
\mu$m in the horizontal direction. This choice is not optimal,
since corresponds to the value $\Omega \sim 10$. The line spread
function for the pinhole camera in the horizontal direction is
illustrated, in this case, in Fig. \ref{pinX10}. Estimation of the
resolution is not easy  because, as already said, is related to
the definition of the width of $l_\mathrm{pc}$, which is quite
subjective in the particular case depicted in Fig. \ref{pinX10}.
However, knowing the profile of the line spread function, we may
use it in order to deconvolve experimental results and extract the
electron beam size. From the same Fig. 2 in \cite{PIN1} we can
also conclude that $2 a \simeq 25 \mu$m in the vertical direction.
Although the definition of the width of $l_\mathrm{pc}$ is
somewhat subjective, we can conclude that the choice made in
\cite{PIN1} is near the optimal value in the vertical direction.
The resolution in the vertical direction turns out to be, from Eq.
(\ref{drs}), $\delta {r} \simeq 14 \mu$m. As a result the
resolution in \cite{PIN1}, that is about $26 \mu$m, is somewhat
underestimated. It should be clear from the previous discussion
that decreasing the pinhole dimension beyond the optimal size not
only will decrease the photon flux, but will also worsen the
camera resolution.

\section{\label{sec:focim} Imaging in the focal plane}

In this Section we will investigate the intensity distribution on
the focal plane due to a quasi-homogeneous source. We will first
consider the case of a Gaussian quasi-homogeneous source ($\hat{N}
\gg 1$, $\hat{D} \gg 1$) and subsequently generalize our
conclusions, as done before for the more comprehensive case of
generic quasi-homogeneous sources. The characteristics of the
intensity distribution in the focal plane can be treated in formal
analogy with the pinhole camera setup treated in Section
\ref{sec:cam} and with the physics of aberrations, in particular
defocusing aberrations, described in Section \ref{sec:abe}. This
may seem counterintuitive, since, at first glance, we are treating
completely different systems from a physical viewpoint.  The
formal analogy between these situations is a demonstration of the
power of the combined Statistical and Fourier Optics approach,
which allows one to unify study cases otherwise completely
distinct. Such unification can be seen from the expression for the
cross-spectral density at any distance from the source in the
presence of a pupil, Eq. (\ref{crspecprop3bis}). As usual, we will
assume that the lens is in the far field, i.e. $\hat{D}\hat{z}_1^2
\gg \hat{N}$. Under this assumption we may use Eq.
(\ref{maintrick2}) to characterize the cross-spectral density at
the lens position. In the image plane, the far field assumption
allows cancellation of all phase factors in the integrand of Eq.
(\ref{crspecprop3bis}) and leads to Eq. (\ref{GPInb1}). This
cancellation does not hold anymore for the focal plane. Use of the
focal plane condition $\hat{f} = \hat{z}_2-\hat{z}_1$ and of Eq.
(\ref{maintrick2})  yields  the intensity ($\Delta \hat{r}_f =0$)

\begin{eqnarray}
\hat{I}(\hat{z}_2,\bar{r}_f) &=& \frac{1}{\mathcal{Z}} \int
d\bar{r}'d\Delta\hat{r}'
\Bigg\{\hat{\mathcal{G}}(0,-\bar{r}',-\Delta\hat{r}') \exp\left[i
\frac{2}{\hat{z}_1}\bar{r}'\Delta \hat{r}' \right] \cr &&  \times
P(\bar{r}'+\Delta\hat{r}')P(\bar{r}'-\Delta \hat{r}')\Bigg\}
\exp\left[-\frac{2i}{\hat{f}}\bar{r}_f\Delta\hat{r}'\right]~.
\label{pineqb2}
\end{eqnarray}
As one can see, the phase factor in the expression for the
cross-spectral density, Eq. (\ref{maintrick2}), survived in Eq.
(\ref{pineqb2}) due to the choice of the focal plane as the
observation plane. From a formal viewpoint Eq. (\ref{pineqb2}) is
identical to Eq. (\ref{pineq}). In fact, as the reader will
remember, the phase $\bar{r}'\Delta \hat{r}'/\hat{d}_i$ in Eq.
(\ref{pineq}) is negligible for $|M| \gg 1$ and can be retained
without change in the formalism in the case $|M|$ is not much
larger than unity by defining $\hat{z}_\mathrm{eff}$ according to
Eq. (\ref{zeff}). As before, the normalization factor
$\mathcal{Z}$ is chosen in such a way that
$\hat{I}(\hat{z}_2,0)=1$.

Under the (for now) accepted assumptions $\hat{N} \gg 1$ and
$\hat{D} \gg 1$, the expression for the Fourier transform of the
cross-spectral density at the virtual source position is given by
Eq.(\ref{crspec0}). Substitution of Eq. (\ref{crspec0}) in Eq.
(\ref{pineqb2}) yields

\begin{eqnarray}
\hat{I}(\hat{z}_f,\bar{r}_f) &=& \frac{1}{\mathcal{Z}}
\int_{-\infty}^{\infty}d\Delta\hat{r}' \exp\left[-\frac{2 \hat{N}
\Delta\hat{r}'^{2}}{\hat{z}_1^2}\right]\exp\left[-\frac{2i}{\hat{f}}\bar{r}_f\Delta\hat{r}'\right]
\cr && \times \Bigg\{\int_{-\infty}^{\infty} d\bar{r}' \exp\left[i
\frac{2}{\hat{z}_1}\bar{r}'\Delta \hat{r}' \right]
P(\bar{r}'+\Delta\hat{r}')P(\bar{r}'-\Delta \hat{r}') \cr &&
\times \exp\left[-\frac{\bar{r}'^2}{2{\hat{z}_1^2
\hat{D}}}\right]\Bigg\} ~. \label{pineqb3}
\end{eqnarray}
Note that when the pupil influence is negligible (that is the case
when $\hat{a}^2 \gg \hat{D} \hat{z}_1^2 \gg \hat{z}_1^2/\hat{N}$),
Eq. (\ref{pineqb3}) reads

\begin{eqnarray}
\hat{I}(\hat{z}_f,\bar{r}_f) &=& \frac{1}{\mathcal{Z}}
\int_{-\infty}^{\infty}d\Delta\hat{r}'
\exp\left[-\frac{2i}{\hat{f}}\bar{r}_f\Delta\hat{r}'\right] \cr &&
\times \Bigg\{\int_{-\infty}^{\infty} d\bar{r}'\exp\left[i
\frac{2}{\hat{z}_1}\bar{r}'\Delta \hat{r}' \right]
\exp\left[-\frac{2 \hat{N} \Delta\hat{r}'^{2}}{\hat{z}_1^2}\right]
\exp\left[-\frac{\bar{r}'^2}{2{\hat{z}_1^2 \hat{D}}}\right]
\Bigg\} ~. \cr && \label{pineqb4}
\end{eqnarray}
The reader may check that evaluation of Eq. (\ref{pineqb4}) gives
back Eq. (\ref{Ingrandefoc}), as it should be. For large
non-limiting apertures, the extra phase imposes an extra Fourier
transformation of the integrand, which gives the usual result. The
intensity in the focal plane is a scaled version of the Fourier
transform of the spectral degree of coherence on the virtual
source plane. Yet, there are situations when one may recover an
image of the virtual source at the focal plane. This happens when
parameters are such that the only influence of a final pupil
aperture is to make the phase factor in parenthesis $\{..\}$ in
Eq. (\ref{pineqb3}) negligible. Looking for a region in parameter
space where this situation is realized is equivalent to what has
been done in Section \ref{sec:cam}. Three conditions were given
such that the phase factor in $\bar{r}'\Delta \hat{r}'$  in Eq.
(\ref{pineq}) could be neglected. In that case, for a pinhole
camera, we had image formation at all positions after the pinhole.

In particular, when $\hat{D} \hat{z}_1^2 \gg \hat{N} \gg \hat{a}^2
\gg \hat{z}_1^2/\hat{N}$ not only we can neglect the phase factor
in Eq. (\ref{pineqb3}) but we can also see that the width of the
exponential function in $\Delta \hat{r}'$ is much narrower than
that of the pupil function $P$. Therefore, the dependence on
$\Delta \hat{r}'$ in $P$ can be neglected. Moreover, the width of
$P$ in $\bar{r}'$ is much narrower than the width of the
exponential function in $\bar{r}'$, so that the latter can be
neglected as well. As a result we simplify Eq. (\ref{pineqb3}) to
obtain

\begin{eqnarray}
\hat{I}(\hat{z}_f,\bar{r}_f) &=& \frac{1}{\mathcal{Z}} \int
d\bar{r}'d\Delta\hat{r}' \exp\left[-\frac{2 \hat{N}
\Delta\hat{r}'^{2}}{\hat{z}_1^2}\right] |P(\bar{r}')|^2
\exp\left[-\frac{2i}{\hat{f}}\bar{r}_f\Delta\hat{r}'\right]~.
\label{pineqb5}
\end{eqnarray}
The integral in $d\bar{r}'$ yields an unessential multiplication
constant to be included in $\mathcal{Z}$, and one is left with

\begin{eqnarray}
\hat{I}(\hat{z}_f,\bar{r}_f) &=& \frac{1}{\mathcal{Z}}
\int_{-\infty}^{\infty} d\Delta\hat{r}' \exp\left[-\frac{2 \hat{N}
\Delta\hat{r}'^{2}}{\hat{z}_1^2}\right]
\exp\left[-\frac{2i}{\hat{f}}\bar{r}_f\Delta\hat{r}'\right]
\label{pineqb6} ~. \end{eqnarray}
Including another unessential constant in the normalization factor
${\mathcal{Z}}$, Eq. (\ref{pineqb6}) can be written as

\begin{eqnarray}
\hat{I}(\hat{z}_f,\bar{r}_f) &=&
\frac{1}{\mathcal{Z}}\exp\left[-\frac{\hat{z}_1^2 \bar{r}_f^{2}}{2
\hat{f}^2 \hat{N} }\right] ~. \label{pineqb7}
\end{eqnarray}
As a result, when $\hat{D} \hat{z}_1^2 \gg \hat{N} \gg \hat{a}^2
\gg \hat{z}_1^2/\hat{N}$, we obtain, in the focal plane, a scaled
image of the virtual source. This is exactly what a pinhole camera
does in the parameter region when conditions 1. to 3. in Section
\ref{sec:cam} are satisfied. A correspondence between completely
different problems has thus been established thanks to the
combined power of Statistical Optics and Fourier Optics approach.

Such a correspondence can be pursued further, up to a complete
formal identification between the pinhole camera and the focal
imaging system. To this purpose we restrict our analysis to the
case $\hat{a}^2 \ll \hat{D} \hat{z}_1^2$. In the opposite limit we
would obtain the already treated result for negligible pupil
influence. In this situation we can neglect the exponential
function in $\bar{r}'^2$ in Eq. (\ref{pineqb3}). Eq.
(\ref{pineqb3}) can thus be simplified to

\begin{eqnarray}
\hat{I}(\hat{z}_f,\bar{r}_f) &=& \frac{1}{\mathcal{Z}}
\int_{-\infty}^{\infty}d\Delta\hat{r}' \exp\left[-\frac{2 \hat{N}
\Delta\hat{r}'^{2}}{\hat{z}_1^2}\right]\exp\left[-\frac{2i}{\hat{f}}\bar{r}_f\Delta\hat{r}'\right]
\cr && \times \Bigg\{\int_{-\infty}^{\infty} d\bar{r}' \exp\left[i
\frac{2}{\hat{z}_1}\bar{r}'\Delta \hat{r}' \right]
P(\bar{r}'+\Delta\hat{r}')P(\bar{r}'-\Delta \hat{r}') \Bigg\} ~.
\label{pineqb3bbb}
\end{eqnarray}
Once the substitutions $\hat{f} \longrightarrow \hat{d}$,
$\hat{z}_f \longrightarrow \hat{z}_2$ and $\bar{r}_f
\longrightarrow \bar{r}$ are made, Eq. (\ref{pineqb3bbb}) is
identical to Eq. (\ref{pineq2}). This means that, in the limit
$\hat{a}^2 \ll \hat{D} \hat{z}_1^2$ studying  the pinhole camera
setup is completely equivalent to studying the focal imaging
detup. As in Eq. (\ref{pineq2}), the quantity in parenthesis
$\{...\}$ constitutes an Optical Transfer Function, and its
Fourier transform yields a line spread function for the system.
Note that in the intermediate region for $\hat{a}^2 \sim \hat{D}
\hat{z}_1^2$ one may retain the same formalism: in this case
though, the exponential function in $\bar{r}'^2$ in Eq.
(\ref{pineqb3}) cannot be neglected and results would be
different, depending also on the source parameter $\hat{D}_y$
which introduces a factor formally identical to lens apodization.

However, in the case  $\hat{a}^2 \ll \hat{D} \hat{z}_1^2$, we can
proceed in perfect parallelism with the study of the pinhole
camera setup in the previous Section \ref{sec:cam}. Starting from
Eq. (\ref{pineqb3bbb}), we may present an expression for the
Optical Transfer Function of the pinhole camera, which is the
quantity in parenthesis $\{...\}$. This quantity can be written as

\begin{eqnarray}
\tilde{P}_{\mathrm{fp}}(\Delta
\hat{r}',\hat{a},\hat{z}_1)=\int_{-\hat{a}}^{\hat{a}} d\bar{r}'
P(\bar{r}'+\Delta \hat{r}') P(\bar{r}'-\Delta \hat{r}')
\exp\left[i 2 \frac{\bar{r}'\Delta \hat{r}'}{\hat{z}_1} \right]~.
\label{accueqf}
\end{eqnarray}
After introduction of $\xi=\bar{r}'/\hat{a}$ and
$\Omega=\hat{a}^2/\hat{z}_1$ we can write Eq. (\ref{accueqf}) as

\begin{eqnarray}
\tilde{P}_{\mathrm{fp}}\left(\frac{\Delta
\hat{r}'}{\hat{a}},\Omega\right)=\int_{-1}^{1} d \xi
P\left(\xi+\frac{\Delta \hat{r}'}{\hat{a}}\right)
P\left(\xi-\frac{\Delta \hat{r}'}{\hat{a}}\right) \exp\left[i 2
\xi \Omega \frac{\Delta \hat{r}'}{\hat{a}} \right]~.
\label{accueq2f}
\end{eqnarray}
Similarly as before, the Fourier transform of Eq. (\ref{accueq2f})
with respect to $\Delta\hat{r}'/\hat{a}$ is the line spread
function for the system. Since it refers to the focal plane it
will be indicated with $l_\mathrm{fp}$. After introduction of $y =
\hat{a} \bar{r}_i/\hat{f}$ and $\eta = \Delta{\hat{r}'}/\hat{a}$
we can write such Fourier transform exactly as  Eq.
(\ref{accueq3}), that is

\begin{eqnarray}
l_\mathrm{fp}\left(\Omega,y\right)=\int_{-1}^{1} d\xi \left \{
\int_{-1}^{1} d \eta P\left(\xi+\eta\right) P\left(\xi-\eta\right)
\exp\left[i 2 \eta \left(\xi \Omega -  y\right)\right]\right\}~,
\label{accueq3f}
\end{eqnarray}
where we accounted for the fact that the Fourier transform
integral in $d\xi$ is limited by the presence of the lens to the
range $[-1,1]$. On the one hand, the integral in parenthesis
$\{...\}$ is the Fourier transform of $P(\xi+\eta) P(\xi-\eta)$
calculated with respect to $\eta$ as a function of  $2 (\xi \Omega
+ y)$. On the other hand, the function $P(\xi+\eta) P(\xi-\eta)$
is a window function similar to the pupil function $P$. It is
equal to unity for values of $|\eta|<1-|\xi|$ and is zero
elsewhere. Therefore, as in the previous Section \ref{sec:cam},
the quantity in parenthesis $\{...\}$ can be calculated
analytically yielding back Eq. (\ref{accueq4}). Moreover, as in
the previous Section \ref{sec:cam} we may use a new variable
$y'=y/\sqrt{\Omega}$ in Eq. (\ref{accueq3f}), thus obtaining the
following final expression for the line spread function:

\begin{eqnarray}
l_\mathrm{fp}\left(\Omega,y'\right)=\int_{-1}^{1} d\xi
\frac{\sin\left[2 (\xi \Omega + y'
\sqrt{\Omega})(1-|\xi|)\right]}{\xi \Omega + y' \sqrt{\Omega}}~.
\label{accueq5f}
\end{eqnarray}
Eq. (\ref{accueq5f}) is identical to Eq. (\ref{accueq5}).
Therefore, the shapes of the function $l_\mathrm{fp}$ and of its
asymptotic limit for different values of $\Omega$ are the same as
those for $l_\mathrm{pc}$ given in Fig. \ref{pinX05}, Fig.
\ref{pinX2}, Fig. \ref{pinX10} and Fig. \ref{pinX100}. As before,
once a definition for the width of $l_\mathrm{fp}$  is chosen, the
optimal operation point for the pinhole camera may be set
requiring that such width be minimal. Defining the width of
$l_\mathrm{fp}$ to be the full width half maximum (FWHM) one
concludes that the optimal lens aperture for the imaging in the
focal plane is given by

\begin{equation}
\hat{a}^2 \simeq 4.5 \hat{z}_1 \label{bestfoc}
\end{equation}
or, in dimensional units

\begin{equation}
{a}^2 \simeq 4.5 \frac{{z}_1 c}{\omega}~. \label{best2foc}
\end{equation}
We can estimate the best resolution in the focal plane $\delta
{r}_f$ in dimensional units requiring that $\delta y'$ be equal to
the minimal FWHM of the line spread function, i.e. $1.5$, and
taking advantage of the definition of $y'$, which gives

\begin{equation}
\delta {r}_f \simeq 1.5 \sqrt{\frac{c}{\omega {z}_1}} f~.
\label{drfoc}
\end{equation}
Note that results in this Section have not been obtained under the
assumption $|\mathrm{M}| \gg 1$ as those in the last Section.
Therefore, substitution of $\hat{z}_1 \longrightarrow
\hat{z}_{\mathrm{eff}}$ is not required in this case.

Similarly as before, we should now discuss the applicability of
Eq. (\ref{accueq5f}). Up to now we discussed about a Gaussian
quasi-homogeneous undulator source, but Eq. (\ref{accueq5f}) can
be applied in a wider variety of cases. In particular, Eq.
(\ref{accueq5f}) can be applied when (a) the source is
quasi-homogeneous, (b) the lens is installed in the far zone and
(c) $\hat{a}^2 \ll \max[1,\hat{D}] \hat{z}_1^2$. When conditions
(a), (b) and (c) are satisfied, the line spread function for the
lens, $l_\mathrm{fp}$ can be calculated from Eq. (\ref{accueq5f}).
Depending on the situation, one may have a good quality or a bad
quality of the image. This is related with how the width of
$l_\mathrm{fp}$ scales with the width of the ideal image. The
intensity profile at the focal plane (located at $\hat{z}=
\hat{z}_1+\hat{f}$) is given, in any case, by the convolution
product of the line spread function $l_\mathrm{fp}$ and the ideal
image. In particular, for an undulator source with Gaussian
intensity profile at the virtual plane, such product is given by
$l_\mathrm{fp} \ast \exp[-\bar{r}^2/(2 \mathrm{M}^2 \hat{N})]$,
where $|\mathrm{M}| = \hat{f}/\hat{z}_1$. When $\sqrt{\hat{N}} \gg
\hat{a} \gg \hat{z}_1/\sqrt{\hat{N}}$ a good quality image is
formed at the observation plane. In other situations instead, the
width of $l_\mathrm{fp}$ is of order of the width of the ideal
image, and a bad quality image will be formed. To be specific, for
values of $\Omega\gg 1$ the line spread function and, therefore,
the intensity distribution at the focal plane tends to a stepped
profile,  while for values of $\Omega \ll 1$ one obtains a
$\mathrm{sinc}^2(\cdot)$ profile. To sum up, in order to calculate
the intensity distribution at the focal plane behind the lens, we
should first calculate $\Omega = \hat{a}^2/\hat{z}_1$. Then use
Eq. (\ref{accueq5f}) to calculate the line spread function
$l_\mathrm{fp}$. Finally, convolve with the ideal image from the
source. The minimal width of the line spread function, which
corresponds to the best image resolution, is when
$\hat{a}^2/\hat{z}_1 \simeq 4.5$. Note that conditions (a), (b)
and (c) above do not require the source to be an undulator source.
In particular, (a), (b) and (c) are automatically satisfied for
thermal sources, or perfectly incoherent objects. As a result, our
theory can be applied for visible light imaging as well. These
remarks simplify our discussion and drastically widen the
applicability region of our results.

\section{\label{sec:imany} A unified theory of incoherent imaging by a single lens}

In the present Section \ref{sec:imany} we will develop a unified
theory which is applicable to imaging in an arbitrary plane behind
the lens. In the previous Section \ref{sec:focim} we discussed the
possibility of imaging the source at the focal plane. There we
recognized that under conditions (a) the source is
quasi-homogeneous  (b) the lens is installed in the far zone and
(c) $\hat{a}^2 \ll \max[1,\hat{D}] \hat{z}_1^2$, the intensity on
the focal plane can be calculated as a convolution product of the
line spread function of the system and the ideal image, which is a
scaled version of the intensity distribution on the source plane.
Moreover, it was shown that the characteristics of the intensity
distribution in the focal plane can be treated in formal analogy
with the pinhole camera setup in Section \ref{sec:cam} and with
the physics of aberrations, in particular defocusing aberrations,
described in Section \ref{sec:abe}. We have seen that when the
lens is in the far field (condition (b))  Eq. (\ref{maintrick2})
can be used to characterize the cross-spectral density at the lens
position. In the image plane, this assumption alone allows
cancellation of all phase factors in the integrand of Eq.
(\ref{crspecprop3bis}) and yields Eq. (\ref{GPInb1}). In the
previous Section \ref{sec:focim} we showed that this cancellation
does not hold anymore for the focal plane. From this viewpoint the
focal plane is not privileged in any way with respect to other
observation planes. It is this last remark which suggests a
generalization of the previous results. For a generic observation
plane positioned at $\hat{z}=\hat{z}_2$ one may use Eq.
(\ref{maintrick2}) to get an expression for the intensity ($\Delta
\hat{r} =0$) from Eq. (\ref{crspecprop3bis}), in analogy with Eq.
(\ref{pineqb2}) and Eq. (\ref{pineq}), that is

\begin{eqnarray}
\hat{I}(\hat{z}_2,\bar{r}) &=& \int d\bar{r}'d\Delta\hat{r}'
\Bigg\{\hat{\mathcal{G}}(0,-\bar{r}',-\Delta\hat{r}') \exp\left[2i
\left(-\frac{1}{f}+\frac{1}{\hat{z}_1}+\frac{1}{\hat{d}}\right)\bar{r}'\Delta
\hat{r}' \right] \cr && \times
P(\bar{r}'+\Delta\hat{r}')P(\bar{r}'-\Delta \hat{r}')\Bigg\}
\exp\left[-\frac{2i}{\hat{d}}\bar{r}\Delta\hat{r}'\right]~,
\label{pineqb2any}
\end{eqnarray}
where $\hat{d} = \hat{z}_2-\hat{z}_1$. As one can see, the phase
factor in $\bar{r}'\Delta \hat{r}' $ is more complicated than
those Eq. (\ref{pineqb2}) and Eq. (\ref{pineq}) due to the
complete arbitrariness of the observation plane. Nevertheless, Eq.
(\ref{pineqb2any}) can be recast to the same form of Eq.
(\ref{pineqb2}) simply with the help of with a new definition of
$\hat{z}_\mathrm{eff}$, that is

\begin{equation}
\frac{1}{\hat{z}_\mathrm{eff}}
=-\frac{1}{f}+\frac{1}{\hat{z}_1}+\frac{1}{\hat{d}}~.
\label{zeffany}
\end{equation}
Accounting for Eq. (\ref{zeffany}) we may rewrite Eq.
(\ref{pineqb2any}) as

\begin{eqnarray}
\hat{I}(\hat{z}_2,\bar{r}) &=& \int d\bar{r}'d\Delta\hat{r}'
\Bigg\{\hat{\mathcal{G}}(0,-\bar{r}',-\Delta\hat{r}')
\exp\left[i\frac{2}{\hat{z}_{\mathrm{eff}}} \bar{r}'\Delta
\hat{r}' \right] \cr && \times
P(\bar{r}'+\Delta\hat{r}')P(\bar{r}'-\Delta \hat{r}')\Bigg\}
\exp\left[-\frac{2i}{\hat{d}}\bar{r}\Delta\hat{r}'\right]~,
\label{pineqb2any2}
\end{eqnarray}
which is equivalent to Eq. (\ref{pineqb2}).

All that is left to do is now follow, step by step, the previous
Section \ref{sec:focim}. This leads to the following results. When
conditions (a), (b) and (c) are satisfied one may introduce, as
before, a line spread function of the system, characteristic of
the observation plane $\hat{z}=\hat{z}_2$. This line spread
function is formally identical to Eq. (\ref{accueq5f}):

\begin{eqnarray}
l_\mathrm{z}\left(\Omega,y'\right)=\int_{-1}^{1} d\xi
\frac{\sin\left[2 (\xi \Omega + y'
\sqrt{\Omega})(1-|\xi|)\right]}{\xi \Omega + y' \sqrt{\Omega}}~,
\label{accueq5fany}
\end{eqnarray}
with $\Omega = \hat{a}^2/\hat{z}_\mathrm{eff}$ and
$y'=y/\sqrt{\Omega}$, $y'$ being defined by $y = \hat{a}
\bar{r}/\hat{d}$. Here similarity techniques have been employed.
The five dimensional parameters $\lambda$, $a$, $f$, $z_1$ and
$z_2$ have been reduced to the only parameters $\Omega$ and $y'$,
so that one is left with the calculation of a dimensionless
function in $y'$ depending on the single parameter $\Omega$,  Eq.
(\ref{accueq5fany}). As before, depending on the situation, one
may have a good quality or a bad quality of the image. This is
related with how the width of $l_\mathrm{z}$ scales with the width
of the ideal image. The intensity profile at the observation plane
located at $\hat{z}= \hat{z}_2$ is given by the convolution
product of the line spread function $l_\mathrm{z}$ and the ideal
image. In particular, for an undulator source with Gaussian
intensity profile at the virtual plane, such product is given by
$l_\mathrm{z} \ast \exp[-\bar{r}^2/(2 \mathrm{M}^2 \hat{N})]$,
where $|\mathrm{M}| = \hat{d}/\hat{z}_1$. When $\sqrt{\hat{N}} \gg
\hat{a} \gg \hat{z}_1/\sqrt{\hat{N}}$ a good quality image is
formed at the observation plane. In other situations instead, the
width of $l_\mathrm{z}$ is of order of the width of the ideal
image, and a bad quality image will be formed. To sum up, we
presented here an algorithm to calculate the intensity
distribution of radiation at any observation plane located at
position $\hat{z}=\hat{z}_2$ behind the lens, given an arbitrary
value of $\hat{a}$ and a system satisfying conditions (a), (b) and
(c). First one should calculate $\Omega =
\hat{a}^2/\hat{z}_\mathrm{eff}$. Then use Eq. (\ref{accueq5fany})
to calculate the line spread function $l_\mathrm{z}$. Finally,
convolve with the ideal image from the source. As before, when
$\Omega\gg 1$ the line spread function tends to a stepped profile
and, as a result, the intensity profile reproduces a stepped
profile too. Note that, $\hat{z}_\mathrm{eff}$ being arbitrary, we
cannot give a relation in terms of $\hat{a}$ and $\hat{z}_1$
corresponding to an optimal line spread function. For instance,
when $\hat{z}_2 \longrightarrow \hat{z}_i$, i.e. when we consider
the asymptotic limit for the image plane, we obtain from Eq.
(\ref{zeffany}) that $\hat{z}_\mathrm{eff} \longrightarrow \infty$
and $\Omega \longrightarrow 0$. Then, from Eq. (\ref{accueq5fany})
one obtains $l_\mathrm{z} \longrightarrow \mathrm{sinc}^2(\hat{a}
\bar{r}_i/\hat{d})$, exactly as in Eq. (\ref{acculim1}). In this
case no defocusing aberration is present, quadratic phase term
having being cancelled by the particular choice of the observation
plane. As a result, we cannot give a criterium for an optimal line
spread function: we only have diffraction effects so that the
larger the aperture $\hat{a}$ (always within the constraint
imposed by condition (c), i.e. $\hat{a}^2 \ll \max[1,\hat{D}]
\hat{z}_1^2$) the better the quality of the image. In closing, it
is worth to mention that the asymptote for $\hat{f}
\longrightarrow \infty$ corresponds to the pinhole camera setup
already discussed in Section \ref{sec:cam}. As a result, this
particular case can be treated in terms of our unified theory as
well.

\section{\label{sec:dof} Depth of focus}

According to \cite{ATTW} "the depth of focus of a lens is the
permitted displacement, away from the focal or image plane, for
which the intensity on axis is diminished by some permissible
amount". In particular, when plane wave illumination is considered
on a perfect circular lens, the focal plane corresponds to the
plane where the radiation assumes the minimal spot size, and the
intensity on axis reaches a maximum at that point. It can be shown
that, in this case, the on-axis intensity decreases of about
$20\%$  when "the observation plane is displaced from the ideal
focal plane [...] by an amount" (see \cite{ATTW})

\begin{equation}
|\Delta' z| = \frac{\lambda}{2 \mathrm{NA}^2} ~,\label{ATTDOF}
\end{equation}
where the quantity $\mathrm{NA}$ indicates the numerical aperture
of the lens. $\mathrm{NA}=\sin\theta$, $\theta$ being the "half
angle measured from the optic axis at the focus back to the lens"
\cite{ATTW}. The concept of depth of focus described in Eq.
(\ref{ATTDOF}), describes a case of coherent illumination of the
lens by a plane wave, and the lens is treated, here, as a
condenser. The object to be imaged is, in fact, the radiation
source itself. Since we are dealing with coherent illumination we
may say that, in this case, the depth of focus is diffraction
limited.

The depth of focus as described in Eq. (\ref{ATTDOF}) is
parametrically related to another concept of depth of focus which
is used, for example, in Optical Lithography \cite{ATTW,BJOR}. In
this case one needs to illuminate a wafer with a demagnified image
of a given pattern on a mask.  Here the lens (actually the
collection of lenses) is no more treated as a condenser: its
function is to produce the demagnified image of the mask. The mask
itself must be illuminated by means of a condenser system instead,
and, as remarked in \cite{ATTW}, "the ability to print fine, high
contrast features is significantly affected by the degree of
coherence within the optical system. If there exists a high degree
of spatial coherence, diffraction from adjacent mask features will
interfere in the image plane, significantly modifying the recorded
pattern". The mask should therefore be illuminated by incoherent
light and should be considered as a quasi-homogeneous source
itself.  For a lithography setup, the resolution $R$ is the width
of the diffraction-limited point spread function of the system,
while the distance $X$ over which the image is in proper focus is,
more quantitatively, the distance over which $R$ is increased by
some permitted amount. Although a lithography setup constitutes  a
completely different setup with respect to the condenser system
illuminated with coherent plane waves, $R$ and $X$ turn out to be
respectively \cite{BJOR}:

\begin{equation}
R  = k_1 \frac{\lambda}{\mathrm{NA}}~;~~~X=k_2
\frac{\lambda}{\mathrm{NA}^2} ~.\label{ReD}
\end{equation}
The distance $X$ is therefore parametrically related to the
distance $|\Delta' z|$ defined in Eq. (\ref{ATTDOF}) and even
though these two quantities refer to very different setups,
involving quite different physics, also $X$ is named, as $|\Delta'
z|$, depth of focus.

To complicate the situation further one is frequently interested
in the depth of focus $|\Delta z|$ for a condenser system when the
lens is not illuminated by plane waves but by other kind of non
quasi-homogeneous or quasi-homogeneous sources, which is a
different situation from both cases considered in the previous
discussion. In this Section we will treat the case of a condenser
lens illuminated by a quasi-homogeneous source characterized by
$\hat{N}_x \gg 1$ and $\hat{D}_x \gg 1$. Separate treatments of
the $x$ and $y$ direction are thus allowed, which simplify to
one-dimensional cases. Our study applies to the horizontal
direction only. However, if also $\hat{N}_y \gg 1$ and $\hat{D}_y
\gg 1$ the same results for the horizontal direction can be
applied to the vertical direction as well. Our definition of depth
of focus will be relative to the plane of smallest spot size of
the radiation. Therefore, in the present Section \ref{sec:dof},
the depth of focus of a lens is defined as the permitted
displacement, away from the waist plane, for which the intensity
on axis is diminished by some permissible amount.

Before starting to discuss about the depth of focus, it is
necessary to derive an expression for the point where the
radiation spot size is the smallest. This can be obtained with the
help of Eq. (\ref{beta}) in Section \ref{sub:imfoge}. In fact,
since $2\sqrt{\hat{\epsilon}\hat{\beta}(\hat{z}_2)}$ is the
characteristic width of the Gaussian radiation spot size at the
observation plane located at $\hat{z}=\hat{z}_2$, it is sufficient
to look at the point $\hat{z}_{2\mathrm{best}}$ where the
derivative of $\hat{\epsilon}\hat{\beta}(\hat{z}_2)$ is zero to
obtain the position of the waist plane. With the help of Eq.
(\ref{beta}) one finds

\begin{eqnarray}
\hat{z}_{2\mathrm{best}} &=&
\frac{\left(\hat{f}+\hat{z}_1\right)\hat{N}
-\hat{D}\hat{f}\hat{z}_1^2+\hat{D}\hat{z}_1^3}{\hat{N}+\hat{D}
\left(\hat{f}-\hat{z}_1\right)^2} =
\frac{\left(\hat{f}+\hat{z}_1\right)\hat{A}\hat{z}_1^2-
\left(\hat{f}-\hat{z}_1\right)\hat{D}\hat{z}_1^2}{\hat{A}
\hat{z}_1^2 +\hat{D} \left(\hat{f}-\hat{z}_1\right)^2 }~.\cr &&
\label{bestf}
\end{eqnarray}
It is interesting to study two limiting cases of Eq. (\ref{bestf})
for $\hat{A} \gg \hat{D}$  and for $\hat{A} \longrightarrow 0$
respectively. Consider first the case when $\hat{A} \gg \hat{D}$.
From Eq. (\ref{bestf})  one may see that, in this first case,
$\hat{z}_{2\mathrm{best}} \longrightarrow \hat{z}_f = \hat{f} +
\hat{z}_1$. This means that the waist plane asymptotically goes to
the focal plane of the lens. Consider now the case when $\hat{A}
\longrightarrow 0$. Again from Eq. (\ref{bestf})  one may see
that, in this second case, $\hat{z}_{2\mathrm{best}}
\longrightarrow \hat{z}_i = \hat{z}_1^2/(\hat{z}_1-\hat{f})$. This
means that the waist plane asymptotically goes to the image plane
of the virtual source (located at $\hat{z}=0$).

From the analysis of the two limiting cases for $\hat{A} \gg
\hat{D}$  and for $\hat{A} \longrightarrow 0$, we conclude that
one always has $\hat{z}_1+\hat{f}<\hat{z}_{2\mathrm{best}}
<\hat{z}_i$. In other words, the waist plane is always located
between the focal and the image plane. This results may seem
counterintuitive. In fact, for our Gaussian virtual source, the
waist is located at $\hat{z}_s = 0$, which is imaged, reversed and
magnified, at the image plane. Therefore, at first glance, the
smallest spot size should be located at the image plane. The
reason why this is not the case is due to the presence of the
magnification factor, which  linearly increases with $\hat{z}_2$.
On the one hand the source has a waist located at $\hat{z}_s = 0$
and its size increases symmetrically as one moves away from
$\hat{z}_s = 0$ in both the positive and the negative direction.
On the other hand the magnification factor increases linearly with
$\hat{z}_2$ and is not characterized by the same symmetric
dependence on the displacement from $\hat{z}_s = 0$.

In the following Section \ref{sub:lapdof} and Section
\ref{sub:sizdof} we study the problem of the depth of focus in a
condenser system with a Gaussian quasi-homogeneous source. Effects
from the pupil width will be neglected in Section
\ref{sub:lapdof}, Finite pupil dimensions will be accounted for in
Section \ref{sub:sizdof}.

\subsection{\label{sub:lapdof} Large non-limiting aperture}

We will discuss, for simplicity, the case when $|\mathrm{M}| =
\hat{d}_i/\hat{z}_1 \ll 1$. In this case, from Eq.
(\ref{lensagain}) we obtain $\hat{d}_i \simeq \hat{f}$. Since we
assumed as a starting point $|\mathrm{M}| \ll 1$ it follows that
$\hat{f} \ll \hat{z}_1$, and that the distance between the image
and the focal plane is $\Delta = \hat{d}_i - \hat{f} \simeq
\hat{f}^2/\hat{z}_1 \simeq |\mathrm{M}| \hat{f}$. Moreover, the
radiation spot size at the image plane is known to be of order
$|\mathrm{M}|\sqrt{\hat{N}}$ from Eq. (\ref{Infrelim}), while the
radiation spot size at the focal plane is obtained from Eq.
(\ref{Infrel}) and is of order $\sqrt{\hat{D}} \hat{f}$. As has
been discussed above, the position of the waist goes from the
focal plane to the image plane as we pass from the near to the far
zone. This can also be seen by comparing the radiation spot sizes
at the image and at the focal plane in the near and in the far
zone. In the near zone $\hat{A} \gg \hat{D}$. It follows that
$\sqrt{\hat{D}} \hat{f} \ll \sqrt{\hat{N}} \hat{f}/\hat{z}_1$,
i.e. the radiation spot size at the focal plane is much smaller
than that at the image plane. On the contrary in the far zone
$\hat{A} \ll \hat{D}$. It follows that $\sqrt{\hat{D}} \hat{f} \gg
\sqrt{\hat{N}} \hat{f}/\hat{z}_1\simeq \sqrt{\hat{N}}
\hat{d}_i/\hat{z}_1$, i.e. the radiation spot size at the focal
plane is much smaller than that at the image plane. We will
analyze these two cases separately.

\subsubsection{Far zone}

In this case $\hat{A} \ll \hat{D}$ and the waist is near the
image. On the one hand, as has been already remarked, the
radiation spot size on the image plane is about $|\mathrm{M}|
\sqrt{\hat{N}}$ and does not depend on $\hat{D}$. On the other
hand, the radiation spot size on the non-limiting pupil aperture
is defined by the beam divergence and can be estimated as
$\sqrt{\hat{D}} \hat{z}_1$. This means that the rate of change of
the radiation spot size from the lens to the waist can be
estimated as $\sqrt{\hat{D}} \hat{z}_1/\hat{f}$, since
$\hat{d}_i\simeq \hat{f}$. The parametric dependence of the depth
of focus can be found requiring that the spot size increase due to
a displacement $\Delta \hat{z}$, that is  $\left(\sqrt{\hat{D}}
\hat{z}_1/\hat{f}\right)\Delta \hat{z}$ be of order of the
radiation spot size at the waist, i.e. $|\mathrm{M}|
\sqrt{\hat{N}}$. This yields the following Geometrical Optics
prediction for the depth of focus:

\begin{equation}
\Delta \hat{z} \simeq \frac{\hat{f}|\mathrm{M}|
\sqrt{\hat{N}}}{\sqrt{\hat{D}} \hat{z}_1} \simeq \mathrm{M}^2
\sqrt{\frac{\hat{N}}{\hat{D}}}~. \label{dof1}
\end{equation}
One may separately estimate the diffraction size related with an
aperture of size $\sqrt{\hat{D}} \hat{z}_1$ (the radiation spot
size at the non-limiting pupil aperture) at the observation plane.
Such estimation yields a diffraction size
$\hat{d}_i/(\sqrt{\hat{D}} \hat{z}_1)\simeq
\hat{f}/(\sqrt{\hat{D}} \hat{z}_1)$. Now, requiring that the spot
size increase due to a displacement $\Delta' \hat{z}$, that is
$\left(\sqrt{\hat{D}} \hat{z}_1/\hat{f}\right)\Delta' \hat{z}$, be
of order of the diffraction size $\hat{f}/(\sqrt{\hat{D}}
\hat{z}_1)$ one may estimate the diffraction limited depth of
focus as

\begin{equation}
\Delta' \hat{z} \simeq \frac{\hat{f}^2}{\hat{D} \hat{z}_1^2}~.
\label{dof2}
\end{equation}
The following comparison between the diffraction limited depth of
focus $\Delta'\hat{z}$ and the Geometrical Optics prediction for
the depth of focus $\Delta \hat{z}$ can be presented:

\begin{equation}
\frac{\Delta' \hat{z}}{\Delta \hat{z}} \simeq
\frac{\hat{f}^2}{\hat{D} \hat{z}_1^2} \cdot \frac{1}{\mathrm{M}^2}
\sqrt{\frac{\hat{D}}{\hat{N}}} \simeq
\frac{1}{\sqrt{\hat{N}\hat{D}}}~. \label{dof3}
\end{equation}
Eq. (\ref{dof3}) allows one to conclude that diffraction effects
can be neglected within the accuracy of the quasi-homogeneous
approximation, and that Eq. (\ref{dof1}) is a correct estimation
of the depth of focus in this case. Note that the definition in
Eq. (\ref{ATTDOF}) has no meaning in the case of a condenser
system with a quasi-homogeneous source. Finally, it is interesting
to remark that the depth of focus is much shorter than the
distance $\Delta$ between the waist (image) and the focal plane.
In fact

\begin{equation}
\frac{\Delta\hat{z}}{\Delta} \simeq \mathrm{M}^2
\sqrt{\frac{\hat{N}}{\hat{D}}} \cdot \frac{1}{|\mathrm{M}|
\hat{f}} \simeq  \sqrt{\frac{\hat{N}}{\hat{D}\hat{z}_1^2}} \ll 1~,
\label{dof4}
\end{equation}
because of the far zone assumption ($\hat{A} \ll \hat{D}$).

\subsubsection{Near zone}

Similar estimations can be made in the near zone, i.e. assuming
$\hat{A} \gg \hat{D}$, when the waist is near the focal plane. On
the one hand, as has been already remarked, the radiation spot
size at the focal plane is about $\sqrt{\hat{D}}\hat{f}$ and does
not depend on $\hat{N}$. On the other hand, the radiation spot
size on the non-limiting pupil aperture is defined by the electron
beam waist and can be estimated as $\sqrt{\hat{N}}$. This means
that the rate of change of the radiation spot size from the lens
to the waist can be estimated as $\sqrt{\hat{N}}/\hat{f}$ since
the waist is near the focal plane. The parametric dependence of
the depth of focus can be found imposing that the spot size
increase due to a displacement $\Delta \hat{z}$, that is
$\left(\sqrt{\hat{N}}/\hat{f}\right)\Delta \hat{z}$ be of order of
the radiation spot size at the waist, i.e.
$\sqrt{\hat{D}}\hat{f}$. This yields the following Geometrical
Optics prediction for the depth of focus:

\begin{equation}
\Delta \hat{z} \simeq \frac{\sqrt{\hat{D}}\hat{f}^2}
{\sqrt{\hat{N}}}~. \label{dof1b}
\end{equation}
One may separately estimate the diffraction size related with an
aperture of size $\sqrt{\hat{N}}$ (the radiation spot size at the
non-limiting pupil aperture) at the observation plane. Such
estimation gives a diffraction size
$\hat{d}_i/\sqrt{\hat{N}}\simeq \hat{f}/\sqrt{\hat{N}}$. Now,
imposing that the spot size increase due to a displacement $\Delta
\hat{z}$, that is $\left(\sqrt{\hat{N}}/\hat{f}\right)\Delta
\hat{z}$ be of order of the diffraction size
$\hat{f}/\sqrt{\hat{N}}$ one may estimate the diffraction limited
depth of focus as

\begin{equation}
\Delta' \hat{z} \simeq \frac{\hat{f}^2}{\hat{N}}~. \label{dof2b}
\end{equation}
The following comparison between the diffraction limited depth of
focus $\Delta'\hat{z}$ and the Geomterical Optics prediction of
the depth of focus $\Delta \hat{z}$:

\begin{equation}
\frac{\Delta' \hat{z}}{\Delta \hat{z}} \simeq
\frac{\sqrt{\hat{N}}}{\sqrt{\hat{D}}\hat{f}^2}
\frac{\hat{f}^2}{\hat{N}}\simeq \frac{1}{\sqrt{\hat{N}\hat{D}}}~
\label{dof3b}
\end{equation}
Eq. (\ref{dof3b}) allows one to conclude that diffraction effects
can be neglected within the accuracy of the quasi-homogeneous
approximation, and that Eq. (\ref{dof1b}) is a correct estimation
for the depth of focus in this case. Finally, it is interesting to
remark that the depth of focus is much shorter than the distance
$\Delta$ between the waist (focal plane) and the image plane. In
fact

\begin{equation}
\frac{\Delta\hat{z}}{\Delta} \simeq \frac{\sqrt{\hat{D}}\hat{f}^2}
{\sqrt{\hat{N}}}\cdot \frac{1}{|\mathrm{M}| \hat{f}} \simeq
\sqrt{\frac{\hat{D}\hat{z}_1^2}{\hat{N}}}  \ll 1 ~,\label{dof4b}
\end{equation}
because of the near zone assumption ($\hat{A} \gg \hat{D}$).

\subsection{\label{sub:sizdof} Effect of aperture size}

We will now analyze the influence of a finite aperture on our
analysis of the depth of focus. For simplicity we will consider
the far field case, as all our examples involving effects of
finite aperture size have been relying on the far field
assumption. At first glance, substitution of the radiation spot
size $\sqrt{\hat{D}}\hat{z}_1$ with the finite aperture size
$\hat{a}$ solves all issues, as it allows estimations of the rate
of change of the radiation spot size and of diffraction effects.
This would lead to the following equations for the Geometrical
Optics prediction of the depth of focus:

\begin{equation}
\Delta \hat{z} \simeq \frac{\hat{f}^2 \sqrt{\hat{N}}}{\hat{z}_1
\hat{a}}~, \label{dof1c}
\end{equation}
for the diffraction limited depth of focus:

\begin{equation}
\Delta' \hat{z} \simeq \frac{\hat{f}^2}{\hat{a}^2}~, \label{dof2c}
\end{equation}
for the ratio between $\Delta' \hat{z}$ and $\Delta \hat{z}$:

\begin{equation}
\frac{\Delta' \hat{z}}{\Delta \hat{z}} \simeq
\frac{\hat{f}^2}{\hat{a}^2} \cdot \frac{\hat{z}_1
\hat{a}}{\hat{f}^2 \sqrt{\hat{N}}} \simeq
\frac{\hat{z}_1}{\hat{a}\sqrt{\hat{N}}}~, \label{dof3c}
\end{equation}
and for the ratio between the depth of focus and the distance
$\Delta$:

\begin{equation}
\frac{\Delta\hat{z}}{\Delta} \simeq \frac{\hat{f}^2
\sqrt{\hat{N}}}{\hat{z}_1 \hat{a}} \cdot \frac{1}{|\mathrm{M}|
\hat{f}} \simeq  \frac{\sqrt{\hat{N}}}{\hat{a}} ~. \label{dof4c}
\end{equation}
We have seen in Section \ref{sec:focim} that there exist
particular situations when the phase factor in parenthesis
$\{...\}$ in Eq. (\ref{pineqb3}) can be neglected. 

In one of these cases, when $\hat{D} \hat{z}_1^2 \gg \hat{N} \gg
\hat{a}^2 \gg \hat{z}_1^2/\hat{N}$ one recovers a scaled image of
the virtual source at the focal plane, where the scaling factor
$\hat{f}/\hat{z}_1 \simeq |\mathrm{M}|$ (see Eq. (\ref{pineqb7})).
Thus, in this situation, it is not correct to state that the
radiation spot size at the image plane is smaller than the
radiation spot size at the focal plane, because in both these
planes we have the same, identical image of the virtual source
with approximatively the same scaling factor. It should be
remarked that in this case the depth of focus $\Delta \hat{z}$ is
longer than the distance between focal and image plane, $\Delta$,
as one may see from Eq. (\ref{dof4c}) when $\hat{N} \gg
\hat{a}^2$. Another critical scale to account for is
$\hat{z}_1^2/\hat{N}$. As we have seen in Section \ref{sec:focim},
when $\hat{a}^2 \ll \hat{z}_1^2/\hat{N}$ one obtains, at the focal
plane, the diffraction pattern of the pupil (aberrated or not,
depending on how $\hat{a}^2$ scales with respect to $\hat{z}_1$).
This case corresponds to coherent illumination of the pupil. In
this situation, the diffraction limited depth of focus $\Delta'
\hat{z}$ is much longer than the Geometrical Optics prediction of
the  depth of focus $\Delta \hat{z}$, as one may see from Eq.
(\ref{dof3c}). In this case, Eq. (\ref{dof2c}) should be used
instead of Eq. (\ref{dof1c}) in order to estimate the depth of
focus of the system. Summing up, Eq. (\ref{dof1c}) can be used
without any remarks in all the cases when $\hat{a}^2 \gg
\hat{z}_1^2/\hat{N}$. When $\hat{a}^2 \ll \hat{N}$ and $\hat{a}^2
\gg \hat{z}_1^2/\hat{N}$ one should remember that the depth of
focus becomes longer than the distance between the image and the
focal plane, this case corresponding the situation when an image
of the virtual source is formed in the focal plane. In all cases
when $\hat{a}^2 \ll \hat{z}_1^2/\hat{N}$ instead, we have coherent
illumination of the pupil.  The diffraction limited depth of focus
becomes longer than the Geometrical Optics prediction of the depth
of focus and, for estimations, one should use Eq. (\ref{dof2c}) in
place of Eq. (\ref{dof1c}).

\section{\label{sec:nonh} Solutions to the image formation problem for non-homogeneous undulator sources}

We will now consider non-homogeneous undulator sources. The next
Section \ref{sub:hqhvn} will deal with the case of a horizontally
quasi-homogenous and vertically non-homogeneous source. Special
emphasis will be given to the study of this situation, which is
relevant for the majority of Synchrotron Radiation sources of the
third generation. In the following Section \ref{sub:ultimaltro} we
will discuss, instead, the case of a horizontally non-homogeneous
and vertically diffraction limited source. These two cases
practically deal with all third generation light sources, from the
soft to the hard X-rays.

Our treatment is based on the assumption that the beta functions
have minima in the center of the undulator. Placing the virtual
source in the center of the undulator, we will obtain a
particularly simple expression for the cross-spectral density of
the source and its Fourier transform. In particular, calculations
will yield real results. Statistical Optics methods conjugated to
Fourier Optics will allow us to give explicit presentations of the
cross-spectral density and its Fourier transform at the virtual
source position. Such presentations will account for all
complexities of the source, intrinsic properties of undulator
radiation and electron-beam phase space distribution. Based on our
previous work \cite{OURS}, we will first calculate the
cross-spectral density $\hat{G}$ in the far zone limit. Then, with
the help of Eq. (\ref{maintrick2}), we will be able to recover the
Fourier transform of the cross-spectral density
$\hat{\mathcal{G}}$ at the virtual source position. By inverse
transforming that expression we will finally recover the
cross-spectral density as well.

In general, we cannot give an explicit expression for the
cross-spectral density at any observation plane. In other words,
as before, Eq. (\ref{crspecprop3bis}) cannot be calculated
explicitly for any value of $\hat{z}_1$ and $\hat{z}_2$. However,
once the center of the undulator is fixed as the virtual source
position, there exists a privileged observation plane. This is the
plane where the center of the undulator is imaged and depends on
the position $\hat{z}_1$ chosen for the lens. Throughout this
paper we simply called it the image plane. Starting from the
expression for the cross-spectral density and its Fourier
transform at the virtual source we will be able to calculate, for
any choice of  $\hat{z}_1$, the cross-spectral density on the
focal ($\hat{z}_2 = \hat{z}_f$) and on the image ($\hat{z}_2 =
\hat{z}_i$) plane. The procedure that we will use to perform these
calculations is similar to what has been proposed in the case of
homogeneous sources and will take advantage of Eq. (\ref{fundfoc})
and Eq. (\ref{fundima}).

As usual, we will begin our investigations neglecting pupil
effects. Results for a large non-limiting aperture on the image
plane will be generalized to include effects of the pupil with the
help of Eq. (\ref{GPIn}). Finally, in Section \ref{sub:gim} we
will critically discuss the assumptions made throughout this paper
and see how and up to what extent they can be relaxed.

\begin{figure}
\begin{center}
\includegraphics*[width=140mm]{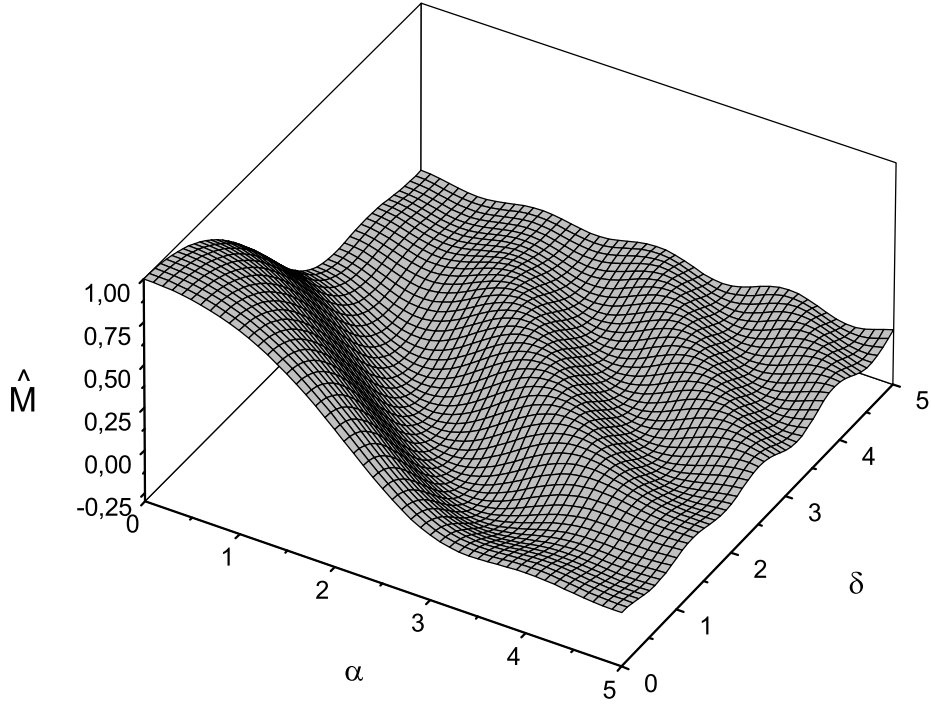}
\caption{\label{M} Plot of the universal function $\hat{M}$, used
to calculate the cross-spectral density at the focal plane when
$\hat{N}_x \gg1$, $\hat{D}_x \gg 1$, $\hat{N}_y$ and $\hat{D}_y$
are arbitrary. }
\end{center}
\end{figure}

\subsection{\label{sub:hqhvn} Horizontally quasi-homogenous and vertically non-homogeneous source}

\subsubsection{Large non-limiting aperture}

Let us first retain the assumption $\hat{D}_x \gg1 $ and
$\hat{N}_x \gg 1$, but allow $\hat{D}_y$ and $\hat{N}_y$ to assume
arbitrary values.  Eq. (\ref{G2Dnewsimplif2y}) allows the
reconstruction of the vertical cross-spectral density in the far
zone. Note that, when both $\hat{N}_y \lesssim 1$ and $\hat{D}_y
\lesssim 1$ the source is non-homogeneous. In this case the far
zone limit is for values $\hat{z}_o\gg1 $, and one obtains

\begin{eqnarray}
\hat{G}(\hat{z}_o,\bar{\theta}_y,\Delta \hat{\theta}_y) &=&
{\exp{\left[i 2\bar{\theta}_y\hat{z}_o\Delta \hat{\theta}_y
\right]}}  \exp{\left[-2\hat{N}_y  \Delta \hat{\theta}_y^2
\right]} \cr &&\times\int_{-\infty}^{\infty} d \hat{\phi}_y
\exp{\left[-\frac{(\bar{\theta}_y+\hat{\phi}_y)^2}{2
\hat{D}_y}\right]}M(\hat{\phi}_y,\Delta \hat{\theta}_y)
~,\label{resu1}
\end{eqnarray}
where the function $\hat{M}(\alpha,\delta)$ is the normalized
version of a universal function first defined in \cite{OURS} and
reads:

\begin{eqnarray}
\hat{M}(\alpha,\delta)= \frac{3}{8 \sqrt{\pi}}
{\int_{-\infty}^{\infty} d \hat{\phi}_x
\mathrm{sinc}{\left[\frac{\hat{\phi}_x^2+(\alpha-\delta)^2}{4}\right]}
\mathrm{sinc}{\left[\frac{\hat{\phi}_x^2+(\alpha+\delta)^2}{4}\right]}}~.\label{Motherofall}
\end{eqnarray}
A plot of the $\hat{M}$ function is given in Fig. \ref{M}. The
$\hat{M}$ function is a real function. Another remarkable property
of  $\hat{M}$  is its invariance for exchange of $\alpha$ with
$\delta$. Also, $\hat{M}$ is invariant for exchange of $\alpha$
with $-\alpha$ (or $\delta$ with $-\delta$). The following
relations between universal functions hold:

\begin{equation}
\hat{I}_S(\alpha) = \hat{M}(\alpha,0) ~, \label{relation1}
\end{equation}
\begin{equation}
\beta(\delta) = \frac{1}{2\pi^2} \int_{-\infty}^{\infty} d \xi
\hat{M}(\xi,\delta) ~\label{relation2}
\end{equation}
and

\begin{equation}
\gamma(x) = \frac{1}{2\pi^2} \int_{-\infty}^{\infty} d \xi \exp
\left [- i 2 x \xi\right] \hat{M}(\xi,0) ~.\label{relation3}
\end{equation}
\begin{figure}
\begin{center}
\includegraphics*[width=140mm]{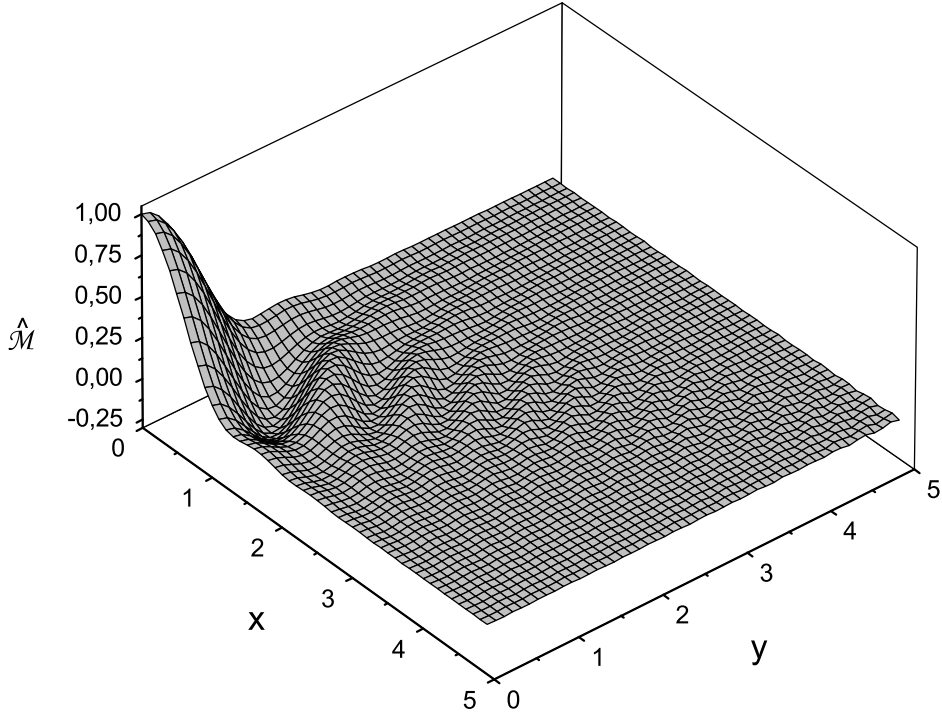}
\caption{\label{FTM} Plot of the universal function
$\hat{\mathcal{M}}$, used to calculate the cross-spectral density
at the image plane when $\hat{N}_x \gg1$, $\hat{D}_x \gg 1$,
$\hat{N}_y$ and $\hat{D}_y$ are arbitrary. }
\end{center}
\end{figure}
Using Eq. (\ref{maintrick2}) and Eq. (\ref{resu1}) we obtain the
Fourier transform of the cross-spectral density at $\hat{z}_o=0$
at the virtual source position

\begin{eqnarray}
\hat{\mathcal{G}}\left(0,{ \bar{u}}, {\Delta{ \hat{u}}}\right) &=&
\exp\left[-2 \hat{N}_y \Delta \hat{u}^2 \right]
\int_{-\infty}^{\infty} d \hat{\phi}_y
\exp{\left[-\frac{(\bar{u}+\hat{\phi}_y)^2}{2
\hat{D}_y}\right]}\hat{M}(\hat{\phi}_y,\Delta \hat{u}) ~. \cr &&
\label{gen1}
\end{eqnarray}
Inverse transforming Eq. (\ref{gen1}) we obtain the cross-spectral
density at the virtual source position

\begin{eqnarray}
\hat{G}(0,\bar{y},\Delta \hat{y}) = \exp\left[-2 \hat{D}_y \Delta
\hat{y}^2 \right]  \int_{-\infty}^{\infty} d \hat{\phi}
\exp\left[-\frac{(\bar{y}+\hat{\phi})^2}{2\hat{N}_y}\right]\hat{\mathcal{M}}(\Delta
\hat{y}, \hat{\phi})~.\label{gen2}
\end{eqnarray}
In analogy with Eq. (\ref{trGtris}), we defined
$\hat{\mathcal{M}}$ as

\begin{eqnarray}
\hat{\mathcal{M}}(x,y) =\frac{8\sqrt{\pi}}{3} \frac{1}{2\pi^2
\mathcal{K}} \int_{-\infty}^{\infty} d\alpha
\int_{-\infty}^{\infty} d\delta \exp\left[-2i(\alpha x+\delta y
)\right] \hat{M}(\alpha,\delta)~, \label{calM}
\end{eqnarray}
where $\mathcal{K}$ is given in Eq. (\ref{BB}). A plot of
$\hat{\mathcal{M}}$ is presented in Fig. \ref{FTM}. Note that the
symmetry of $\hat{M}$ for exchange of $\alpha$ with $-\alpha$ or
$\delta$ with $-\delta$ implies that $\hat{\mathcal{M}}$ is a real
function. This fact, together with Eq. (\ref{gen2}), implies that
$\hat{G}$ at the virtual plane, that is an ensemble-averaged field
product, is real as well. This is a consequence of the fact that
single particle sources produce a field characterized by a plane
wavefront at the virtual source position. It can be seen comparing
Eq. (\ref{cpact5}) with Eq. (\ref{cpact61a}) or, directly, by
inspecting Eq. (\ref{virsof}). Moreover, note that
$\hat{\mathcal{M}}$ is invariant for the exchange of $x$ with $y$.

From Eq. (\ref{fundfoc}) and Eq. (\ref{gen1}) we obtain the
cross-spectral density on the focal plane in the vertical
direction, that is

\begin{eqnarray}
\hat{G}( \hat{z}_f,{ \bar{y}}_f,\Delta{\hat{y}}_f) &=&
\exp\left[\frac{2i}{\hat{f}^2} \left(\hat{f}-\hat{z}_1\right){
\bar{y}}_f\Delta{\hat{y}}_f\right]\exp\left[- \frac{2\hat{N}_y
\Delta{\hat{y}}_f^2 }{ \hat{f}^2}\right] \cr && \times
\int_{-\infty}^{\infty} d \hat{\phi}_y
\exp{\left[-\frac{(\bar{y}_f/\hat{f}+\hat{\phi}_y)^2}{2 \hat{D}_y
}\right]}\hat{M}\left(\hat{\phi}_y,\frac{\Delta
\hat{y}_f}{\hat{f}} \right).\label{focino}
\end{eqnarray}
In the vertical direction, the relative intensity on the focal
plane is therefore given by

\begin{eqnarray}
\hat{I}(\hat{z}_f,\bar{y}_f) &=&~\int_{-\infty}^{\infty} d
\hat{\phi}_y
\exp{\left[-\frac{(\bar{y}_f/\hat{f}+\hat{\phi}_y)^2}{2 \hat{D}_y
}\right]}\hat{I}_S\left(\hat{\phi}_y \right)\cr && \times
\left\{\int_{-\infty}^{\infty} d \hat{\phi}_y
\exp{\left[-\frac{\hat{\phi}_y^2}{2 \hat{D}_y
}\right]}\hat{I}_S\left(\hat{\phi}_y \right)\right\}^{-1}
,\label{Ifocino}
\end{eqnarray}
because $\hat{I}_S(\hat{\phi}_y)=\hat{M}(\hat{\phi}_y,0)$. In the
limit  $\hat{D}_y \ll 1$, Eq. (\ref{Ifocino}) reduces to

\begin{eqnarray}
\hat{I}(\hat{z}_f,\bar{y}_f)=
\hat{I}_S\left(\frac{\bar{y}_f}{\hat{f}}
\right).\label{Ifocinolim}
\end{eqnarray}

\begin{figure}
\begin{center}
\includegraphics*[width=140mm]{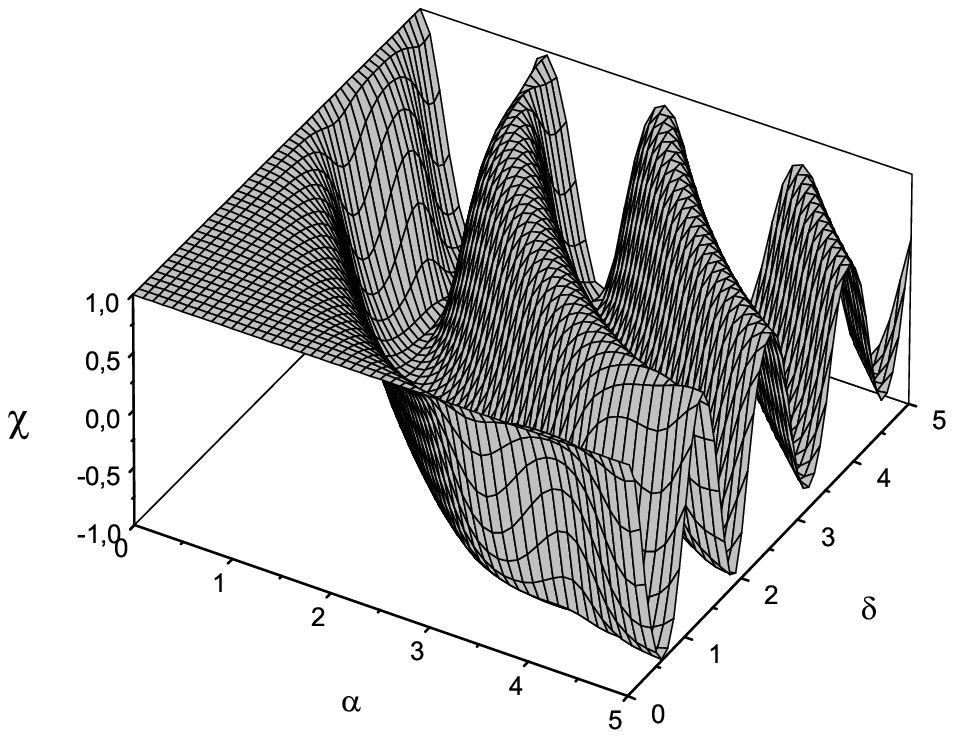}
\caption{\label{chi} Plot of the universal function $\hat{\chi}$,
used to calculate the modulus of the spectral degree of coherence
on the focal plane when $\hat{N}_x \gg1$, $\hat{D}_x \gg 1$,
$\hat{N}_y \ll 1$ and $\hat{D}_y \ll 1$. }
\end{center}
\end{figure}
The modulus of the spectral degree of coherence on the focal plane
in the vertical direction reads

\begin{eqnarray}
|g(\hat{z}_f,\bar{y}_f, \Delta \hat{y}_f)| &=&  \exp\left[-2
\frac{\hat{N}_y }{\hat{f}^2} \Delta \hat{y}_f^2\right] \cr
&&\times \int_{-\infty}^{\infty} d \hat{\phi}_y
\exp{\left[-\frac{(\bar{y}_f/\hat{f}+\hat{\phi}_y)^2}{2
\hat{D}_y}\right]}\hat{M}\left(\hat{\phi}_y,\frac{\Delta
\hat{y}_f}{\hat{f}}\right)\cr && \times \left\{
\int_{-\infty}^{\infty} d \hat{\phi}_y
\exp{\left[-\frac{(\bar{y}_f/\hat{f}+\Delta
\hat{y}_f/\hat{f}+\hat{\phi}_y)^2}{2
\hat{D}_y}\right]}\hat{I}_S(\hat{\phi}_y)\right\}^{-1/2}\cr &&
\times \left\{ \int_{-\infty}^{\infty} d \hat{\phi}_y
\exp{\left[-\frac{(\bar{y}_f/\hat{f}-\Delta
\hat{y}_f/\hat{f}+\hat{\phi}_y)^2}{2
\hat{D}_y}\right]}\hat{I}_S(\hat{\phi}_y)\right\}^{-1/2}~.\cr &&
\label{gfoc}
\end{eqnarray}
In the limiting case  for $\hat{D}_y \ll 1$ one obtains the
simplified expression

\begin{eqnarray}
|g( \hat{z}_f,\bar{y}_f, \Delta \hat{y}_f)| &=& \exp\left[-2
\frac{\hat{N}_y }{\hat{f}^2} \Delta \hat{y}_f^2\right]
\chi{\left(\frac{\bar{y}_f}{\hat{f}},\frac{\Delta
\hat{y}_f}{\hat{f}}\right)} ~, \label{gfoclim1}
\end{eqnarray}
where, as in Eq. (145) of \cite{OURS},

\begin{equation}
\chi{\left(\alpha,\delta\right)} =
\frac{\hat{M}(\alpha,\delta)}{\left[\hat{I}_S(\alpha-\delta)\right]^{1/2}
\left[\hat{I}_S(\alpha+\delta)\right]^{1/2}}~. \label{chidef}
\end{equation}
A plot of $\chi$ is given in Fig. \ref{chi}.  Eq. (\ref{gfoc})
reduces even more in the case when both $\hat{D}_y \ll 1$ and
$\hat{N}_y \ll 1$. In this case we have

\begin{eqnarray}
|g( \hat{z}_f,\bar{y}_f, \Delta \hat{y}_f)| &=&
\chi{\left(\frac{\bar{y}_f}{\hat{f}},\frac{\Delta
\hat{y}_f}{\hat{f}}\right)} ~.\label{gfoclim2}
\end{eqnarray}
On the image plane, Eq. (\ref{fundima}) and Eq. (\ref{gen2}) give
the following cross-spectral density in the vertical direction:

\begin{eqnarray}
\hat{G}(\hat{z}_i,{ \bar{y}}_i,\Delta{\hat{y}}_i)  &=&
\exp\left[\frac{i \mathrm{m}(\mathrm{m}+1) { \bar{y}}_i
\Delta{\hat{y}}_i }{2 \hat{z}_1} \right] \int_{-\infty}^{\infty} d
\hat{\phi} \exp\left[-\frac{(\mathrm{m}\bar{y}_i+
\hat{\phi})^2}{2\hat{N}_y}\right]\cr &&\times
\hat{\mathcal{M}}(\mathrm{m}\Delta \hat{y}_i,
\hat{\phi})\exp\left[-2 \hat{D}_y \mathrm{m}^2\Delta \hat{y}_i^2
\right] ~,\cr && \label{genimaM}
\end{eqnarray}
corresponding to a relative intensity on the image plane

\begin{eqnarray}
\hat{I}( \hat{z}_i,{ \bar{y}}_i)  &=& \int_{-\infty}^{\infty} d
\hat{\phi}
\exp\left[-\frac{(\mathrm{m}\bar{y}_i+\hat{\phi})^2}{2\hat{N}_y}\right]\hat{\mathcal{M}}(0,
\hat{\phi})\cr && \times \left\{\int_{-\infty}^{\infty} d
\hat{\phi}
\exp\left[-\frac{\hat{\phi}^2}{2\hat{N}_y}\right]\hat{\mathcal{M}}(0,
\hat{\phi})\right\}^{-1}~. \label{IgenimaM}
\end{eqnarray}
With the help of Eq. (\ref{calM}), Eq. (\ref{relation2}) and Eq.
(\ref{BB}) one sees that

\begin{eqnarray}
\hat{\mathcal{M}}(0,y) &=& \frac{1}{2\pi^2 \mathcal{K}}
\int_{-\infty}^{\infty} d\delta \left[\exp\left[i (-2 \delta) y
\right] \int_{-\infty}^{\infty} d\alpha
\hat{M}(\alpha,\delta)\right]   =  \hat{\mathcal{B}}(y) ~.
\label{unpoinsie}
\end{eqnarray}
The solution of the image formation problem is thus constituted by
a convolution product between a Gaussian function and the
universal function $ \hat{\mathcal{B}}$, which admits the
analytical representation given in Eq. (\ref{ftbrepr}).  The
intensity on the image plane is independent of the value of
$\hat{D}_y$. In the limit $\hat{N}_y \gg 1$ Eq. (\ref{IgenimaM})
gives back Eq. (\ref{Ingandeima}). Instead, in the limit
$\hat{N}_y \ll 1$ we obtain

\begin{eqnarray}
\hat{I}(\hat{z}_i,{ \bar{y}}_i)  &=& \hat{\mathcal{M}}(0,
\mathrm{m}\bar{y}_i) =  \hat{\mathcal{B}}(\mathrm{m}\bar{y}_i) ~.
\label{IgenimaMlim}
\end{eqnarray}
The modulus of the spectral degree of coherence in the vertical
direction is given by

\begin{figure}
\begin{center}
\includegraphics*[width=140mm]{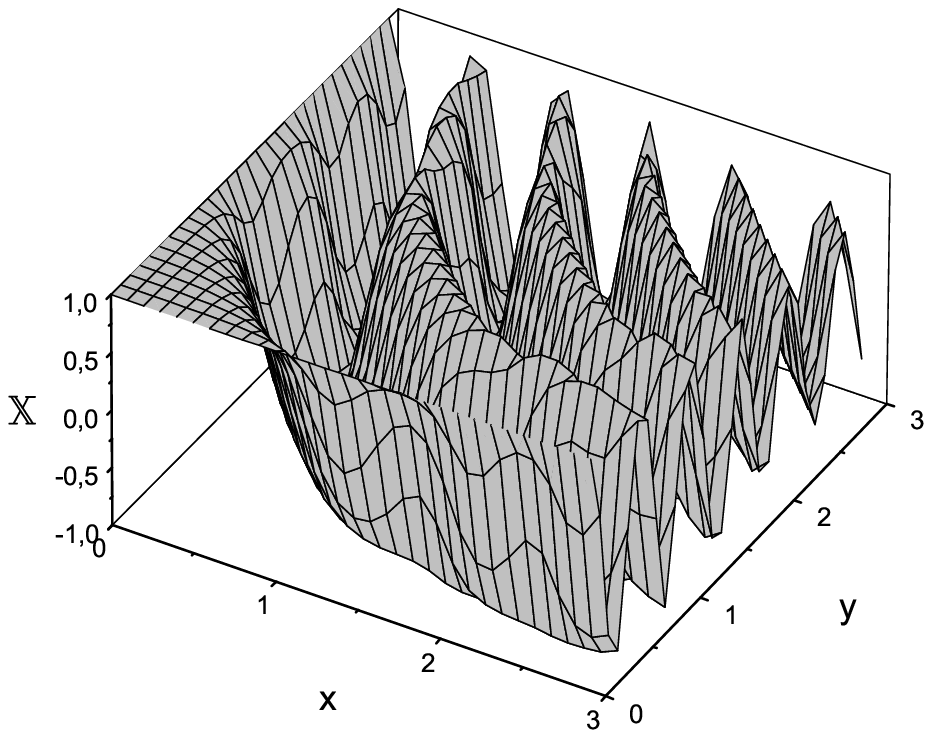}
\caption{\label{ftchi} Plot of the universal function
$\mathbb{X}$, used to calculate the modulus of the spectral degree
of coherence on the image plane when $\hat{N}_x \gg1$, $\hat{D}_x
\gg 1$, $\hat{N}_y \ll 1$ and $\hat{D}_y \ll 1$. }
\end{center}
\end{figure}
\begin{eqnarray}
|{g}( \hat{z}_i, \bar{y}_i, \Delta{\hat{y}}_i)|  &=&
\int_{-\infty}^{\infty} d \hat{\phi}
\exp\left[-\frac{(\mathrm{m}\bar{y}_i+\hat{\phi})^2}{2\hat{N}_y}\right]\cr
&&\times \hat{\mathcal{M}}(\mathrm{m}\Delta \hat{y}_i, \hat{\phi})
\exp\left[-2 \hat{D}_y \mathrm{m}^2\Delta \hat{y}_i^2 \right] \cr
&&\times \left\{\int_{-\infty}^{\infty} d \hat{\phi}
\exp\left[-\frac{(\mathrm{m}\bar{y}_i+\mathrm{m} \Delta \hat{y}_y
+\hat{\phi})^2}{2\hat{N}_y}\right]\hat{\mathcal{M}}(0,
\hat{\phi})\right\}^{-1/2} \cr &&\times
\left\{\int_{-\infty}^{\infty} d \hat{\phi}
\exp\left[-\frac{(\mathrm{m}\bar{y}_i-\mathrm{m} \Delta \hat{y}_y
+\hat{\phi})^2}{2\hat{N}_y}\right]\hat{\mathcal{M}}(0,
\hat{\phi})\right\}^{-1/2}. \cr && \label{spdngandeimab}
\end{eqnarray}
In the case $\hat{D}_y \ll1 $ and $\hat{N}_y \ll 1$ one obtains

\begin{eqnarray}
|{g}( \hat{z}_i,  \bar{y}_i,\Delta{\hat{y}}_i)|  &=&
{\mathbb{X}}\left(\mathrm{m}\Delta \hat{y}_i,\mathrm{m}
\bar{y}_i\right)~ \label{spdngandeimalim}
\end{eqnarray}
where, in analogy with Eq. (\ref{chidef}), we define

\begin{equation}
\mathbb{X}(x,y) = \frac{\hat{\mathcal{M}}(x,y)}
{\left[\hat{\mathcal{B}}(x-y)\right]^{1/2}
\left[\hat{\mathcal{B}}(x+y)\right]^{1/2}}~. \label{mathcalchi}
\end{equation}
Note that when $\hat{N}_y \gg 1$ and $\hat{D}_y$ assumes arbitrary
values, the modulus of the spectral degree of coherence, Eq.
(\ref{spdngandeimab}), simplifies to Eq. (\ref{spdngandeima}). A
plot of the universal function $\mathbb{X}$ is given in Fig.
\ref{ftchi}.

\subsubsection{Effect of aperture size}

We will now include, with the help of Eq. (\ref{GPIn}), the
effects of pupil in the one-dimensional case. The pupil function
and $\hat{\mathcal{P}}$ are given by Eq. (\ref{Pyfunc}) and Eq.
(\ref{verpy}). The $r$-direction should be now substituted with
the $y$-direction.

We can use condition (\ref{farlens}) and Eq. (\ref{genimaM}) to
describe the case when the lens is in the far zone. From Eq.
(\ref{genimaM}) we can estimate the order of the source dimension,
that is $\max[1,\sqrt{\hat{N}_y}]$, and the order of the coherence
length, that is $\min[1/\sqrt{\hat{D}_y},1]$. The lens is in the
far zone when $\max[1,\sqrt{\hat{N}_y}]/\hat{z}_1 \ll
\max[\sqrt{\hat{D}_y},1]$. Note that several particular cases of
interest are automatically included in this condition: the case
for $\hat{D}_y \gg 1$ and $\hat{N}_y \gg 1$, that gives $\hat{z}_1
\gg \sqrt{\hat{N}_y/\hat{D}_y}$, the case for $\hat{D}_y \gg 1$
and $\hat{N}_y \lesssim 1$, that gives $\hat{z}_1 \gg
1/\sqrt{\hat{D}_y}$, as well as the case for $\hat{D}_y \lesssim
1$ and $\hat{N}_y \gg 1$, that gives $\hat{z}_1 \gg
\sqrt{\hat{N}_y}$. All these situations have been previously
discussed in Section \ref{sec:qhso}. The new situation left to
consider is for $\hat{N}_y \sim 1$ and $\hat{D}_y \sim 1$. In this
case, the lens is in the far field when $\hat{z}_1 \gg 1$. From
Eq. (\ref{GPIn}) and Eq. (\ref{genimaM}) we obtain

\begin{eqnarray} \hat{G}_P(\hat{z}_i,{\bar{y}}_i, \Delta {\hat{y}}_i)&=&4 \hat{a}^2
\exp\left[\frac{2i\mathrm{m}}{\hat{z}_1}{\bar{y}}_i\Delta{\hat{y}}_i\right]
\int d{\bar{u}} ~d \Delta {\hat{u}}~  \exp\left[-2 \hat{D}_y
\hat{z}_1^2\Delta \hat{u}^2 \right] \cr &&\times
\int_{-\infty}^{\infty} d\hat{\phi}
\exp\left[-\frac{\left(\hat{\phi}+\hat{z}_1 \bar{u}\right)^2}{2
\hat{N}_y} \right] \hat{\mathcal{M}}(\hat{z}_1\Delta \hat{u},
\hat{\phi})\cr &&\times\mathrm{sinc}\left\{\hat{a}\left[\frac{
\mathrm{m}}{\hat{z}_1}\left({\bar{y}}_i+ \Delta
{\hat{y}}_i\right)-{\bar{u}}- \Delta {\hat{u}}\right]\right\}\cr
&& \times\mathrm{sinc}\left\{\hat{a}\left[\frac{
\mathrm{m}}{\hat{z}_1}\left({\bar{y}}_i-\Delta{\hat{y}}_i\right)-{\bar{u}}+\Delta{\hat{u}}\right]\right\}
~. \label{GPInbqhb11d}
\end{eqnarray}
According to Eq. (\ref{resu1}), $\hat{z}_1^2 \max[1,\hat{D}_y]$ is
of order of the square of the radiation spot size on the pupil,
while $\hat{z}_1^2/\max[\hat{N}_y,1]$ is of order of the square of
the coherence length on the pupil. It is interesting to see that
if $\hat{D}_y \gg 1$ the exponential function in $\Delta
{\hat{u}}$ has a very narrow characteristic length with respect to
unity, and one can make the substitution
$\hat{\mathcal{M}}(\hat{z}_1\Delta \hat{u}, \hat{\phi}_y)
\longrightarrow \hat{\mathcal{M}}(0, \hat{\phi}_y)
=\hat{\mathcal{B}}(\hat{\phi}_y)$, thus getting back Eq.
(\ref{GPInb1bc}). If $\hat{N}_y \gg 1$ instead, $\hat{\phi}_y$ can
be set to zero in the exponential function in
$\hat{\phi}_y+\hat{z}_1 \bar{u}$, and the entire exponential
function can be taken out of the integral in $d \hat{\phi}_y$.
Then, it is possible to show that the surviving integral in $d
\hat{\phi}_y$ is equal to $\gamma(\hat{z}_1 \Delta \hat{u})$,
giving back Eq. (\ref{GPInb1b}). Finally, if both $\hat{N}_y \gg
1$ and $\hat{D}_y \gg 1$ we get back Eq. (\ref{GPInb1}).

Eq. (\ref{GPInbqhb11d}) should be considered as a limiting case
when no aberrations are present. More generally, when aberrations
and apodization are present, Eq. (\ref{GPInbqhb11d}) should be
substituted by the following expression:

\begin{eqnarray} \hat{G}_P(\hat{z}_i,{\bar{y}}_i, \Delta {\hat{y}}_i)&=&4 \hat{a}^2
\exp\left[\frac{2i\mathrm{m}}{\hat{z}_1}{\bar{y}}_i\Delta{\hat{y}}_i\right]
\int d{\bar{u}} ~d \Delta {\hat{u}}~  \exp\left[-2 \hat{D}_y
\hat{z}_1^2\Delta \hat{u}^2 \right] \cr &&\times
\int_{-\infty}^{\infty} d\hat{\phi}
\exp\left[-\frac{\left(\hat{\phi}+\hat{z}_1 \bar{u}\right)^2}{2
\hat{N}_y} \right] \hat{\mathcal{M}}(\hat{z}_1\Delta \hat{u},
\hat{\phi})\cr &&\times\hat{\mathcal{P}_a}\left[\frac{
\mathrm{m}}{\hat{z}_1}\left({\bar{y}}_i+ \Delta
{\hat{y}}_i\right)-{\bar{u}}- \Delta {\hat{u}}\right]\cr &&
\times\hat{\mathcal{P}_a}^*\left[\frac{
\mathrm{m}}{\hat{z}_1}\left({\bar{y}}_i-\Delta{\hat{y}}_i\right)-{\bar{u}}+\Delta{\hat{u}}\right]
~. \label{ultimasp}
\end{eqnarray}
Eq. (\ref{ultimasp}) is very general and is valid under the only
assumption that the lens is in the far field region. The use of
old coordinates $\hat{y}_{i1}$ and $\hat{y}_{i2}$ instead of
$\bar{y}_i$ and $\Delta \hat{y}_i$ somewhat clarifies  the meaning
of Eq. (\ref{ultimasp}). Eq. (\ref{ultimasp}) states that the
cross-spectral density accounting for the pupil influence
(aberrations and diffraction effects) is a double convolution
product of Eq. (\ref{genimaM}), i.e. the cross-spectral density in
the ideal case, and $\hat{\mathcal{P}}_a$, which can be written as

\begin{eqnarray}
\hat{G}_P(\hat{z}_i,{\hat{y}}_{i1},{\hat{y}}_{i2}) &=& \left [
\left<\hat{E}(\hat{u}_1)\hat{E}^*(\hat{u}_2)\right> \ast
\hat{\mathcal{P}_a} (\hat{u}_1) \ast \hat{\mathcal{P}}^*_a
(\hat{u}_2) \right]({\hat{y}}_{i1},{\hat{y}}_{i2}) \cr &=&
\left<\left[\hat{E}\ast \hat{\mathcal{P}_a}
\right](\hat{y}_{i1})\left[\hat{E}^*\ast \hat{\mathcal{P}}^*_a
\right](\hat{y}_{i2})\right> ~.\label{ultimasp2}
\end{eqnarray}
The intensity at the image plane is found setting
$\hat{y}_{i1}=\hat{y}_{i2}=\hat{y}_i$, which gives

\begin{eqnarray}
\hat{I}_P(\hat{z}_i,{\hat{y}}_{i}) &=& \left [
\left<\hat{E}(\hat{u}_1)\hat{E}^*(\hat{u}_2)\right> \ast
\hat{\mathcal{P}_a} (\hat{u}_1) \ast \hat{\mathcal{P}}^*_a
(\hat{u}_2) \right]({\hat{y}}_{i},{\hat{y}}_{i}) \cr &=&
\left<\left|\hat{E}\ast
\hat{\mathcal{P}_a}\right|^2(\hat{y}_{i})\right>
~.\label{ultimasp3}
\end{eqnarray}
In particular, in the case of a completely coherent source, we may
neglect the ensemble average.

Note that in order to evaluate the intensity at the image plane,
it is no more enough to know the ideal intensity and to convolve
with a line spread function. One has to know the cross-spectral
density in the ideal case and convolve twice with
$\hat{\mathcal{P}_a}$, which is known as the amplitude line spread
function of the system (or the amplitude point spread function in
the two-dimensional case) and is a more general identifier of the
system characteristics. Even in the non quasi-homogeneous case one
may continue to use, for evaluating the intensity at the image
plane, an algorithm based on the calculation of ideal
characteristics and further convolution with a function
characterizing the system. The difference with respect to the
quasi-homogeneous case is that the amplitude line spread function
must be used in place of the line spread function, and that the
cross-spectral density must be used in place of the intensity.

Note that, in our approach, the Wigner function plays no role, as
it constitutes an artificial quantity in the image formation
problem, whereas the natural quantity to consider is the
cross-spectral density. Even if the virtual source can be
characterized by a phase space distribution (i.e. by a positive
Wigner function) one has no simplification in the imaging problem.
The only simplification of Eq. (\ref{ultimasp3}) takes place in
the quasi-homogeneous case when, due to the separability of the
cross-spectral density variables and to a short coherence length
(compared with the size of the source), one obtains the usual
incoherent line spread function formalism.

\subsection{\label{sub:ultimaltro} Horizontally non-homogeneous and vertically diffraction limited source}

We will now relax the assumption of a large horizontal electron
beam size and divergence and deal with the non-homogeneous case
when $\hat{N}_x$ assumes arbitrary values, while $\hat{D}_x \ll
1$. This is a rather exotic range of parameters, and we will
discuss the case for a large non-limiting aperture only. In this
case, for third generation light sources we have automatically
$\hat{N}_y \ll 1$ and $\hat{D}_y \ll 1$ because ${\epsilon}_y \ll
{\epsilon}_x$\footnote{This treatment can be easily generalized to
the more exotic situation for ${\hat{N}_x}$ arbitrary,
${\hat{D}_x\ll 1 }$, ${\hat{N}_y}$ arbitrary and ${\hat{D}_y \ll
1}$. However, here we will be concerned with third generation
light sources only. Assuming reasonable values for ${\beta}_x
\lesssim 10 ~L_w$ and ${\epsilon}_y = \chi {\epsilon}_x$ with
$\chi \simeq 10^{-2}$ we see that $\hat{D}_x \ll 1$ implies both
$\hat{N}_y \ll 1$ and $\hat{D}_y \ll 1$. Therefore we will avoid
to make generalizations which are not pertinent to the case under
study, e.g. exotic case for ${\beta}_x > 10~L_w$, ${\beta}_y <
10^{-1}~L_w$ or ${\beta}_y
> 10~L_w$. }. In this case, in the
limit $\hat{z}_o \gg 1$, from Eq. (\ref{G2D}) we obtain

\begin{eqnarray}
\hat{G} &=&{\exp{\left(i 2 \bar{\theta}_x\Delta \hat{\theta}_x
\hat{z}_o\right)}} \exp{\left[-{ 2 \hat{N}_x  \Delta
\hat{\theta}_x^2 }\right]} {\exp{\left(i 2 \bar{\theta}_y\Delta
\hat{\theta}_y \hat{z}_o\right)}} \cr && \times\mathrm{sinc}
{\left[\frac{(\bar{\theta}_x-\Delta
\hat{\theta}_x)^2+(\bar{\theta}_y-\Delta
\hat{\theta}_y)^2}{4}\right]} \mathrm{sinc}
{\left[\frac{(\bar{\theta}_x+\Delta
\hat{\theta}_x)^2+(\bar{\theta}_y+\Delta
\hat{\theta}_y)^2}{4}\right]} ~.\cr &&\label{Gdis2}
\end{eqnarray}
Using Eq. (\ref{maintrick2}) and Eq. (\ref{Gdis2}) one obtains the
Fourier transform of the cross-spectral density at $\hat{z}_o=0$,
i.e. at the virtual-source position, that is given by

\begin{eqnarray}
\hat{\mathcal{G}}\left(0,{ \bar{\theta}_x}, {\Delta{
\hat{\theta}}_x},{ \bar{\theta}_y}, {\Delta{
\hat{\theta}}_y}\right)&=& \exp{\left[-{ 2 \hat{N}_x \Delta
\hat{\theta}_x^2 }\right]} \mathrm{sinc}
{\left[\frac{(\bar{\theta}_x-\Delta
\hat{\theta}_x)^2+(\bar{\theta}_y-\Delta
\hat{\theta}_y)^2}{4}\right]} \cr &&\times \mathrm{sinc}
{\left[\frac{(\bar{\theta}_x+\Delta
\hat{\theta}_x)^2+(\bar{\theta}_y+\Delta
\hat{\theta}_y)^2}{4}\right]} .\cr && \label{xsmallftg}
\end{eqnarray}
Inverse transforming Eq. (\ref{xsmallftg}) it is possible to
express the cross-spectral density at the virtual source position
as

\begin{eqnarray}
\hat{G}(0,\bar{x},\Delta \hat{x},\bar{y},\Delta \hat{y}) &=&
\int_{-\infty}^{\infty} d \bar{\theta}_x d\Delta \hat{\theta}_x d
\bar{\theta}_y d\Delta \hat{\theta}_y \exp{\left[-{ 2 \hat{N}_x
\Delta \hat{\theta}_x^2 }\right]} \cr && \times  \mathrm{sinc}
{\left[\frac{(\bar{\theta}_x-\Delta
\hat{\theta}_x)^2+(\bar{\theta}_y-\Delta
\hat{\theta}_y)^2}{4}\right]} \cr  && \times \mathrm{sinc}
{\left[\frac{(\bar{\theta}_x+\Delta
\hat{\theta}_x)^2+(\bar{\theta}_y+\Delta
\hat{\theta}_y)^2}{4}\right]} \cr && \times  \exp{\left[2i (\Delta
\hat{\theta}_x \bar{x} + \bar{\theta}_x \Delta \hat{x})\right]
}\exp{\left[2i (\Delta \hat{\theta}_y \bar{y} + i\bar{\theta}_y
\Delta \hat{y})\right] }~. \cr &&\label{gen2cambio}
\end{eqnarray}
From Eq. (\ref{fundfoc}) and Eq. (\ref{xsmallftg}) we obtain the
cross-spectral density on the focal plane

\begin{eqnarray}
\hat{G}( \hat{z}_f,{ \bar{x}}_f,\Delta{\hat{x}}_f,{
\bar{y}}_f,\Delta{\hat{y}}_f) &=& \exp\left[\frac{2i}{\hat{f}^2}
\left(\hat{f}-\hat{z}_1\right){
\bar{x}}_f\Delta{\hat{x}}_f\right]\exp{\left[-\frac{ 2
\hat{N}_x}{\hat{f}^2} \Delta{\hat{x}}_f^2 \right]} \cr && \times
\exp\left[\frac{2i}{\hat{f}^2} \left(\hat{f}-\hat{z}_1\right){
\bar{y}}_f\Delta{\hat{y}}_f\right]\cr && \times \mathrm{sinc}
{\left[\frac{(\bar{x}_f-\Delta \hat{x}_f)^2+(\bar{y}_f-\Delta
\hat{y}_f)^2}{4 \hat{f}^2}\right]} \cr && \times
 \mathrm{sinc} {\left[\frac{(\bar{x}_f+\Delta
\hat{x}_f)^2+(\bar{y}_f+\Delta \hat{y}_f)^2}{4\hat{f}^2}\right]}
.\label{focino2}
\end{eqnarray}
The relative intensity on the focal plane is therefore given by

\begin{eqnarray}
\hat{I}(\hat{z}_f,\bar{x}_f,\bar{y}_f) &=&\mathrm{sinc}^2
{\left[\frac{\bar{x}_f^2+\bar{y}_f^2}{4
\hat{f}^2}\right]}.\label{Ifocino2}
\end{eqnarray}
This is just the relative intensity on the focal plane from a
single electron, i.e. Eq. (\ref{Icpact31}). It is interesting to
note that the modulus of the spectral degree of coherence on the
focal plane depends on $\Delta \hat{x}_f$ only, and can be written
as

\begin{eqnarray}
|g( \hat{z}_f, \bar{x}_f, \Delta \hat{x}_f, \bar{y}_f, \Delta
\hat{y}_f)| &=& \exp{\left[-\frac{ 2 \hat{N}_x}{\hat{f}^2}
\Delta{\hat{x}}_f^2 \right]}  ~.\cr && \label{gfoc2}
\end{eqnarray}
In the limit $\hat{N}_x \ll 1$ one recovers the deterministic case
of a single particle. In this limit $|g|$ reduces to unity, and
the wavefront is perfectly coherent.

As regards the image plane, we should note that Eq.
(\ref{gen2cambio}) is not easy to manipulate analytically in the
most general case. However, when $\Delta \hat{x}=0$ and $\Delta
\hat{y}=0$, one can calculate the intensity of the virtual source.
With the help of Eq. (\ref{G2D2perDFgenApp}) and  Eq.
(\ref{appliedapp}) one obtains

\begin{eqnarray}
\hat{I}(0,\bar{x},\bar{y}) &=& \int_{-\infty}^{\infty} d
\hat{\phi}_x d \exp{\left[-\frac{(\bar{x}+\hat{\phi}_x)^2}{ 2
\hat{N}_x}\right]} \tilde{\Psi}(\hat{\phi}_x,\bar{y})~,
\label{gen3b}
\end{eqnarray}
where we have set

\begin{equation}
\tilde{\Psi}(x,y) = \Psi\left(\sqrt{x^2+y^2}\right)~.
\label{psipsi}
\end{equation}
The function $\Psi$ was already defined in Eq. (\ref{psiuni}).
Using Eq. (\ref{fundima}) and Eq. (\ref{gen3b}) we can now give
the following expression for the relative intensity:

\begin{eqnarray}
\hat{I}(\hat{z}_i,\bar{x}_i,{ \bar{y}}_i)  &=&
\int_{-\infty}^{\infty} d \hat{\phi}_x
\exp{\left[-\frac{(\mathrm{m}\bar{x}_i+\hat{\phi}_x)^2}{ 2
\hat{N}_x}\right]} \tilde{\Psi}(\hat{\phi}_x,\mathrm{m}
\bar{y}_i)\cr && \times \left\{\int_{-\infty}^{\infty} d
\hat{\phi}_x \exp{\left[-\frac{\hat{\phi}_x^2}{ 2
\hat{N}_x}\right]} \tilde{\Psi}(\hat{\phi}_x,0)\right\}^{-1}~.\cr
&& \label{IgenimaMmm}
\end{eqnarray}
In the limit  $\hat{N}_x \ll 1$  we  have

\begin{eqnarray}
\hat{I}(\hat{z}_i, \bar{x}_i,{ \bar{y}}_i)  &=& {\mathrm{m}^2}
\tilde{\Psi}(\mathrm{m}\bar{x}_i,\mathrm{m}\bar{y}_i)~,
\label{IgenimaMlim2}
\end{eqnarray}
in agreement with Eq. (\ref{cpact61}).

\subsection{\label{sub:gim} General imaging considerations}

In the present Section \ref{sub:gim} we discuss a general
algorithm for the solution to the image formation problem for
undulator sources based on our Statistical Optics approach.

Eq. (\ref{ultimasp}) is an expression for the cross-spectral
density on the image plane in the case of a non-homogeneous
undulator source and of a lens with an arbitrary pupil function
(i.e. a lens with aberrations, apodization and finite aperture
size). However, we assumed that  the electron beam has (i) a
Gaussian transverse profile and (ii) a large horizontal emittance
compared with the radiation wavelength ($\hat{N}_x \gg 1$ and
$\hat{D}_x \gg 1$), that (iii) the radiation frequency is tuned at
perfect resonance with the fundamental frequency of the undulator,
i.e. $\hat{C} \ll 1$\footnote{This means that monochromatization
is good enough to neglect finite bandwidth of the radiation around
the fundamental frequency, as well as electron beam energy
spread.}, that (iv) the minimal beta functions in both horizontal
and vertical directions are located at $\hat{z}=0$, that (v) there
is no influence of focusing inside the undulator, that (vi) the
lens is placed in the far zone and, finally, that (vii) the
observation plane is located at position $\hat{z}=\hat{z}_i$,
where the virtual source (that we assume at $\hat{z}=0$) is
imaged. Assumptions (i), (ii), (iii), (iv) and (v) are related to
the form of the cross-spectral density at the virtual source
plane, Eq. (\ref{gen2}). They are very often, but not always
verified. Moreover they do not depend on the particular imaging
setup related with a given photon beamline. Assumptions (vi) and
(vii) instead, are related with the imaging setup i.e. with how
the lens and the observation points are positioned.

The majority of the assumptions from (i) to (vii) are often
verified for third generation light sources. As a matter of fact,
our theory is specifically built to deal with third generation
light sources. However, a generalization to include the case of
spontaneous undulators installed in XFEL facilities (see e.g.
\cite{AUDE}, \cite{CORN}) is certainly desirable. Of all
restrictions from (i) to (vii), (v) is the more difficult to be
relaxed. In order to do so, one needs to modify the expression for
the single particle field to account for the influence of focusing
inside the undulator.  The other assumptions may be more easily
relaxed, to give a more general algorithm for the calculation of
the cross-spectral density in the case of arbitrary position of
the lens (near or far zone). The case of spontaneous undulators
installed in XFEL facilities is a particular one when it is needed
to deal with both horizontal and vertical electron beam emittances
comparable or smaller than the radiation wavelength. In fact,
$\epsilon_x \simeq \epsilon_y \simeq 0.3 \mathrm{\AA}$. Results
based on a Gaussian model of the electron beam with generic
$\hat{N}_{x,y}$ and $\hat{D}_{x,y}$ may be presented. To this
purpose one may use Eq. (\ref{G2D}) to calculate the
cross-spectral density in free space. As one may see inspecting
Eq. (\ref{G2D}), it is no more possible to separate the horizontal
and the vertical direction. Eq. (\ref{G2D}) is still subject to
assumptions (i), (iii), (iv) and (v). Our theory is built by
exploiting many simplifying assumptions. When some of them fail,
one should go back to the point where the invalid simplification
is exploited, and use a more generic expression in its place. In
particular, while assumption (i) is quite realistic in the case if
storage ring sources, it is to be regarded as a conventional
assumption when treating sources based on linear accelerators. Our
most generic expression, Eq. (\ref{Gzlarge2}) should be used in
place of Eq. (\ref{G2D}) if one wishes to relax assumption (i), as
well as (iii) and (iv). Once the cross-spectral density in free
space is known through Eq. (\ref{G2D}) or Eq. (\ref{Gzlarge2}),
one may account for a generic observation plane, thus relaxing
assumption (vii). To this purpose one needs to put attention on
the fact that any observation plane located at position
$\hat{z}=\hat{z}_2$ is related with a certain plane at position
$\hat{z}=\hat{z}_s$ in front of the lens through the lens-maker
equation, Eq. (\ref{lens}). The next step in our algorithm
consists in finding the cross-spectral density at
$\hat{z}=\hat{z}_s$, which will be imaged at our chosen
observation plane, located at $\hat{z}=\hat{z}_2$. To this
purpose, Eq. (\ref{ftGprop}) may be used. In fact, Eq.
(\ref{ftGprop}) describes the propagation of the Fourier transform
of the cross-spectral density, $\hat{\mathcal{G}}$, in free space.
As we have seen, the fact that the lens is placed in the far field
allows cancellation of one phase factor in the relation between
the cross-spectral density on the image plane and the
cross-spectral density on the virtual source plane. This can be
seen using condition (\ref{farlens}) in Eq. (\ref{fundima}). Such
phase factor should be retained if the lens is in the near zone
(that is when condition (\ref{farlens}) is not satisfied). In this
way, assumption (vi) can be relaxed as well. At this point, the
cross-spectral density $\hat{G}(\hat{z}_s, \vec{\bar{r}}, \Delta
\vec{\hat{r}})$ is known. The final step consists in the
calculation of the amplitude point spread function
$\hat{\mathcal{P}}_a$ of the system. The cross-spectral density at
position $\hat{z} =\hat{z}_2$ is found convolving twice the
product of the cross-spectral density at the source plane and the
extra phase-factor due to the failure of condition (\ref{farlens})
with the amplitude point spread function. This gives

\begin{eqnarray} \hat{G}_P(\hat{z}_2,\vec{\bar{r}}, \Delta \vec{\hat{r}})&=&4 \hat{a}^2
\exp\left[\frac{2i\mathrm{m}}{\hat{z}_1}\vec{\bar{r}}\cdot \Delta
\vec{\hat{r}}\right] \int d\vec{\bar{u}} ~d \Delta
\vec{\hat{u}}~\cr && \times \exp\left[2 i \hat{z}_1
\vec{\bar{u}}\cdot \Delta \vec{\hat{u}} \right]
\hat{G}\left(\hat{z}_s,-\hat{z}_1\vec{\bar{u}},-\hat{z}_1 \Delta
\vec{\hat{u}}\right) \cr && \times
\hat{\mathcal{P}_a}~\left[\frac{\mathrm{m}}{\hat{z}_1}\left(\vec{\bar{r}}+
\Delta \vec{\hat{r}}\right)-\vec{\bar{u}}- \Delta
\vec{\hat{u}}\right]\cr &&\times
\hat{\mathcal{P}_a}^*\left[\frac{\mathrm{m}}{\hat{z}_1}\left(\vec{\bar{r}}-
\Delta \vec{\hat{r}}\right)-\vec{\bar{u}}+ \Delta
\vec{\hat{u}}\right]~, \label{ultimasp4}
\end{eqnarray}
where $\mathrm{m}$ is now defined as $\mathrm{m} = (\hat{z}_1
-\hat{z}_s) / (\hat{z}_2-\hat{z}_1)$, according to Eq.
(\ref{Mdefined}).

It should be noted that, in this paper, together with a theory for
third generation light sources we also developed a particular
language that can be applied for a wider range of problems. Up to
now, research works dealing with transverse coherence properties
of Synchrotron Radiation have used the standard language developed
to treat Statistical Optics problems. Such a language has a very
limited scope because Statistical Optics has mainly dealt with
thermal-like sources. As a result, a time domain approach has
often been used. Quasi-stationary approximation and ergodicity are
usually assumed, so that time averages are used instead of
ensemble averages. Then, the concept of cross-spectral purity
\cite{MAND,GOOD} must be forcefully evoked in order to separate
longitudinal and transverse coherence effects, which are described
through the mutual intensity function. Such language though, is
not suitable to describe Synchrotron Radiation experiments, where
many radiation pulses are collected and results are averaged over
an ensemble. Our approach starts from the very foundation of
Statistical Optics, thus avoiding inconvenient assumptions. A
language based on a frequency domain description and on ensemble
averages over an ensemble of radiation pulses (each corresponding
to a different electron bunch) has been developed. In our paper we
aimed at presenting a satisfactory description of the physics
involved in the characterization of light sources from a
statistical viewpoint, trying to exhaustively explain where the
main ideas come from and where they lead to. We restricted our
attention to a quantitative treatment of third-generation light
sources but we also laid the foundations to describe other kind of
radiation sources, e.g. spontaneous undulator sources installed in
XFEL facilities. In fact, many of the features of the relatively
specialized setup considered in this work are common in the
general theory of undulator sources.

\section{\label{sec:supp}Supplementary remarks on quasi-homogeneous undulator
source asymptotes}

In the last Section \ref{sec:nonh} we presented quite general
results for vertically non-homogeneous sources. Under the only
assumption of a large horizontal emittance ($\hat{D}_x \gg 1$ and
$\hat{N}_x \gg 1$), Eq. (\ref{resu1}) specifies the cross-spectral
density of the radiation in the far zone and in the vertical
direction without constraints on $\hat{N}_y$ and $\hat{D}_y$.
Under the same conditions, Eq. (\ref{gen2}) specifies the
cross-spectral density of the virtual source in the vertical
direction. Note that Eq. (\ref{gen2}) is a convolution. When the
dependence of $\hat{G}$ on $\Delta{\hat{y}}$ can be isolated in a
single factor, the source is forcefully quasi-homogeneous, while
there are no cases when it is described by a more generic kind of
Shell model (i.e. $\hat{G} \sim \sqrt{I(y_1)}\sqrt{I(y_2)}g(\Delta
\hat{y})$). It is important to stress this fact in connection with
several research works \cite{COI1, COI2} devoting particular
attention to the relation between the Gaussian-Shell model and
undulator sources. As we remarked in Section \ref{sub:imfoge},
Shell models (and, in particular, Gaussian-Shell models) may
certainly be useful for describing light sources other than
undulator-based ones and for educational purposes, but they do not
describe any practical realization of undulator radiation sources.

From Eq. (\ref{resu1}), setting  $\Delta \hat{\theta}_y =0$,  one
obtains the intensity distribution in the far zone

\begin{eqnarray}
\hat{I}(\bar{\theta}_y) = \int_{-\infty}^{\infty} d\hat{\phi}_y
\exp\left[-\frac{(\bar{\theta}_y+\hat{\phi}_y)^2}{2\hat{D}_y}
\right] \hat{I}_S(\hat{\phi}_y) ~,\label{unogen}
\end{eqnarray}
having used Eq. (\ref{relation1}). From Eq. (\ref{gen2}), setting
$\Delta \hat{y} =0$, one gets the intensity distribution of the
virtual source in the center of the undulator\footnote{ We are
always assuming that the minimal beta function of the electron
beam is located at the center of the undulator.},

\begin{eqnarray}
\hat{I}(0,\bar{y}) = \int_{-\infty}^{\infty} d\hat{\phi}
\exp\left[-\frac{(\bar{y}_y+\hat{\phi})^2}{2\hat{N}_y} \right]
\hat{\mathcal{B}}(\hat{\phi}) ~,\label{duegen}
\end{eqnarray}
having taken advantage of Eq. (\ref{unpoinsie}). $\hat{I}_S$ in
Eq. (\ref{unogen}) and $\hat{\mathcal{B}}$ in Eq. (\ref{duegen})
are the universal functions given in Eq. (\ref{bingo2}) and Eq.
(\ref{ftbrepr}). Both Eq. (\ref{unogen}) and Eq. (\ref{duegen})
are convolutions, and are valid regardless  the values of
$\hat{N}_y$ and $\hat{D}_y$, i.e. regardless the fact that the
source is quasi-homogeneous or not. In the case of a large
non-limiting aperture and an ideal lens, other two exact results
can be found in Section \ref{sec:nonh}, which are independent of
the values of $\hat{N}_y$ and $\hat{D}_y$. In fact, Eq.
(\ref{Ifocino}) and Eq. (\ref{IgenimaM}) prove that the exact
expression for the intensity distribution is, both on the focal
and on the image plane, a convolution between Gaussian and
universal functions.

When the source is quasi-homogeneous, with an accuracy scaling as
the inverse number of modes,
$1/\sqrt{\max[\hat{N}_y,1]\max[\hat{D}_y,1]}$, we may take the
approximation $\hat{\mathcal{M}}(\Delta\hat{y}, \hat{\phi}) \simeq
\gamma(\Delta\hat{y}) \mathcal{B}(\hat{\phi})$ in Eq.
(\ref{gen2}).

This fact may be demonstrated as follows. First, let us introduce
a normalized version of the one-dimensional inverse Fourier
transform of the function ${M}$, that is

\begin{eqnarray}
M'(\Delta \hat{y},\Delta \hat{\theta}_y) &=&
\frac{1}{\mathcal{A}}\int_{-\infty}^{\infty}
\hat{M}\left(\bar{\theta}_y, \Delta \hat{\theta}_y\right)
\exp\left[-2 i \Delta \hat{y} \bar{\theta}_y \right] d
\bar{\theta}_y \cr &=&
\frac{1}{\mathcal{A}}\int_{-\infty}^{\infty}
\hat{\mathcal{M}}\left(\Delta \hat{y},\bar{y}\right) \exp\left[2 i
\Delta \hat{\theta}_y \bar{y} \right] d \bar{y}~.\label{WIG2}
\end{eqnarray}
The normalization factor $\mathcal{A}$ in Eq. (\ref{WIG2}) is
defined as

\begin{eqnarray}
\mathcal{A} &=& \int_{-\infty}^{\infty}
\hat{\mathcal{M}}\left(0,\bar{y}\right)  d
\bar{y}~,\label{WIG2norm}
\end{eqnarray}
so that $M'(0,0)=1$. The cross-spectral density  in Eq.
(\ref{gen2}) can therefore be written as

\begin{eqnarray}
\hat{G}(0,\bar{y},\Delta \hat{y}) &=& \exp\left[-2 \hat{D}_y
\Delta \hat{y}^2 \right]  \cr &&\times \int_{-\infty}^{\infty} d
\hat{u} \exp\left[-2 i \hat{u}\bar{y}\right] \exp\left[-2\hat{N}_y
\hat{u}^2\right] M' (\Delta \hat{y}, \hat{u})~,\label{gen2sepa}
\end{eqnarray}
having used the convolution theorem. Under the quasi-homogeneous
assumption, we can approximate $M' (\Delta \hat{y}, \hat{u})
\simeq M' (\Delta \hat{y}, 0)M' (0, \hat{u})$. To show this, let
us represent $M'(x,y)$ using a Taylor expansion around the point
$(0,0)$. One obtains

\begin{eqnarray}
M'(x,y) &=& 1 + \sum_{k=1}^{\infty} \frac{1}{k!} \left[x^k
\frac{\partial^k M'(x,0)}{\partial x^k}\Bigg|_{x=0}+y^k
\frac{\partial^k M'(0,y)}{\partial y^k}\Bigg|_{y=0}\right] \cr &&
+ O(xy) ~, \label{Mpexp}
\end{eqnarray}
where the normalization relation $M'(0,0)=1$ has been taken
advantage of. Similarly, one may consider the following
representation of the product $M'(x,0)$ $M'(0,y)$ also obtained by
means of a Taylor expansion:

\begin{eqnarray}
M'(x,0)M'(0,y) &=& \left[M'(0,0)+ \sum_{k=1}^{\infty}
\frac{x^k}{k!} \frac{d^k M'(x,0)}{dx^k}\Bigg|_{x=0}\right]\cr
&\times& \left[M'(0,0)+ \sum_{j=1}^{\infty} \frac{y^j}{j!}
\frac{d^j M'(0,y)}{dy^j}\Bigg|_{y=0}\right] \cr &=& 1 +
\sum_{n=1}^{\infty} \frac{1}{n!} \left[x^n \frac{d^n
M'(x,0)}{dx^n}\Bigg|_{x=0} \right.\cr && \left. +y^n \frac{d^n
M'(0,y)}{dy^n}\Bigg|_{y=0}\right] + O(xy) ~,\label{Mpexp2}
\end{eqnarray}
having used $M'(0,0)=1$. Comparison of the last equality in
(\ref{Mpexp2}) with the right hand side of Eq. (\ref{Mpexp}) shows
that $M'(x,y)\simeq M'(x,0)M'(0,y)$ up to corrections of order $xy
\sim 1/\sqrt{\max[\hat{N}_y,1]\max[\hat{D}_y,1]}$, that is the
quasi-homogeneous accuracy. Using this approximation in Eq.
(\ref{gen2sepa}) yields

\begin{eqnarray}
\hat{G}(0,\bar{y},\Delta \hat{y}) &=& \exp\left[-2 \hat{D}_y
\Delta \hat{y}^2 \right] M' (\Delta \hat{y}, 0) \cr &&\times
\int_{-\infty}^{\infty} d \hat{u} \exp\left[-2 i
\hat{u}\bar{y}\right] \exp\left[-2\hat{N}_y \hat{u}^2\right] M'
(0, \hat{u})~.\label{gen2sepabis}
\end{eqnarray}
Finally, recalling the definitions of $\gamma$ and $\mathcal{B}$
we can write Eq. (\ref{gen2sepabis}) as

\begin{eqnarray}
\hat{G}(0,\bar{y},\Delta \hat{y}) = \exp\left[-2 \hat{D}_y \Delta
\hat{y}^2 \right]  \gamma(\Delta \hat{y}) \int_{-\infty}^{\infty}
d \hat{\phi}
\exp\left[-\frac{(\bar{y}+\hat{\phi})^2}{2\hat{N}_y}\right]\hat{\mathcal{B}}(\hat{\phi})~.\label{gen2qhapp}
\end{eqnarray}
Eq. (\ref{gen2qhapp}) is valid in any quasi-homogeneous case.

Note that Eq. (\ref{gen2qhapp}) accounts for diffraction effects
through the universal functions $\gamma$ and $\mathcal{B}$. This
may be traced back to the use of the inhomogeneous wave equation
to calculate the cross-spectral density for the virtual source,
from which Eq. (\ref{gen2qhapp}) follows. Deriving Eq.
(\ref{gen2qhapp}), we assume a large number of modes, and this
justifies the use of phase space representation as an alternative
characterization of the source, in place of the cross-spectral
density (i.e. Eq. (\ref{gen2qhapp}) itself).

Setting $\Delta \hat{y}=0$, Eq. (\ref{gen2qhapp}) gives the exact
intensity distribution at the virtual source, i.e. Eq.
(\ref{duegen}). The spectral degree of coherence on the virtual
source is then recovered using the definition of quasi-homogeneous
source $\hat{G} = \hat{I}(\bar{y}) g(\Delta \hat{y})$. Since the
source is quasi-homogeneous, the Fourier transform of the spectral
degree of coherence $g(\Delta \hat{y})$ yields the intensity in
the far zone. Remembering that $\gamma$ and $I_S$ form a Fourier
pair, we conclude that, starting from Eq. (\ref{gen2qhapp}) it is
possible to reproduce the exact result for the intensity in the
far zone, Eq. (\ref{unogen}). Quite remarkably, Eq.
(\ref{gen2qhapp}), which is derived under the quasi-homogeneous
approximation and is related to an accuracy
$1/\sqrt{\max[\hat{N}_y,1]\max[\hat{D}_y,1]}$, yields back two
results, Eq. (\ref{unogen}) and Eq. (\ref{duegen}) which are valid
regardless the fact that the source is quasi-homogeneous or not.
Moreover, in the case of perfect optics and non-limiting pupil
aperture, and independently of the quasi-homogeneous assumption,
the intensity profile in the virtual plane reproduces the
intensity profile in the image plane, while the intensity profile
in the far zone reproduces the intensity profile in the focal
plane. Therefore, we can also conclude that Eq. (\ref{gen2qhapp})
gives both the intensity in the focal and in the image plane for
an ideal lens. Note again that also these two results,  Eq.
(\ref{Ifocino}) and Eq. (\ref{IgenimaM}),  have perfect accuracy.
They are exact and are not subject to the quasi-homogeneous
accuracy $1/\sqrt{\max[\hat{N}_y,1]\max[\hat{D}_y,1]}$.

Let us now consider the particular quasi-homogeneous case when
both  $\hat{N}_y \gg 1$ and $\hat{D}_y \gg 1$. In this case, the
number of modes along the virtual source is of order
$\sqrt{\hat{N}_y\hat{D}_y}$ and  the normalized coherence length
can be estimated as $\hat{\xi}_c \sim 1/\sqrt{\hat{D}_y} \ll 1$.
In the case of non-ideal optics, once a line spread function $l$
for the system is found, one may obtain the intensity distribution
of the radiation by convolving $l$ with the ideal image. The
accuracy of these calculations is now the accuracy of the
quasi-homogeneous assumption, $1/\sqrt{\hat{N}_y\hat{D}_y}$. In
the case a pupil with aperture $\hat{a}$ is present, we concluded
in Section \ref{sub:qhpup} that it makes sense to account for
diffraction effects when $\hat{a}\ll \sqrt{\hat{D}} \hat{z}_1$. In
this case in fact, the ratio between the width of the line spread
function and the width of the ideal image is of order
$\hat{z}_1/(\hat{a} \sqrt{\hat{N}_y}) \gg
1/\sqrt{\hat{N}_y\hat{D}_y}$. As a result, when Eq.
(\ref{gen2qhapp}) is used to calculate the ideal intensity, it
makes sense to account for diffraction effects. Worsening the
accuracy in the calculation of the cross-spectral density of the
source, we may reduce Eq. (\ref{gen2qhapp}) to

\begin{eqnarray}
\hat{G}(0,\bar{y},\Delta \hat{y}) = \exp\left[-2 \hat{D}_y \Delta
\hat{y}^2 \right]
\exp\left[-\frac{\bar{y}^2}{2\hat{N}_y}\right]~,\label{gen2qhappworse}
\end{eqnarray}
that is Eq. (\ref{crossx1l}). Note that neglecting the product
with the $\gamma$ function can be done with an accuracy
$1/\sqrt{\hat{D}_y}$, while extraction of the exponential function
in $\bar{y}$ from the convolution product with the $\mathcal{B}$
function can be done with an accuracy $1/\sqrt{\hat{N}_y}$. In our
study case when $\hat{D}_y \gg 1$ and $\hat{N}_y \gg 1$, the
overall accuracy of Eq. (\ref{gen2qhappworse}) (or Eq.
(\ref{crossx1l})) can be estimated as
$\max(1/\sqrt{\hat{D}_y},1/\sqrt{\hat{N}_y})$, that is the
accuracy of the Gaussian approximation. Such accuracy is much
worse than that of the quasi-homogeneous assumption in Eq.
(\ref{gen2qhapp}), that is $1/\sqrt{\hat{N}_y\hat{D}_y}$. This
fact has interesting consequences.  In fact, Eq.
(\ref{gen2qhappworse}) can first be used to calculate the ideal
intensity on the image plane and, then, it may be convolved with
the line spread function of the lens to give a characterization of
the intensity distribution with a reduced accuracy. If, for
instance, $\hat{a}\ll \sqrt{\hat{D}} \hat{z}_1$ we can have
situation when $\hat{z}_1/(\hat{a} \sqrt{\hat{N}_y}) \gg
1/\sqrt{\hat{N}_y\hat{D}_y}$ but $\hat{z}_1/(\hat{a}
\sqrt{\hat{N}_y}) \ll 1/\sqrt{\hat{N}_y}$ and, as a result,
accounting for diffraction effects would not modify the intensity
with accuracy $1/\sqrt{\hat{N}_y}$. In spite of this, going back
to Eq. (\ref{gen2qhapp}) to calculate the ideal intensity with
better accuracy $1/\sqrt{\hat{N}_y\hat{D}_y}$, diffraction effects
will appreciably modify the intensity within the accuracy
$1/\sqrt{\hat{N}_y\hat{D}_y}$.

When $\hat{N}_y \gg 1$ and $\hat{D}_y \simeq 1$ the accuracy of
the quasi-homogeneous approximation becomes $1/\sqrt{\hat{N}_y
\max[1,\hat{D}_y]}$. When $\hat{N}_y \simeq 1$ and $\hat{D}_y \gg
1$ it becomes, instead, $1/\sqrt{\max[1,\hat{N}_y]\hat{D}_y }$. In
these cases, the accuracy of the quasi-homogeneous approximation
is comparable to the accuracy of the Gaussian approximation. To be
specific, when $\hat{N}_y \gg 1$ and $\hat{D}_y \simeq 1$ the
accuracy of the quasi-homogeneous approximation is of order
$1/\sqrt{\hat{N}_y}$ (note that the coherence length at the pupil
is $\hat{\xi}_c \simeq 1$, that is the diffraction size)  and Eq.
(\ref{gen2qhapp}) can be substituted with

\begin{eqnarray}
\hat{G}(0,\bar{y},\Delta \hat{y}) =
\exp\left[-\frac{\bar{y}^2}{2\hat{N}_y}\right] \exp\left[-2
\hat{D}_y \Delta \hat{y}^2 \right]  \gamma(\Delta \hat{y})
~\label{gen2qhappN}
\end{eqnarray}
without loss of accuracy, because the relative accuracy of the
convolution is of order $1/\sqrt{\hat{N}_y}$ as the accuracy of
the quasi-homogenous approximation. Eq. (\ref{gen2qhappN}) is just
Eq. (\ref{Ngrande}). A similar reasoning can be done when
$\hat{D}_y \gg 1$ and $\hat{N}_y \simeq 1$. In this case the
accuracy of the quasi-homogeneous approximation is of order
$1/\sqrt{\hat{D}_y}$ (note that the coherence length at the pupil
is $\hat{\xi}_c \ll 1$, that is much smaller than the diffraction
size), and Eq. (\ref{gen2qhapp}) can be substituted with

\begin{eqnarray}
\hat{G}(0,\bar{y},\Delta \hat{y}) = \exp\left[-2 \hat{D}_y \Delta
\hat{y}^2 \right] \int_{-\infty}^{\infty} d \hat{\phi}
\exp\left[-\frac{(\bar{y}+\hat{\phi})^2}{2\hat{N}_y}\right]\hat{\mathcal{B}}(\hat{\phi})~.\label{gen2qhappD}
\end{eqnarray}
without loss of accuracy. In fact, neglecting the $\gamma$
function in Eq. (\ref{gen2qhapp}) is equivalent to approximate the
convolution in Eq. (\ref{unogen}) with a Gaussian distribution,
which can be done with an accuracy of order $1/\sqrt{\hat{D}_y}$,
the same of the quasi-homogenous approximation. Eq.
(\ref{gen2qhappD}) is just Eq. (\ref{Ngr3}).

In closing this Section we should stress  that, in the case of
ideal lenses, the intensity on the image plane, that is a scaled
version of Eq. (\ref{duegen}), is exactly given by a convolution
of a known universal function and the electron beam profile. In
the case the electron beam profile is unknown, one may measure the
intensity and deconvolve Eq. (\ref{duegen}) in order to find back
the electron beam profile. In literature (see, for example,
\cite{HOFF}) it is usually accepted that the resolution of any
electron beam size $\sigma$ inferred from the measurement of the
radiation intensity distribution on the image plane is limited (in
the case of an ideal lens) by the diffraction size of the single
particle undulator radiation, i.e. the resolution is of order
$\sqrt{\lambda L_w}/(2\pi \sigma)$. However, we have seen that Eq.
(\ref{duegen}) is an exact result. Therefore, any measurement of
$\sigma$ obtained by deconvolution of Eq. (\ref{duegen}) is only
limited by the finite accuracy of the detector.

\section{\label{sec:conc} Conclusions}

As has been remarked in \cite{HOWE}: "[...] it is very desirable
to have a way to model the performance of undulator beamlines with
significant partial coherent effects, and such modelling would,
naturally, start with the source. The calculation would involve
the knowledge of the partial coherence properties of the source
itself and of how to propagate partially coherent fields through
space and through the optical components used in the beamline.
[...] it is important to recognize that, although most of these
calculations are, in principle, straightforward applications of
conventional coherence theory (Born and Wolf, 1980; Goodman,
1985), there is not much current interest in the visible optics
community. [...] For example, even for the rather simple problem
of diffraction by an open aperture with partially coherent
illumination, we have found published solutions only for circular
and slit-shaped apertures and only for sources consisting of an
incoherently illuminated aperture of similar shape to the
diffracting aperture. Thus, there is no counterpart in these types
of Fourier Optics problem to the highly developed art of ray
tracing in geometrical optics, not is there anything as simple as
a ray to which an exact system response can be calculated.". This
program of development of Synchrotron Radiation theory was
formulated more than ten years ago. Operation of third generation
light sources also started in this period. This demonstrates that
when third generation light sources were born, it was immediately
recognized that the usual theory of Synchrotron Radiation was not
adequate to describe them. Yet, up to now, no theoretical progress
has been made in that direction. The present paper, as well as our
previous work \cite{OURS} are devoted to the realization of the
before mentioned program of development of Synchrotron Radiation
theory.

In \cite{OURS} we described spatial coherence propertied of
undulator radiation from third-generation light sources in free
space. In this paper we aim at an extension of \cite{OURS}.
Previous scientific works and textbooks postulate that the
cross-spectral density at the virtual undulator source can be
described in terms of a Gaussian-Shell model or, even more
restrictively, that undulator sources are perfectly incoherent.
Such assumption is not adequate when treating third generation
light sources, because the vertical emittance is comparable or
even much smaller than the radiation wavelength (i.e.
${\epsilon}_y \ll \lambda/(2\pi)$) in a wide spectral interval
extending from the $\mathrm{\AA}$mstrong wavelength range up to
the soft X-rays.

In this work we combined Statistical Optics methods with Fourier
Optics techniques in order to describe in an analytical way the
propagation of the cross-spectral density of Synchrotron Radiation
through a lens. In particular, we focused our attention on the
problem of finding both the intensity and the spectral degree of
coherence of undulator radiation at the focal and at the image
plane of the lens. Although our paper is not limited to this
situation alone, our main result deals with the quite generic case
of a large normalized horizontal emittance ${\epsilon}_x \gg
\lambda/(2\pi)$ and an arbitrary vertical emittance
${\epsilon}_y$.

Our paper provides physical understanding of a setup of general
interest and we expect it to be useful for practical estimations
in almost all range of the parameter space for third generation
light sources. We expect that, in the future, numerical codes
fully capable of dealing with transverse coherence properties of
Synchrotron Radiation will also be developed, and will be capable
of providing detailed analysis of particular experimental setups.
Our theory will be of help to developers of these codes because it
provides both benchmarks and partially manipulated equations for
the field correlation, simpler to treat numerically than first
principle calculations and still reasonably generic.

Two basic non-restrictive assumptions made in our theory are the
paraxial approximation and the resonance approximation. The first
is justified by the fact that we are treating an
ultra-relativistic system. The second means that we are working
with an undulator composed of a large number of period $N_w \gg 1$
and that we are interested in frequencies near the fundamental.
This allows to neglect the vertical polarization component of the
field and to treat the field within a scalar theory.

Analytical studies also required the introduction of some
restrictive assumptions introduced in our theory, to be relaxed in
the future. First, we assumed that the beta functions in both
directions have their minima in the center of the undulator.
However, beta functions are (or may be) often tuned around the
center of the undulator in cases of practical interest. Second, we
assumed a single converging thin lens with none or very specific
pupil functions. However, other shapes of the lens can be
accounted for by means of numerical convolutions between a more
complicated complex pupil function and our fundamental results
referring to the case when influence of the pupil is negligible.
Note that the situation of a thin lens is often met in practice in
the case of X-ray radiation, as grazing incidence reflective
optics is quite frequently used for image formation in this
spectral range. Third, we assumed that monochromatization is good
enough to neglect finite bandwidth of the radiation around the
fundamental and also electron beam energy spread. Both these
effects may be taken into account by an extension of this theory.
However, in practical cases of interest these restrictions are
often met. For instance, a monochromator relative bandwidth of
$10^{-3}$ is sufficient to guarantee a reasonably narrow bandwidth
for undulators up to about $40$ periods. Also, with this number of
undulator periods, such a monochromator guarantees small
corrections to our results for electron beams with a relative
energy spread of $10^{-3}$ or better.

The three before mentioned assumptions are somewhat restrictive.
However, they are often met in practice and our theory can be
extended to the case when they are not met. A more restrictive
assumption is that radiation frequency is tuned at perfect
resonance with the fundamental frequency of the undulator.
Although this last assumption is also often met in practice, our
theory cannot be easily extended to account for the situation when
it is not satisfied because the starting point for the calculation
of the cross-spectral density is given by an expression for the
electric field around the fundamental frequency.

As a final remark, it should be said that we chose not to deal
with bending magnet sources in this paper. This is left for future
investigations. However, some estimates based on dimensional
analysis suggest that, as discussed in Section \ref{sub:imfoge},
Geometrical Optics treatments may be sufficient to describe in a
satisfactory way bending magnet radiation from third generation
light sources.

To conclude, our paper constitutes, to our knowledge, the first
satisfactory theory describing imaging of undulator sources by a
single non-ideal lens. We restrict ourselves to the analysis of a
single lens for simplicity, the results for more complicated
optical systems involving a larger quantity of optical elements
being a straightforward extension of the present work.

\newpage

\section*{Acknowledgements}

The authors wish to thank Hermann Franz and Petr Ilinski for many
useful discussions, Jochen Schneider and Edgar Weckert for their
interest in this work.

\newpage

\listoffigures

\newpage

\end{document}